\documentclass[a4paper,11pt]{article}
\pdfoutput=1 
\usepackage{jcappub}
\usepackage{amsmath}
\usepackage{graphicx}
\usepackage{latexsym}
\usepackage{xspace}
\usepackage{color}
\usepackage{hyperref} 
\usepackage{bm}
\usepackage{relsize}
\usepackage{tabularx}
\usepackage{multirow}
\usepackage{amssymb}
\usepackage[table]{xcolor}
\usepackage{slashbox}
\usepackage{braket}
\usepackage{comment}
\usepackage[utf8]{inputenc}
\usepackage{slashed}

\makeatletter
\gdef\@fpheader{}
\g@addto@macro\bfseries{\boldmath}
\makeatother

\newcommand{\ds}{\displaystyle}

\newcommand{\ie}{\textsl{i.e.~}}

\newcommand{\eg}{\textsl{e.g.~}}

\newcommand{\etc}{\textsl{etc.~}}

\newcommand{\order}[1]{\mathcal{O}\!\left(#1\right)}

\newcommand{\dd}{\mathrm{d}}
\newcommand{\ee}{e}

\newcommand{\sss}[1]{{\scriptscriptstyle{#1}}}

\newcommand{\uPl}{\mathrm{Pl}}

\newcommand{\ucl}{\mathrm{cl}}

\newcommand{\usssPl}{\sss{\uPl}}

\newcommand{\Rea}{\Re \mathrm{e}\,}

\newcommand{\Mp}{M_\usssPl}

\newcommand{\beq}{\begin{equation}}
\newcommand{\eeq}{\end{equation}}
\newcommand{\bea}{\begin{eqnarray}}
\newcommand{\eea}{\end{eqnarray}}

\newlength{\wsingfig}
\newlength{\wdblefig}
\newlength{\wquadfig}
\newlength{\wtriplefig}
\newcommand{\Eq}[1]{Eq.~(\ref{#1})}
\newcommand{\Eqs}[1]{Eqs.~(\ref{#1})}
\newcommand{\Fig}[1]{Fig.~{\ref{#1}}}

\newcommand{\Refa}[1]{Ref.~{\cite{#1}}}
\newcommand{\Refs}[1]{Refs.~{\cite{#1}}}
\newcommand{\Sec}[1]{Sec.~\ref{#1}}
\newcommand{\Secs}[1]{Secs.~\ref{#1}}
\newcommand{\App}[1]{Appendix~\ref{#1}}
\newcommand{\Apps}[1]{Appendices~\ref{#1}}

\subheader{}

\title{Canonical transformations and squeezing formalism in cosmology}

\author[a]{Julien Grain,}
\affiliation[a]{Institut d'Astrophysique Spatiale, UMR8617, CNRS, Univ. Paris Sud, Universit\'e Paris-Saclay, Bt. 121, Orsay, France, F-91405}
\emailAdd{julien.grain@ias.u-psud.fr}

\author[b]{Vincent Vennin}
\affiliation[b]{Laboratoire Astroparticule et Cosmologie, Universit\'e Denis Diderot Paris 7, 10 rue
Alice Domon et L\'eonie Duquet, 75013 Paris, France}
\emailAdd{vincent.vennin@apc.univ-paris7.fr}

\date{today}

\begin{document}
\sloppy

\abstract{Canonical transformations are ubiquitous in Hamiltonian mechanics, since they not only describe the fundamental invariance of the theory under phase-space reparameterisations, but also generate the dynamics of the system. In the first part of this work we study the symplectic structure associated with linear canonical transformations. After reviewing salient mathematical properties of the symplectic group in a pedagogical way, we introduce the squeezing formalism, and show how any linear dynamics can be cast in terms of an invariant representation. In the second part, we apply these results to the case of cosmological perturbations, and focus on scalar field fluctuations during inflation. We show that different canonical variables select out different vacuum states, and that this leaves an ambiguity in observational predictions if initial conditions are set at a finite time in the past. We also discuss how the effectiveness of the quantum-to-classical transition of cosmological perturbations depends on the set of canonical variables used to describe them.}



\maketitle

\flushbottom

\section{Introduction}
\label{sec:intro}

The physical description of a given system should not depend on the way it is parametrised. This is why canonical transformations play a fundamental role in physics, as they relate different parameterisations of phase space. In some sense, they play the same role for phase space as diffeomorphisms do for space-time in general relativity. They are a fundamental invariance of any Hamiltonian (or Lagrangian) theory. 

Canonical transformations are defined as follows~\cite{goldstein2002classical}. Let phase space be parametrised by the phase-space variables $q_i$ and $p_i$, with Hamiltonian $H(\lbrace q_i \rbrace , \lbrace p_i \rbrace,t)$, where $i$ runs over the number of degrees of freedom. Hamilton's equations of motion are given by
\bea
\label{eq:Hamilton}
{\dot{q}_i}=\dfrac{\partial H}{\partial p_i}\, , \quad\quad\quad 
{\dot{p}_i}=-\dfrac{\partial H}{\partial q_i}\, ,
\eea
where a dot denotes derivation with respect to time. A transformation of the phase-space coordinates 
\bea
q_i &\rightarrow \tilde{q}_i\left(\lbrace q_j \rbrace , \lbrace p_j \rbrace,t\right)\\
p_i &\rightarrow \tilde{p}_i\left(\lbrace q_j \rbrace , \lbrace p_j \rbrace,t\right)
\eea
is said to be \emph{canonical} if $\tilde{q}_i$ and $\tilde{p}_i$ are canonical variables, \ie if $\tilde{p}_i$ is the momentum canonically conjugate to $\tilde{q}_i$. In other words, there must exist a Hamiltonian form $\tilde{H}(\lbrace \tilde{q}_i\rbrace,\lbrace \tilde{p}_i\rbrace,t)$ such that Hamilton's equations are satisfied for $\tilde{q}_i$ and $\tilde{p}_i$, 
\bea
\dot{\tilde{q}}_i =\dfrac{ \partial \tilde{H}}{\partial \tilde{p}_i}\, , \quad\quad\quad 
\dot{\tilde{p}}_i = -\dfrac{\partial \tilde{H}}{\partial \tilde{q}_i}\, .
\eea 
This can also be formulated at the Lagrangian level. Hamilton's equations of motion for $q_i$ and $p_i$ follow from variation of the Lagrangian, $\delta\int_{t_1}^{t_2} \left[p_i \dot{q}_i-H(\lbrace q_j \rbrace , \lbrace p_j \rbrace,t)\right]\dd t=0$, where an implicit summation over $i$ is implied. If $\tilde{q}_i$ and $\tilde{p}_i$ are canonical variables, there must exist a Hamiltonian form $\tilde{H}$ such that, similarly, their equation of motion is obtained from varying the Lagrangian, $\delta\int_{t_1}^{t_2} \left[\tilde{p}_i \dot{\tilde{q}}_i-\tilde{H}(\lbrace \tilde{q}_j \rbrace , \lbrace \tilde{p}_j \rbrace,t)\right]\dd t=0$. For these two formulas to be equivalent, the two Lagrangians should differ at most by a pure time derivative function, and an overall constant. In other words, $\tilde{q}_i$ and $\tilde{p}_i$ are canonical variables if there exist two functions $\tilde{H}(\lbrace \tilde{q}_i \rbrace , \lbrace \tilde{p}_i \rbrace,t)$  and $F(\lbrace \tilde{q}_i \rbrace , \lbrace \tilde{p}_i \rbrace,t)$ and a constant $\lambda$ such that 
\bea
\label{eq:canonical:Lagrangian}
\lambda \left[p_i \dot{q}_i-H(\lbrace q_j \rbrace , \lbrace p_j \rbrace,t)\right] = \tilde{p}_i \dot{\tilde{q}}_i-\tilde{H}(\lbrace \tilde{q}_j \rbrace , \lbrace \tilde{p}_j \rbrace,t) + \frac{\dd}{\dd t} F(\lbrace \tilde{q}_j \rbrace , \lbrace \tilde{p}_j \rbrace,t)\, ,
\eea
where $F$ is called the generating function of the canonical transformation.

The states of a system at two different times are also related by a canonical transformation. Indeed, let us consider the solution of the equations of motion~(\ref{eq:Hamilton}), with initial conditions $\lbrace q_i, p_i\rbrace $ at time $t$. When evaluated at time $t+\Delta t$, this solution returns $\lbrace \tilde{q}_i, \tilde{p}_i\rbrace $. The transformation $\lbrace q_i, p_i\rbrace \rightarrow \lbrace \tilde{q}_i, \tilde{p}_i\rbrace$ defined in that way is canonical by construction, and the new Hamiltonian reads $\tilde{H}(\lbrace \tilde{q}_i \rbrace , \lbrace \tilde{p}_i \rbrace,t)=H[\lbrace q_i(\lbrace \tilde{q}_j, \tilde{p}_j \rbrace)\rbrace , \lbrace p_i(\lbrace \tilde{q}_j, \tilde{p}_j \rbrace)\rbrace,t+\Delta t]$. This is why canonical transformations do not only describe a fundamental invariance of the theory, they also generate its dynamics. They therefore provide valuable insight into the physics of a given system. 

If different canonical variables can be equivalently used to describe the state and the evolution of a physical system, at the quantum mechanical level, they may not lead to equivalent definitions of the ``vacuum state''. This has important implications in inflationary cosmology, where initial conditions are often set in the ``vacuum'', under the assumption that all pre-existing classical fluctuations are red-shifted by the accelerated expansion. This implies that, at first sight, different initial states for the universe need to be considered for different choices of canonical variables one uses to describe it. This makes canonical transformations of even greater importance in the context of cosmology, where the present work studies various of their aspects. This paper contains materials of different kinds, and is, as such, a rather hybrid object: while it reviews in a pedagogical format standard results about canonical transformations discussed in general (\ie outside the context of cosmology), it also derives new formal results (for instance composition laws in $\mathrm{SU}(1,1)$, configuration representations of states in the invariant representation, generic initial conditions setting in the invariant representation), it investigates new observational signatures of non Bunch-Davies initial states in the context of inflationary cosmology (coming from ambiguities in the choice of canonical variables), and it offers new insight in describing the quantum-to-classical transition of cosmological fluctuations.

This work focuses on linear canonical transformations, for systems endowed with linear dynamics. The reason is that, at leading order in perturbation theory, the dynamics of cosmological perturbations is indeed linear. Moreover, linear canonical transformations can be studied with linear algebraic techniques and can therefore be more thoroughly examined. More precisely, they possess a symplectic structure, which can be shown as follows. Using matricial notations, let us cast the phase-space coordinates into the vector 
\bea
\boldsymbol{z}=(q_1,\cdots,q_n;p_1,\cdots,p_n)^\mathrm{T}\, ,
\eea
where ``$\mathrm{T}$'' stands for the transpose. The dynamics of $\boldsymbol{z}$ being linear, it is generated by a quadratic Hamiltonian $H$, that can be expressed in terms of a symmetric $(2n\times 2n)$-real matrix $\boldsymbol{H}$ as
\bea
H=\frac{1}{2} \boldsymbol{z}^\mathrm{T}\boldsymbol{H}\boldsymbol{z}\, .
\eea
Hamilton's equations of motion~(\ref{eq:Hamilton}) for $\boldsymbol{z}$ can then be recast as
\bea
\dot{\boldsymbol{z}} = \boldsymbol{\Omega}\boldsymbol{H}\boldsymbol{z}\, ,
\eea
where the matrix $\boldsymbol{\Omega}$ is given by
\bea
\label{eq:Omega:def}
	\boldsymbol{\Omega}=\left(\begin{array}{cc}
		0 & \boldsymbol{I}_n \\
		-\boldsymbol{I}_n & 0
	\end{array}\right).
\eea
In this expression, $\boldsymbol{I}_n$ is the $n\times n$ identity matrix. One can check that the matrix $\boldsymbol{\Omega}$ is such that $\boldsymbol{\Omega}^\mathrm{T} = \boldsymbol{\Omega}^{-1} = -\boldsymbol{\Omega}$. Let us now introduce a linear canonical transformation, parametrised by the $(2n\times 2n)$-real matrix  $\boldsymbol{M}$,
\bea
\label{eq:canonical:transformation}
 \boldsymbol{z}\rightarrow \tilde{\boldsymbol{z}} = \boldsymbol{M} \boldsymbol{z}\, .
\eea
The dynamics of $\tilde{\boldsymbol{z}}$ is still linear, so if $\boldsymbol{M}$ describes a canonical transformation, there should exist a symmetric real matrix $\tilde{\boldsymbol{H}}$ such that $\dot{\tilde{\boldsymbol{z}}} = \boldsymbol{\Omega}\tilde{\boldsymbol{H}}\tilde{\boldsymbol{z}}$. If this is the case, differentiating \Eq{eq:canonical:transformation} with respect to time leads to $\tilde{\boldsymbol{H}} = -\boldsymbol{\Omega} \dot{\boldsymbol{M}} \boldsymbol{M}^{-1} - \boldsymbol{\Omega} \boldsymbol{M} \boldsymbol{\Omega} \boldsymbol{H} \boldsymbol{M}^{-1} $. The only requirement that is left is that $\tilde{\boldsymbol{H}}$ must be symmetric. Imposing that $\tilde{\boldsymbol{H}}-\tilde{\boldsymbol{H}}^\mathrm{T}=0$, and multiplying the expression one obtains by $\boldsymbol{M}^\mathrm{T}$ on the left and by $\boldsymbol{M}$ on the right, one has
\bea
\label{eq:canonical:condition}
\boldsymbol{H}\boldsymbol{\Omega} \boldsymbol{M}^\mathrm{T}\boldsymbol{\Omega}\boldsymbol{M} - \boldsymbol{M}^\mathrm{T}\boldsymbol{\Omega} \boldsymbol{M} \boldsymbol{\Omega}\boldsymbol{H}-\frac{\dd}{\dd t}\left(\boldsymbol{M}^\mathrm{T}\boldsymbol{\Omega} \boldsymbol{M} \right)=0\, .
\eea
For $\boldsymbol{M}$ to define a generic canonical transformation of phase space, independently of the dynamics it is endowed with, the above should be valid for any Hamiltonian $\boldsymbol{H}$. Therefore, the sum of the first two terms, which involve $\boldsymbol{H}$, and the third term, which does not, should vanish independently. The requirement that the sum of the two first terms vanishes can be written as $\boldsymbol{H} \boldsymbol{A} = \boldsymbol{A}^\mathrm{T} \boldsymbol{H}$ for any $\boldsymbol{H}$, with $\boldsymbol{A} = \boldsymbol{\Omega} \boldsymbol{M}^\mathrm{T}\boldsymbol{\Omega}\boldsymbol{M} $. This imposes that $\boldsymbol{A} $ is proportional to the identity matrix, $\boldsymbol{A}=-\lambda \boldsymbol{I}_{2n}$ (where the minus sign is introduced for convenience, in order to match the notation of \Eq{eq:canonical:Lagrangian}]. This translates into the condition
\bea
\label{eq:canonical:symplectic:lambda}
\boldsymbol{M}^\mathrm{T}\boldsymbol{\Omega}\boldsymbol{M} = \lambda \boldsymbol{\Omega}\, ,
\eea
which is such that the third term in \Eq{eq:canonical:condition} also trivially vanishes.
The multiplicative constant $\lambda$ can be absorbed through a particularly simple type of canonical transformations called scale transformations. Indeed, if one changes the ``units'' in which $q_i$ and $p_i$ are measured, $\tilde{q}_i = \lambda_q q_i$ and $\tilde{p}_i = \lambda_p p_i$, it is clear that \Eq{eq:canonical:Lagrangian} is satisfied, or that, equivalently, \Eq{eq:canonical:symplectic:lambda} is satisfied, if $\lambda=\lambda_q \lambda_p$. Any extended canonical transformation (\ie a canonical transformation with $\lambda\neq 1$) can therefore always be decomposed as the product of a scale transformation with a canonical transformation having $\lambda=1$. This is why the term ``canonical transformations'' usually refers to those transformations having $\lambda=1$ only, and for which
\bea
\label{eq:canonical:symplectic}
\boldsymbol{M}^\mathrm{T}\boldsymbol{\Omega}\boldsymbol{M} =  \boldsymbol{\Omega}\, .
\eea
Real $(2n\times 2n)$ matrices satisfying this condition define the symplectic group $\mathrm{Sp}(2n,\mathbb{R})$, the properties of which are at the basis of the analysis of linear canonical transformations.

This article is organised as follows. In \Sec{sec:group}, we first derive the main properties of Sp($2,\mathbb{R}$), and of ${\mathrm{SU}}(1,1)$ to which it is isomorphic. We also study their Lie algebras. In \Sec{sec:classical}, we apply these tools to the study of canonical transformations for classical scalar field degrees of freedom and their linear dynamics. In \Sec{sec:Quantum}, the case of quantum scalar fields is analysed. In particular, we show how different sets of canonical variables select out different vacuum states, and how they are related. The linear dynamics of quantum scalar field is described in the squeezing formalism, and the connection between canonical transformations and squeezing operators is established. In \Sec{ssec:invariant}, we show that any quadratic Hamiltonian can be cast under the form of a standard harmonic oscillator by a suitable canonical transformation. The choice of the corresponding canonical variables is called the ``invariant representation'', of which we study various aspects. Let us stress that these 5 first sections (as well as the appendices, see below) are independent of cosmology, and may therefore be useful to a reader interested in canonical transformations but with no particular interest in cosmology.

In \Sec{sec:inflation}, we then apply our findings to the context of cosmology, and consider the case of scalar fields in an inflating background. In particular, we show how different ``natural'' choices of canonical variables lead to different observational predictions when initial conditions are set at a finite time. Our conclusions are presented in \Sec{sec:conclusion}, and the paper then ends with a number of appendices. They contain additional material that is not directly essential to follow the main text but that will allow the interested reader to go deeper into various aspects of this work. In \App{app:composition}, the composition laws on ${\mathrm{SU}}(1,1)$ are derived. To our knowledge, this represents a new result. In \App{app:representation}, quantum representations of $\frak{su}(1,1)$ and $\frak{sp}(2,\mathbb{R})$ are constructed, in terms of the field operators and in terms of the creation and annihilation operators. In \App{app:realvariables}, the one-mode representation needed in the use of the Wigner-Weyl function is introduced, and mapped to the two-mode representation mainly used throughout this article. \Apps{app:rho} and~\ref{app:wave} complement \Sec{ssec:invariant} in investigating the invariant representation. \App{app:rho} shows how the dynamics can be solved in the invariant representation, and as an application, the case where inflation is followed by an era of radiation is considered. In \App{app:wave}, the wavefunction of the Fock states and of the quasiclassical states is derived in the invariant representation.
\section{Preliminaries on $\mathrm{Sp}(2,\mathbb{R})$ and ${\mathrm{SU}}(1,1)$ groups}
\label{sec:group}

\subsection{$\mathrm{Sp}(2,\mathbb{R})$ and its link with ${\mathrm{SU}}(1,1)$}
\subsubsection{Symplectic matrices} 
As explained around \Eq{eq:canonical:symplectic}, the group of symplectic $(2n\times2n)$-matrices, denoted $\mathrm{Sp}(2n,\mathbb{R})$, is formed by the real matrices $\boldsymbol{M}$ satisfying
\bea
	\boldsymbol{M}^\mathrm{T}\boldsymbol{\Omega}\boldsymbol{M}=\boldsymbol{\Omega}, \label{eq:symp}
\eea
with $\boldsymbol{\Omega}$ given in \Eq{eq:Omega:def}. Since $\det \boldsymbol{\Omega}=1$, \Eq{eq:symp} implies that $\det \boldsymbol{M}=\pm 1$.
This group is a non-compact Lie group and, as explained in \Sec{sec:intro}, it is of specific interest for Hamiltonian dynamics since symplectic transformations generate both the dynamics of the system, and the canonical transformations. Quadratic Hamiltonians indeed lead to linear evolutions encoded in a Green's matrix which is symplectic, \ie $\boldsymbol{z}(t)=\boldsymbol{G}(t,t_0)\boldsymbol{z}(t_0)$ with $\boldsymbol{G}\in\mathrm{Sp}(2n,\mathbb{R})$ (this will be shown below in \Sec{sssec:complexdynamics}). Similarly, any linear canonical transformation of the phase space is generated by a symplectic matrix, \ie the variables $\boldsymbol{\widetilde{z}}=\boldsymbol{M}\boldsymbol{z}$ with $\boldsymbol{M}\in\mathrm{Sp}(2n,\mathbb{R})$ are also described by a Hamiltonian (note that $\boldsymbol{M}$ can be time dependent). The Green's matrix $\boldsymbol{\widetilde{G}}$ generating the evolution of $\boldsymbol{\widetilde{z}}$ is also an element of $\mathrm{Sp}(2n,\mathbb{R})$ and 
\bea
\label{eq:Green:canonical:transformation}
\boldsymbol{\widetilde{G}}(t,t_0)=\boldsymbol{M}(t)\boldsymbol{G}(t,t_0)\boldsymbol{M}^{-1}(t_0).
\eea

In what follows, we consider the simple case of $n=1$, which for a scalar field would correspond to one of its infinitely many $\vec{k}$-modes. From now on, $\boldsymbol{\Omega}$ stands for 
\bea
\label{eq:Omega:2D:def}
	\boldsymbol{\Omega}=\left(\begin{array}{cc}
		0 & 1 \\
		-1 & 0
	\end{array}\right),
\eea
and $\boldsymbol{I}$ stands for $\boldsymbol{I}_2$. In this restricted case, $\boldsymbol{M}$ being symplectic is equivalent to $\det(\boldsymbol{M})=1$, \ie $\boldsymbol{M}\in\mathrm{Sp}(2,\mathbb{R})$ if and only if $\boldsymbol{M}\in\mathrm{SL}(2,\mathbb{R})$. Any matrix of Sp(2,$\mathbb{R}$) can be parametrised with 2 angles, $\gamma$ and $\gamma'$, and a squeezing amplitude $d\in\mathbb{R}$, through the Bloch-Messiah decomposition\footnote{We note that the Bloch-Messiah decomposition can be used for any value of $n$.}~\cite{Bloch:1962zj}
\bea
	\boldsymbol{M}(\gamma,d,\gamma')=\boldsymbol{R}(\gamma)\left(\begin{array}{cc}
		e^d & 0 \\
		0 & e^{-d}
	\end{array}\right)\boldsymbol{R}(\gamma') .
	\label{eq:BlochMessiah:Sp2R}
\eea
In this expression, $\boldsymbol{R}(\alpha)\in\mathrm{SO}(2)$ is a rotation in the phase-space plane,
\bea
	\boldsymbol{R}(\alpha)=\left(\begin{array}{cc}
		\cos\alpha & -\sin\alpha \\
		\sin\alpha & \cos\alpha
	\end{array}\right) .
	\label{eq:R(alpha):real}
\eea
From this decomposition, it is easy to get the inverse of the matrix, $\left[\boldsymbol{M}(\gamma,d,\gamma')\right]^{-1}=\boldsymbol{M}(-\gamma',-d,-\gamma)$, while the identity matrix is obtained for $\gamma+\gamma'=d=0$.

The reason why $d$ is called a squeezing amplitude is the following. Consider a trajectory in the phase space given by a unit circle, \ie $\left|\boldsymbol{z}\right|^2=q^2+p^2=1$ (where $q$ and $p$ are normalised in such a way that they share the same physical dimension), and apply a canonical transformation of the form $\boldsymbol{M}(0,d,0)$. The trajectory with the new set of canonical variables is thus given by an ellipse squeezed along the configuration axis $\widetilde{q}$ for $d>0$, or along the momentum axis $\widetilde{p}$ for $d<0$, with eccentricity $\sqrt{1-e^{-4\vert d\vert}}$.

\subsubsection{From $\mathrm{Sp}(2,\mathbb{R})$ to ${\mathrm{SU}}(1,1)$}
\label{sec:HelicityBasis:SU11}

Instead of working with the real-valued variables $\boldsymbol{z}^\mathrm{T}=(q,p)$, one can introduce the set of complex variables $\boldsymbol{a}^\dag=\left(a^*,a\right)$ with $a=\left(q+ip\right)/\sqrt{2}$ (``$\dag$'' stands for the conjugate transpose). From the forthcoming perspective of a quantum scalar field, this set of variables corresponds to a set of creation/annihilation operators. Such a transformation will be referred to as working in the ``helicity basis'' in what follows. This change of variables is generated by the unitary matrix
\bea
\label{eq:U:def}
	\boldsymbol{U}=\frac{1}{\sqrt{2}}\left(\begin{array}{cc}
		1 & i \\
		1 & -i
	\end{array}\right),
\eea
\ie $\boldsymbol{a}=\boldsymbol{U}\boldsymbol{z}$, and $\boldsymbol{U}\boldsymbol{U}^\dag=\boldsymbol{U}^\dag\boldsymbol{U}=\boldsymbol{I}$. If one performs a canonical transformation generated by $\boldsymbol{M}$, then it is easy to check that $\boldsymbol{\widetilde{a}}=\boldsymbol{\mathcal{M}}\boldsymbol{a}$ with 
\bea
\label{eq:calM:def}
\boldsymbol{\mathcal{M}}=\boldsymbol{U}\boldsymbol{M}\boldsymbol{U}^\dag.
\eea
Similarly, the dynamics in the helicity basis is obtained from the Green's matrix $\boldsymbol{\mathcal{G}}(t,t_0)=\boldsymbol{U}\boldsymbol{G}(t,t_0)\boldsymbol{U}^\dag$, with $\boldsymbol{a}(t)=\boldsymbol{\mathcal{G}}(t,t_0)\boldsymbol{a}(t_0)$. The relation between the Green's matrices of two such sets of complex variables is then $\boldsymbol{\widetilde{\mathcal{G}}}(t,t_0)=\boldsymbol{\mathcal{M}}(t)\boldsymbol{\mathcal{G}}(t,t_0)\boldsymbol{\mathcal{M}}^{-1}(t_0)$.

It is easy to check that in the helicity basis, canonical transformations and the dynamics are now generated by matrices in the $\mathrm{SU}(1,1)$ group. First, since $\det(\boldsymbol{M})=1$ and $\boldsymbol{U}$ is unitary, one has $\det(\boldsymbol{\mathcal{M}})=1$. Second, \Eq{eq:symp} leads to
\bea
\label{eq:eqdef:SU11}
\boldsymbol{\mathcal{M}}^\dag\boldsymbol{\mathcal{J}}\boldsymbol{\mathcal{M}}=\boldsymbol{\mathcal{J}},
\eea  
with
\bea
\label{eq:J:def}
	\boldsymbol{\mathcal{J}}=\boldsymbol{U}\boldsymbol{\Omega}\boldsymbol{U}^\dag=(-i)\left(\begin{array}{cc}
		1 & 0 \\
		0 & -1
	\end{array}\right).
\eea
These are the two conditions for the matrix $\boldsymbol{\mathcal{M}}$ to be an element of the $\mathrm{SU}(1,1)$ group (see \Refa{Martin-Dussaud:2019ypf} for a more formal presentation of the relations between $\mathrm{Sp}(2,\mathbb{R})$ and ${\mathrm{SU}}(1,1)$).
\subsection{SU(1,1) toolkit}
\subsubsection{General aspects}
\label{sec:SU11:General}
From \Eq{eq:eqdef:SU11} and the condition $\det(\boldsymbol{M})=1$, matrices of the $\mathrm{SU}(1,1)$ group read
\bea
	\boldsymbol{\mathcal{M}}=\left(\begin{array}{cc}
		\alpha & \beta \\
		\beta^* & \alpha^*
	\end{array}\right)
	\quad\text{with}\quad 
	\left|\alpha\right|^2-\left|\beta\right|^2=1.
	\label{eq:su11mat}
\eea
Since in the following we will mainly work in the helicity basis, let us provide other useful ways of writing these matrices. Many aspects of $\mathrm{SU}(1,1)$ are discussed \eg in \Refs{Chiribella2006,doi:10.1063/1.525254,puri2001mathematical,perelomov1986generalized}.

From the Bloch-Messiah decomposition~(\ref{eq:BlochMessiah:Sp2R}) of symplectic matrices, one first obtains the following decomposition of matrices in $\mathrm{SU}(1,1)$:
\begin{equation}\begin{aligned}
\label{eq:bloch}
	\boldsymbol{\mathcal{M}}(\gamma,d,\gamma')&=\left(\begin{array}{cc}
		e^{i\gamma} & 0 \\
		0 & e^{-i\gamma}
	\end{array}\right)\left(\begin{array}{cc}
		\cosh d & \sinh d \\
		\sinh d & \cosh d
	\end{array}\right)\left(\begin{array}{cc}
		e^{i\gamma'} & 0 \\
		0 & e^{-i\gamma'}
	\end{array}\right),  \\
	&=\left(\begin{array}{cc}
		e^{i(\gamma+\gamma')}\cosh d & e^{i(\gamma-\gamma')}\sinh d \\
		e^{-i(\gamma-\gamma')}\sinh d & e^{-i(\gamma+\gamma')}\cosh d
	\end{array}\right). 
\end{aligned}\end{equation} 
The relation with the expression~(\ref{eq:su11mat}) is given by $\alpha=e^{i(\gamma+\gamma')}\cosh d$ and $\beta=e^{i(\gamma-\gamma')}\sinh d$. This way of writing $\mathrm{SU}(1,1)$ matrices will also be dubbed ``Bloch-Messiah decomposition'' in the rest of this paper. As with \Eq{eq:BlochMessiah:Sp2R}, this decomposition gives rise to the inverse $\left[\boldsymbol{\mathcal{M}}(\gamma,d,\gamma')\right]^{-1}=\boldsymbol{\mathcal{M}}(-\gamma',-d,-\gamma)$. One can also derive the following identity from the Bloch-Messiah decomposition: 
\bea
	\boldsymbol{\mathcal{M}}(\widetilde\gamma,-d,\widetilde\gamma')=\boldsymbol{\mathcal{M}}(\gamma,d,\gamma'), \label{eq:sym}
\eea 
where the angles have to satisfy $\widetilde\gamma=\gamma+(n+1/2)\pi$ and $\widetilde\gamma'=\gamma'-(m+1/2)\pi$ with $n,m\in\mathbb{Z}$ and $n$ and $m$ have the same parity.

Such matrices also admit the left-polar decomposition given by $\boldsymbol{\mathcal{M}}=\boldsymbol{\mathcal{S}}\boldsymbol{\mathcal{R}}$ with $\boldsymbol{\mathcal{S}}=\sqrt{\boldsymbol{\mathcal{M}}\boldsymbol{\mathcal{M}}^\dag}$ a positive-definite matrix, and $\boldsymbol{\mathcal{R}}$ a unitary matrix. The first of these two matrices is obtained by diagonalising the matrix $\boldsymbol{\mathcal{M}}\boldsymbol{\mathcal{M}}^\dag$ (which is also positive definite) starting from the Bloch-Messiah decomposition~(\ref{eq:bloch}), and then taking the square root of it. The second is derived by calculating $\boldsymbol{\mathcal{S}}^{-1}\boldsymbol{\mathcal{M}}$, making use of the Bloch-Messiah decomposition for $\boldsymbol{\mathcal{M}}$ again. This gives 
\bea
	\boldsymbol{\mathcal{M}}(\gamma,d,\gamma')=\left(\begin{array}{cc}
		\cosh d & e^{2i\gamma}\sinh d \\
		e^{-2i\gamma}\sinh d & \cosh d
	\end{array}\right)\left(\begin{array}{cc}
		e^{i(\gamma+\gamma')} & 0 \\
		0 & e^{-i(\gamma+\gamma')}
	\end{array}\right).
\eea
We note that one can instead use the right-polar decomposition $\boldsymbol{\mathcal{M}}=\boldsymbol{\mathcal{R}}\boldsymbol{\mathcal{S}'}$ with $\boldsymbol{\mathcal{S}'}=\sqrt{\boldsymbol{\mathcal{M}}^\dag\boldsymbol{\mathcal{M}}}$,\bea
	\boldsymbol{\mathcal{M}}(\gamma,d,\gamma')=\left(\begin{array}{cc}
		e^{i(\gamma+\gamma')} & 0 \\
		0 & e^{-i(\gamma+\gamma')}
	\end{array}\right)\left(\begin{array}{cc}
		\cosh d & e^{-2i\gamma'}\sinh d \\
		e^{2i\gamma'}\sinh d & \cosh d
	\end{array}\right).
\eea
We stress that the unitary matrix $\boldsymbol{\mathcal{R}}$ for the left-polar and the right-polar decompositions are the same.

The three above decompositions of $\mathrm{SU}(1,1)$ elements can be wrapped up introducing the two types of matrices 
\bea
	\boldsymbol{\mathcal{R}}(\theta)=\left(\begin{array}{cc}
		e^{i\theta} & 0 \\
		0 & e^{-i\theta}
	\end{array}\right) &\quad\text{and}\quad& 
	\boldsymbol{\mathcal{S}}(r,\varphi)=\left(\begin{array}{cc}
		\cosh r & e^{2i\varphi}\sinh r \\
		e^{-2i\varphi}\sinh r & \cosh r
	\end{array}\right),
\label{eq:polar:RS:explicit}
\eea
which will be referred to as {\it rotation} and {\it squeezing} matrices respectively. It is rather obvious that $\boldsymbol{\mathcal{R}}$ operates a rotation of the phase space in the helicity basis, \ie $\boldsymbol{\mathcal{R}}(\alpha)= \boldsymbol{U} \boldsymbol{{R}}(\alpha) \boldsymbol{U}^\dagger$, where $\boldsymbol{{R}}(\alpha)$ was defined in \Eq{eq:R(alpha):real}. In full generality, $\boldsymbol{\mathcal{S}}$ operates both a squeezing and a rotation in the phase space. We will however keep the term ``squeezing matrices'' since, as it will be clear hereafter, the so-called squeezing operator introduced in quantum optics is the unitary representation of the action of the matrices $\boldsymbol{\mathcal{S}}$ on the complex variables. The inverses of these matrices are respectively $\left[\boldsymbol{\mathcal{R}}(\theta)\right]^{-1}=\boldsymbol{\mathcal{R}}(-\theta)$ and $\left[\boldsymbol{\mathcal{S}}(r,\varphi)\right]^{-1}=\boldsymbol{\mathcal{S}}(-r,\varphi)=\boldsymbol{\mathcal{S}}(r,\varphi+\frac{\pi}{2})$. The following composition laws apply: $\boldsymbol{\mathcal{R}}(\theta_1)\boldsymbol{\mathcal{R}}(\theta_2)=\boldsymbol{\mathcal{R}}(\theta_1+\theta_2)$ and $\boldsymbol{\mathcal{S}}(d_1,\varphi)\boldsymbol{\mathcal{S}}(d_2,\varphi)=\boldsymbol{\mathcal{S}}(d_1+d_2,\varphi)$. 
From these relations, its is easy to obtain the three compact forms of the elements of $\mathrm{SU}(1,1)$:
\bea
\label{eq:bloch:left:right}
	\boldsymbol{\mathcal{M}}(\gamma,d,\gamma')&=&\boldsymbol{\mathcal{R}}(\gamma)\boldsymbol{\mathcal{S}}(d,0)\boldsymbol{\mathcal{R}}(\gamma')\\ 
	&=&\boldsymbol{\mathcal{S}}(d,\gamma)\boldsymbol{\mathcal{R}}(\gamma+\gamma')=\boldsymbol{\mathcal{R}}(\gamma+\gamma')\boldsymbol{\mathcal{S}}(d,-\gamma'), \nonumber
\eea
with the first line corresponding to the Bloch-Messiah decomposition, and the second to the left- and right-polar decompositions.  
\subsubsection{Composition} 
\label{sec:SU11:Composition}
From the expression~(\ref{eq:su11mat}), one can easily show that the composition of two elements of $\mathrm{SU}(1,1)$ leads to another element of $\mathrm{SU}(1,1)$, \ie 
\bea
\label{eq:com:su11:formal}
\boldsymbol{\mathcal{M}}(\gamma_1,d_1,\gamma'_1)\boldsymbol{\mathcal{M}}(\gamma_2,d_2,\gamma'_2)=\boldsymbol{\mathcal{M}}(\gamma,d,\gamma').
\eea 
The relation between the two sets of parameters of the matrices to be composed in the left-hand side of \Eq{eq:com:su11:formal}, and the set of parameters of the resulting matrix in the right-hand side, can be obtained as follows.

First, we note that any matrix of $\mathrm{SU}(1,1)$ can be decomposed on the basis $\left\{\boldsymbol{J}_a\right\}_{a=0,\cdots,4}=\left\{\boldsymbol{I},\boldsymbol{J}_x,\boldsymbol{J}_y,\boldsymbol{J}_z\right\}$, with $\boldsymbol{J}_{x,y,z}$ the Pauli matrices,\footnote{The Pauli matrices are given by
\bea
\boldsymbol{J}_x =
\left(\begin{array}{cc}
		0 & 1 \\
		1& 0
	\end{array}\right)
	,\quad
\boldsymbol{J}_y =
\left(\begin{array}{cc}
		0 & -i \\
		i& 0
	\end{array}\right)	
		,\quad
\boldsymbol{J}_z =
\left(\begin{array}{cc}
		1 & 0 \\
		0 & -1
	\end{array}\right).	
\eea
\label{footnote:Pauli}} as follows
\bea
\label{eq:SU11:exp:Pauli}
	\boldsymbol{\mathcal{M}}=T\boldsymbol{I}+X\boldsymbol{J}_x+Y\boldsymbol{J}_y+iZ\boldsymbol{J}_z .
\eea
In this expression, $T$, $X$, $Y$ and $Z$ are four real numbers satisfying $T^2+Z^2-X^2-Y^2=1$ because of \Eq{eq:su11mat}.\footnote{The components are simply given by $T=\mathrm{Re}(\alpha)=\cos(\gamma+\gamma')\cosh d$, $X=\mathrm{Re}(\beta)=\cos(\gamma-\gamma')\sinh d$, $Y=-\mathrm{Im}(\beta)=-\sin(\gamma-\gamma')\sinh d$, and $Z=\mathrm{Im}(\alpha)=\sin(\gamma+\gamma')\cosh d$.} We note that, in practice, these numbers can be computed by making use of the relation $\frac{1}{2}\mathrm{Tr}\left[\boldsymbol{J}_a\boldsymbol{J}_b\right]=\delta_{a,b}$. By writing $\boldsymbol{\mathcal{M}}=\ds\sum_{a=0}^4\mathcal{M}_a\boldsymbol{J}_a$, it is indeed straightforward to show that $\mathcal{M}_a=\frac{1}{2}\mathrm{Tr}\left[\boldsymbol{\mathcal{M}}\boldsymbol{J}_a\right]$.

One then decomposes both sides of \Eq{eq:com:su11:formal} on the basis $\left\{\boldsymbol{J}_a\right\}$, which yields 
\bea
	\frac{1}{2}\mathrm{Tr}\left[\boldsymbol{\mathcal{M}}\boldsymbol{J}_a\right]=\frac{1}{2}\mathrm{Tr}\left[\boldsymbol{\mathcal{M}}_1\boldsymbol{\mathcal{M}}_2\boldsymbol{J}_a\right].
\eea
By plugging the Bloch-Messiah decomposition~(\ref{eq:bloch}) in the above, one obtains a set of four equations relating the set $(\gamma,d,\gamma')$ to the two sets $(\gamma_1,d_1,\gamma'_1)$ and $(\gamma_2,d_2,\gamma'_2)$,
\bea
	\cos(\gamma+\gamma')\cosh d&=&\cos(\gamma_1+\gamma'_2)\cos(\gamma'_1+\gamma_2)\cosh(d_1+d_2) \nonumber \\ \label{eq:comp:BM:1}
	&&-\sin(\gamma_1+\gamma'_2)\sin(\gamma'_1+\gamma_2)\cosh(d_1-d_2), \\
	\cos(\gamma-\gamma')\sinh d&=&\cos(\gamma_1-\gamma'_2)\cos(\gamma'_1+\gamma_2)\sinh(d_1+d_2) \nonumber \\
	&&+\sin(\gamma_1-\gamma'_2)\sin(\gamma'_1+\gamma_2)\sinh(d_1-d_2), \\
	\sin(\gamma-\gamma')\sinh d&=&\sin(\gamma_1-\gamma'_2)\cos(\gamma'_1+\gamma_2)\sinh(d_1+d_2) \nonumber \\
	&&-\cos(\gamma_1-\gamma'_2)\sin(\gamma'_1+\gamma_2)\sinh(d_1-d_2), \\
	\sin(\gamma+\gamma')\cosh d&=&\sin(\gamma_1+\gamma'_2)\cos(\gamma'_1+\gamma_2)\cosh(d_1+d_2) \nonumber \\
	&&+\cos(\gamma_1+\gamma'_2)\sin(\gamma'_1+\gamma_2)\cosh(d_1-d_2).
	\label{eq:comp:BM:4}
\eea
As it is, the above set of constraints cannot be simply inverted to get close expressions of $(\gamma,d,\gamma')$ as functions of $(\gamma_1,d_1,\gamma'_1)$ and $(\gamma_2,d_2,\gamma'_2)$. However, one can derive close formulas for the hyperbolic sine and cosine of $d$, and the sines and cosines of linear combinations of $\gamma$ and $\gamma'$ (full expressions for $(\gamma,d,\gamma')$ are obtained below in two specific cases). 

The first step consists in rewriting the group composition law~(\ref{eq:com:su11:formal}) as
\bea
	\boldsymbol{\mathcal{M}}(0,d,0)=\boldsymbol{\mathcal{M}}(\gamma_1-\gamma,d_1,\gamma'_1)\boldsymbol{\mathcal{M}}(\gamma_2,d_2,\gamma'_2-\gamma'),
\eea
which is easily obtained by operating \Eq{eq:com:su11:formal} with $\boldsymbol{\mathcal{R}}(-\gamma)$ on the left and $\boldsymbol{\mathcal{R}}(-\gamma')$ on the right. Denoting 
\begin{equation}\begin{aligned}
\label{eq:gamma+-12:def}
	\gamma_+&=\gamma_1-\gamma+\gamma'_2-\gamma', \\
	\gamma_-&=\gamma_1-\gamma-\gamma'_2+\gamma',  \\
	\gamma_{1,2}&=\gamma'_1+\gamma_2,
\end{aligned}\end{equation}
the four constraints~(\ref{eq:comp:BM:1})-(\ref{eq:comp:BM:4}) then read
\bea
	\cosh d&=&\cos(\gamma_+)\cos(\gamma_{1,2})\cosh(d_1+d_2)-\sin(\gamma_+)\sin(\gamma_{1,2})\cosh(d_1-d_2), \label{eq:c1}\\
	\sinh d&=&\cos(\gamma_-)\cos(\gamma_{1,2})\sinh(d_1+d_2)+\sin(\gamma_-)\sin(\gamma_{1,2})\sinh(d_1-d_2), \label{eq:c2}\\
	0&=&\sin(\gamma_-)\cos(\gamma_{1,2})\sinh(d_1+d_2)-\cos(\gamma_-)\sin(\gamma_{1,2})\sinh(d_1-d_2), \label{eq:c3}\\
	0&=&\sin(\gamma_+)\cos(\gamma_{1,2})\cosh(d_1+d_2)+\cos(\gamma_+)\sin(\gamma_{1,2})\cosh(d_1-d_2). \label{eq:c4}
\eea
The unknowns are $d$ and $\gamma_\pm$,\footnote{It is clear that $\gamma$ and $\gamma'$ can easily be obtained from $\gamma_\pm$, namely $\gamma=\gamma_1-\frac{1}{2}(\gamma_++\gamma_-)$ and $\gamma'=\gamma_2'-\frac{1}{2}(\gamma_+-\gamma_-)$.\label{footnote:gamma:gammapm}} which should be expressed as functions of the two squeezing parameters, $d_1$ and $d_2$, and one single angle $\gamma'_1+\gamma_2$. We note that it is not a surprise that only $\gamma'_1+\gamma_2$ is involved, since from the Bloch-Messiah decomposition~(\ref{eq:bloch:left:right}), one has, using the composition law for rotation matrices given above,
\bea
	\boldsymbol{\mathcal{M}}(\gamma_1,d_1,\gamma'_1)\boldsymbol{\mathcal{M}}(\gamma_2,d_2,\gamma'_2)=\boldsymbol{\mathcal{R}}(\gamma_1)\boldsymbol{\mathcal{S}}(d_1,0)\boldsymbol{\mathcal{R}}(\gamma'_1+\gamma_2)\boldsymbol{\mathcal{S}}(d_2,0)\boldsymbol{\mathcal{R}}(\gamma'_2).
\eea

Solving the above system for $(d,\gamma_\pm)$ is done case by case in \App{app:composition}. Here we just list our final results.
\begin{itemize}
\item For $\gamma'_1+\gamma_2=n\pi$ with $n\in\mathbb{Z}$:
\bea
	\gamma=\left\{\begin{array}{c}
		\gamma_1-n\pi \\
		\gamma_1-\left(n+\frac{1}{2}\right)\pi
	\end{array}\right., &~~~	d=\left\{\begin{array}{c}
		d_1+d_2 \\
		-d_1-d_2
	\end{array}\right., &~~~	\gamma'=\left\{\begin{array}{c}
		\gamma'_2 \\
		\gamma'_2+\frac{\pi}{2}
	\end{array}\right..
\eea
The two choices (upper and lower rows) yield the same final matrix because of the symmetry~(\ref{eq:sym}).
\item For $\gamma'_1+\gamma_2=(n+1/2)\pi$ with $n\in\mathbb{Z}$:
\bea
	\gamma=\left\{\begin{array}{c}
		\gamma_1-(n+1)\pi \\
		\gamma_1-\left(n-\frac{1}{2}\right)\pi
	\end{array}\right. , &~~~ d=\left\{\begin{array}{c}
		d_1-d_2 \\
		-d_1+d_2
	\end{array}\right. , &~~~	\gamma'=\left\{\begin{array}{c}
		\gamma'_2-\frac{\pi}{2} \\
		\gamma'_2
	\end{array}\right..
\eea
Here again, the two solutions are related by the symmetry~(\ref{eq:sym}).
\item If $\gamma_1'$ and $\gamma_2$ are such that neither of the two above conditions apply, one can derive close expressions for $\cosh d$, $\sinh d$, $\cos\gamma_\pm$ and $\sin\gamma_\pm$, which are sufficient to unequivocally determine the set $(\gamma,d,\gamma')$. For the squeezing amplitude, one has
\bea
	\cosh d&=&\sqrt{\cos^2\left(\gamma_{1,2}\right)\cosh^2(d_1+d_2)+\sin^2\left(\gamma_{1,2}\right)\cosh^2\left(d_1-d_2\right)}, \\
	\sinh d&=&\sqrt{\cos^2\left(\gamma_{1,2}\right)\sinh^2\left(d_1+d_2\right)+\sin^2\left(\gamma_{1,2}\right)\sinh^2\left(d_1-d_2\right)},
\eea
while for the angles, one finds
\bea
	\cos\gamma_+&=&\mathrm{sign}\left(\cos\gamma_{1,2}\right)\sqrt{\frac{\cos^2\left(\gamma_{1,2}\right)\cosh^2(d_1+d_2)}{\cos^2\left(\gamma_{1,2}\right)\cosh^2(d_1+d_2)+\sin^2\left(\gamma_{1,2}\right)\cosh^2(d_1-d_2)}}, \quad\quad\quad\quad \\
	\sin\gamma_+&=&-\mathrm{sign}\left(\sin\gamma_{1,2}\right)\sqrt{\frac{\sin^2\left(\gamma_{1,2}\right)\cosh^2(d_1-d_2)}{\cos^2\left(\gamma_{1,2}\right)\cosh^2(d_1+d_2)+\sin^2\left(\gamma_{1,2}\right)\cosh^2(d_1-d_2)}},\\
	\cos\gamma_-&=&\mathrm{sign}\left(\cos\gamma_{1,2}\right)\sqrt{\frac{\cos^2\left(\gamma_{1,2}\right)\sinh^2(d_1+d_2)}{\cos^2\left(\gamma_{1,2}\right)\sinh^2(d_1+d_2)+\sin^2\left(\gamma_{1,2}\right)\sinh^2(d_1-d_2)}}, \\
	\sin\gamma_-&=&\mathrm{sign}\left(\sin\gamma_{1,2}\right)\sqrt{\frac{\sin^2\left(\gamma_{1,2}\right)\sinh^2(d_1-d_2)}{\cos^2\left(\gamma_{1,2}\right)\cosh^2(d_1+d_2)+\sin^2\left(\gamma_{1,2}\right)\cosh^2(d_1-d_2)}}.
\eea
\end{itemize}

This closes the composition rule of elements of $\mathrm{SU}(1,1)$. Up to our knowledge, the above formulas relating the parameters of the composed matrix to the two sets of parameters describing the two matrices to be composed, were not presented in the literature (except in restricted cases, see \Refa{Cervero2002}). We also mention that since $\mathrm{SU}(1,1)$ is isomorphic to $\mathrm{Sp}(2,\mathbb{R})$, the above results equally apply to the symplectic group of dimension 2.

\subsubsection{Generators and Lie algebra}
\label{sec:LieAlgebra}
Since $\mathrm{SU}(1,1)$ is a Lie group, one can write some of the above relations using the generators of its associated Lie algebra. The ${\mathfrak{su}}(1,1)$ Lie algebra is a three-dimensional vector space whose basis vectors satisfy the following commutation relations 
\begin{equation}\begin{aligned}
\label{eq:Lie:commutation}
	\left[\boldsymbol{\mathcal{K}}_+,\boldsymbol{\mathcal{K}}_-\right]&=-2\boldsymbol{\mathcal{K}}_z, \\
	\left[\boldsymbol{\mathcal{K}}_z,\boldsymbol{\mathcal{K}}_\pm\right]&=\pm\boldsymbol{\mathcal{K}}_\pm.
\end{aligned}\end{equation}
The generators are given by $\boldsymbol{\mathcal{K}}_\pm=\frac{i}{2}\left(\boldsymbol{J}_x\pm i \boldsymbol{J}_y\right)$ and $\boldsymbol{\mathcal{K}}_z=\frac{1}{2}\boldsymbol{J}_z$, where the Pauli matrices $\boldsymbol{J}_a$ were given in footnote~\ref{footnote:Pauli}. One can alternatively make use of the generators $\boldsymbol{\mathcal{K}}_x=\frac{i}{2}\boldsymbol{J}_x$ and $\boldsymbol{\mathcal{K}}_y=\frac{i}{2}\boldsymbol{J}_y$, leading to $\boldsymbol{\mathcal{K}}_\pm=\boldsymbol{\mathcal{K}}_x\pm i\boldsymbol{\mathcal{K}}_y$. Any vector of the $\mathfrak{su}(1,1)$ algebra can be written as $\boldsymbol{m}=d\left(-n_x\boldsymbol{\mathcal{K}}_x-n_y\boldsymbol{\mathcal{K}}_y+n_z\boldsymbol{\mathcal{K}}_z\right)$ where $d\in\mathbb{R}^+$ is the ``norm'' of the vector defined such that $-n_x^2-n_y^2+n^2_z$ equal either to $0$ or $\pm1$. For any $\boldsymbol{m}\in\mathfrak{su}(1,1)$, $\boldsymbol{\mathcal{M}}=\exp\left(i\boldsymbol{m}\right)$ is an element of $\mathrm{SU}(1,1)$. However, we note that the exponential map is not surjective, meaning that there exist elements of  $\mathrm{SU}(1,1)$ that cannot be written as $\exp(i\boldsymbol{m})$.\footnote{We refer the interested reader to Sec. II B of \Refa{Chiribella2006} for details on the exponential maps of $\mathrm{SU}(1,1)$.}

By making use of $\left(\boldsymbol{J}_a\right)^2=\boldsymbol{I}$, leading to $\left(\boldsymbol{J}_a\right)^{2n}=\boldsymbol{I}$ and $\left(\boldsymbol{J}_a\right)^{2n+1}=\boldsymbol{J}_a$, one can explicitly check that 
$\exp\left(2i\gamma\boldsymbol{\mathcal{K}}_z\right)\exp\left(-2id\boldsymbol{\mathcal{K}}_x\right)\exp\left(2i\gamma'\boldsymbol{\mathcal{K}}_z\right)$ leads to the right-hand side of the Bloch-Messiah decomposition~(\ref{eq:bloch}). Hence, any matrix of $\mathrm{SU}(1,1)$ can be written as
\bea	
\label{eq:calM:calK:xyz}
	\boldsymbol{\mathcal{M}}(\gamma,d,\gamma')=e^{2i\gamma\boldsymbol{\mathcal{K}}_z}e^{-2id\boldsymbol{\mathcal{K}}_x}e^{2i\gamma'\boldsymbol{\mathcal{K}}_z}.
\eea
Similarly, the polar decompositions can be written using the generators of $\mathrm{SU}(1,1)$. Introducing $\xi(r,\varphi)=-i r e^{2i\varphi}$, one has\footnote{It is straightforward to establish that \Eq{eq:polar:Lie:R} matches the first expression in \Eq{eq:polar:RS:explicit}. For \Eq{eq:polar:Lie:S}, one first notes that $\xi\boldsymbol{\mathcal{K}}_+-\xi^*\boldsymbol{\mathcal{K}}_-=2ir\left[-\cos(2\varphi)\boldsymbol{\mathcal{K}}_x+\sin(2\varphi)\boldsymbol{\mathcal{K}}_y\right]$. Following \Refa{Chiribella2006}, this leads to $\exp\left(\xi\boldsymbol{\mathcal{K}}_+-\xi^*\boldsymbol{\mathcal{K}}_-\right)=\cosh(r)\boldsymbol{I}_2+2i\sinh(r)\left[-\cos(2\varphi)\boldsymbol{\mathcal{K}}_x+\sin(2\varphi)\boldsymbol{\mathcal{K}}_y\right]$. From the expressions given in footnote~\ref{footnote:Pauli} for the generators as Pauli matrices, it is easy to check that this last expression matches \Eq{eq:polar:RS:explicit}. }
\bea
\label{eq:polar:Lie:S}
\boldsymbol{\mathcal{S}}(r,\varphi)&=&\exp\left(\xi\boldsymbol{\mathcal{K}}_+-\xi^*\boldsymbol{\mathcal{K}}_-\right) ,\\
\label{eq:polar:Lie:R}
\boldsymbol{\mathcal{R}}(\theta)&=&\exp\left(2i\theta\boldsymbol{\mathcal{K}}_z\right).
\eea
Let us finally mention that the matrix $\boldsymbol{\mathcal{S}}(r,\varphi)$ can be decomposed using the Baker-Campbell-Haussdorff formula as follows \cite{puri2001mathematical,perelomov1986generalized,barnett2002methods}
\bea
	\boldsymbol{\mathcal{S}}(r,\varphi)&=&\exp\left(-ie^{2i\varphi}\tanh r\boldsymbol{\mathcal{K}}_+\right)\left(\frac{1}{\cosh r}\right)^{2\boldsymbol{\mathcal{K}}_z}\exp\left(-ie^{-2i\varphi}\tanh r\boldsymbol{\mathcal{K}}_-\right) \\
	&=&\exp\left(-ie^{-2i\varphi}\tanh r\boldsymbol{\mathcal{K}}_-\right)\left({\cosh r}\right)^{2\boldsymbol{\mathcal{K}}_z}\exp\left(-ie^{2i\varphi}\tanh r\boldsymbol{\mathcal{K}}_+\right). \nonumber
\eea
These two expressions are particularly useful if the squeezing matrix is to be operated on a vector belonging to the null space of either $\boldsymbol{\mathcal{K}}_-$ or $\boldsymbol{\mathcal{K}}_+$, which are respectively given by the set of vectors $\boldsymbol{v}^\dag_-=\left(\lambda^*_-,0\right)$ and $\boldsymbol{v}^\dag_+=\left(0,\lambda^*_+\right)$ where $\lambda_\pm$ are complex numbers. This indeed simplifies to $\exp\left(-ie^{\mp i\varphi}\tanh r\boldsymbol{\mathcal{K}}_\mp\right)\boldsymbol{v}_\mp=\boldsymbol{v}_\mp$ by using the series expansion of the exponential of a matrix. 

\subsection{Final remarks on $\mathrm{Sp}(2,\mathbb{R})$}

We finally mention that all the results derived for $\mathrm{SU}(1,1)$ can be directly transposed to $\mathrm{Sp}(2,\mathbb{R})$ since any matrix of the latter group is similar to a matrix belonging to the former group via $\boldsymbol{M}=\boldsymbol{U}^\dag\boldsymbol{\mathcal{M}}\boldsymbol{U}$. In particular, this allows one to write the Bloch-Messiah decomposition and the polar decomposition of elements of the symplectic group by means of the Lie generators.\footnote{Remind that $\mathrm{Sp}(2,\mathbb{R})$ coincides with $\mathrm{SL}(2,\mathbb{R})$, as noted below \Eq{eq:Omega:2D:def}. Its Lie algebra is denoted $\frak{sl}(2,\mathbb{R})$.} For the Bloch-Messiah decomposition, this gives
\bea
	\boldsymbol{M}(\gamma,d,\gamma')&=&e^{2i\gamma\boldsymbol{U}^\dag\boldsymbol{\mathcal{K}}_z\boldsymbol{U}}e^{-2id\boldsymbol{U}^\dag\boldsymbol{\mathcal{K}}_x\boldsymbol{U}}e^{2i\gamma'\boldsymbol{U}^\dag\boldsymbol{\mathcal{K}}_z\boldsymbol{U}},
	\label{eq:M:calK}
\eea
where we use the fact that for any matrix $\boldsymbol{X}$ one has $\boldsymbol{U}^\dag e^{\boldsymbol{X}}\boldsymbol{U}=e^{\boldsymbol{U}^\dag \boldsymbol{X}\boldsymbol{U}}$ since $\boldsymbol{U}$ is unitary. For the polar decomposition, the squeezing and the rotation matrices read 
\begin{equation}\begin{aligned}
\label{eq:polar:calK}
	\boldsymbol{{S}}(r,\varphi)&=\exp\left(\xi\boldsymbol{U}^\dag\boldsymbol{\mathcal{K}}_+\boldsymbol{U}-\xi^*\boldsymbol{U}^\dag\boldsymbol{\mathcal{K}}_-\boldsymbol{U}\right), \\
	\boldsymbol{R}(\theta)&=\exp\left(2i\theta\boldsymbol{U}^\dag\boldsymbol{\mathcal{K}}_z\boldsymbol{U}\right).
\end{aligned}\end{equation}
The generators of the symplectic group are therefore given by
\begin{equation}\begin{aligned}
	\boldsymbol{K_x}&=\boldsymbol{U}^\dag\boldsymbol{\mathcal{K}}_x\boldsymbol{U}=\frac{i}{2}\boldsymbol{J}_z, \\
	\boldsymbol{K_y}&=\boldsymbol{U}^\dag\boldsymbol{\mathcal{K}}_y\boldsymbol{U}=\frac{-i}{2}\boldsymbol{J}_x, \\
	\boldsymbol{K_z}&=\boldsymbol{U}^\dag\boldsymbol{\mathcal{K}}_z\boldsymbol{U}=\frac{-1}{2}\boldsymbol{J}_y.
\end{aligned}\end{equation}
From the unitarity of $\boldsymbol{U}$, it is easy to verify that the generators of $\mathrm{Sp}(2,\mathbb{R})$ have the same commutation relations~(\ref{eq:Lie:commutation}) as the ones of $\mathrm{SU}(1,1)$.

Finally, since $\boldsymbol{K_x}$, $ \boldsymbol{K_y}$ and $\boldsymbol{K_z}$ are pure imaginary from the previous relations, it is clear that the matrices belonging to $\mathrm{Sp}(2,\mathbb{R})$ are real because of \Eq{eq:M:calK}, as they should. To make this more obvious, it is sometimes convenient to introduce the generators 
\begin{equation}\begin{aligned}
\boldsymbol{K}_1&=2 i \boldsymbol{K}_y ,\\
\boldsymbol{K}_2&=-2 i \boldsymbol{K}_z ,\\
\boldsymbol{K}_3&=-2 i \boldsymbol{K}_x ,
\label{eq:K123}
\end{aligned}\end{equation}
which are real. Matrices belonging to $\mathrm{SU}(1,1)$ can then be expanded according to $\boldsymbol{\mathcal{M}}=t\boldsymbol{I}+x\boldsymbol{K}_1+y\boldsymbol{K}_2+z\boldsymbol{K}_3$ with $(t,x,y,z)\in\mathbb{R}^4$ satisfying $t^2-x^2+y^2-z^2=1$.\footnote{This is nothing but the decomposition~(\ref{eq:SU11:exp:Pauli}) with the identification $t=T$, $x=-Y$, $y=-Z$ and $z=X$.} The commutation relations of these Lie generators are given by $\left[\boldsymbol{K}_1,\boldsymbol{K}_2\right]=-2\boldsymbol{K}_3$, $\left[\boldsymbol{K}_1,\boldsymbol{K}_3\right]=-2\boldsymbol{K}_2$ and $\left[\boldsymbol{K}_2,\boldsymbol{K}_3\right]=-2\boldsymbol{K}_1$ and they satisfy $\left(\boldsymbol{\Omega K}_a\right)^\mathrm{T}=\boldsymbol{\Omega K}_a$ for $a=1$, $2$, $3$. With these matrices, the Bloch-Messiah decomposition~(\ref{eq:M:calK}) of matrices in $\mathrm{Sp}(2,\mathbb{R})$ reads $\boldsymbol{M}(\gamma,d,\gamma')=e^{-\gamma\boldsymbol{K}_2}e^{d\boldsymbol{K}_3}e^{-\gamma'\boldsymbol{K}_2}$, and the polar decomposition~(\ref{eq:polar:calK}) is obtained by $\boldsymbol{{S}}(r,\varphi)=\exp\left[r \sin(2\varphi)\boldsymbol{K}_1+r \cos(2\varphi)\boldsymbol{K}_3\right]=\exp\left[r\left(e^{2i\varphi}\boldsymbol{K}_++e^{-2i\varphi}\boldsymbol{K}_-\right)\right] $ and $\boldsymbol{R}(\theta)=\exp(-\theta\boldsymbol{K}_2)$, where $\boldsymbol{K}_\pm=\left(\boldsymbol{K}_3\mp i\boldsymbol{K}_1\right)/2$.
\section{Classical scalar field}
\label{sec:classical}
\subsection{Dynamics}
\label{sec:classical:dynamics}
We consider the case of a real-valued, test scalar field, $\phi(\vec{x})$, and we denote its canonically conjugate momentum $\pi(\vec{x})$. By {\it test} field, we mean that the scalar field does not backreact on the space-time dynamics, which we assume is homogeneous and isotropic (having in mind applications to cosmology, that will come in \Sec{sec:inflation}). Let us arrange the set of canonical variables into the vector $\boldsymbol{z}^\mathrm{T}(\vec{x})=\left(\phi(\vec{x}),\pi(\vec{x})\right)$. We consider a free field for which the Hamiltonian is a local and quadratic form
\bea
	H=\frac{1}{2}\ds\int\dd^3x\boldsymbol{z}^\mathrm{T}(\vec{x})\boldsymbol{H}(\tau)\boldsymbol{z}(\vec{x}),
\eea
with the symmetric Hamiltonian kernel, $\boldsymbol{H}$, admitting the general form
\bea
\label{eq:Hamiltonian:kernel}
	\boldsymbol{H}=\left(\begin{array}{ccc}
		f_\Delta(\tau)\delta^{ij}\overleftarrow{\partial_i}~\overrightarrow{\partial_j}+f_m(\tau)& {}^{} & f_\times(\tau) \\
		f_\times(\tau)&  & f_\pi(\tau)
	\end{array}\right) ,
\eea
where the notation $\overleftarrow{\partial_i}~\overrightarrow{\partial_j}$ is to be understood as $F(\vec{x})\overleftarrow{\partial_i}~\overrightarrow{\partial_j}G(\vec{x})\equiv \left(\partial_i F\right)\left(\partial_j G\right)$. We note that apart from the gradients, there is no space dependence thanks to homogeneity and isotropy of the background. The equations of motion are obtained using the Poisson bracket, \ie
\bea
	\left\{F,G\right\}=\ds\int\dd^3x\left[\frac{\delta F}{\delta \phi(\vec{x})}\frac{\delta G}{\delta\pi(\vec{x})}-\frac{\delta F}{\delta\pi(\vec{x})}\frac{\delta G}{\delta \phi(\vec{x})}\right],
\eea
which leads to
\bea
\label{eq:commutator:z}
	\left\{\boldsymbol{z}(\vec{x}),\boldsymbol{z}^\mathrm{T}(\vec{y})\right\}=\boldsymbol{\Omega}\delta^3\left(\vec{x}-\vec{y}\right).
\eea
The notation in this equation has to understood as follows. We introduce the greek indices $\mu,~\nu,~\cdots$ spanning the 2-dimensional phase space at a given point, \ie $\boldsymbol{z}_\mu(\vec{x})=\phi(\vec{x})$ or $\pi(\vec{x})$ for $\mu=1$ and $\mu=2$ respectively. \Eq{eq:commutator:z} has to be read as $\left\{\boldsymbol{z}_\mu(\vec{x}),\boldsymbol{z}_\nu(\vec{y})\right\}=\boldsymbol{\Omega}_{\mu\nu}\delta^3(\vec{x}-\vec{y})$.
The dynamics of any function of the phase-space field variables is finally $\dot{F}(\phi,\pi)=\left\{F(\phi,\pi),H\right\}$ where an overdot means differentiation with respect to the time variable, denoted $t$ hereafter.\footnote{Note that, in the context of cosmology, $t$ can be any time variable: cosmic time, conformal time, {\etc}. Changing the time variable results in rescaling the $f_i$ functions entering the Hamiltonian kernel~(\ref{eq:Hamiltonian:kernel}).} 

The above relations can equivalently be written in Fourier space. We introduce the Fourier transform of the canonical variables
\bea
\label{eq:def:zk}
	\boldsymbol{z}_{\vec{k}}:=\left(\begin{array}{c}
		\phi_{\vec{k}} \\
		\pi_{\vec{k}}
	\end{array}\right)=\ds\int\frac{\dd^3x}{(2\pi)^{3/2}}\boldsymbol{z}(\vec{x})e^{-i\vec{k}\cdot\vec{x}}.
\eea
The inverse transform is
\bea
\label{eq:z(x):Fourrier}
	\boldsymbol{z}(\vec{x})=\ds\int\frac{\dd^3k}{(2\pi)^{3/2}}\boldsymbol{z}_{\vec{k}}e^{i\vec{k}\cdot\vec{x}},
\eea
with the normalisation $\ds\int\dd^3{x}e^{i(\vec{k}-\vec{q})\cdot\vec{x}}=(2\pi)^3\delta^3(\vec{k}-\vec{q})$ and $\ds\int\dd^3{k}e^{i\vec{k}\cdot(\vec{x}-\vec{y})}=(2\pi)^3\delta^3(\vec{x}-\vec{y})$. We also introduce the vector
\bea
\label{eq:z:def}
	\boldsymbol{z}^*_{\vec{k}}:=\left(\begin{array}{c}
		\phi^*_{\vec{k}} \\
		\pi^*_{\vec{k}}
	\end{array}\right),
\eea
which should not be mislead with the conjugate transpose, \ie $\boldsymbol{z}^\dag_{\vec{k}}=\left(\phi^*_{\vec{k}},\pi^*_{\vec{k}}\right)=\ds\int\frac{\dd^3x}{(2\pi)^{3/2}}\boldsymbol{z}^\mathrm{T}(\vec{x})e^{i\vec{k}\cdot\vec{x}}$. 

The field is real-valued, leading to the constraint in its Fourier representation $\boldsymbol{z}^*_{\vec{k}}=\boldsymbol{z}_{-\vec{k}}$. In order to avoid double counting of the degrees of freedom, the integration over Fourier modes is thus split into two parts, $\mathbb{R}^{3+} \equiv \mathbb{R}^2\times \mathbb{R}^+$ and $\mathbb{R}^{3-} \equiv \mathbb{R}^2\times \mathbb{R}^-$, and the integral over $\mathbb{R}^{3-}$ is written as an integral over $\mathbb{R}^{3+}$ using the above constraint, leading to
\bea
\label{eq:Hamiltonian:Fourrier}
	H=\int_{\mathbb{R}^{3+}}\dd^3k\boldsymbol{z}_{\vec{k}}^\dag\boldsymbol{H}_k\boldsymbol{z}_{\vec{k}},
\eea
where $\boldsymbol{H}_k$ is a time-dependent matrix given by the background evolution and the Fourier transform of the Laplacian (hence the $k$ dependence). We note that this matrix depends only on the norm of $\vec{k}$ thanks to the isotropy of space-time. It is easily obtained by replacing $\delta^{ij}\overleftarrow{\partial_i}~\overrightarrow{\partial_j}$ by $k^2$ in \Eq{eq:Hamiltonian:kernel}.\footnote{Note that with the splitting, the gradients apply to $\delta^{ij}(\partial_ie^{i\vec{k}\cdot\vec{x}})(\partial_je^{-i\vec{q}\cdot\vec{x}})$.} The Poisson brackets of the Fourier-transformed field variables are readily given by 
\bea
\label{eq:PoissonBracket:Fourrier}
\left\{\boldsymbol{z}_{\vec{k}},\boldsymbol{z}^\dag_{\vec{q}}\right\}=\boldsymbol{\Omega}\delta^3(\vec{k}-\vec{q}),
\eea
which, as before, is a shorthand notation for $\left\{\boldsymbol{z}_{\vec{k},\mu},\boldsymbol{z}^*_{\vec{q},\nu}\right\}=\boldsymbol{\Omega}_{\mu\nu}\delta^3(\vec{k}-\vec{q})$. We can thus define it directly through functional derivatives with respect to the Fourier-transformed variables
\bea
\label{eq:Poisson:Fourier}
	\left\{F,G\right\}=\ds\int\dd^3k\left(\frac{\delta F}{\delta \phi_{\vec{k}}}\frac{\delta G}{\delta\pi^*_{\vec{k}}}-\frac{\delta F}{\delta\pi_{\vec{k}}}\frac{\delta G}{\delta \phi^*_{\vec{k}}}\right).
\eea
In Fourier space, the equations of motion are then simply given by 
\bea
\label{eq:eom:Fourrier}
\dot{\boldsymbol{z}}_{\vec{k}}=\left(\boldsymbol{\Omega}\boldsymbol{H}_k\right)\boldsymbol{z}_{\vec{k}}.
\eea
Solutions of this can be obtained using a Green's matrix formalism, see \eg Appendix A of \Refa{Grain:2017dqa}. For a given set of initial conditions, $\boldsymbol{z}^{(0)}_{\vec{k}}:=\boldsymbol{z}_{\vec{k}}(t_0)$, the solution is $\boldsymbol{z}_{\vec{k}}(t)=\boldsymbol{G}_k(t,t_0)\boldsymbol{z}^{(0)}_{\vec{k}}$. The Green's matrix $\boldsymbol{G}_k$ (parametrised by the wavenumber only, and not the full wavevector, since the Hamiltonian kernel is isotropic) is an element of $\mathrm{Sp}(2,\mathbb{R})$ and is a solution of $\dot{\boldsymbol{G}}_k=\left(\boldsymbol{\Omega}\boldsymbol{H}_k\right)\boldsymbol{G}_k+\boldsymbol{I}\delta(t-t_0)$.\footnote{The presence of the term $\boldsymbol{I}\delta(t-t_0)$ in the differential equation for $\boldsymbol{G}_k$ ensures that the initial condition $\boldsymbol{G}_k(t_0,t_0)=\boldsymbol{I}$ is satisfied.}
\subsection{Linear canonical transformation}
\label{ssec:lincantrans}
We now consider a linear, real-valued and isotropic canonical transformation that operates on the Fourier transform variables,
\bea
\label{eq:canonical:transform:Fourrier}
\widetilde{\boldsymbol{z}}_{\vec{k}}=\boldsymbol{M}_{{k}}(t)\boldsymbol{z}_{\vec{k}} .
\eea
In this expression, $\boldsymbol{M}_k$ is a time-dependent symplectic matrix, parametrised by the wavenumber only (isotropy hypothesis). Thanks to \Eq{eq:symp}, the Poisson bracket~(\ref{eq:PoissonBracket:Fourrier}) is preserved through canonical transformations, {\ie} $\left\{\boldsymbol{\widetilde{z}}_{\vec{k}},\boldsymbol{\widetilde{z}}^\dag_{\vec{q}}\right\}=\left\{\boldsymbol{{z}}_{\vec{k}},\boldsymbol{{z}}^\dag_{\vec{q}}\right\}=\boldsymbol{\Omega}\delta^3\left(\vec{k}-\vec{q}\right)$. This also ensures that the Poisson bracket between two arbitrary phase-space functions, calculated with the $\boldsymbol{z}_{\vec{k}}$ variables as in \Eq{eq:Poisson:Fourier}, coincides wit the one calculated with the $\widetilde{\boldsymbol{z}}_{\vec{k}}$ variables (\ie replacing $\phi_{\vec{k}}$ and $\pi_{\vec{k}}$ by $\tilde{\phi}_{\vec{k}}$ and $\tilde{\pi}_{\vec{k}}$ in \Eq{eq:Poisson:Fourier}).

Starting from \Eq{eq:eom:Fourrier}, making use of the fact that $\boldsymbol{\Omega}^2=-\boldsymbol{I}$ as can be checked from \Eq{eq:Omega:def}, and since \Eq{eq:symp} gives rise to $\boldsymbol{M}^{-1}=-\boldsymbol{\Omega}\boldsymbol{M}^\mathrm{T}\boldsymbol{\Omega}$, one can verify that the equations of motion for the new set of canonical variables are also given by Hamilton equations, with the new Hamiltonian reading
\bea
\label{eq:Hamiltonian:canonical:transform}
	\widetilde{H}&=&\int_{\mathbb{R}^{3+}}\dd^3k\widetilde{\boldsymbol{z}}_{\vec{k}}^\dag\underbrace{\left[\left(\boldsymbol{M}_k^{-1}\right)^\mathrm{T}\boldsymbol{H}_k\boldsymbol{M}_k^{-1}+\left(\boldsymbol{M}_k^{-1}\right)^\mathrm{T}\boldsymbol{\Omega}\frac{\dd\boldsymbol{M}^{-1}_k}{\dd t}\right]}_{\boldsymbol{\widetilde{H}}_k}\widetilde{\boldsymbol{z}}_{\vec{k}}, 
\eea
which defines the Hamiltonian kernel $\boldsymbol{\widetilde{H}}_k$ for the $\boldsymbol{\widetilde{z}}$ variables. The inverse relation is easily obtained as $\boldsymbol{H}_k=\boldsymbol{M}_k^\mathrm{T}\boldsymbol{\widetilde{H}}_k\boldsymbol{M}_k+\boldsymbol{M}_k^\mathrm{T}\boldsymbol{\Omega}\frac{\dd\boldsymbol{M}_k}{\dd t}$. The new Hamiltonian is obviously quadratic if the initial one is. 

One can also show that the new Hamiltonian kernel remains symmetric, $(\boldsymbol{\widetilde{H}}_k)^{\mathrm{T}}=\boldsymbol{\widetilde{H}}_k$, if the initial one is. In fact, in \Sec{sec:intro}, the symplectic structure was introduced precisely to ensure that this is the case. The first term defining $\boldsymbol{\widetilde{H}}_k$ in \Eq{eq:Hamiltonian:canonical:transform} is indeed obviously symmetric if $\boldsymbol{H}_k$ is. For the second term, since  $\boldsymbol{\Omega}^\mathrm{T}=-\boldsymbol{\Omega}$, one first notes that $[(\boldsymbol{M}_k^{-1})^\mathrm{T}\boldsymbol{\Omega}\frac{\dd}{\dd t}\boldsymbol{M}^{-1}_k]^\mathrm{T} = - \frac{\dd}{\dd t}(\boldsymbol{M}_k^{-1})^\mathrm{T}\boldsymbol{\Omega}\boldsymbol{M}^{-1}_k$. Second, since $\boldsymbol{M}_k^{-1}$ is a symplectic matrix, it satisfies \Eq{eq:symp}. Differentiating this expression with respect to time yields $(\boldsymbol{M}_k^{-1})^\mathrm{T}\boldsymbol{\Omega}\frac{\dd}{\dd t}\boldsymbol{M}^{-1}_k +\frac{\dd}{\dd t}(\boldsymbol{M}_k^{-1})^\mathrm{T}\boldsymbol{\Omega}\boldsymbol{M}^{-1}_k =0$, hence the second term is also symmetric. 

Finally, Hamilton's equations for the new set of canonical variables can also be solved using the Green's matrix formalism, {\ie} $\boldsymbol{\widetilde{z}}_{\vec{k}}(t)=\boldsymbol{\widetilde{G}}_k(t,t_0)\boldsymbol{\widetilde{z}}^{(0)}_{\vec{k}}$, where the new Green's matrix is given by $\boldsymbol{\widetilde{G}}_k(t,t_0)=\boldsymbol{M}_k(t)\boldsymbol{G}_k(t,t_0)\boldsymbol{M}^{-1}_k(t_0)$, as already noted in \Eq{eq:Green:canonical:transformation}.
\subsection{Helicity basis}
As explained in \Sec{sec:HelicityBasis:SU11}, one can introduce the set of complex variables
\bea
\label{eq:a(z)}
	\boldsymbol{a}_{\vec{k}}=\boldsymbol{U}\boldsymbol{D}_k\boldsymbol{z}_{\vec{k}},
\eea
where $\boldsymbol{U}$ was defined in \Eq{eq:U:def}, and which we refer to as working in the helicity basis. Compared to \Sec{sec:HelicityBasis:SU11}, a new degree of freedom has been introduced in the transformation, $\boldsymbol{D}_k$, which can be any $k$-dependent (and potentially time-dependent) symplectic matrix.  The $k$-dependence is necessary since the configuration and momentum variables do not share the same dimension. It can therefore be convenient to write 
\bea
\boldsymbol{D}_k = \boldsymbol{M}_k 
\left(\begin{array}{cc}
		\sqrt{k} & 0 \\
		0 & \frac{1}{\sqrt{k}}
	\end{array}\right),
\eea
where $\boldsymbol{M}_k $ is a \emph{dimensionless} symplectic matrix, and one can check that $\mathrm{diag}(\sqrt{k},1/\sqrt{k})$ is of course symplectic (it corresponds to a ``scale transformation'' as introduced in  \Sec{sec:intro}). 

The entries of the vector $\boldsymbol{a}_{\vec{k}}$ can be written as
\bea
\label{eq:a:components}
	\boldsymbol{a}_{\vec{k}}:=\left(\begin{array}{c}
		a_{\vec{k}} \\
		a^*_{-\vec{k}}
	\end{array}\right),
\eea
where it is worth noticing that the second entry is for $(-\vec{k})$. 
The reason why $\boldsymbol{a}_{\vec{k}}$ can be written as above can be understood as follows. Let us first notice that \Eq{eq:a(z)} can be inverted as $\boldsymbol{z}_{\vec{k}} =\boldsymbol{D}_k^{-1}\boldsymbol{U}^\dagger \boldsymbol{a}_{\vec{k}}$, hence $\boldsymbol{z}_{-\vec{k}} =\boldsymbol{D}_k^{-1}\boldsymbol{U}^\dagger \boldsymbol{a}_{-\vec{k}}$ and $\boldsymbol{z}_{\vec{k}}^* =\boldsymbol{D}_k^{-1}\boldsymbol{U}^{\mathrm{T}} \boldsymbol{a}_{\vec{k}}^*$. Identifying $\boldsymbol{z}_{-\vec{k}}$ and $\boldsymbol{z}_{\vec{k}}^*$ (since, as explained above, this is required by the condition that $\boldsymbol{z}(\boldsymbol{x})$ is a real field), one obtains $\boldsymbol{a}_{-\vec{k}} = \boldsymbol{U} \boldsymbol{U}^\mathrm{T} \boldsymbol{a}_{\vec{k}}^*$, \ie
\bea
\boldsymbol{a}_{-\vec{k}} =
\left(\begin{array}{cc}
		0&1 \\
		1&0
	\end{array}\right)
\boldsymbol{a}_{\vec{k}}^*\, .
\eea
If one introduces the two components of $\boldsymbol{a}_{\vec{k}}$ as $\alpha_{\vec{k}}$ and $\beta_{\vec{k}}$, the above relation implies that $\alpha_{-\vec{k}}=\beta_{\vec{k}}^*$ and $\beta_{-\vec{k}}=\alpha_{\vec{k}}^*$, hence the notation of \Eq{eq:a:components}.

\subsubsection{Canonical transformations}
\label{sssec:canonicalcomplex}
\begin{figure}[t]
\begin{center}
\includegraphics[width=0.4\textwidth]{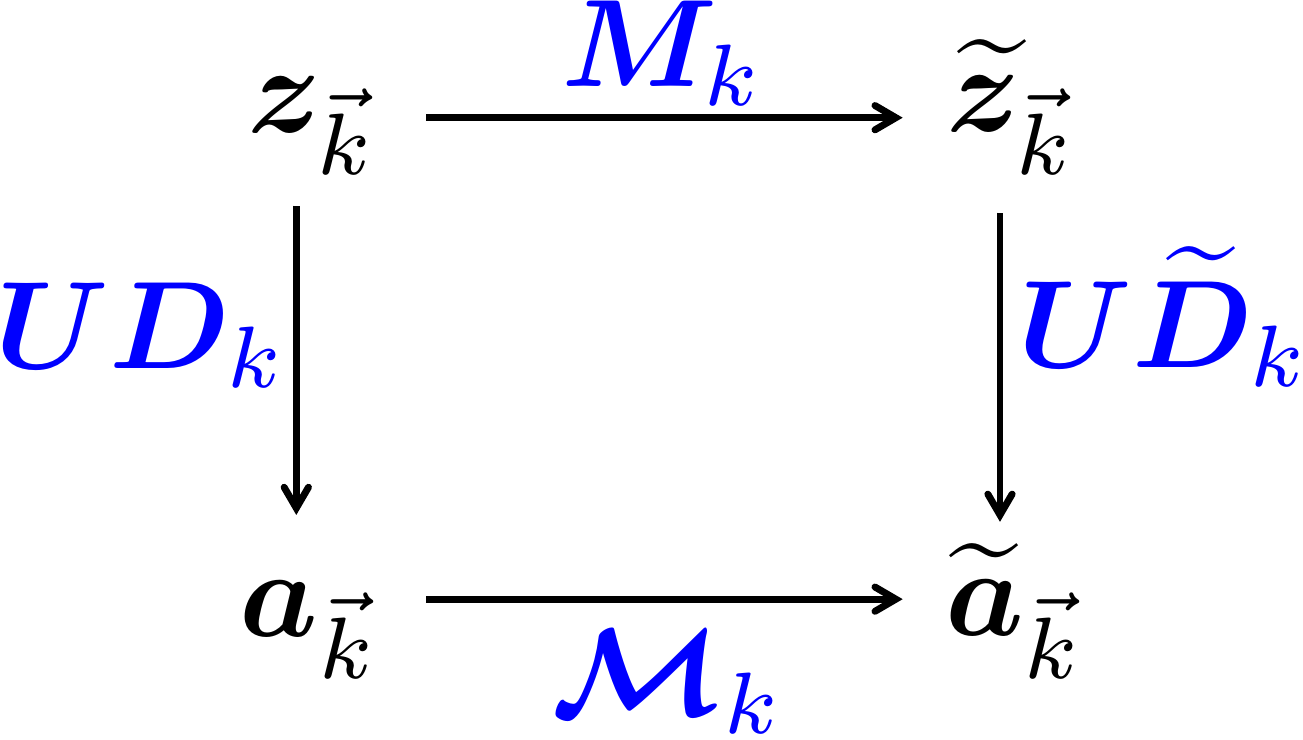}
\includegraphics[width=0.15\textwidth]{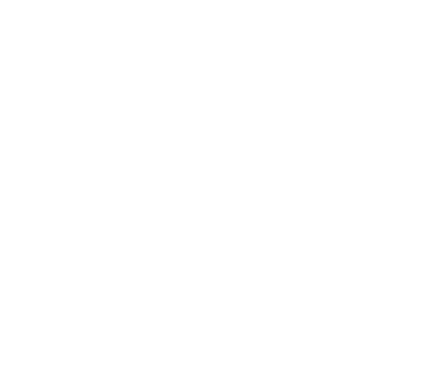}
\includegraphics[width=0.37\textwidth]{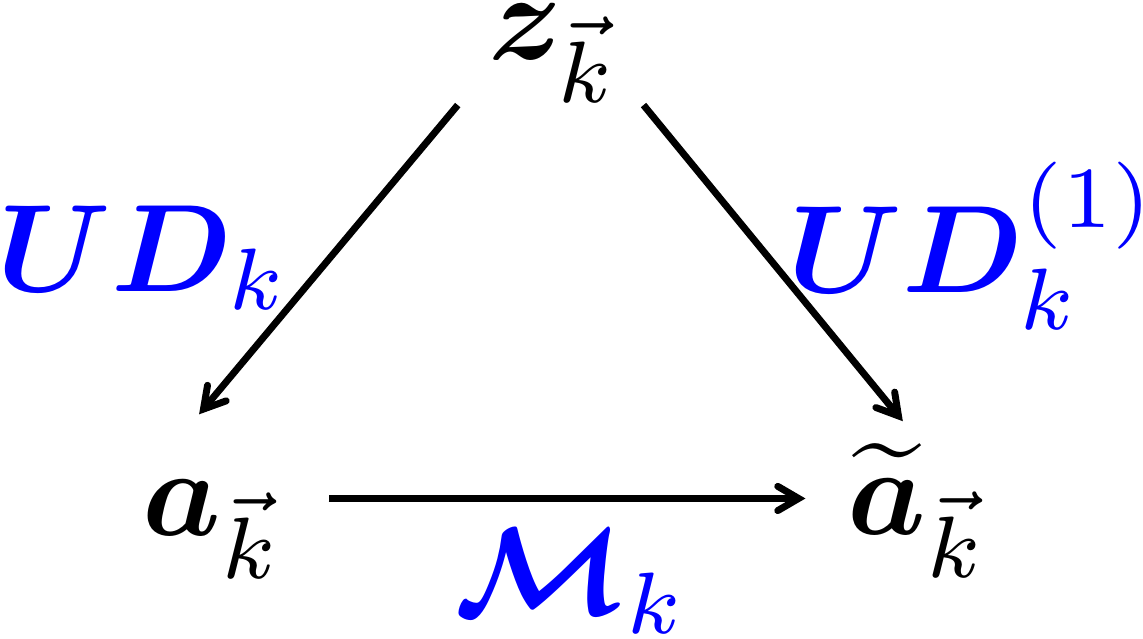}
\caption{Canonical transformations in the field and helicity variables. The matrices $\boldsymbol{M}_k$, $\boldsymbol{D}_k$, $\boldsymbol{\widetilde{D}}_k$ and $\boldsymbol{D}_k^{(1)}$ belong to $\mathrm{Sp}(2,\mathbb{R})$, while $\boldsymbol{\mathcal{M}}_k$ belongs to $\mathrm{SU}(1,1)$. In the case depicted in the left diagram, it is given by \Eq{eq:calMk:def}, while it is given by \Eq{eq:calM:D:D1:def} in the case represented in the right diagram.}  
\label{fig:diagram}
\end{center}
\end{figure}
Let us consider a canonical transformation of the form~(\ref{eq:canonical:transform:Fourrier}). A new set of helicity variables can be defined from $\boldsymbol{\widetilde{z}}_{\vec{k}}$ through $\boldsymbol{\widetilde{a}}_{\vec{k}}=\boldsymbol{U}\boldsymbol{\widetilde{D}}_k\boldsymbol{\widetilde{z}}_{\vec{k}}$, where $\boldsymbol{\widetilde{D}}_k$ is a priori different from $\boldsymbol{D}_k$. The situation is depicted in \Fig{fig:diagram}. It is easy to check that
\bea
	\boldsymbol{\widetilde{a}}_{\vec{k}}=\boldsymbol{\mathcal{M}}_k\boldsymbol{a}_{\vec{k}},
\eea
with
\bea
\label{eq:calMk:def}
	\boldsymbol{\mathcal{M}}_k=\boldsymbol{U}\left(\boldsymbol{\widetilde{D}}_k\boldsymbol{M}_k\boldsymbol{D}^{-1}_k\right)\boldsymbol{U}^\dag\, ,
\eea
which generalises \Eq{eq:calM:def}.
Since $\boldsymbol{\widetilde{D}}_k$, $\boldsymbol{M}_k$, and $\boldsymbol{D}^{-1}_k$ all belong to $\mathrm{Sp}(2,\mathbb{R})$, it is obvious that $\boldsymbol{\mathcal{M}}_k\in \mathrm{SU}(1,1)$. 

These matrices retain some degeneracy. For instance, suppose that one performs the helicity transform through the matrix $\boldsymbol{D}^{(1)}_k$ instead of $\boldsymbol{D}_k$, {\ie} $\boldsymbol{a}^{(1)}_{\vec{k}}=\boldsymbol{U}\boldsymbol{D}^{(1)}_k\boldsymbol{z}_{\vec{k}}$. It is clear that $\boldsymbol{a}^{(1)}_{\vec{k}}=\boldsymbol{\mathcal{M}}^{(1)}_k\boldsymbol{a}_{\vec{k}}$, where 
\bea
\label{eq:calM:D:D1:def}
\boldsymbol{\mathcal{M}}^{(1)}_k=\boldsymbol{U}\boldsymbol{D}^{(1)}_k\boldsymbol{D}^{-1}_k\boldsymbol{U}^\dag
\eea
is also an element of $\mathrm{SU}(1,1)$. From the viewpoint of the helicity variables, a different choice of the matrix $\boldsymbol{D}_k$ is thus equivalent to a canonical transformation, identifying $\boldsymbol{D}^{(1)}_k$ with $\boldsymbol{\widetilde{D}}_k\boldsymbol{M}_k$. Conversely, any redefinition of the complex variables can be interpreted as working with a {\it canonically-transformed} set of fields variables.

In the following, by ``canonical transformation of the helicity variables'', we will refer to any transformation generated by an element of $\mathrm{SU}(1,1)$, regardless of whether this is due to a canonical transformation ($\boldsymbol{M}_k$) of the field variables, or to different definitions ($\boldsymbol{{D}}_k$) of the helicity variables -- or both.

\subsubsection{Dynamics}
\label{sssec:complexdynamics}
The dynamics for the helicity variables is generated by the quadratic Hamiltonian
\bea
	\mathcal{H}=\ds\int_{\mathbb{R}^{3+}}\dd^3k\boldsymbol{a}_{\vec{k}}^\dag\boldsymbol{\mathcal{H}}_k\boldsymbol{a}_{\vec{k}},
\eea
where, making use of \Eq{eq:Hamiltonian:canonical:transform}, the kernel $\boldsymbol{\mathcal{H}}_k$ reads
\bea
\label{eq:Hamiltonian:helicity}
	\boldsymbol{\mathcal{H}}_k=\boldsymbol{U}\left[\left(\boldsymbol{D}_k^{-1}\right)^\mathrm{T}\boldsymbol{H}_k\boldsymbol{D}_k^{-1}+\left(\boldsymbol{D}_k^{-1}\right)^\mathrm{T}\boldsymbol{\Omega}\frac{\dd\left(\boldsymbol{D}_k^{-1}\right)}{\dd t}\right]\boldsymbol{U}^\dag.
\eea
The Poisson bracket for the helicity variables is\footnote{Making use of \Eq{eq:a(z)} one has $\left\{\boldsymbol{a}_{\vec{k}},\boldsymbol{a}^\dag_{\vec{q}}\right\}= \boldsymbol{U}\boldsymbol{D}_k\   \left\{\boldsymbol{z}_{\vec{k}},\boldsymbol{z}^\dag_{\vec{q}}\right\} \boldsymbol{D}_k^\dagger \boldsymbol{U}^\dagger = \boldsymbol{U}\boldsymbol{D}_k\  \boldsymbol{\Omega} \boldsymbol{D}_k^\dagger \boldsymbol{U}^\dagger \delta^3\left(\vec{k}-\vec{q}\right) $ where \Eq{eq:PoissonBracket:Fourrier} has been inserted in the second equality. Since $\boldsymbol{D}_k\in\mathrm{Sp}(2,\mathbb{R})$, it is a real matrix hence $\boldsymbol{D}_k^\dagger=\boldsymbol{D}_k^{\mathrm{T}}$, and it satisfies \Eq{eq:symp}. This gives rise to $\left\{\boldsymbol{a}_{\vec{k}},\boldsymbol{a}^\dag_{\vec{q}}\right\}= \boldsymbol{U} \boldsymbol{\Omega}  \boldsymbol{U}^\dagger \delta^3\left(\vec{k}-\vec{q}\right)  = \boldsymbol{\mathcal{J}}  \delta^3\left(\vec{k}-\vec{q}\right) $ where \Eq{eq:J:def} has been used. }  
\bea
\label{eq:PoissonBracket:helicity}
\left\{\boldsymbol{a}_{\vec{k}},\boldsymbol{a}^\dag_{\vec{q}}\right\}=\boldsymbol{\mathcal{J}}\delta^3\left(\vec{k}-\vec{q}\right),
\eea
and any function of the helicity phase-space variables evolves according to $\dot{F}\left(\boldsymbol{a}_{\vec{k}}\right)=\left\{F\left(\boldsymbol{a}_{\vec{k}}\right),\mathcal{H}\right\}$. As for the field variables, \Eq{eq:PoissonBracket:helicity} has to be understood as $\left\{\boldsymbol{a}_{\vec{k},\mu},\boldsymbol{a}^*_{\vec{q},\nu}\right\}=\boldsymbol{\mathcal{J}}_{\mu\nu}\delta^3(\vec{k}-\vec{q})$ with the greek indices now running over $a_{\vec{k}}$ for $\mu=1$ and $a^*_{-\vec{k}}$ for $\mu=2$. The Poisson bracket of the helicity variables is also preserved through canonical transformations, {\ie} $\left\{\boldsymbol{\widetilde{a}}_{\vec{k}},\boldsymbol{\widetilde{a}}^\dag_{\vec{q}}\right\}=\left\{\boldsymbol{a}_{\vec{k}},\boldsymbol{a}^\dag_{\vec{q}}\right\}=\boldsymbol{\mathcal{J}}\delta^3\left(\vec{k}-\vec{q}\right)$, as a result of \Eq{eq:eqdef:SU11}.

Hamilton's equations providing the evolution of the helicity variables read 
\bea
\label{eq:Hamilton:a}
\dot{\boldsymbol{a}}_{\vec{k}}=\boldsymbol{\mathcal{JH}}_k\boldsymbol{a}_{\vec{k}}\, .
\eea
Thanks to the remark made below \Eq{eq:Hamiltonian:canonical:transform}, since $\boldsymbol{D}_k$ is a symplectic matrix, the quantity $\left(\boldsymbol{D}_k^{-1}\right)^\mathrm{T}\boldsymbol{H}_k\boldsymbol{D}_k^{-1}+\left(\boldsymbol{D}_k^{-1}\right)^\mathrm{T}\boldsymbol{\Omega}\frac{\dd\left(\boldsymbol{D}_k^{-1}\right)}{\dd t}$ is symmetric and real. From \Eq{eq:Hamiltonian:helicity} it is then easy to see that $\boldsymbol{\mathcal{JH}}_k$ has the following structure
\bea
\label{eq:Hamiltonian:helicity:FRTheta}
	\boldsymbol{\mathcal{JH}}_k&=&(-i)\left(\begin{array}{cc}
		F_k(t) & R_k(t)e^{i\Theta_k(t)} \\
		-R_k(t)e^{-i\Theta_k(t)} & -F_k(t)
	\end{array}\right) \nonumber \\
	&=&(-2i)\left[R_k(t)\sin\Theta_k(t)\boldsymbol{\mathcal{K}}_x+R_k(t)\cos\theta_k(t)\boldsymbol{\mathcal{K}}_y+F_k(t)\boldsymbol{\mathcal{K}}_z\right],
\eea
where $F_k$, $R_k$ and $\Theta_k$ are real functions given by combinations of the (Fourier transformed) $f_i$'s functions entering $\boldsymbol{H}_k$ in \Eq{eq:Hamiltonian:kernel} and the coefficients of the matrix $\boldsymbol{D}_k$.  This means that $\boldsymbol{\mathcal{JH}}_k=i\boldsymbol{m}(t)$ with $\boldsymbol{m}(t)$ a vector in the $\mathfrak{su}(1,1)$ Lie algebra, following the notations of \Sec{sec:LieAlgebra}. A formal solution to Hamilton's equations~(\ref{eq:Hamilton:a}) reads $\boldsymbol{a}_{\vec{k}}(t)=\exp\left[\boldsymbol{A}(t,t_0)\right]\boldsymbol{a}_{\vec{k}}(t_0)$ where the matrix $\boldsymbol{A}(t,t_0)$ is obtained from the Magnus expansion \cite{Magnus:1954zz}, which can be expressed using a continuous version of the Baker-Campbell-Haussdorff formula. Since $i \boldsymbol{\mathcal{JH}}_k\in\mathfrak{su}(1,1)$, solutions using the Magnus expansion is in the exponential map of $\mathfrak{su}(1,1)$, hence an element of the group $\mathrm{SU}(1,1)$ (see \Refa{BLANES2009151} for a review). 

We thus search for a Green's matrix solution, $\boldsymbol{\mathcal{G}}_k(t,t_0)$, belonging to $\mathrm{SU}(1,1)$. Following the notation of \Eq{eq:su11mat}, this matrix is denoted
\bea
	\boldsymbol{\mathcal{G}}_k(t,t_0)=\left(\begin{array}{cc}
		\alpha_k(t) & \beta_k(t) \\
		\beta^*_k(t) & \alpha^*_k(t)
	\end{array}\right),
\label{eq:G:Bogoliubov}
\eea
where the coefficients $\alpha_k(t)$ and $\beta_k(t)$ satisfy $\vert \alpha_k(t) \vert^2 - \vert \beta_k(t) \vert^2=1 $ and are often called the ``Bogolyubov coefficients'' in the context of quantum field theory in curved spaces, see \Sec{Sec:Dynamics:Creation:Annihilation}. \Eq{eq:Hamilton:a} give rise to
\bea
	i\dot{\alpha}_k(t)&=&F_k(t)\alpha_k+R_k(t)e^{i\Theta_k(t)}\beta^*_k, \label{eq:alpha}\\
	i\dot{\beta}_k(t)&=&F_k(t)\beta_k+R_k(t)e^{i\Theta_k(t)}\alpha^*_k, \label{eq:beta}
\eea
with initial conditions $\alpha_k(t_0)=1$ and $\beta_k(t_0)=0$, such that $\boldsymbol{\mathcal{G}}_k(t_0,t_0)=\boldsymbol{I}$. 

The Green's matrix does not depend on the initial conditions for the helicity variables, $\boldsymbol{a}_{\vec{k}}(t_0)$. It however depends on the initial time which is chosen, \ie $\boldsymbol{\mathcal{G}}_k(t,t'_0)\neq\boldsymbol{\mathcal{G}}_k(t,t_0)$ for $t'_0\neq t_0$. It is nevertheless possible to generate the entire set of Green's matrix from a single one. Suppose that the dynamics can be coherently formulated from the infinite past, $t_0\to-\infty$, at least conceptually. With initial conditions set in the infinite past, the dynamics is generated by the Green's matrix $\boldsymbol{\mathcal{G}}_k(t,-\infty)$ and we denote its Bogolyubov coefficients by $\alpha_k^{(-\infty)}(t)$ and $\beta^{(-\infty)}_k(t)$. By construction, such a Green's matrix satisfies the group composition law 
\bea
	\boldsymbol{\mathcal{G}}_k(t,-\infty)=\boldsymbol{\mathcal{G}}_k(t,t_0)\boldsymbol{\mathcal{G}}_k(t_0,-\infty),
\eea
with $t_0<t$. Hence for any arbitrary choice of the initial time $t_0$, one can generate the Green's matrix $\boldsymbol{\mathcal{G}}_k(t,t_0)$ from $\boldsymbol{\mathcal{G}}_k(t,-\infty)$ as 
\bea
	\boldsymbol{\mathcal{G}}_k(t,t_0)=\boldsymbol{\mathcal{G}}_k(t,-\infty)\left[\boldsymbol{\mathcal{G}}_k(t_0,-\infty)\right]^{-1}.
\eea
The interpretation is rather simple: evolving from $t_0$ to $t$ is equivalent to first evolving {\it backward} from $t_0$ to the infinite past, and then evolving {\it forward} from the infinity past to $t$ since here we are dealing with time-reversible systems. The Bogolyubov coefficients of $\boldsymbol{\mathcal{G}}_k(t,t_0)$ are functions of the ones entering $\boldsymbol{\mathcal{G}}_k(t,-\infty)$. They read
\bea
	\alpha_k(t)&=&\alpha^{(-\infty)}_k(t)\left[\alpha^{(-\infty)}_k(t_0)\right]^*-\beta^{(-\infty)}_k(t)\left[\beta^{(-\infty)}_k(t_0)\right]^*, \\
	\beta_k(t)&=&-\alpha^{(-\infty)}_k(t)\beta^{(-\infty)}_k(t_0)+\beta^{(-\infty)}_k(t)\alpha^{(-\infty)}_k(t_0).
\eea
\\

Let us finally mention that the evolution could alternatively be recast as dynamical equations on the parameters determining $\mathrm{SU}(1,1)$ elements. Using the Bloch-Messiah decomposition of $\boldsymbol{\mathcal{G}}_k=\boldsymbol{\mathcal{R}}(\gamma_k)\boldsymbol{\mathcal{S}}(d_k,0)\boldsymbol{\mathcal{R}}(\gamma'_k)$ this leads to dynamical equations on $(\gamma_k,d_k,\gamma'_k)$. The initial conditions ensuring $\boldsymbol{\mathcal{G}}_k(t_0,t_0)=\boldsymbol{I}$ are $d_k(t_0)=0$ and $\gamma_k(t_0)+\gamma'_k(t_0)=2n\pi$ with $n\in\mathbb{Z}$. Using instead the left-polar decomposition, \ie $\boldsymbol{\mathcal{G}}_k=\boldsymbol{\mathcal{S}}(r_k,\varphi_k)\boldsymbol{\mathcal{R}}(\theta_k)$, dynamics will be formulated in terms of $(r_k,\varphi_k,\theta_k)$ with initial conditions $r_k(t_0)=0$, $\theta_k(t_0)=2n\pi$ with $n\in\mathbb{Z}$, and $\varphi_k(t_0)$ can admit any value. [Note that in terms of parameters, two constraints are sufficient to target the identity matrix in $\mathrm{SU}(1,1)$.]
\section{Quantum scalar field}
\label{sec:Quantum}
\subsection{Quantisation in a nutshell}
\subsubsection{Hamiltonian and Fock space}
Let us now consider the case of a quantum scalar field. The configuration and momentum variables are promoted to operators acting on a Hilbert space, $\boldsymbol{\widehat{z}}^\dag_{\vec{k}}=\left(\widehat\phi^\dag_{\vec{k}},\widehat{\pi}^\dag_{\vec{k}}\right)$. These operators satisfy the canonical commutation relation 
\bea
\label{eq:commutator:hatz}
\left[\widehat{\boldsymbol{z}}_{\vec{k}},\widehat{\boldsymbol{z}}^{\dag}_{\vec{q}}\right]=i\boldsymbol{\Omega}\delta^3(\vec{k}-\vec{q}),
\eea 
which is the quantum analogue of \Eq{eq:PoissonBracket:Fourrier}. The Hamiltonian operator~(\ref{eq:Hamiltonian:Fourrier}) now reads
\bea
	\widehat{H}=\int_{\mathbb{R}^{3+}}\dd^3k\widehat{\boldsymbol{z}}_{\vec{k}}^\dag\boldsymbol{H}_k\widehat{\boldsymbol{z}}_{\vec{k}},
\eea
and any function of the quantum phase-space field variables evolves according to the Heisenberg equation
\bea
\label{eq:Heisenberg}
\dot{F}\left(\widehat{\boldsymbol{z}}_{\vec{k}}\right) = -i \left[F\left(\widehat{\boldsymbol{z}}_{\vec{k}}\right) ,\widehat{H}\right].
\eea 
In particular, this gives rise to $\dot{\widehat{\boldsymbol{z}}}_{\vec{k}}=\left(\boldsymbol{\Omega H}_k\right)\widehat{\boldsymbol{z}}_{\vec{k}}$, which directly transposes \Eq{eq:eom:Fourrier}.

One can then introduce the creation and annihilation operators through the same transformation as \Eq{eq:a(z)}, namely
\bea
\label{eq:a(z):quantum}
	\boldsymbol{\widehat{a}}_{\vec{k}}:=\left(\begin{array}{c}
		\widehat{a}_{\vec{k}} \\
		\widehat{a}^\dag_{-\vec{k}}
	\end{array}\right)=\boldsymbol{U}\boldsymbol{D}_k\widehat{\boldsymbol{z}}_{\vec{k}},
\eea
where we recall that $\boldsymbol{D}_k$ can be any symplectic matrix. The canonical commutation relation then reads 
\bea
\label{eq:commutator:hat:a}
\left[\boldsymbol{\widehat{a}}_{\vec{k}},\boldsymbol{\widehat{a}}^\dag_{\vec{q}}\right]=i\boldsymbol{\mathcal{J}}\delta^3(\vec{k}-\vec{q}),
\eea
which is the analogue of \Eq{eq:PoissonBracket:helicity}. Following the notation of \Eq{eq:Hamiltonian:helicity:FRTheta}, the Hamiltonian operator in the helicity basis reads
\bea
\label{eq:H:F:R:Theta}
	\widehat{\mathcal{H}}&=&\int_{\mathbb{R}^{3+}}\dd^3k\boldsymbol{\widehat{a}}^\dag_{\vec{k}}\boldsymbol{\mathcal{H}}_k\boldsymbol{\widehat{a}}_{\vec{k}}, \\
	&=&\int_{\mathbb{R}^{3+}}\dd^3k\left\{F_k(t)\left[\widehat{a}^\dag_{\vec{k}}\widehat{a}_{\vec{k}}+\widehat{a}^\dag_{-\vec{k}}\widehat{a}_{-\vec{k}}+1\right]+R_k(t)\left[e^{i\Theta_k(t)}\widehat{a}^\dag_{\vec{k}}\widehat{a}^\dag_{-\vec{k}}+e^{-i\Theta_k(t)}\widehat{a}_{\vec{k}}\widehat{a}_{-\vec{k}}\right]\right\}. \nonumber
\eea
The part multiplied par $F_k$ can be viewed as a \emph{free} Hamiltonian, and the term multiplied by $R_ke^{\pm i\Theta_k}$ can be interpreted as an \emph{interaction} Hamiltonian caused by the curvature of the background, which couples two modes of opposite wavevectors, $\vec{k}$ and $-\vec{k}$.

To define a Hilbert space as a Fock space, one must check that the creation and annihilation operators are ladder operators for the Hamiltonian. This is however not generically true for the situation presently at hand. Indeed, one has $\left[\widehat{\mathcal{H}},\boldsymbol{\widehat{a}}_{\vec{k}}\right]=-i\boldsymbol{\mathcal{J}}\boldsymbol{\mathcal{H}}_k\boldsymbol{\widehat{a}}_{\vec{k}}$, while for proper ladder operators, the commutator should read $\left[\widehat{\mathcal{H}},\boldsymbol{\widehat{a}}_{\vec{k}}\right]=-ic_k\boldsymbol{\mathcal{J}}\boldsymbol{\widehat{a}}_{\vec{k}}$ with $c_k\in\mathbb{R}$.\footnote{Ladder operators are usually defined as $\left[\widehat{H},\widehat{a}\right]=-c~\widehat{a}$ and $\left[\widehat{H},\widehat{a}^\dag\right]=c~\widehat{a}^\dag$ with $c\in\mathbb{R}$. Since $-i\boldsymbol{\mathcal{J}}=\mathrm{diag}(-1,1)$, see \Eq{eq:J:def}, this can be written in the compact matricial form given in the main text.} From \Eq{eq:Hamiltonian:helicity:FRTheta}, one can see that the former cannot be written in the form of the latter as soon as $R_k\neq 0$, \ie as soon as the interaction term is present. In some cases however, it is possible to neglect the interaction Hamiltonian at some initial times, $t_0$, and/or for some range of wavenumbers (as \eg $k\gg |a H|$ in an inflationary background, see \Sec{sec:inflation}). Then, the initial creation and annihilation operators, $\hat{a}_{\vec{k}}(t_0)$ and $\hat{a}^\dag_{-\vec{k}}(t_0)$, are ladder operators for $\left[\widehat{\mathcal{H}},\boldsymbol{\widehat{a}}_{\vec{k}}(t_0)\right]=-iF_k(t_0)\boldsymbol{\mathcal{J}}\boldsymbol{\widehat{a}}_{\vec{k}}(t_0)$.
In terms of number operators, $\widehat{N}_{\vec{k}}=\widehat{a}^\dag_{\vec{k}}\widehat{a}_{\vec{k}}$ and $\widehat{N}_{-\vec{k}}=\widehat{a}^\dag_{-\vec{k}}\widehat{a}_{-\vec{k}}$, the raising and lowering values associated to the ladder operators is $1$. It is thus straightforward to built the Fock basis of the {\it initial} Hilbert space,
\bea
\mathcal{E}(\tau_0)=\ds\prod_{\vec{k}\in\mathbb{R}^{3+}}\mathcal{E}_{\vec{k}}(\tau_0)\otimes\mathcal{E}_{-\vec{k}}(\tau_0),
\eea 
which is a bipartite system. The vacuum state is the state annihilated by all annihilation operators, and we denote it 
\bea
\label{eq:vacuum:NumberOfParticle:def}
\left|\slashed{0}(t_0)\right>=\ds\prod_{\vec{k}\in\mathbb{R}^{3+}}\left|0_{\vec{k}},0_{-\vec{k}}(t_0)\right>.
\eea
The rest of the Fock basis is simply built by successive applications of the creation operators, namely 
\bea
\left|n_{\vec{k}},m_{-\vec{k}}(t_0)\right>=\frac{1}{n!}\left[\widehat{a}^\dag_{\vec{k}}(t_0)\right]^n\frac{1}{m!}\left[\widehat{a}^\dag_{-\vec{k}}(t_0)\right]^m\left|0_{\vec{k}},0_{-\vec{k}}(t_0)\right>.
\eea
\subsubsection{Dynamics of the creation and annihilation operators}
\label{Sec:Dynamics:Creation:Annihilation}
The dynamics of the creation and annihilation operators is provided by the Heisenberg equation~(\ref{eq:Heisenberg}), which gives rise to
\bea
\label{eq:Heisenberg:hat:a}
\dot{\boldsymbol{\widehat{a}}}_{\vec{k}}=-i \left[\boldsymbol{\widehat{a}}_{\vec{k}},\widehat{\mathcal{H}}\right].
\eea
It is usually solved by means of a so-called ``Bogolyubov transform'', which recasts the problem of differential equations for operators into a problem of differential equations for functions, \ie
\bea
\label{eq:Bogoljubov:a:1}
	\widehat{a}_{\vec{k}}(t)&=&\alpha_{\vec{k}}(t)\widehat{a}_{\vec{k}}(t_0)+\beta_{\vec{k}}(t)\widehat{a}^\dag_{-\vec{k}}(t_0), \label{eq:transport1} \\
	\widehat{a}^\dag_{-\vec{k}}(t)&=&\alpha^*_{\vec{k}}(t)\widehat{a}^\dag_{-\vec{k}}(t_0)+\beta^*_{\vec{k}}(t)\widehat{a}_{\vec{k}}(t_0) \label{eq:transport2} .
\label{eq:Bogoljubov:a:2}
\eea
The two Bogolyubov coefficients $(\alpha_{\vec{k}},\beta_{\vec{k}})$, which are \emph {functions}, initially satisfy $\alpha_k(t_0)=1$ and $\beta_k(t_0)=0$, and since the commutator $\left[\boldsymbol{\widehat{a}}_{\vec{k}},\boldsymbol{\widehat{a}}^\dag_{\vec{q}}\right]$ is preserved, see \Eq{eq:commutator:hat:a}, they are such that $\left|\alpha_{\vec{k}}\right|^2-\left|\beta_{\vec{k}}\right|^2=1$ at every time. It is obvious that such a way to solve for the dynamics is no more than the Green's matrix approach introduced in \Sec{sssec:complexdynamics} for the classical helicity variables.  Indeed, the Bogolyubov equations~(\ref{eq:transport1}) and~(\ref{eq:transport2}) can be recast as 
\bea
\label{eq:a:evol:Green}
\widehat{\boldsymbol{a}}_{\vec{k}}(t)={\boldsymbol{\mathcal{G}}(t,t_0)}\widehat{\boldsymbol{a}}_{\vec{k}}(t_0),
\eea 
where ${\boldsymbol{\mathcal{G}}(t,t_0)}$ has been defined in \Eq{eq:G:Bogoliubov}, and plugging this into \Eq{eq:Heisenberg:hat:a}, it is easy to see that the coefficients of ${\boldsymbol{\mathcal{G}}(t,t_0)}$ are solutions of \Eqs{eq:alpha} and~(\ref{eq:beta}). 

Let us further comment on the initial values of the Bogolyubov coefficients. As mentioned above, strictly speaking, they have to be set as $\alpha_k(t_0)=1$ and $\beta_k(t_0)=0$. In some specific cases however, it is sufficient to constrain $\alpha_{\vec{k}}$ up to a global phase, \ie $\alpha_{\vec{k}}=e^{i\theta_k(t_0)}$. Supposing that one initially sets $\widetilde{\alpha}_{\vec{k}}(t_0)=e^{i\theta_{\vec{k}}(t_0)}$ and $\widetilde{\beta}_k(t_0)=0$, the transport equations~(\ref{eq:transport1}) and~(\ref{eq:transport2}) are then rewritten as 
\bea
	\widehat{\widetilde{a}}_{\vec{k}}(t)&=&\alpha_{\vec{k}}(t)e^{i\theta_k(t_0)}\widehat{a}_{\vec{k}}(t_0)+\beta_{\vec{k}}(t)e^{-i\theta_k(t_0)}\widehat{a}^\dag_{-\vec{k}}(t_0),\\
	\widehat{\widetilde{a}}^\dag_{-\vec{k}}(t)&=&\alpha^*_{\vec{k}}(t)e^{-i\theta_k(t_0)}\widehat{a}^\dag_{-\vec{k}}(t_0)+\beta^*_{\vec{k}}(t)e^{i\theta_k(t_0)}\widehat{a}_{\vec{k}}(t_0), 
\eea
with $\alpha_{\vec{k}}$ and $\beta_{\vec{k}}$ constrained to be initially equal to 1 and 0, respectively. This is equivalent to a redefinition of the initial creation and annihilation operators by applying a rotation, \ie $\widehat{\widetilde{a}}_{\vec{k}}(t_0)=e^{i\theta_k(t_0)}\widehat{a}_{\vec{k}}(t_0)$ and $\widehat{\widetilde{a}}^\dag_{-\vec{k}}(t_0)=e^{-i\theta_k(t_0)}\widehat{a}^\dag_{-\vec{k}}(t_0)$. Rephrasing this in the Schr\"odinger picture leads to a new basis for the Fock space $\left|\cdots \widetilde{n}_{\vec{k}}\cdots\right>=e^{i\int\dd^3k(n_{\vec{k}}+1)\theta_k}\left|\cdots n_{\vec{k}}\cdots\right>$, while operators are now left unchanged. This  defines the same initial vacuum state, and the initial $(n_{\vec{k}},m_{-\vec{k}})$-particles states are just multiplied by a global phase. Hence, if the initial state is {\it not} a superposition of many-particles states (\ie in the restricted case of $\left|\Psi(t_0)\right>=\left|\cdots n_{\vec{k}}\cdots\right>$), one obtains the same prediction regardless of the initial phase chosen for $\alpha_{\vec{k}}$ since the evolution does not couple different modes (homogeneous and isotropic hypothesis).\footnote{Note that this is not the case anymore for quantum states that are superpositions of Fock states (though it remains so for the mixed states that are fully decohered in the number of particles basis). Considering indeed an operator $\boldsymbol{\widehat{\mathcal{O}}}$ whose matrix elements in the initial Fock basis are $\boldsymbol{\mathcal{O}}_{n,m}$, its matrix element in the ``rotated'' basis are $\boldsymbol{\widetilde{\mathcal{O}}}_{n,m}=e^{i(n-m)\theta}\boldsymbol{\mathcal{O}}_{n,m}$.}
\subsubsection{Initial conditions}
\label{sssec:initialcond}
The last step consists in choosing the initial conditions, \ie the initial $\boldsymbol{D}_k$ matrix such that a specific vacuum state is selected, and a specific initial basis in the Hilbert space is defined.\footnote{Note that setting the initial values of the Bogolyubov coefficients is not what is meant by ``setting the initial conditions'', since they are constrained to be $1$ and $0$ by construction, see \Sec{Sec:Dynamics:Creation:Annihilation}.} To this end, one first notes that according to \Eq{eq:a(z):quantum}, $\widehat{\boldsymbol{z}}_{\vec{k}}$ can be expanded onto the entries of $\boldsymbol{\widehat{a}}_{\vec{k}}$, namely $\widehat{a}_{\vec{k}} $ and $\widehat{a}^\dag_{-\vec{k}}$, which, in turn, can be written as combinations of $\widehat{a}_{\vec{k}}(t_0)$ and $\widehat{a}^\dag_{-\vec{k}}(t_0)$, see \Eqs{eq:Bogoljubov:a:1} and~(\ref{eq:Bogoljubov:a:2}). We write this expansion as $\widehat{\boldsymbol{z}}_{\vec{k}} = \boldsymbol{z}_k(t) \widehat{a}_{\vec{k}}(t_0) + \boldsymbol{z}_k^*(t) \widehat{a}_{-\vec{k}}^\dagger(t_0)  $, leading to
\bea
\label{eq:field:decomposition:quantum:z:a}
	\boldsymbol{\widehat{z}}(\vec{x},t)=\ds\int_{\mathbb{R}^{3+}}\frac{\dd^3k}{(2\pi)^{3/2}}\left[\widehat{a}_{\vec{k}}(t_0)\boldsymbol{z}_k(t)e^{-i\vec{k}\cdot\vec{x}}+\widehat{a}^\dag_{\vec{k}}(t_0)\boldsymbol{z}^*_k(t)e^{i\vec{k}\cdot\vec{x}}\right],
\eea
where the $\boldsymbol{z}_{\vec{k}}$'s are \emph{classical} functions. In order to recover the commutator~(\ref{eq:commutator:hatz}) for the field variables from the commutator~(\ref{eq:commutator:hat:a}) for the ladder operators, these {\it mode} functions have to satisfy $\boldsymbol{z}_{\vec{k}} \boldsymbol{z}_{\vec{k}}^{*\mathrm{T}}-\boldsymbol{z}_{\vec{k}}^* \boldsymbol{z}_{\vec{k}}^\mathrm{T} = i \boldsymbol{\Omega}$,\footnote{This means that the mode functions, $\boldsymbol{z}_k(t)e^{-i\vec{k}\cdot\vec{x}}$, are normalised through the Klein-Gordon product, which is preserved through Hamiltonian evolutions (see \eg \Refa{birrell1982} for a Lagrangian viewpoint and \eg \Refa{Grain:2017dqa} for a Hamiltonian viewpoint).\label{footnote:KleinGordonProduct}} \ie
\bea
\phi_{\vec{k}} \pi_{\vec{k}}^*  - \phi_{\vec{k}}^* \pi_{\vec{k}} = i\, .
\label{eq:Wronskian}
\eea
In the following, this constraint is referred to as the ``Wronskian'' condition. The mode functions have two complex degrees of freedom and the Wronskian condition provides one constraint, hence one is free to select different initial conditions for the mode functions. Depending on this choice, different initial vacuum states are selected (since different initial mode functions lead to having the creation and annihilation operators creating and annihilating different kinds of excitations, \ie different kinds of mode functions).\footnote{In what follow, we loosely use the notation $\widehat{a}_{\vec{k}}$. In principle, one should however write $\widehat{a}_{z_{\vec{k}}(t)}$, in the sense that operators are creating and annihilating entire mode functions (each of them carrying a given wave vector). If, for a given wave vector, two different initial conditions are chosen, then one has two different mode functions, hence two different sets of creation and annihilation operators.} 

The question is thus: given a choice of initial conditions for the mode functions, what should the matrix $\boldsymbol{D}_k(t)$ appearing in \Eq{eq:a(z):quantum} be such that the creation and annihilation operators create and annihilate the desired type of quantas? The key point is to ensure that the two Bogolyubov coefficients entering the Green's matrix~(\ref{eq:G:Bogoliubov}) are related to the classical mode functions~(\ref{eq:z:def}) via the same transform~(\ref{eq:a(z):quantum}) that relates the operators, \ie 
\bea	
	\left(\begin{array}{c}
		\alpha_{\vec{k}}(t) \\
		\beta^*_{\vec{k}}(t)
	\end{array}\right)=\boldsymbol{UD}_k(t)\left(\begin{array}{c}
		\phi_{\vec{k}}(t) \\
		\pi_{\vec{k}}(t)
	\end{array}\right).
\eea
The requirement $\alpha_{\vec{k}}(t_0)=1$ and $\beta_{\vec{k}}(t_0)=0$ then leads to a constraint relating the matrix $\boldsymbol{D}_k(t_0)$ and the initial mode functions, namely
\bea
	\boldsymbol{e}_\alpha \equiv \left(\begin{array}{c}
		1 \\
		0
	\end{array}\right)=\boldsymbol{UD}_k(t_0)\left(\begin{array}{c}
		\phi_{\vec{k}}(t_0) \\
		\pi_{\vec{k}}(t_0)
	\end{array}\right). \label{eq:initialD}
\eea
This relation can be used in two ways. On the one hand, one can chose an explicit expression for the matrix $\boldsymbol{D}_k(t)$ which unequivocally selects a given initial state by selecting the mode functions initially satisfying $\boldsymbol{z}_{\vec{k}}(t_0)=\boldsymbol{D}^{-1}_k(t_0)\boldsymbol{U}^\dag\boldsymbol{e}_\alpha$. Choosing this matrix thus selects a specific initial basis of the Hilbert space. On the other hand, one can start from a specific vacuum state and initial basis of the initial Hilbert space, encoded in a specific choice for the initial mode functions, $\boldsymbol{z}_{\vec{k}}(t_0)$, and determine the entries of $\boldsymbol{D}_k(t_0)$ according to \Eq{eq:initialD}.\footnote{Let us note that \Eq{eq:initialD} relates complex numbers and thus yields four constraints, while the matrix $\boldsymbol{D}_k$ is characterised by three parameters only, making the problem a priori over-constrained. The field variables should however satisfy the Wronskian condition~(\ref{eq:Wronskian}) which fixes the remaining degree of freedom.}

\subsection{Squeezing formalism}
The squeezing formalism is a common approach in quantum optics to study time-dependent quadratic Hamiltonians, and it has also been applied to the case of cosmological perturbations in an inflationary background (see e.g. \Refa{Zeldovich:1971mw,Polarski:1995jg,Lesgourgues:1996jc,Martin:2007bw} and references therein). 

We first describe the squeezing formalism from a simply technical perspective, before adopting a more formal approach. The Green's matrix introduced below \Eq{eq:calM:def} is an element of $\mathrm{SU}(1,1)$, hence it can be decomposed along the left-polar decomposition~(\ref{eq:polar:RS:explicit})-(\ref{eq:bloch:left:right}), \ie
\bea
	\boldsymbol{\mathcal{G}}_k(t,t_0)=\boldsymbol{\mathcal{S}}(r_k,\varphi_k)\boldsymbol{\mathcal{R}}_k(\theta_k)\, .
\eea
In this expression, $r_k$ is the squeezing amplitude and $\varphi_k$ the squeezing angle ($r_k$ and $\varphi_k$ are referred to as the ``squeezing parameters'' below), and $\theta_k$ is a rotation angle. Making use of \Eq{eq:polar:RS:explicit}, the Bogolyubov coefficients entering the Green's matrix~(\ref{eq:G:Bogoliubov}) read 
\bea
\label{eq:alpha:squeezing}
\alpha_k(t)&=&e^{i\theta_k(t)}\cosh\left[ r_k(t) \right],\\
\beta_k(t)&=&e^{-i\theta_k(t)+2i\varphi_k(t)}\sinh \left[r_k(t)\right].
\label{eq:beta:squeezing}
\eea
This is no more than the usual rewriting of the Bogolyubov coefficients in the squeezing formalism, which allows one to recast the dynamics in terms of evolution equations for the parameters $(r_k(t),\varphi_k(t),\theta_k(t))$, which are solved with initial conditions $r_k(t_0)=\varphi_k(t_0)=\theta_k(t_0)=0$.

Let us now introduce the squeezing formalism from a more formal perspective. Since the dynamics of the creation and annihilation operators is generated by $\mathrm{SU}(1,1)$ elements, the evolution operator, $\widehat{\mathcal{U}}(t,t_0)$, should be given by a representation of the group $\mathrm{SU}(1,1)$. In the left-polar decomposition, one therefore has
\bea
\label{eq:Uevol:def}
	\widehat{\mathcal{U}}\left(t,t_0\right)=\exp\left\lbrace\ds\int_{\mathbb{R}^{3+}}\dd^3k\left[\xi_k(t)\widehat{\mathcal{K}}^{(\vec{k})}_+-\xi^*_k(t)\widehat{\mathcal{K}}^{(\vec{k})}_-\right]\right\rbrace\exp\left[2i\ds\int_{\mathbb{R}^{3+}}\dd^3k\theta_k(t)\widehat{\mathcal{K}}^{(\vec{k})}_z\right],
\eea
where we recall that $\xi_k(t)=-ir_k(t)e^{2i\varphi_k(t)}$ has been introduced above \Eq{eq:polar:Lie:R}, and the $\widehat{\mathcal{K}}^{(\vec{k})}_i$'s operators should be built from the creation and annihilation operators and represent the Lie algebra of $\mathrm{SU}(1,1)$ (note that the parameters depend on the wavenumber only because of the isotropic assumption, while the generator operators depend on the wave vectors). As noticed below \Eq{eq:H:F:R:Theta}, the dynamics thus couples the two modes $\vec{k}$ and $-\vec{k}$. In \App{app:representation}, we show that a two-mode representation of the generator of $\mathrm{SU}(1,1)$ is given by
\bea
\label{eq:K+:a}
	\widehat{\mathcal{K}}^{(\vec{k})}_+&=&i~\widehat{a}^\dag_{\vec{k}}(t_0)~\widehat{a}^\dag_{-\vec{k}}(t_0), \\
\label{eq:K-:a}
	\widehat{\mathcal{K}}^{(\vec{k})}_-&=&-i~\widehat{a}_{\vec{k}}(t_0)~\widehat{a}_{-\vec{k}}(t_0), \\
\label{eq:Kz:a}
	\widehat{\mathcal{K}}^{(\vec{k})}_z&=&\frac{1}{2}\left[\widehat{a}^\dag_{\vec{k}}(t_0)~\widehat{a}_{\vec{k}}(t_0)+\widehat{a}^\dag_{-\vec{k}}(t_0)~\widehat{a}_{-\vec{k}}(t_0)+1\right],
\eea
and we present a systematic way to built such a representation, which is mainly adapted from the approach developed in \Refs{PhysRevA.36.3868, SIMON1987223, Arvind:1995ab}. This is done both for the field operators (\App{sec:QuantumRepresentation:FieldOperators}), and for the creation and annihilation operators (\App{sec:QuantumRepresentation:CreationAnnihilationOperators}). 

In the Heisenberg picture, any operator $\widehat{\mathcal{O}}_{\vec{k}}$ evolves according to $\widehat{\mathcal{O}}_{\vec{k}}(t)=\widehat{\mathcal{U}}^\dag(t,t_0)\widehat{\mathcal{O}}_{\vec{k}}(t_0)\widehat{\mathcal{U}}(t,t_0)$. Using operator ordering \cite{barnett2002methods}, one can show that 
\bea
\label{eq:transport:Heisenberg:a}
\kern-2em
	\widehat{\mathcal{U}}^\dag(t,t_0)\widehat{a}_{\vec{k}}(t_0)\widehat{\mathcal{U}}(t,t_0)&=&e^{i\theta_k(t)}\cosh r_k(t)\widehat{a}_{\vec{k}}(t_0)+e^{-i\theta_k(t)+2i\varphi_k(t)}\sinh r_k(t)\widehat{a}^\dag_{-\vec{k}}(t_0), \\
\label{eq:transport:Heisenberg:adag}
	\kern-2em
	\widehat{\mathcal{U}}^\dag(t,t_0)\widehat{a}^\dag_{-\vec{k}}(t_0)\widehat{\mathcal{U}}(t,t_0)&=&e^{-i\theta_k(t)}\cosh r_k(t)\widehat{a}^\dag_{-\vec{k}}(t_0)+e^{i\theta_k(t)-2i\varphi_k(t)}\sinh r_k(t)\widehat{a}_{\vec{k}}(t_0).
\eea
Making use of \Eqs{eq:alpha:squeezing} and~(\ref{eq:beta:squeezing}), the transport equations~(\ref{eq:transport1}) and~(\ref{eq:transport2}), derived from the Heisenberg equations, are thus exactly recovered.

In the Schr\"odinger picture instead, the evolution of quantum states is given by $\left|\Psi(t)\right>=\widehat{\mathcal{U}}(t,t_0)\left|\Psi(t_0)\right>$. Starting for example from the vacuum state, which, from \Eqs{eq:K-:a} and~(\ref{eq:Kz:a}) is in the null space of $\widehat{\mathcal{K}}^{(\vec{k})}_-$ and $\widehat{\mathcal{K}}^{(\vec{k})}_z-1/2$, one easily obtains that it evolves towards the two-mode squeezed state
\bea
\label{eq:TMSS}
	\left|\slashed{0}(t)\right>=\ds\prod_{\vec{k}\in\mathbb{R}^{3+}}\left[\frac{e^{i\theta_k}}{\cosh r_k}\ds\sum_{n=0}^\infty(-1)^ne^{2in\varphi_k}\tanh^nr_k\left|n_{\vec{k}}(t_0),n_{-\vec{k}}(t_0)\right>\right],
\eea
where in the above $n_{\vec{k}}=n_{-\vec{k}}$.\footnote{We note the additional phase factor $e^{i\theta_k}$ as compared to usual results found in the literature (see \eg \Refa{Martin:2015qta}). This is because the evolution operator~(\ref{eq:Uevol:def}) can alternatively be defined as $\widehat{\mathcal{U}}(t,t_0)=\exp\left[\ds\int\dd^3k\left(\xi_k\widehat{\mathcal{K}}^{(\vec{k})}_+-\xi^*_k\widehat{\mathcal{K}}^{(\vec{k})}_-\right)\right]\exp\left[2i\ds\int\dd^3k\theta_k\left(\widehat{\mathcal{K}}^{(\vec{k})}_z-1/2\right)\right]$. This leads to the same transport equations~(\ref{eq:transport:Heisenberg:a}) and~(\ref{eq:transport:Heisenberg:adag}) in the Heisenberg picture~\cite{Martin:2015qta}, but removes the phase factor $\ee^{i\theta_k}$ from \Eq{eq:TMSS} in the Schr\"odinger picture. This overall phase factor is obviously not an issue from the viewpoint of observables since {\it all} quantum states acquire the very same phase, hence making it unobservable by measurements of hermitic observables (recall that, here, there is no coupling here between modes with different wavenumbers).\label{footnote:phase:factor}}

Let us finally note that \Eqs{eq:polar:Lie:S} and~(\ref{eq:polar:Lie:R}) lead to the following representation for the squeezing and rotation matrices:
\bea
\label{eq:quantum:operator:S}
	\widehat{\mathcal{S}}\left(r_k,\varphi_k\right)&=&\exp\left[\ds\int_{\mathbb{R}^{3+}}\dd^3kr_k\left(e^{2i\varphi_k}\widehat{a}^\dag_{\vec{k}}~\widehat{a}^\dag_{-\vec{k}}-e^{-2i\varphi_k}\widehat{a}_{\vec{k}}~\widehat{a}_{-\vec{k}}\right)\right],\\
	\widehat{\mathcal{R}}\left(\theta_k\right)&=&\exp\left[i\ds\int_{\mathbb{R}^{3+}}\dd^3k\theta_k\left(\widehat{a}^\dag_{\vec{k}}~\widehat{a}_{\vec{k}}+\widehat{a}^\dag_{-\vec{k}}~\widehat{a}_{-\vec{k}}+1\right)\right].
\label{eq:quantum:operator:R}
\eea
Any quantum operator representing an element of $\mathrm{SU}(1,1)$, such as the evolution operator, can thus be written using the left-polar decomposition~(\ref{eq:bloch:left:right}) as $\widehat{\mathcal{S}}(r,\varphi)~\widehat{\mathcal{R}}(\theta)$, as done explicitly in \Eq{eq:Uevol:def}. \\

From the above considerations, it is straightforward to show that a two-mode squeezed states evolves into a two-mode squeezed state. Suppose the initial state to be any two-mode squeezed state, $\left|\Psi_{d,\varpi}(t_0)\right>$, where $d$ and $\varpi$ are the squeezing parameters that characterise it. (Note that the vacuum state is the specific two-mode squeezed state obtained by setting $d=0$.) Such a state is realised by applying the squeezing operator $\widehat{\mathcal{S}}(d,\varpi)$ on the vacuum state, \ie $\left|\Psi_{d,\varpi}(t_0)\right>=\widehat{\mathcal{S}}(d,\varpi)\left|\slashed{0}(t_0)\right>$. This state thus evolves according to
\bea
	\left|\Psi_{d,\varpi}(t)\right>=\widehat{\mathcal{U}}(t,t_0)\left|\Psi_{d,\varpi}(t_0)\right>=\widehat{\mathcal{U}}(t,t_0)\widehat{\mathcal{S}}(d,\varpi)\left|\slashed{0}(t_0)\right>,
\eea
where the evolution operator is characterised by some squeezing parameters $(r,\varphi)$ and a rotation angle $\theta$, all of them being functions of time. Since both $\widehat{\mathcal{U}}(t,t_0)$ and $\widehat{\mathcal{S}}(d,\varphi)$ are elements of $\mathrm{SU}(1,1)$, the product operator $\widehat{\mathcal{U}}(t,t_0)\widehat{\mathcal{S}}(d,\varpi)$ is also an element of $\mathrm{SU}(1,1)$. As such it can be written using the left-polar decomposition leading to
\bea
	\widehat{\mathcal{U}}(t,t_0)\widehat{\mathcal{S}}(d,\varpi)=\widehat{\mathcal{S}}(\widetilde{r},\widetilde{\varphi})\widehat{\mathcal{R}}(\widetilde{\theta}),
\eea
where the parameters $(\widetilde{r},\widetilde{\varphi})$ and $\widetilde{\theta}$ are obtained from $(d,\varpi)$ and $(r,\varphi,\theta)$ by using the composition law of $\mathrm{SU}(1,1)$ derived in \Sec{sec:SU11:Composition}. Using the fact that the vacuum state is invariant under rotations of the phase space, \ie $\widehat{\mathcal{R}}(\widetilde{\theta})\left|\slashed{0}(t_0)\right>=\left|\slashed{0}(t_0)\right>$, one gets for the evolved two-mode squeezed state 
\bea
	\left|\Psi_{d,\varpi}(t)\right>=\widehat{\mathcal{S}}(\widetilde{r},\widetilde{\varphi})\left|\slashed{0}(t_0)\right>,
\eea
which is indeed a two-mode squeezed state itself.
\subsection{Canonical transformations}
Let us consider a canonical transformations encoded in an $\mathrm{SU}(1,1)$ matrix, $\widehat{\boldsymbol{a}}_{\vec{k}}\to\widehat{\boldsymbol{\widetilde{a}}}_{\vec{k}}=\boldsymbol{\mathcal{M}}_k(t)\widehat{\boldsymbol{a}}_{\vec{k}}$. As explained in \Sec{sssec:canonicalcomplex}, see \Fig{fig:diagram}, this can be interpreted in two ways: either as the effect of a canonical transformation $\boldsymbol{M}_k$ of the field operators, hence $\boldsymbol{\mathcal{M}}_k$ is given by \Eq{eq:calMk:def}, or as the result of having two different choices for the matrix $\boldsymbol{D}_k$ used to define the creation and annihilation operators starting from the same field operators, hence $\boldsymbol{\mathcal{M}}_k$ is given by \Eq{eq:calM:D:D1:def}.
\subsubsection{Canonical transformations as squeezing}
\label{sec:canonical:transf:as:squeezing}
Since $\boldsymbol{\mathcal{M}}_k$ is an element of $\mathrm{SU}(1,1)$, the quantum operator that generates it has to be given by a representation of that transformation group, and in the present case it should be a two-mode representation (see also \Refa{Cervero2002} for a similar analysis in the case of one-mode representations). In the same way that, for the Green's matrix generating the dynamical evolution and belonging to $\mathrm{SU}(1,1)$, the corresponding evolution operator is built from a rotation operator and a squeezing operator, the unitary transformation that generates $\boldsymbol{\mathcal{M}}_k$ is also given by a rotation operator and a squeezing operator. 

Denoting $d_k(t)$ and $\varpi_k(t)$ the squeezing amplitude and the squeezing angle of the canonical transformation respectively, and $\vartheta_k(t)$ its rotation angle, the unitary transformation that generates $\boldsymbol{\mathcal{M}}_k$ is given by
\bea
	\widehat{\mathcal{M}}=e^{\int_{\mathbb{R}^{3+}}\dd^3k\left(d_ke^{2i\varpi_k}\widehat{a}^\dag_{\vec{k}}\widehat{a}^\dag_{-\vec{k}}-d_ke^{-2i\varpi_k}\widehat{a}_{\vec{k}}\widehat{a}_{-\vec{k}}\right)}e^{i\int_{\mathbb{R}^{3+}}\dd^3k\vartheta_k\left(\widehat{a}^\dag_{\vec{k}}\widehat{a}_{\vec{k}}+\widehat{a}^\dag_{-\vec{k}}\widehat{a}_{-\vec{k}}+1\right)}, \label{eq:canonicalquantum}
\eea
according to \Eqs{eq:quantum:operator:S} and~(\ref{eq:quantum:operator:R}). The quantum operator $\widehat{\mathcal{M}}$ acts on operators $\widehat{\mathcal{O}}$ and quantum states $\left|\Psi\right>$ as follows,
\bea
	\mathrm{Heisenberg\,picture:}&&\widehat{\mathcal{O}}\to\widehat{\widetilde{\mathcal{O}}}=\widehat{\mathcal{M}}^\dag\widehat{\mathcal{O}}\widehat{\mathcal{M}}, \\
	\mathrm{Schrodinger\,picture:}&&\left|\Psi\right>\to\left|\widetilde{\Psi}\right>=\widehat{\mathcal{M}}\left|\Psi\right>.
	\label{eq:Schrodinger:picture}
\eea
The unitary representation~\eqref{eq:canonicalquantum} has the same form as the evolution operator~(\ref{eq:Uevol:def}) since both transformations, canonical and dynamical, are generated by the action of $\mathrm{SU}(1,1)$. One can however note that in the evolution operator, the {\it initial} creation and annihilation operators are involved, while in the canonical transformation, the creation and annihilation operators {\it at the time} $t$ are involved. This is because, despite the formal similarity, there are key conceptual differences between them.

On the one hand, linear canonical transformations generated by the symplectic group correspond to a {\it symmetry} of the Hamiltonian system, which implies that one can describe the dynamics using any set of canonical variables, related one to another by a canonical transformation. These transformations relate two different sets of canonical variables evaluated at the {\it same} time (and for any time). This explains why the creation and annihilation operators at the time $t$ enter in $\widehat{\mathcal{M}}(t)$. The choice of canonical variables is conventional, and one is free to chose the values and time dependence of $(\varpi_k,d_k,\vartheta_k)$ and work with any set of canonical variables. 

On the other hand, the Hamiltonian dynamics is the result of the action of the symplectic transformations generated by the Green's matrix. This action has a different meaning than the first one: it relates the same set of canonical variables evaluated at two different times, \ie it transforms the initial values of phase-space variables into their final values. This is why the initial creation and annihilation operators are involved in $\widehat{\mathcal{U}}(t,t_0)$. In addition, the symplectic matrix encoding this evolution, \ie the parameters $(\varphi_k,r_k,\theta_k)$, is not conventional but determined by the considered, peculiar dynamics (\ie the peculiar Hamiltonian describing the system). 

In the following, the squeezing parameters related to canonical transformations will be denoted by $(\varpi_k,d_k,\vartheta_k)$, while the ones related to the Green's matrix (or the evolution operator) will be denoted by $(\varphi_k,r_k,\theta_k)$. Despite their identical mathematical nature, we introduce these two different notations to make obvious the type of squeezing being discussed, either canonical or dynamical.
\subsubsection{Vacuum states transformation and equivalent representations}
\label{sec:EquivalentRepresentations}
We denote by $\left|\slashed{0}(t)\right>$ and $\left|\widetilde{\slashed{0}}(t)\right>$ the vacuum states generated by the set of operators $\widehat{\boldsymbol{a}}_{\vec{k}}$ and $\widehat{\boldsymbol{\widetilde{a}}}_{\vec{k}}$, respectively. From what precedes, the $\widetilde{a}$-vacuum, $\left|\widetilde{\slashed{0}}(t)\right>$, can be expressed as a two-mode squeezed state of $a$-quantas,\footnote{The converse is also true, \ie the $a$-vacuum is a two-mode squeezed state of $\widetilde{a}$-quantas with parameters given by the inverse canonical transformation.}
\bea
	\left|\widetilde{\slashed{0}}(t)\right>=\ds\prod_{\vec{k}\in\mathbb{R}^{3+}}\left[\frac{e^{i\vartheta_k}}{\cosh d_k}\ds\sum_{n=0}^\infty(-1)^ne^{2in\varpi_k}\tanh^nd_k\left|n_{\vec{k}}(t_0),n_{-\vec{k}}(t_0)\right>\right],
\eea
see \Eq{eq:TMSS}. Two sets of canonical variables therefore lead to two different vacuum states, hence different initial conditions, hence different physical predictions. 

A sufficient and necessary condition for the two vacua to be identical is that the squeezing amplitude of the canonical transformation, $d_k$, vanishes. In this case, the canonical transformation reduces to a rotation in the phase space, which, as explained in footnote~\ref{footnote:phase:factor}, leaves the vacuum state unchanged. Working with the same vacuum thus constrains the canonical transformation to be of the form $\boldsymbol{\mathcal{M}}_k=\boldsymbol{\mathcal{R}}[\vartheta_k(t)]$, with $\vartheta_k(t)$ any angle. If one starts from two different sets of field operators as depicted in the left diagram of \Fig{fig:diagram}, related one to the other by a canonical transformation $\boldsymbol{M}_k$, this condition translates into
\bea
	\boldsymbol{\widetilde{D}}_k(t)=\boldsymbol{R}[\vartheta_k(t)]\boldsymbol{D}_k(t)\boldsymbol{M}^{-1}_k(t)\, , \label{eq:rotconstraint}
\eea
where \Eq{eq:calMk:def} has been used. The choice of $\boldsymbol{D}_k(t)$ and $\boldsymbol{\widetilde{D}}_k(t)$ unequivocally defines the two sets of creation and annihilation operators $\widehat{\boldsymbol{a}}_{\vec{k}}$ and $\widehat{\boldsymbol{\widetilde{a}}}_{\vec{k}}$. From \Eq{eq:rotconstraint}, the choice of $\boldsymbol{\widetilde{D}}_k(t)$ knowing $\boldsymbol{D}_k(t)$ and $\boldsymbol{M}_k(t)$ is constrained (and vice-versa), which specifies how the creation and annihilation operators $\widehat{\boldsymbol{a}}_{\vec{k}}$ and $\widehat{\boldsymbol{\widetilde{a}}}_{\vec{k}}$ have to be {\it coherently} defined in order to select the same vacuum state.

If one instead starts from the same set of field operators, but uses two different $\boldsymbol{D}_k$ matrices, as in the right diagram of \Fig{fig:diagram}, \Eq{eq:rotconstraint} imposes the two matrices to differ by a rotation only. Stated otherwise, the vacuum state is defined up to rotations of the phase space. 

Finally, if one imposes that the entire basis states of the Hilbert space are the same, as explained at the end of \Sec{Sec:Dynamics:Creation:Annihilation}, the remaining rotation is constrained to be the identity, since the basis states with non-zero particle numbers are not rotationally invariant, but are instead transform through a global phase $e^{i(n_k+1)\vartheta_k(t)}$.  In this case, one imposes 
\bea
	\boldsymbol{\widetilde{D}}_k(t)=\boldsymbol{D}_k(t)\boldsymbol{M}^{-1}_k(t). \label{eq:fullconstraint}
\eea

If this relation is not satisfied, one works with two different representations and there are two usual circumstances under which this may happen. First, one may define creation and annihilation operators always in the same way, \ie $\boldsymbol{\widetilde{D}}_k(t)=\boldsymbol{D}_k(t)$, but work with different field variables, $\boldsymbol{M}_k(t)\neq\boldsymbol{I}$. Second, one may start from the same field variables, $\boldsymbol{M}_k(t) = \boldsymbol{I}$, but define creation and annihilation operators in different ways, $\boldsymbol{\widetilde{D}}_k(t)\neq\boldsymbol{D}_k(t)$. 
As mentioned below \Eq{eq:calM:D:D1:def}, these two ways of building different representations are equivalent from the perspective of the helicity variables, and this is why in the following we will discuss only the first situation, \ie $\boldsymbol{\widetilde{D}}_k(t)=\boldsymbol{D}_k(t)$ but $\boldsymbol{M}_k(t)\neq\boldsymbol{I}$.
\subsubsection{Relating observational predictions}
\label{sec:RelatingObservationalPredictions}
Since different sets of canonical variables select different vacuum states initially, the dynamical evolution being entirely determined by the Hamiltonian, they give rise to different final states, hence different observational predictions. In this section, we study how they are related. Let us note that if the restriction~(\ref{eq:rotconstraint}) is imposed at initial time, the same initial state is selected and one expects to recover the same predictions (if one chooses to start from the vacuum state, otherwise the more restrictive condition~(\ref{eq:fullconstraint}) must be required).

We consider an operator $\widehat{\mathcal{O}}$ that represents a given observable, that we write as some function $\mathcal{O}$ of the phase-space variables $\boldsymbol{z}$, \ie $\widehat{\mathcal{O}}= \mathcal{O}(\widehat{\boldsymbol{z}})$. In terms of the phase-space variables $\widetilde{\boldsymbol{z}} = \boldsymbol{M}(t)\boldsymbol{z} $, it is given by  $\widehat{\mathcal{O}}= \mathcal{O}[\boldsymbol{M}^{-1}(t)\widehat{\widetilde{\boldsymbol{z}}}] \equiv \widetilde{\mathcal{O}}(\widehat{\widetilde{\boldsymbol{z}}})$, which defines the function $\widetilde{\mathcal{O}}$. 

\paragraph{Wigner-Weyl transform --}  A convenient tool to discuss how expectation values of operators change under canonical transformations is the Wigner-Weyl transform~\cite{1927ZPhy...46....1W,1946Phy....12..405G,1949PCPS...45...99M}. This is defined by first decomposing the two complex field variables $\phi_{\vec{k}}$ and $\pi_{\vec{k}}$ into four real variables $\mathfrak{q}_{\vec{k}}$, $\mathfrak{q}_{-\vec{k}}$, $\mathfrak{p}_{\vec{k}}$ and $\mathfrak{p}_{-\vec{k}}$,
 \bea
 \label{eq:Wigner:Canonical:Transform:phi}
 \phi_{\vec{k}} &=& \frac{1}{2}\left[\mathfrak{q}_{\vec{k}}+\mathfrak{q}_{-\vec{k}}+\frac{i}{k}\left(\mathfrak{p}_{\vec{k}}-\mathfrak{p}_{-\vec{k}}\right)\right] ,\\
 \pi_{\vec{k}} &=& \frac{1}{2i}\left[k\left(\mathfrak{q}_{\vec{k}}-\mathfrak{q}_{-\vec{k}}\right)+i\left(\mathfrak{p}_{\vec{k}}+\mathfrak{p}_{-\vec{k}}\right)\right] ,
 \label{eq:Wigner:Canonical:Transform:pi}
 \eea
which can easily be inverted to give $\mathfrak{q}_{\vec{k}}$, $\mathfrak{q}_{-\vec{k}}$, $\mathfrak{p}_{\vec{k}}$ and $\mathfrak{p}_{-\vec{k}}$ in terms of $\phi_{\vec{k}}$, $\phi_{-\vec{k}}=\phi_{\vec{k}}^*$, $\pi_{\vec{k}}$ and $\pi_{-\vec{k}}=\pi_{\vec{k}}^*$. One then defines
 \bea
 \kern-2em 
\mathcal{O}_{(W)}\left(\mathfrak{q}_{\vec{k}},\mathfrak{p}_{\vec{k}},\mathfrak{q}_{-\vec{k}},\mathfrak{p}_{-\vec{k}}\right) =\int \dd x \dd y e^{-i \mathfrak{p}_{\vec{k}} x - i \mathfrak{p}_{-\vec{k}} y }\left\langle \mathfrak{q}_{\vec{k}} + \frac{x}{2},  \mathfrak{q}_{-\vec{k}} + \frac{y}{2} \right\vert \widehat{\mathcal{O}} \left\vert \mathfrak{q}_{\vec{k}} - \frac{x}{2},  \mathfrak{q}_{-\vec{k}} - \frac{y}{2}\right\rangle.
 \eea
This allows one to map $\widehat{\mathcal{O}}$ into a classical function of the phase-space variables, and provides one with a phase-space representation of $\widehat{\mathcal{O}}$. Note that since the Wigner-Weyl transform of an operator is a function, $\mathcal{O}_{1(W)}\mathcal{O}_{2(W)}=\mathcal{O}_{2(W)}\mathcal{O}_{1(W)}$. However, $(\mathcal{O}_1 \mathcal{O}_2)_{(W)} \neq (\mathcal{O}_2 \mathcal{O}_1)_{(W)}$ in general. The Wigner-Weyl transform of an operator need also not be equal to its classical counterpart, \ie $\mathcal{O}$ and $\mathcal{O}_{(W)}$ are a priori two different functions. 

It is shown in \App{app:realvariables} that the transformation~\eqref{eq:Wigner:Canonical:Transform:phi}-\eqref{eq:Wigner:Canonical:Transform:pi}, that relates the four-dimensional spaces $(\phi_{\vec{k}},\phi_{-\vec{k}},\pi_{\vec{k}},\pi_{-\vec{k}})$ and $(\mathfrak{q}_{\vec{k}},\mathfrak{q}_{-\vec{k}},\mathfrak{p}_{\vec{k}},\mathfrak{p}_{-\vec{k}})$, is a canonical transformation. Moreover, we show that if a canonical transformation $\boldsymbol{M}_{{k}}$ is isotropic, \ie if it acts identically on $(\phi_{\vec{k}},\pi_{\vec{k}})$ and $(\phi_{-\vec{k}},\pi_{-\vec{k}})$, it is mapped through \Eqs{eq:Wigner:Canonical:Transform:phi}-\eqref{eq:Wigner:Canonical:Transform:pi} to a canonical transformation on $(\mathfrak{q}_{\vec{k}},\mathfrak{q}_{-\vec{k}},\mathfrak{p}_{\vec{k}},\mathfrak{p}_{-\vec{k}})$. Therefore, since the dynamics of $(\phi_{\vec{k}},\pi_{\vec{k}})$ and $(\phi_{-\vec{k}},\pi_{-\vec{k}})$ is generated by the same Green's matrix, it is an isotropic canonical transformation, and the dynamics on the four-dimensional space $(\mathfrak{q}_{\vec{k}},\mathfrak{q}_{-\vec{k}},\mathfrak{p}_{\vec{k}},\mathfrak{p}_{-\vec{k}})$ is also generated by a $4\times4$ Green's matrix. These mappings are summarised in \App{subapp:mappingcomplexreal}, in which it is also shown that they preserve many properties of the initial group $\mathrm{Sp}(2,\mathbb{R})$. For instance, the four-dimensional extended Green's matrices and the canonical transformations still have a unit determinant.

One of the reasons why the Wigner-Weyl transform is handy in the context of canonical transformations is that it transforms according to a simple rule: considering a canonical transformation encoded in the matrix $\boldsymbol{M}$, any operator $\widehat{\mathcal{A}}$  with Wigner-Weyl transforms $\mathcal{A}_{(W)}$ transform according to~\cite{PhysRevA.36.3868,SIMON1987223,Arvind:1995ab}
\bea
\label{eq:A:tranform:M}
	\widehat{\mathcal{A}}&\to&\widehat{\widetilde{\mathcal{A}}}=\widehat{\mathcal{M}}^\dag\widehat{\mathcal{A}}\widehat{\mathcal{M}}, \\
	\mathcal{A}_{(W)}(\boldsymbol{z})&\to&\widetilde{\mathcal{A}}_{(W)}(\boldsymbol{z})=\mathcal{A}_{(W)}\left[\boldsymbol{M}\boldsymbol{z}\right] .
\label{eq:AW:tranform:M}
\eea

For the specific case of the density matrix operator, $\widehat{\rho}(t):=\left|\Psi(t)\right>\left<\Psi(t)\right|$, the Wigner-Weyl transform is the well-known Wigner function \cite{1932PhRv...40..749W,doi:10.1119/1.2957889}, denoted $\mathcal{W}(\boldsymbol{z},t)$ in the following. The Wigner function can be interpreted as a {\it quasi-distribution function} in the sense that expectation values of operators can be expressed as (we work here in the Schr\"odinger representation)
\bea
	\left<\Psi(t)\right|\widehat{\mathcal{O}}\left|\Psi(t)\right>=\mathrm{Tr}\left[\widehat{\rho}(t)\widehat{\mathcal{O}}\right]=\ds\int\dd\boldsymbol{z}\mathcal{O}_{(W)}(\boldsymbol{z})\mathcal{W}(\boldsymbol{z},t). \label{eq:expectationweyl}
\eea
The density matrix evolves as $\widehat{\rho}(t)=\widehat{\mathcal{U}}(t,t_0)\widehat{\rho}(t_0)\widehat{\mathcal{U}}^\dag(t,t_0)$, where the evolution operator $\widehat{\mathcal{U}}$ corresponds to the classical canonical transformation generated by the Green's matrix solving for the classical evolution, $\boldsymbol{G}(t,t_0)$. Hence, from \Eq{eq:AW:tranform:M}, one can relate the evolved Wigner distribution with the initial one,
\bea
\label{eq:Wigner:evol}
\mathcal{W}\left(\boldsymbol{z},t\right)=\mathcal{W}\left[\boldsymbol{G}^{-1}(t,t_0)\boldsymbol{z},t_0\right]\, .
\eea
In the following, we will denote $\mathcal{W}_0(\boldsymbol{z}):=\mathcal{W}\left(\boldsymbol{z},t_0\right)$ the initial Wigner distribution, defined as the Wigner-Weyl transform of the initial density matrix. Combining the two previous relations, one obtains
\bea
	\left<\Psi(t)\right|\widehat{\mathcal{O}}\left|\Psi(t)\right>=\ds\int\dd\boldsymbol{z}\,\mathcal{O}_{(W)}\left[\boldsymbol{G}(t,t_0)\boldsymbol{z}\right]\,\mathcal{W}_0(\boldsymbol{z}), \label{eq:expectationevolvedweyl}
\eea
where we have performed a change of integration variables $\boldsymbol{z}\to\boldsymbol{G}^{-1}(t,t_0)\boldsymbol{z}$, noting that Green's matrices are unimodular since they are elements of $\mathrm{Sp}(2,\mathbb{R})$. The above is easily interpreted as follows: expectation values are obtained by averaging a function on the {\it evolved} phase space, with a measure given by the {\it initial} Wigner distribution. 

\paragraph{Expectation values of observables and canonical transformations --} Let us now compute the expectation value of $\widehat{\mathcal{O}}$ if phase space is parametrised with $\widetilde{\boldsymbol{z}}$ instead of $\boldsymbol{z}$,
\bea
	\left<\widetilde{\Psi(t)}\right|\widehat{\mathcal{O}}\left|\widetilde{\Psi(t)}\right>=\int\dd\widetilde{\boldsymbol{z}}\,\widetilde{\mathcal{O}}_{(W)}\left[\widetilde{\boldsymbol{G}}(t,t_0)\widetilde{\boldsymbol{z}}\right]\,\widetilde{\mathcal{W}}_0(\widetilde{\boldsymbol{z}}) .
\eea
In this expression, by construction, $\widetilde{\mathcal{O}}_{(W)}$ and $\mathcal{O}_{(W)}$ are related in the same way that $\widetilde{\mathcal{O}}$ and $\mathcal{O}$ are, \ie $\widetilde{\mathcal{O}}_{(W)}(\cdot) =  \mathcal{O}_{(W)}[\boldsymbol{M}^{-1}(t)\cdot]$. For the Green function, $\widetilde{\boldsymbol{G}}(t,t_0)$ is related to $\boldsymbol{G}(t,t_0)$ via \Eq{eq:Green:canonical:transformation}.  For the initial Wigner function, if the two sets of ladder operator are constructed in the same way, \ie if $\boldsymbol{D}_k=\boldsymbol{\widetilde{D}}_k$ in the notations of \Fig{fig:diagram}, selecting the vacuum state leads to a uniquely defined functional form for the Wigner function $\mathcal{W}_{\mathrm{vacuum}}(\cdot)$ One then has $\widetilde{\mathcal{W}}_0(\widetilde{\boldsymbol{z}}) =\mathcal{W}_0(\widetilde{\boldsymbol{z}})  =  \mathcal{W}_{\mathrm{vacuum}}(\widetilde{\boldsymbol{z}}) $. Notice that ${\mathcal{W}}_0(\cdot)$ and $\widetilde{\mathcal{W}}_0(\cdot)$ would also be identical (as functions of dummy parameters) if the prescription for the initial state was chosen differently, say as the state containing one particle in a certain mode $\vec{k}$, \etc. In fact, a prescription for the initial state exactly corresponds to a given functional form for the initial Wigner function. One then obtains
\bea
\label{eq:expectation:value:tilde:1}
	\left<\widetilde{\Psi(t)}\right|\widehat{\mathcal{O}}\left|\widetilde{\Psi(t)}\right>&=&\int\dd\boldsymbol{z}\,\mathcal{O}_{(W)}\left[\boldsymbol{G}(t,t_0)\boldsymbol{M}^{-1}(t_0){\boldsymbol{z}}\right]\,\mathcal{W}_0\left(\boldsymbol{z}\right)\\
	&=&\int\dd\boldsymbol{z}\,\mathcal{O}_{(W)}\left[\boldsymbol{G}(t,t_0){\boldsymbol{z}}\right]\,\mathcal{W}_0\left[\boldsymbol{M}(t_0)\boldsymbol{z}\right],
	\label{eq:expectation:value:tilde}
\eea
where in the second line we have performed the change of integration variable $\boldsymbol{z}\rightarrow \boldsymbol{M}(t_0){\boldsymbol{z}}$, and used the fact that $\det(\boldsymbol{M})=1$. Several remarks are in order. 

First, if the two sets of canonical variables initially coincide, $\boldsymbol{M}(t_0) =\boldsymbol{I} $, then \Eqs{eq:expectation:value:tilde} and~(\ref{eq:expectationevolvedweyl}) are identical, and the two expectation values are the same. This confirms that, if one starts from the same initial state, observable predictions do not depend on the choice of canonical variables.

Second, if the initial state is symmetrical with respect to the transformation $\boldsymbol{M}(t_0)$, namely if $\mathcal{W}_0\left[\boldsymbol{M}(t_0)\boldsymbol{z}\right]=\mathcal{W}_0\left(\boldsymbol{z}\right)$, then the same predictions are again obtained. This is for instance the case for the situation described in \Eq{eq:rotconstraint} where the initial state is the vacuum state and $\boldsymbol{M}(t_0)$ is a pure phase-space rotation. 

Third, if the two sets of canonical variables select out two different initial states, the differences in their observable predictions lies only in $\boldsymbol{M}(t_0)$ and does not involve $\boldsymbol{M}(t>t_0)$. In that case, working with the phase-space variables $\widetilde{{\boldsymbol{z}}}$ instead of ${\boldsymbol{z}}$ can either be viewed as working with a different phase-space measure, see \Eq{eq:expectation:value:tilde}, or as working with a different dynamical evolution, see \Eq{eq:expectation:value:tilde:1}.\\
\section{The invariant representation}
\label{ssec:invariant}
As we have made clear in \Sec{sec:Quantum}, different sets of canonical variables select out different vacuum states, hence they give rise to different predictions if initial conditions are to be set in the vacuum. One may therefore wonder to what extent universal predictions can be made within a given theory, if they depend on the way the system is parametrised. In this section, we show that there exists in fact one particular choice of canonical variables, for which the Hamiltonian becomes the one of a standard harmonic oscillator. This is dubbed the ``invariant representation'', and may be used as a well-defined reference canonical frame, in which uniquely defined initial conditions can be set.

We first show in \Sec{eq:sec:parametric:oscillator} how a generic quadratic action for a scalar field can be recast into the one of a standard parametric oscillator, and then explain in \Sec{sssec:invariant} how this can be, in turn, formulated in terms of a standard harmonic oscillator. In \Sec{sec:inflation:invariant:representation}, we illustrate the use of the invariant representation by solving for the dynamics of parametric oscillators. 

Let us note that hereafter, in order to prepare for the notations of \Sec{sec:inflation}, time is denoted $\eta$, and differentiation with respect to time is denoted with a prime. In this section however, $\eta$ is simply a dummy time variable and plays the same role as $t$ before.
\subsection{From arbitrary quadratic hamiltonians to parametric oscillators}
\label{eq:sec:parametric:oscillator}
Let us start from a generic quadratic action for a scalar field $\phi$,
\bea
	S_\phi=\frac{1}{2}\ds\int\dd \eta\dd^3k\left[\gamma^2_k(\eta)\left\vert\phi'_{\vec{k}}\right\vert^2-\mu^2_k(\eta)\left\vert \phi_{\vec{k}}\right\vert^2-\lambda_k(\eta)\left(\phi_{\vec{k}}{\phi'_{\vec{k}}}^*+\phi_{\vec{k}}^*\phi'_{\vec{k}}\right)\right], \label{eq:squad}
\eea
where $\gamma_k,~\mu_k$ and $\lambda_k$ are arbitrary time-dependent functions, and show that it can be recast into the form of a parametric oscillator
\bea
	S_v=\frac{1}{2}\ds\int\dd \eta\dd^3k\left[\left\vert v'_{\vec{k}}\right\vert^2-\omega^2_k(\eta)\left\vert v_{\vec{k}}\right\vert^2\right]. \label{eq:sho}
\eea
In \Refs{goldstein2002classical,Cervero2002}, this is done by introducing a specific generating function. Here, we obtain the result by means of canonical transformations only. The first step is to introduce the variable $v_{\vec{k}}=\gamma_k(\eta)\phi_{\vec{k}}$ in \Eq{eq:squad}. The second step consists in adding to the Lagrangian the following {\it total} derivative
\bea
\label{eq:total:derivative}
	\frac{\dd F}{\dd \eta}=\left[\frac{1}{2}\left(\frac{\gamma'_k}{\gamma_k}+\frac{\lambda_k}{\gamma_k^2}\right)\left\vert v_{\vec{k}}\right\vert^2\right]',
\eea
which does not change the equation of motion. This gives rise to the action~(\ref{eq:sho}), with the time-dependent frequency reading 
\bea
\label{eq:omega:from:squad}
	\omega^2_k(\eta)=\frac{\mu^2_k}{\gamma^2_k}-\frac{\gamma''_k}{\gamma_k}-\frac{\lambda'_k}{\gamma_k^2}.
\eea 

The canonical analysis of the action $S_\phi$ gives $\pi_{\vec{k}}=\gamma^2_k\phi'_{\vec{k}}-\lambda_k\phi_{\vec{k}}$. The Hamiltonian is then obtained from performing a Legendre transformation, and one gets
\bea
	H_\phi=\int_{\mathbb{R}^{3+}}\dd^3k\left[\frac{\left\vert\pi_{\vec{k}}\right\vert^2}{\gamma^2_k}+\left(\mu^2_k+\frac{\lambda_k^2}{\gamma^2_k}\right)\left\vert\phi_{\vec{k}}\right\vert^2+\frac{\lambda_k}{\gamma^2_k}\left(\phi_{\vec{k}}\pi_{\vec{k}}^*+\phi_{\vec{k}}^*\pi_{\vec{k}}\right)\right].
	\label{eq:Hquad}
\eea
Similarly, for the action $S_v$, one obtains $\Pi_{\vec{k}}=v'_{\vec{k}}$ with the Hamiltonian of a parametric oscillator
\bea
\label{eq:Hv}
	H_v=\int_{\mathbb{R}^{3+}}\dd^3k\left[\left\vert \Pi_{\vec{k}}\right\vert^2+\omega^2_k(\eta)\left\vert v_{\vec{k}}\right\vert^2\right].
\eea
The canonical transformation relating the two sets of canonical variables, \ie $\boldsymbol{v}_{\vec{k}}=\left(v_{\vec{k}}^*,\Pi^*_{\vec{k}}\right)^\dag$ as a function of $\boldsymbol{z}_{\vec{k}}=\left(\phi_{\vec{k}}^*,\pi_{\vec{k}}^*\right)^\dag$, simply reads
\bea
	\left(\begin{array}{c}
		v_{\vec{k}} \\
		\Pi_{\vec{k}}
	\end{array}\right)=\left(\begin{array}{cc}
		\gamma_k(\eta) & 0 \\
		\ds\gamma'_k(\eta)+\frac{\lambda_k(\eta)}{\gamma_k(\eta)} & \ds\frac{1}{\gamma_k(\eta)}
	\end{array}\right)\left(\begin{array}{c}
		\phi_{\vec{k}} \\
		\pi_{\vec{k}}
	\end{array}\right) . \label{eq:candiag}
\eea
This is the general canonical transformation that recasts any quadratic Hamiltonian, $H_\phi$, into the specific form of a time-dependent harmonic oscillator, $H_v$.\footnote{Starting directly from the most general quadratic Hamiltonian, 
\bea
	H_\phi=\frac{1}{2}\ds\int\dd^3k\left[T^2_k(\eta)\pi_{\vec{k}}^2+M^2_k(\eta)\phi^2_{\vec{k}}+2C_k(\eta)\pi_{\vec{k}}\phi_{\vec{k}}\right],
\eea
the canonical transformation leading to \Eq{eq:Hv} is obtained  by identifying $T_k(\eta)=1/\gamma_k(\eta)$, $M_k(\eta)=\sqrt{\mu_k^2(\eta)+\lambda^2_k(\eta)/\gamma^2_k(\eta)}$, and $C_k(\eta)=\lambda_k(\eta)/\gamma^2_k(\eta)$ in Eq. (\ref{eq:candiag}) [and similarly for the frequency $\omega_k(\eta)$].} This canonical transformation does not depend on the effective mass, $\mu_k$, of the initial configuration variable $\phi_k$.\footnote{Note also that matrices of the form encountered in \Eq{eq:candiag}, \ie
\bea
	\boldsymbol{M}=\left(\begin{array}{cc}
		f & 0 \\
		g & f^{-1}
	\end{array}\right)
\eea
with $f\neq0$ and $g\in\mathbb{R}$, form the (lower triangular) Borel subgroup of $\mathrm{Sp}(2,\mathbb{R})$. This specific subclass of linear canonical transformation will play an important role in the following, see \eg \Eq{eq:ztovutov}.}
\subsection{From parametric to harmonic oscillators}
\label{sssec:invariant}
Here we closely follow the approach developed in \Refs{PhysRevLett.18.510, PhysRevLett.18.636.2, doi:10.1063/1.1664532, Lewis:1968tm}. To lighten up notations, we consider a single degree of freedom, for which the Hamiltonian reads
\begin{equation}
\label{eq:H:parma:osc}
	\hat{H}=\frac{1}{2}\left[\hat{p}^2+\omega^2(\eta)\hat{q}^2\right] .
\end{equation}
Here, $(q,p)$ stands for the two real canonical variables, and $\omega(\eta)$ is the time-dependant, possibly complex-valued, frequency. The dynamics, either at the classical or quantum level, can be worked out by making use of a quadratic and selfadjoint invariant, $I(\eta)$, such that ~\cite{PhysRevLett.18.510,PhysRevLett.18.636.2,doi:10.1063/1.1664532}
\bea
\label{eq:invariant:def}
\frac{\dd \widehat{I}}{\dd \eta}=\frac{\partial\widehat{I}}{\partial{\eta}}-i\left[\widehat{I},\widehat{H}\right]=0\, .
\eea
Thanks to this invariant (an explicit expression will be given below), a complete quantum theory of parametric oscillators has been developed in \Refa{Lewis:1968tm} that we now briefly review. The goal is to express solutions to the Schr\"odinger equation,
\bea
i \frac{\partial}{\partial \eta}\left|\Psi(\eta)\right>_S = \hat{H}\left|\Psi(\eta)\right>_S\, ,
\eea
in terms of the eigenstates of the $\hat{I}$ operator,
\bea
\label{eq:invariant:eigenstates}
\hat{I}(\eta)\left\vert n,\kappa\right\rangle =n \left\vert n,\kappa\right\rangle\, ,
\eea
where the eigenvalues of  $\hat{I}(\eta)$ are denoted by $n$, $\left\vert n,\kappa\right\rangle$ are the eigenvectors associated with a given $n$, and $\kappa$ represents all of the quantum numbers other than $n$ that are necessary to specify the eigenstates. We choose the eigenstates to be orthonormalised, \ie $\left\langle \bar{n},\bar{\kappa}\right\vert\left. n,\kappa\right\rangle=\delta_{n,\bar{n}}\delta_{\kappa,\bar{\kappa}}$.

 A first remark is that, thanks to \Eq{eq:invariant:def}, if $\left|\Psi(\eta)\right>_S$ is a solution to the Schr\"odinger equation, then $\hat{I}(\eta)\left|\Psi(\eta)\right>_S$ is also a solution. Let us then show that the eigenvalues $n$'s are independent of time. By differentiating \Eq{eq:invariant:eigenstates} with respect to time, one has
\bea
\label{eq:invariant:rep:eom:1}
\frac{\partial\hat{ I}}{\partial \eta}\left\vert n,\kappa\right\rangle
+\hat{I} \frac{\partial}{\partial \eta}\left\vert n,\kappa\right\rangle
 = {n}' \left\vert n,\kappa\right\rangle+ n \frac{\partial}{\partial \eta}\left\vert n,\kappa\right\rangle\, .
\eea
By taking the scalar product of this expression with $\left\langle n,\kappa\right\vert$, one obtains
\bea
\label{eq:invariant:rep:ndot}
n' = \left\langle n,\kappa\right\vert\frac{\partial\hat{ I}}{\partial \eta}\left\vert n,\kappa\right\rangle\, .
\eea
The right-hand side of this expression can be evaluated by acting \Eq{eq:invariant:def} onto a state $\left\vert n,\kappa\right\rangle$,
\bea
\frac{\partial\widehat{I}}{\partial{\eta}}\left\vert n,\kappa\right\rangle
-i \hat{I} \hat{H} \left\vert n,\kappa\right\rangle
+i n   \hat{H} \left\vert n,\kappa\right\rangle
=0\, ,
\eea
and by taking the scalar product of this expression with $\left\langle \bar{n},\bar{\kappa}\right\vert$, which gives rise to
\bea
\label{eq:invariant:rep:eom:2}
\left\langle \bar{n},\bar{\kappa}\right\vert\frac{\partial\widehat{I}}{\partial{\eta}}\left\vert n,\kappa\right\rangle=
i \left(\bar{n}-n\right) \left\langle \bar{n},\bar{\kappa}\right\vert \hat{H} \left\vert n,\kappa\right\rangle\, .
\eea
When $n=\bar{n}$, this formula indicates that the right-hand side of \Eq{eq:invariant:rep:ndot} vanishes, hence $n$ is indeed constant. 

Let us now see how the eigenstates of $\hat{I}$ evolve in time. By taking the scalar product of \Eq{eq:invariant:rep:eom:1} with $\left\langle \bar{n},\bar{\kappa}\right\vert$ and using \Eq{eq:invariant:rep:eom:2}, one finds
\bea
\label{eq:invariant:rep:eom:3}
i \left(\bar{n}-n\right) \left\langle \bar{n},\bar{\kappa}\right\vert \hat{H} \left\vert n,\kappa\right\rangle
 =  \left(n-\bar{n}\right) \left\langle \bar{n},\bar{\kappa}\right\vert\frac{\partial}{\partial \eta}\left\vert n,\kappa\right\rangle\, ,
\eea
where we have used that $n'=0$. Two different situations arise whether $n=\bar{n}$ or not. If $n\neq \bar{n}$, then \Eq{eq:invariant:rep:eom:3} implies that $ i \left\langle \bar{n},\bar{\kappa}\right\vert\frac{\partial}{\partial \eta}\left\vert n,\kappa\right\rangle=  \left\langle \bar{n},\bar{\kappa}\right\vert \hat{H} \left\vert n,\kappa\right\rangle
$. If $n=\bar{n}$, nothing can be deduced from \Eq{eq:invariant:rep:eom:3}.\footnote{If the relation  $ i \left\langle \bar{n},\bar{\kappa}\right\vert\frac{\partial}{\partial \eta}\left\vert n,\kappa\right\rangle=  \left\langle \bar{n},\bar{\kappa}\right\vert \hat{H} \left\vert n,\kappa\right\rangle$ held also for $n=\bar{n}$, then $\left\vert n,\kappa\right\rangle$ would satisfy the Schr\"odinger equation and our task would be accomplished.} Let us however notice that the phase of the eigenvectors $\left\vert n,\kappa\right\rangle$ has been left unspecified so far. Let us then introduce a new set of eigenvectors of $\hat{I}(\eta)$ related to the current one by
\bea
\label{eq:eigenstates:I:time:dependent}
\left\vert n,\kappa\right\rangle_\vartheta = \ee^{i \vartheta_{n,\kappa}\left(\eta\right)}\left\vert n,\kappa\right\rangle\, ,
\eea
where $\vartheta_{n,\kappa}(\eta)$ are arbitrary free functions of time. All the properties derived above for $\left\vert n,\kappa\right\rangle$ also hold for $\left\vert n,\kappa\right\rangle_\vartheta$. However, $\vartheta_{n,\kappa}$ can be chosen such that the relation $ i \left\langle \bar{n},\bar{\kappa}\right\vert\frac{\partial}{\partial \eta}\left\vert n,\kappa\right\rangle=  \left\langle \bar{n},\bar{\kappa}\right\vert \hat{H} \left\vert n,\kappa\right\rangle$ also hold for $n=\bar{n}$, which is the case provided 
\bea
{\vartheta}_{n,\kappa}^\prime \delta_{\kappa,\bar{\kappa}}=\left\langle n,\bar{\kappa}\right\vert i \frac{\partial}{\partial \eta}-\hat{H}\left\vert n,\kappa\right\rangle\, .
\eea
When $\kappa\neq\bar{\kappa}$, the states $\left\vert n,\kappa\right\rangle$ must be chosen such that the right-hand side of the above vanishes, which is always possible since the operator $i \frac{\partial}{\partial \eta}-\hat{H}$ is Hermitian. One is then simply left with a first-order differential equation for the phases, 
\bea
\label{eqphase}
{\vartheta}_{n,\kappa}^\prime =\left\langle n,\kappa\right\vert i \frac{\partial}{\partial \eta}-\hat{H}\left\vert n,\kappa\right\rangle\, ,
\eea
which can always be solved, and which guarantees that the states $\left\vert n,\kappa\right\rangle_\vartheta$ are solutions to the Schr\"odinger equation. A generic solution can thus be decomposed according to
\bea
\label{ItoS}
\left|\Psi(\eta)\right>_S = \sum_{n,\kappa} c_{n,\kappa}\ee^{i\vartheta_{n,\kappa}(\eta)}\left\vert n,\kappa\right\rangle\, .
\eea
Resolving the quantum dynamics then proceeds along three steps: first, find a hermitian invariant, second, quantise it, and third, make use of the two above equations.

For the parametric oscillator~(\ref{eq:H:parma:osc}), a quadratic invariant has been proposed in \Refa{Lewis:1968tm}, 
\begin{equation}
\label{eq:invariant:quadratic:hamiltonian}
	\widehat{I}(\eta)=\frac{1}{2}\left[\frac{{\widehat{q}}^2}{\rho^2}+\left(\rho\widehat{p}-\rho'\widehat{q}\right)^2\right],
\end{equation}
where $\rho$ is a real function, solution of the non-linear, differential equation 
\begin{equation}
	\rho''+\omega^2(\eta)\rho=\frac{1}{\rho^3}
	\label{eqrho}
\end{equation}
that is obtained by plugging \Eq{eq:invariant:quadratic:hamiltonian} into \Eq{eq:invariant:def}. Two remarks are in order at this stage. 

First, one may question the relevance of the present approach since it requires to solve \Eq{eqrho}, which is more complicated than the usual differential equation involved in the second quantisation approach of parametric oscillators and given by 
\begin{equation}
	\ddot{q}_{\mathrm{cl}}+\omega^2(\eta)q_{\mathrm{cl}}=0\, .
	\label{eqqcl}
\end{equation}
However, in \App{app:rho}, it is shown how to express the solutions \Eq{eqrho} in terms of the solutions of \Eq{eqqcl}, such that solving the two differential equations is technically equivalent (we also refer the interested reader to the original articles on the subject, see \Refs{PhysRevLett.18.510,PhysRevLett.18.636.2,doi:10.1063/1.1664532,Lewis:1968tm}).

Second, since there is an infinite set of solutions to \Eq{eqrho}, an infinite set of invariants can be constructed. However, the observable predictions derived in this formalism are obviously independent of the choice of the invariant~\cite{Lewis:1968tm}. Besides, as will be made clearer when studying quasi-classical states below, there is a single prescription for the choice of $\rho$ that leads to a natural interpretation of the Schr\"odinger states.

Let us now quantise the $\widehat{I}$ invariant. This is done by performing the canonical transformation $\widehat{q}\to\widehat{Q}=\widehat{q}/\rho$ and $\widehat{p}\to\widehat{P}=\rho\widehat{p}-\rho'\widehat{q}$, which leads to a standard harmonic oscillator structure for the invariant,
\begin{equation}
\label{eq:I:P:Q}
	\widehat{I}(t)=\frac{1}{2}\left({\widehat{P}}^2+{\widehat{Q}}^2\right),
\end{equation}
with $\left[\widehat{Q},\widehat{P}\right]=i$. It is worth mentioning that the above invariant is {\it not} the Hamiltonian generating the dynamics of the new variables $(Q,P)$. Using \Eq{eq:Hamiltonian:canonical:transform}, one instead obtains that the Hamiltonian for the $(Q,P)$ variables is given by $\widehat{H}_{(Q,P)}=\widehat{I}/\rho^2(\eta)$, where \Eq{eqrho} has been employed. The invariant can then be quantised making use of creation and annihilation operators, defined as
\begin{eqnarray}
\label{eq:d:QP:def}
	\widehat{d}&=&\frac{1}{\sqrt{2}}\left(\widehat{Q}+i\widehat{P}\right), \\
	\widehat{d}^\dag&=&\frac{1}{\sqrt{2}}\left(\widehat{Q}-i\widehat{P}\right),
\label{eq:d:dagger:QP:def}
\end{eqnarray}
which satisfy the canonical commutation relation at any time. As a function of the creation and annihilation operators, the invariant reads as
\bea
	\widehat{I}(t)=\left(\widehat{d}^\dag\widehat{d}+\frac{1}{2}\right).
\eea
Its eigenstates being non-degenerate, the label $\kappa$ can be dropped and they can be simply denoted $\left| n\right>$, with $n=\left<n\right|d^\dag{d}\left|n\right>$ (and we recall that $n$ is time-independent, see the discussion below \Eq{eq:invariant:rep:ndot}). They define a Fock space, and the associated wave functions are obtained~\cite{Pal:2011xx, Balajani:2018qnn} by using the same approach as for the time-independent harmonic oscillator, as explained in \App{app:wave}. Plugging those states into \Eq{eqphase} gives rise to the following phases:
\bea
\label{eq:vartheta:sol}
	\vartheta_n(\eta)=-\left(n+\frac{1}{2}\right)\displaystyle\int^\eta\frac{d\bar{\eta}}{\rho^2(\bar{\eta})}.
\eea
The Schr\"odinger states are finally given by \Eq{ItoS} where the $c_n$ constants, being time-independent, are derived from the initial conditions $c_ne^{i\vartheta_n(-\infty)}=\left<n\right|\left.\Psi(-\infty)\right>_S$. 
\subsection{Dynamics of parametric oscillators in the invariant representation}
\label{sec:inflation:invariant:representation}
The invariant representation provides us with a systematic way to study the dynamics generated by quadratic Hamiltonians. The above approach can indeed be reformulated in terms of canonical transformations, which allows us to generalise the prescriptions for initial conditions usually studied in the literature to non-necessarily adiabatic vacua (see \App{app:iniinv}). As an illustration, let us consider the case of a set of parametric oscillators (describing \eg the Fourier modes of a scalar field), where the Hamiltonian is of the form~(\ref{eq:Hv}) (as will be explained in \Sec{sec:inflation}, this can be directly applied to free scalar fields in inflationary space-times, see footnote~\ref{footnote:inflation:invariant:representation} below). Following the approach of \Sec{sssec:invariant}, the first step is to introduce the canonical transformation
\bea
\label{eq:Q:S}
	\boldsymbol{Q}_{\vec{k}}=\boldsymbol{R}(\vartheta_{\vec{k}})\underbrace{\left(\begin{array}{cc}
		1/\rho_{\vec{k}} & 0 \\
		-\rho'_{\vec{k}} & \rho_{\vec{k}}
	\end{array}\right)}_{\boldsymbol{S}_{\vec{k}}}\boldsymbol{v}_{\vec{k}},
\eea
where the $\rho_{\vec{k}}$'s are real-valued functions and solutions of \Eq{eqrho} (where $\omega$ needs to be replaced with $\omega_{\vec{k}}$). The angle $\vartheta_{\vec{k}}$ will be chosen later on to further simplify the resulting Hamiltonian.  Using $\boldsymbol{R}^\mathrm{T}(\vartheta)=\boldsymbol{R}^{-1}(\vartheta)=\boldsymbol{R}(-\vartheta)$ and $\boldsymbol{R}'(\vartheta)=-\vartheta'\boldsymbol{\Omega}\boldsymbol{R}(\vartheta)$, see \Eq{eq:R(alpha):real}, and introducing the inverse of $\boldsymbol{S}_{\vec{k}}$, \ie
\bea
\label{eq:invariant:rep:S:inverse}
\boldsymbol{S}^{-1}_{\vec{k}}=\left(\begin{array}{cc}
	\rho_{\vec{k}} & 0 \\
	\rho'_{\vec{k}} & 1/\rho_{\vec{k}}
\end{array}\right),
\eea
the Hamiltonian kernel for the new canonical variables, $\boldsymbol{Q}_{\vec{k}}$, can be obtained from \Eq{eq:Hamiltonian:canonical:transform} and reads
\bea
	\boldsymbol{H}_{Q_{\vec{k}}}&=&\boldsymbol{R}\left(\vartheta_{\vec{k}}\right)\left[\left(\boldsymbol{S}_{\vec{k}}^{-1}\right)^\mathrm{T}\boldsymbol{H}_{v_{\vec{k}}}\boldsymbol{S}_{\vec{k}}^{-1}+\left(\boldsymbol{S}_{\vec{k}}^{-1}\right)^\mathrm{T}\boldsymbol{\Omega}\frac{\dd\boldsymbol{S}_{\vec{k}}^{-1}}{\dd\eta}-\vartheta_{\vec{k}}'\boldsymbol{I}\right]\boldsymbol{R}(-\vartheta_{\vec{k}})\, .
\eea
In this expression, we have used the fact that $\boldsymbol{S}_{\vec{k}}$ is a symplectic matrix, hence $\boldsymbol{S}_{\vec{k}}^\mathrm{-1}$ is a symplectic matrix and $(\boldsymbol{S}_{\vec{k}}^\mathrm{-1})^\mathrm{T}\boldsymbol{\Omega}\boldsymbol{S}_{\vec{k}}^\mathrm{-1}=\boldsymbol{S}_{\vec{k}}^\mathrm{-1}$, see \Eq{eq:symp}; and that $\boldsymbol{\Omega}^2=-\boldsymbol{I}$ as can be checked from \Eq{eq:Omega:def}.
Making use of \Eq{eqrho}, one can show that the term inside the brackets reads $\left(1/\rho_{\vec{k}}^2-\vartheta_{\vec{k}}'\right)\boldsymbol{I}$. The Hamiltonian for the variables $Q$ and $P$ hence simplifies to
\bea
	H_Q=\int_{\mathbb{R}^{3+}}\dd^3k\left(\frac{1}{\rho^2_{\vec{k}}}-\vartheta'_{\vec{k}}\right)\left(\left\vert Q_{\vec{k}}\right\vert ^2+\left\vert P_{\vec{k}}\right\vert^2\right)\, .
\eea 
A convenient choice for the angle is such that $1/\rho^2_{\vec{k}}-\vartheta'_{\vec{k}}$ is a non-vanishing constant, say $1/\rho^2_{\vec{k}}-\vartheta'_{\vec{k}}=\lambda_{\vec{k}}$. The angle then reads 
\bea
\vartheta_{\vec{k}}(\eta)=-\lambda_{\vec{k}}(\eta-\eta_0)+\int^\eta_{\eta_0}\frac{\dd\bar{\eta}}{\rho^2_{\vec{k}}(\bar{\eta})}+\vartheta_{\vec{k}}(\eta_0),
\label{eq:vartheta:lambda}
\eea
and the dynamics reduces to the one of a standard (\ie time-independent) harmonic oscillator.  Choosing for instance $\lambda_{\vec{k}}=1$ for all wavenumbers, one can see that the Hamiltonian reduces to the invariant $I$ introduced in \Sec{sssec:invariant}. Alternatively, one can choose $\lambda_{\vec{k}}=k$, which makes the Hamiltonian describe a collection of harmonic oscillators with frequency $k$. Finally, if one sets $\lambda_{\vec{k}}=0$, the Hamiltonian $H_Q$ vanishes, and $Q$ and $P$ become integrals of motion. This is at the basis of the so-called ``canonical perturbation theory''~\cite{goldstein2002classical}.

The quantisation then proceeds as follows. One first defines the Fock states associated to $\widehat{H}_Q$ by introducing the creation and annihilation operators~(\ref{eq:d:QP:def}) and (\ref{eq:d:dagger:QP:def}) [where $\hat{d}$ needs to be replaced with $\hat{d}_{\vec{k}}$ and $\hat{d}^\dagger$ with $\hat{d}_{-\vec{k}}^\dagger$, see \Eq{eq:a:components}], which with matrix notations read $\boldsymbol{\widehat{d}}_{\vec{k}}=\boldsymbol{U\widehat{Q}}_{\vec{k}}$. The Hamiltonian is now given by 
\bea
\widehat{H}_Q=\int_{\mathbb{R}^{3+}}\dd^3k \frac{\lambda_{\vec{k}}}{2}\left(\widehat{d}^\dag_{\vec{k}}\widehat{d}_{\vec{k}}+\widehat{d}_{-\vec{k}}\widehat{d}^\dag_{-\vec{k}}\right).
\eea
The Heisenberg equation~(\ref{eq:Heisenberg}) for these operators gives $ (\widehat{d}_{\vec{k}})^\prime = -i \lambda_{\vec{k}} \widehat{d}_{\vec{k}}$ and $ ({\widehat{d}_{\vec{k}}^\dagger})^\prime = i \lambda_{\vec{k}} \widehat{d}_{\vec{k}}^\dagger$, where we have used the canonical commutation relations~(\ref{eq:commutator:hat:a}). One thus obtains $\widehat{d}_{\vec{k}}(\eta)=e^{-i\lambda_{\vec{k}}(\eta-\eta_0)}\widehat{d}_{\vec{k}}(\eta_0)$ and $\widehat{d}^\dag_{\vec{k}}(\eta)=e^{i\lambda_{\vec{k}}(\eta-\eta_0)}\widehat{d}^\dag_{\vec{k}}(\eta_0)$. Otherwise stated, the Green's matrix, introduced in \Eq{eq:a:evol:Green}, is given by a rotation operator with the angle $-\lambda_{\vec{k}}(\eta-\eta_0)$, see \Eq{eq:polar:RS:explicit}. Making use of \Eq{eq:Uevol:def}, the evolution of states in the Schr\"odinger picture is thus generated by
\bea
	\widehat{\mathcal{U}}\left(\eta,\eta_0\right)=\exp\left[{-i\int_{\mathbb{R}^{3+}}\dd^3k\lambda_{\vec{k}}(\eta-\eta_0)(\widehat{d}^\dag_{\vec{k}}\widehat{d}_{\vec{k}}+\widehat{d}^\dag_{-\vec{k}}\widehat{d}_{-\vec{k}}+1)}\right]
	\label{eq:Qevol}
\eea
where, hereafter, the creation and annihilation operators are the initial ones.

Denoting by $\left|\cdots n_{\vec{k}}\cdots(\eta_0)\right>_Q$ the basis of the Fock space defined by the initial creation and annihilation operators, the evolution of these states is thus given by $\left|\cdots n_{\vec{k}}\cdots(\eta)\right>_Q = \widehat{\mathcal{U}}\left(\eta,\eta_0\right)\left|\cdots n_{\vec{k}}\cdots(\eta_0)\right>_Q$. Any other choice for the basis of the Hilbert space can then be generated from the above expressions. Suppose for instance that the initial states are chosen according to some creation and annihilation operators defined by $\widehat{\boldsymbol{a}}_{\vec{k}}=\boldsymbol{U}\boldsymbol{D}_{\vec{k}}\hat{\boldsymbol{v}}_{\vec{k}}$, where we use the notation introduced in \Eq{eq:a(z)}. These operators are related to $\widehat{d}_{\vec{k}}$ and $\widehat{d}^\dag_{-\vec{k}}$ by a canonical transformation,  $\widehat{\boldsymbol{a}}_{\vec{k}}=\boldsymbol{\mathcal{M}}_{\vec{k}} \widehat{\boldsymbol{d}}_{\vec{k}}$, generated by the matrix $\boldsymbol{\mathcal{M}}_{\vec{k}}=\boldsymbol{UD}_{\vec{k}}\boldsymbol{S}^{-1}_{\vec{k}}\boldsymbol{R}(-\vartheta_{\vec{k}})\boldsymbol{U}^\dag$, as explained in \Eq{eq:calMk:def}. The $\boldsymbol{v}$-states are then related to the $\boldsymbol{Q}$-states using the operator associated to the canonical transformation $\boldsymbol{\mathcal{M}}_{\vec{k}}$, \ie $\left|\Psi(t)\right>_v=\widehat{\mathcal{M}}_{\vec{k}}\left|\Psi(t)\right>_Q$, see \Eq{eq:Schrodinger:picture}. Suppose that the matrices are initially chosen such as $\boldsymbol{\mathcal{M}}_{\vec{k}}(\eta_0)$ is the identity matrix. This is obtained by an appropriate choice for the initial function $\rho_{\vec{k}}(\eta_0)$ and the angle $\vartheta_{\vec{k}}(\eta_0)$ given the matrix $\boldsymbol{D}_{\vec{k}}(\eta_0)$, as explicitly shown in \App{app:iniinv}.\footnote{We note that the arbitrary parameter $\lambda_{\vec{k}}$ is not involved in making $\boldsymbol{\mathcal{M}}_{\vec{k}}(\eta_0)$ equal to $\boldsymbol{I}$. The explicit expressions derived in \App{app:iniinv} show that neither $\rho_{\vec{k}}(\eta_0)$, $\rho'_{\vec{k}}(\eta_0)$, nor $\vartheta_{\vec{k}}(\eta_0)$ depend on $\lambda_{\vec{k}}$.\label{footnote:ind:lambda}} The creation and annihilation operators then coincide, \ie $\widehat{\boldsymbol{a}}_{\vec{k}}=\widehat{\boldsymbol{d}}_{\vec{k}}$, and the initial states in the Hilbert space are equal, $\left|\cdots n_{\vec{k}}\cdots(\eta_0)\right>_v=\left|\cdots n_{\vec{k}}\cdots(\eta_0)\right>_Q$ (below we drop the subscripts $Q$ and $v$ since the two {\it initial} states are identical, which we stress does not depend on the choice of $\lambda_{\vec{k}}$, see footnote~\ref{footnote:ind:lambda}). Since we work in the Schr\"odinger picture, the equality between the two sets of creation and annihilation operators holds at any time. The canonical transformation can be written as $\widehat{\mathcal{M}}_{\vec{k}}=\widehat{\mathcal{N}}_{\vec{k}}(\rho_{\vec{k}})\widehat{\mathcal{R}}_{\vec{k}}(-\vartheta_{\vec{k}})$ where $\vartheta_{\vec{k}}$ is given by \Eq{eq:vartheta:lambda}, with
\bea
	\widehat{\mathcal{R}}_{\vec{k}}(-\vartheta_{\vec{k}})=\exp\left\{i\ds\int_{\mathbb{R}^{3+}}\dd^3\vec{k}\left[\lambda_{\vec{k}}(\eta-\eta_0)-\vartheta_{\vec{k}}(\eta_0)-\int^\eta_{\eta_0}\frac{\dd\bar{\eta}}{\rho^2_{\vec{k}}(\bar{\eta})}\right]\left(\widehat{d}^\dag_{\vec{k}}\widehat{d}_{\vec{k}}+\widehat{d}^\dag_{-\vec{k}}\widehat{d}_{-\vec{k}}+1\right)\right\},
	\nonumber\\
\eea
see \Eq{eq:quantum:operator:R}, and where $\widehat{\mathcal{N}}_{\vec{k}}(\rho_{\vec{k}})$ is the quantum operator representing the canonical transformation generated by $\boldsymbol{\mathcal{N}}_{\vec{k}}=\boldsymbol{UD}_{\vec{k}}\boldsymbol{S}^{-1}_{\vec{k}}\boldsymbol{U}^\dag$. Let us stress that one can equivalently use the operators $\widehat{\boldsymbol{a}}_{\vec{k}}$ in that expression. The arbitrary parameter $\lambda_{\vec{k}}$ is involved in $\widehat{\mathcal{R}}_{\vec{k}}$ but not in $\widehat{\mathcal{N}}_{\vec{k}}$, since none of the matrices $\boldsymbol{D}_{\vec{k}}$ and $\boldsymbol{S}_{\vec{k}}$ depend on it. Using \Eq{eq:Qevol}, the dynamics of the $\boldsymbol{v}$-states thus reads
\bea
	\left|\Psi(\eta)\right>_v=\left\{\ds\prod_{\vec{k}\in\mathbb{R}^{3+}}\widehat{\mathcal{N}}_{\vec{k}}(\rho_{\vec{k}})\exp\left[-i\left(\vartheta_{\vec{k}}(\eta_0)+\int^\eta_{\eta_0}\frac{\dd\bar{\eta}}{\rho^2_{\vec{k}}(\bar{\eta})}\right)\left(\widehat{d}^\dag_{\vec{k}}\widehat{d}_{\vec{k}}+\widehat{d}^\dag_{-\vec{k}}\widehat{d}_{-\vec{k}}+1\right)\right]\right\}\left|\Psi(\eta_0)\right>\, ,
	\nonumber \\ \label{eq:Vevol}
\eea
where the arbitrary parameter $\lambda_{\vec{k}}$ has now totally disappeared. Since the initial state can be chosen independently of $\lambda_{\vec{k}}$, the outcome of the calculation in the invariant representation is therefore independent of these arbitrary integration constants, as it should.

Let us briefly interpret the above results in terms of particles content. Following \App{sec:connect:harmonic:oscillators}, we consider the case where the time-dependent frequency is constant in both the asymptotic past  and the asymptotic future, with the same asymptotic value that (without loss of generality) we set equal to $k$, \ie $\omega_k(\eta\to-\infty)=\omega_k(\eta\to+\infty)=k$. In the infinite past, it is thus convenient to define the operators $\widehat{\boldsymbol{a}}_{\vec{k}}$ associated to $\boldsymbol{v}_{\vec{k}}$ using $\boldsymbol{D}_k=\mathrm{diag}\left(\sqrt{k},1/\sqrt{k}\right)$.  For $\omega_k=k$, an obvious peculiar solution to \Eq{eqrho} is $\rho_k=1/\sqrt{k}$. Choosing $\rho_k(\eta\to-\infty)=1/\sqrt{k}$ leads to $\boldsymbol{S}_{\vec{k}}(\eta\to-\infty)=\boldsymbol{D}_k$, see \Eq{eq:Q:S}, hence $\boldsymbol{\mathcal{N}}_{\vec{k}}(\eta\to-\infty)=\boldsymbol{I}$ and this guarantees that in the infinite past, the two sets of creation and annihilation operators only differ by a rotation. Since vacuum states are rotationally invariant (see the discussion of \Sec{sec:EquivalentRepresentations}), this implies that the initial vacuum in terms of $\boldsymbol{v}$-quantas, dubbed $\left|\slashed{0}(-\infty)\right>_v$, equals the vacuum of $\boldsymbol{Q}$-quantas, $\left|\slashed{0}(-\infty)\right>_Q$. The subscript labelling canonical variables can thus be dropped from the initial state, \ie we use the notation $\left|\slashed{0}(-\infty)\right>$.

The evolution of $\boldsymbol{Q}$-states is given by \Eq{eq:Qevol}. For the vacuum state, this leads to $\left|\slashed{0}(\eta)\right>_Q=e^{-i\int\dd^3k\lambda_{\vec{k}}(\eta-\eta_0)}\left|\slashed{0}(-\infty)\right>$. As expected for a quantum harmonic oscillator (which here describes the evolution of $\boldsymbol{Q}$-states), the particle content is unchanged by the evolution. For the $\boldsymbol{v}$-states, the evolution~(\ref{eq:Vevol}) leads to
\bea
	\left|\slashed{0}(\eta)\right>_v=\left(\ds\prod_{\vec{k}\in\mathbb{R}^{3+}}\exp\left\lbrace-i\left[\vartheta_{\vec{k}}(\eta_0)+\int^\eta_{-\infty}\frac{\dd\bar{\eta}}{\rho^2_{\vec{k}}(\bar{\eta})}\right]\right\rbrace\widehat{\mathcal{N}}_{\vec{k}}(\rho_{\vec{k}})\right)\left|\slashed{0}(-\infty)\right>.
\eea
As shown in \App{app:rho}, the function $\rho_k$ in the infinite future evolves towards $\rho_k(\eta\to+\infty)=\left[\sqrt{1+A^2}+A^2\sin(2k\eta+\varphi)\right]/\sqrt{k}$, where $A$ and $\varphi$ are two integration constants that depend on the details of the function $\omega_{\vec{k}}(\eta)$ at finite times. If $A=0$, $\rho_{\vec{k}}(\eta\to+\infty)=1/\sqrt{k}$, $\boldsymbol{S}_{\vec{k}}(\eta\to+\infty)=\boldsymbol{D}_k$, hence $\boldsymbol{\mathcal{N}}_{\vec{k}}(\eta\to-\infty)=\boldsymbol{I}$ and in the infinite future, the two sets of creation and annihilation operators only differ by a rotation. In that case the evolution is adiabatic: the $\boldsymbol{v}$-states in the asymptotic future match the $\boldsymbol{Q}$-states up to a phase, and the number of $\boldsymbol{v}$-quanta remains unchanged. If $A\neq 0$ however, the evolution is not adiabatic and the operator $\widehat{\mathcal{N}}(\rho_k)$ has a non-zero squeezing amplitude. In that case, the number of $\boldsymbol{v}$-quanta changes throughout the evolution and the final $\boldsymbol{v}$-states do not coincide with the final $\boldsymbol{Q}$-states anymore (a concrete example of this effect is given in \App{sec:InvariantRep:inf:plus:rad}).
\section{The example of a scalar field in an inflationary space-time}
\label{sec:inflation}
We now apply the formalism developed in the previous sections to the specific case of a free scalar field evolving in a spatially flat Friedmann-Lema\^itre-Robertson-Walker space-time, for which the metric is given by
\bea
\dd s^2= -\dd t^2 +a^2(t)\dd \vec{x}\cdot\dd\vec{x}\, ,
\eea
where $a(t)$ is the scale factor and $t$ is the cosmic time. In the following we will rather work with conformal time $\eta$, defined by $\dd t = a \dd\eta$, and primes will denote differentiation with respect to conformal time. Using the notations of \Sec{sec:classical:dynamics}, the Hamiltonian of a free scalar field $\phi$ reads
\bea
\label{eq:Hamiltonian:phi}
	H_\phi=\frac{1}{2}\ds\int\dd^3k\boldsymbol{z}^\dag_{\vec{k}}\left(\begin{array}{cc}
		k^2a^2+m^2a^4 & 0 \\
		 0 & 1/a^2
		\end{array}\right)\boldsymbol{z}_{\vec{k}}\, ,
\eea
where $\boldsymbol{z}^\dag_{\vec{k}}=\left(\phi^*_{\vec{k}},~\pi^*_{\vec{k}}\right)$ are the canonical variables introduced in \Eq{eq:def:zk}. Two other sets of canonical variables are of particular interest for what follows. The first one is 
\bea
	\boldsymbol{u}_{\vec{k}}=\left(\begin{array}{cc}
		a & 0 \\
		0 & 1/a
	\end{array}\right)\boldsymbol{z}_{\vec{k}}\, , \label{eq:ztou}
\eea
for which, using \Eq{eq:Hamiltonian:canonical:transform}, the Hamiltonian reads
\bea
\label{eq:Hu}
	H_u=\frac{1}{2}\ds\int\dd^3k\boldsymbol{u}^\dag_{\vec{k}}\left(\begin{array}{cc}
		k^2+m^2a^2 & a'/a \\
		 a'/a & 1
		\end{array}\right)\boldsymbol{u}_{\vec{k}}\, .
\eea
The second one is
\bea
	\boldsymbol{v}_{\vec{k}}=\left(\begin{array}{cc}
		a & 0 \\
		a' & 1/a
	\end{array}\right)\boldsymbol{z}_{\vec{k}}=\left(\begin{array}{cc}
		1 & 0 \\
		a'/a & 1
	\end{array}\right)\boldsymbol{u}_{\vec{k}}\, , \label{eq:ztovutov}
\eea
for which the Hamiltonian is given by\footnote{Let us note that the Hamiltonian~(\ref{eq:Hv:inf}) can be seen as the result of the application of the invariant representation technique presented in \Sec{eq:sec:parametric:oscillator}, where an arbitrary quadratic Hamiltonian is recast into the one of a parametric oscillator. Indeed, the Hamiltonian~(\ref{eq:Hamiltonian:phi}) can be written in the parametrisation~(\ref{eq:squad}) with $\gamma^2_k=a^2$, $\mu^2_k=k^2a^2+m^2a^4$, and $\lambda_k=0$, see \Eq{eq:Hquad}. Introducing $\boldsymbol{v}_{\vec{k}}$ as in \Eq{eq:ztovutov} is then equivalent to \Eq{eq:candiag}, and by adding the total derivative~(\ref{eq:total:derivative}), \Eq{eq:omega:from:squad} gives rise to the time-dependent frequency $\omega_k^2=k^2+m^2a^2-a''/a$. The techniques presented in \Sec{sec:inflation:invariant:representation} can then be used to study the system.\label{footnote:inflation:invariant:representation}}
\bea
\label{eq:Hv:inf}
	H_v=\frac{1}{2}\ds\int\dd^3k\boldsymbol{v}^\dag_{\vec{k}}\left(\begin{array}{cc}
		k^2+m^2a^2-a''/a & 0 \\
		 0 & 1
		\end{array}\right)\boldsymbol{v}_{\vec{k}}\, .
\eea
From a Lagrangian viewpoint, the action for $u_{\vec{k}}$ is obtained from the action for $\phi_{\vec{k}}$ by simply introducing $u_{\vec{k}}=a\phi_{\vec{k}}$, while the action for $v_{\vec{k}}$ is derived from the action of $u_{\vec{k}}$ by adding the total derivative $[(a'/a)u_{\vec{k}}^2]'$. We note that the configuration variable is the same for $\boldsymbol{u}_{\vec{k}}$ and $\boldsymbol{v}_{\vec{k}}$, but that their respective canonically conjugate momentum differ (see \Refs{Peter:2005hm, Martin:2007bw} for further analyses of the canonical transformation between $\boldsymbol{u}_{\vec{k}}$ and $\boldsymbol{v}_{\vec{k}}$).

These canonical variables are introduced for their formal analogy with the so-called Mukhanov-Sasaki variable and the curvature perturbation in the context of inflationary cosmological perturbations~\cite{Mukhanov:1990me}. Setting the mass $m$ of the scalar field to zero, and denoting by $\zeta_{\vec{k}}$ the curvature perturbation, the above is analogous to the dynamics of cosmological scalar perturbations by replacing $\phi_{\vec{k}}$ with $\zeta_{\vec{k}}$, and $a(\eta)$ with $z_{{}_\mathrm{S}}(\eta)=a(\eta)\sqrt{\epsilon_1(\eta)}$ where $\epsilon_1=1-(a'/a)'/(a'/a)^2$ is the first slow-roll parameter \cite{Schwarz:2001vv}. Similarly, the configuration variables associated to either $\boldsymbol{u}_{\vec{k}}$ or $\boldsymbol{v}_{\vec{k}}$ are analogous to the Mukhanov-Sasaki variable~\cite{Mukhanov:1981xt,doi:10.1143/PTPS.78.1} (with different canonically conjugate momentum). The correspondences between the different sets of variables are summarised in Table~\ref{tab:variables}.

\begin{table}
\begin{center}
\begin{tabular}{lc|ccc} \hline\hline
 	&& $\boldsymbol{z}_{\vec{k}}$ & $\boldsymbol{u}_{\vec{k}}$ &$\boldsymbol{v}_{\vec{k}}$  \\ \hline
	\bf{Free scalar field} &&&& \\
	\hspace*{0.5cm}Configuration variable && $\phi_{\vec{k}}$ & $u_{\vec{k}}=a\phi_{\vec{k}}$ & $v_{\vec{k}}=a\phi_{\vec{k}}$ \\
	\hspace*{0.5cm}Momentum variable && $\pi_{\vec{k}}=a^2\phi'_{\vec{k}}$ & $p_{\vec{k}}=u'_{\vec{k}}-(a'/a)u_{\vec{k}}$ & $\kern-4em \Pi_{\vec{k}}=v'_{\vec{k}}$\\
		\hspace*{0.5cm}  &&  & $\kern-3em =a\phi'_{\vec{k}}$ & $\kern1.3em=a\phi'_{\vec{k}}+a'\phi_{\vec{k}}$\\	
	&&&& \\
	\bf{Curvature perturbations} &&&& \\
	\hspace*{0.5cm}Configuration variable && $\zeta_{\vec{k}}$ & $u_{\vec{k}}=z_{{}_\mathrm{S}}\zeta_{\vec{k}}$ & $v_{\vec{k}}=z_{{}_\mathrm{S}}\zeta_{\vec{k}}$ \\
	\hspace*{0.5cm}Momentum variable && $\pi_{\vec{k}}=z_{{}_\mathrm{S}}^2\zeta'_{\vec{k}}$ & $p_{\vec{k}}=u'_{\vec{k}}-(z_{{}_\mathrm{S}}'/z_{{}_\mathrm{S}})u_{\vec{k}}$ & $\kern-4em\Pi_{\vec{k}}=v'_{\vec{k}}$ \\
	\hspace*{0.5cm}  &&   & $ \kern-3.2em=z_{{}_\mathrm{S}}\zeta'_{\vec{k}}$ & $\kern1.5em=z_{{}_\mathrm{S}}\zeta'_{\vec{k}}+z_{{}_\mathrm{S}}'\zeta_{\vec{k}}$ \\
	 \hline\hline
\end{tabular}
\caption{Sets of canonical variables used in the analysis of \Sec{sec:inflation}. If the mass of the field is set to zero and the scale factor $a$ is replaced with $z_{{}_\mathrm{S}}=a\sqrt{1-(a'/a)'/(a'/a)^2}$, a simple analogy can be drawn with the canonical variables commonly used in the study of curvature perturbations in an inflationary background, \ie the Mukhanov-Sasaki variable and the curvature perturbation.}
\label{tab:variables} 
\end{center}
\end{table}

\subsection{Choice of the initial state}
Let us now discuss how canonical transformations (or equivalently, the choice of canonical variables) may affect the initial quantum state and the resulting predictions. To this end, we consider the two sets of canonical variables $\boldsymbol{u}_{\vec{k}}$ and $\boldsymbol{v}_{\vec{k}}$. Usually, in this context, the two corresponding sets of creation and annihilation operators are defined through \Eq{eq:a(z):quantum} using the same symplectic matrix $\boldsymbol{D}_k=\mathrm{diag}(\sqrt{k},1/\sqrt{k})$, \ie
\bea
\label{eq:twovacua:def}
	\widehat{\boldsymbol{a}}_{\vec{k}}=\boldsymbol{U}\left(\begin{array}{cc}
		\sqrt{k} & 0 \\
		0 & \frac{1}{\sqrt{k}}
	\end{array}\right)\widehat{\boldsymbol{u}}_{\vec{k}}&~~~~\mathrm{and}~~~~& \widehat{\boldsymbol{d}}_{\vec{k}}=\boldsymbol{U}\left(\begin{array}{cc}
		\sqrt{k} & 0 \\
		0 & \frac{1}{\sqrt{k}}
	\end{array}\right)\widehat{\boldsymbol{v}}_{\vec{k}}\, .
\eea
These two sets of operators are related through the canonical transformation $\widehat{\boldsymbol{d}}_{\vec{k}}=\boldsymbol{\mathcal{M}}_k(\eta)\widehat{\boldsymbol{a}}_{\vec{k}}$, with $\boldsymbol{\mathcal{M}}_k$ given by \Eq{eq:calMk:def} where $\boldsymbol{{M}}_k$ can be read off from \Eq{eq:ztovutov}, leading to
\bea
	\boldsymbol{\mathcal{M}}_k(\eta)=\left(\begin{array}{cc}
		1+iaH/(2k) & iaH/(2k) \\
		-iaH/(2k) & 1-iaH/(2k)
	\end{array}\right)=\boldsymbol{I}-\left(\frac{aH}{2k}\right)\boldsymbol{J}_y+i\left(\frac{aH}{2k}\right)\boldsymbol{J}_z\, , \label{eq:canomatapp}
\eea
where $\boldsymbol{J}_y$ and $\boldsymbol{J}_z$ are the Pauli matrices defined in footnote~\ref{footnote:Pauli}. In this expression, we have introduced the Hubble parameter, $H=a'/a^2$. The matrix $\boldsymbol{\mathcal{M}}_k$ differs from the identity because the two sets of variables have different momenta, despite sharing the same configuration variable. The departure from the identity is controlled by the ratio $aH/k$, that is the ratio between the physical wavelength of the mode considered, $a/k$, and the Hubble radius, $H^{-1}$. The canonical transformation thus tends to the identity for sub-Hubble modes, \ie for wavenumbers such that $aH/k \ll 1$, but strongly departs from the identity for super-Hubble modes, for which $aH/k \gg 1$. 

Making use of the techniques introduced in \Sec{sec:SU11:General}, the matrix $\boldsymbol{\mathcal{M}}_k$ can be written in the left-polar decomposition of \Eq{eq:bloch:left:right}, $\boldsymbol{\mathcal{M}}_k=\boldsymbol{\mathcal{S}}(d_k,\varpi_k)\boldsymbol{\mathcal{R}}(\vartheta_k)$, with the squeezing amplitude, squeezing angle and rotation angle reading
\bea
	d_k&=&\ln\left[\frac{aH}{2k}+\sqrt{1+\left(\frac{aH}{2k}\right)^2}\right], \\
	\varpi_k&=&\frac{\pi}{4}+\frac{1}{2}\arctan\left(\frac{aH}{2k}\right), \\
	\vartheta_k&=&\arctan\left(\frac{aH}{2k}\right).
	\label{eq:vartheta:u:v}
\eea
For sub-Hubble modes, these parameters can be approximated by $d_k\simeq aH/(2k)$, $\varpi_k=\pi/4+aH/(4k)$, and $\vartheta_k=aH/(2k)$.

In an inflationary universe, the inverse of the comoving Hubble horizon behaves as $aH\simeq(-1+\epsilon_1)/\eta$, where $\epsilon_1\ll 1$ for quasi de Sitter (\ie quasi exponential) expansion. For an infinitely long inflationary phase (both in the past and in the future), the conformal time varies from $-\infty$ in the remote past, to $0^-$ in the asymptotic future. The canonical transformation~\eqref{eq:canomatapp} thus asymptotically tends to the identity matrix in the remote past, $\boldsymbol{\mathcal{M}}_k(\eta\to-\infty)\to\boldsymbol{I}$, since all modes are asymptotically sub-Hubble in that limit. Hence the initial vacuum states are the same using either the $\boldsymbol{u}_{\vec{k}}$ or $\boldsymbol{v}_{\vec{k}}$ canonical variables, if they are set in the infinite past, $\eta\to-\infty$. 

However, there are situations where setting initial conditions in the infinite past may not be possible. This is for instance the case if inflation does not last infinitely long but rather extends over a finite period of time. Another context in which this issue arises is the ``trans-Planckian problem'' (see \Refa{Brandenberger:2012aj} and references therein), namely the fact that if one traces inflation backwards in time, one inevitably meets a point where the physical wavelength of a given Fourier mode becomes smaller than the Planck length. Beyond that point, one enters the realm of quantum gravity where physics remains elusive. This is why one may prefer not to set initial conditions in that regime, but rather at its boundary, \ie at a finite time $\eta_0$ such that the physical wavelength of the mode under consideration equals the Planck length, \ie $a(\eta_0)/k=\ell_{{}_\mathrm{Pl}}$, or any other UV physical cutoff. This gives rise to a value for $\eta_0$ that depends on $k$, though one may prefer to set initial conditions for all modes at the same time, in which case one can choose $\eta_0$ to be the time when \eg the largest observable wavelength crosses out the UV scale. Let us also note that a trans-Planckian problem can exist at the background level as well, in models (such as ``large-field'' inflation) where the energy density grows above the Planck scale when one traces inflation backwards in time. In principle, beyond that point, quantum gravity effects come into play, and could for instance give rise to a bounce (see \eg \Refs{Brandenberger:2016vhg,Barrau:2013ula}). The onset of inflation would then occur at a time $\eta_0$ that depends on the details of the theory of quantum gravity (in loop quantum cosmology for example, it is roughly given by $\rho(\eta_0)\lesssim 10^2\Mp^4$ since for smaller energy densities, corrections to the modified Friedmann and Raychauduri equations becomes negligible). 

For all these reasons, we now consider the case where initial conditions are set at a finite time $\eta_0$, that may or may not depend on $k$. The operators $\widehat{\boldsymbol{a}}_{\vec{k}}(\eta_0)$ and $\widehat{\boldsymbol{d}}_{\vec{k}}(\eta_0)$ introduced above then define two different vacua, 
and working with the variables $\boldsymbol{u}_{\vec{k}}$ or the variables $\boldsymbol{v}_{\vec{k}}$ thus leads to different predictions. In order to illustrate this point, in what follows, we compute the two-point correlation functions of the $\boldsymbol{u}_{\vec{k}}$-field variables, if initial conditions are set in the vacuum state annihilated by the set of $\widehat{a}_{\vec{k}}(\eta_0)$ operators (dubbed the $\boldsymbol{u}$-vacuum), or in the vacuum state annihilated by the $\widehat{d}_{\vec{k}}(\eta_0)$ operators (dubbed the $\boldsymbol{v}$-vacuum). 

Before turning to this calculation, let us notice that these two vacua play the role for each other of the so-called generalised $\alpha$-vacua~\cite{Einhorn:2003xb}. More precisely, the $\boldsymbol{v}$-vacuum is an $\alpha$-vacuum from the viewpoint of states defined by the operators $\widehat{\boldsymbol{a}}_{\vec{k}}$~\cite{PhysRevD.31.754,PhysRevD.32.3136,Danielsson:2002kx,Danielsson:2002qh}, and the $\boldsymbol{u}$-vacuum is an $\alpha$-vacuum from the viewpoint of states defined by the operators $\widehat{\boldsymbol{d}}_{\vec{k}}$. These $\alpha$-vacua can be generically defined as follows. Starting from a given set of creation and annihilation operators, $\widehat{\boldsymbol{a}}_f$, which  create and annihilate a given mode function $f_{\vec{k}}(\eta,\vec{x})$ (for example~\cite{Danielsson:2002kx} the mode functions $f_{\vec{k}}(\eta,\vec{x})=e^{-ik\eta+i\vec{k}\vec{x}}(1-i/k\eta)/\sqrt{2k}$, corresponding to the Bunch-Davies vacuum in de Sitter space), the class of $\alpha$-vacua associated to $\widehat{\boldsymbol{a}}_f$ is obtained by operating canonical transformations at $\eta_0$ on the vacuum state defined by the $\widehat{a}_f$ operator. 
When doing so, any vacuum state can a priori be reached (since any two vacuum states are always related through a canonical transform), but in \Refa{Danielsson:2002kx}, only those minimising Heisenberg's uncertainty principle are considered (which amounts to fixing the squeezing angle of the canonical transform, see footnote~\ref{footnote:Heisenberg:Uncertainty:Principle} below), effectively selecting a sub-class of these vacua called $\alpha$-vacua. Besides, $\alpha$-vacua are such that the canonical transformation generating them is independent of $k$. In \Refa{Brandenberger:2012aj}, the case of more general Bogolyubov coefficients is considered, and in \Refa{Einhorn:2003xb}, $\alpha$-vacua are also extended to $k$-dependent canonical transformations. In what follows, we consider fully generic canonical transforms, and we use the langage of ``$\alpha$-vacua'' not for the specific vacuum states they describe when taken in the restricted sense, but rather for the techniques commonly employed to study them that will prove useful for the problem at hand. 

We thus first provide general results for $\alpha$-vacua, leaving the squeezing parameters and the rotation angle that characterise the canonical transformation unspecified. In that respect, the results presented in \Secs{sec:2pt:uvac} and~\ref{sec:2pt:alphavac} are generic: the $\boldsymbol{u}$-vacuum can be thought of as any selected vacuum of a quadratic Hamiltonian and the $\boldsymbol{v}$-vacuum as any $\alpha$-vacuum obtained from the $\boldsymbol{u}$-vacuum. In \Sec{sec:2pt:vvac}, the calculation is specified to the case of the $\boldsymbol{u}$- and $\boldsymbol{v}$-vacua defined above for a scalar field in de Sitter, and related through \Eq{eq:canomatapp}.

\subsubsection{Two-point correlations in the original vacuum}
\label{sec:2pt:uvac}
The two-point correlation functions of the $\boldsymbol{u}_{\vec{k}}$-field variables are encoded in the hermitic matrix 
\bea
\boldsymbol{\Sigma}_{\vec{k},\vec{q}}(\eta)=\left<\slashed{0}_u\right|\widehat{\boldsymbol{u}}_{\vec{k}}(\eta)\widehat{\boldsymbol{u}}^\dag_{\vec{q}}(\eta)\left|\slashed{0}_u\right>
\eea
where $\left|\slashed{0}_u\right>$ is the vacuum state defined according to the $\boldsymbol{u}_{\vec{k}}$ variables, \ie the state such that $\widehat{a}_{\vec{k}}(\eta_0)\left|\slashed{0}_u\right>=0$ for all $\vec{k}$, and where hereafter the Heisenberg picture is used. Since modes are uncoupled for a free field, one can write 
\bea
\boldsymbol{\Sigma}_{\vec{k},\vec{q}}(\eta)=\boldsymbol{\Sigma}_k(\eta)\delta^3\left(\vec{k}-\vec{q}\right).
\eea 
Denoting by $\boldsymbol{\mathcal{G}_u}(\eta,\eta_0)$ the Green's matrix associated to the evolution of the $\widehat{\boldsymbol{a}}_{\vec{k}}$-operators, see \Eq{eq:a:evol:Green}, \Eq{eq:a(z):quantum} leads to
\bea
\label{eq:Sigma:uvacuum:def}
		\boldsymbol{\Sigma}_k(\eta)=\boldsymbol{D}^{-1}_k\boldsymbol{U}^\dag\boldsymbol{\mathcal{G}_u}(\eta,\eta_0)\left<\slashed{0}_u\right|\widehat{\boldsymbol{a}}_{\vec{k}}(\eta_0)\widehat{\boldsymbol{a}}^\dag_{\vec{k}}(\eta_0)\left|\slashed{0}_u\right>\boldsymbol{\mathcal{G}_u}^\dag(\eta,\eta_0)\boldsymbol{U}\left(\boldsymbol{D}_k^{-1}\right)^{\dag}\, ,
\eea
where we remind that the matrix $\boldsymbol{D}_k$ does not depend on time here. An explicit computation leads to\footnote{From the definition~(\ref{eq:a:components}), the expectation values of $\widehat{\boldsymbol{a}}_{\vec{k}}(\eta_0)\widehat{\boldsymbol{a}}^\dag_{\vec{k}}(\eta_0)$ on the $\boldsymbol{u}$-vacuum can be readily calculated and one finds
\bea
	\left<\slashed{0}_u\right|\widehat{\boldsymbol{a}}_{\vec{k}}(\eta_0)\widehat{\boldsymbol{a}}^\dag_{\vec{k}}(\eta_0)\left|\slashed{0}_u\right>=\left(\begin{array}{cc}
		1 & 0 \\
		0 & 0
	\end{array}\right).
\label{eq:aa:u}
\eea} 
\bea
	\boldsymbol{\Sigma}_k(\eta)&=&\frac{1}{2k}\left|\alpha_k^{(u)}+\beta_k^{(u)*}\right|^2\boldsymbol{I}_++\frac{k}{2}\left|\alpha_k^{(u)}-\beta_k^{(u)*}\right|^2\boldsymbol{I}_- +\mathrm{Im}\left[\alpha^{(u)}_k\beta^{(u)}_k\right]\boldsymbol{J}_x-\frac{1}{2}\boldsymbol{J}_y\, ,
\eea
where $\alpha_k^{(u)}(\eta)$ and $\beta^{(u)}_k(\eta)$ are the two Bogolyubov coefficients of $\boldsymbol{\mathcal{G}_u}(\eta,\eta_0)$ defined in \Eq{eq:G:Bogoliubov}, the matrices $\boldsymbol{I}_\pm$ are defined as $\boldsymbol{I}_\pm=\left(\boldsymbol{I}\pm\boldsymbol{J}_z\right)/2$,\footnote{This decomposition is obtained from the expansion of any $2\times2$ hermitic matrix over the basis $\left\{\boldsymbol{J}_a\right\}\equiv\left\{\boldsymbol{I},\boldsymbol{J}_x,\boldsymbol{J}_y,\boldsymbol{J}_z\right\}$ with real coefficients, \ie $\boldsymbol{\Sigma}=\sum_a\sigma_a\boldsymbol{J}_a$ with each real coefficients obtained using $\sigma_a=\mathrm{Tr}\left[\boldsymbol{\Sigma J}_a\right]/2$, as explained below \Eq{eq:SU11:exp:Pauli}.\label{footnote:decomposition:IJxJyJz}} and the Wronskian condition given below \Eq{eq:G:Bogoliubov}, $\left\vert \alpha_k^{(u)} \right\vert^2-\left\vert \beta_k^{(u)} \right\vert^2 =1$, has been used. The two first terms in the above expression, \ie the ones multiplying $\boldsymbol{I}_+$ and $\boldsymbol{I}_-$, respectively correspond to the power spectrum of the configuration operator, \ie $P_{uu}(k)=\left<\widehat{u}_{\vec{k}}\widehat{u}^\dag_{\vec{k}}\right>$, and the power spectrum of the momentum operator, $P_{pp}(k)=\left<\widehat{p}_{\vec{k}}\widehat{p}^\dag_{\vec{k}}\right>$. The term multiplying $\boldsymbol{J}_x$ is the cross-power spectrum defined as $P_{up}(k)=\left<\frac{1}{2}\left(\widehat{u}_{\vec{k}}\widehat{p}^\dag_{\vec{k}}+\widehat{p}_{\vec{k}}\widehat{u}^\dag_{\vec{k}}\right)\right>$. The last term proportional to $\boldsymbol{J}_y$ comes from the commutator of the configuration and momentum operators~(\ref{eq:commutator:hatz}), $[\widehat{\boldsymbol{u}}_{\vec{k}}(\eta),\widehat{\boldsymbol{u}}_{\vec{q}}(\eta)]=i\boldsymbol{\Omega}= \boldsymbol{J}_y$, and can be seen as arising from the decomposition $\langle \widehat{\boldsymbol{u}}_{\vec{k}}(\eta)\widehat{\boldsymbol{u}}^\dag_{\vec{q}}(\eta)\rangle = \langle \lbrace \widehat{\boldsymbol{u}}_{\vec{k}}(\eta),\widehat{\boldsymbol{u}}^\dag_{\vec{q}}(\eta)\rbrace\rangle/2-\langle [ \widehat{\boldsymbol{u}}_{\vec{k}}(\eta),\widehat{\boldsymbol{u}}^\dag_{\vec{q}}(\eta)]\rangle/2= \langle \lbrace \widehat{\boldsymbol{u}}_{\vec{k}}(\eta),\widehat{\boldsymbol{u}}^\dag_{\vec{q}}(\eta)\rbrace\rangle/2-\boldsymbol{J}_y/2$. One thus has
\bea
\label{eq:2ptcorr:2msq}
	\boldsymbol{\Sigma}_k(\eta)=P_{uu}(k)\boldsymbol{I}_++P_{pp}(k)\boldsymbol{I}_-+P_{up}(k)\boldsymbol{J}_x-\frac{1}{2}\boldsymbol{J}_y\, .
\eea

In terms of the squeezing parameters, the Bogolyubov coefficients are given by \Eqs{eq:alpha:squeezing} and~(\ref{eq:beta:squeezing}) and the above considerations lead to
\bea
	P_{uu}(k,\eta)&=&\frac{1}{2k}\left[\cosh(2r_k)+\sinh(2r_k)\cos(2\varphi_k)\right], \label{eq:Puu} \\
	P_{pp}(k,\eta)&=&\frac{k}{2}\left[\cosh(2r_k)-\sinh(2r_k)\cos(2\varphi_k)\right], \label{eq:Ppp} \\
	P_{up}(k,\eta)&=&\frac{1}{2}\sinh(2r_k)\sin(2\varphi_k), \label{eq:Pup}
\eea
where the dynamical squeezing parameters are evaluated at the time $\eta$ (hereafter the explicit time dependence will be dropped from our notations except when it plays an important role). Note that the power spectra do not depend on the rotation angle, $\theta_k$, since the vacuum is rotationally invariant.

The correlation between the two phase-space variables is given by
\bea
\label{eq:up:correlator}
\frac{P_{up}(k)}{\sqrt{P_{uu}(k)P_{pp}(k)}}=\frac{\sinh(2r_k)\sin(2\varphi_k)}{\sqrt{1+\sinh^2(2r_k)\sin^2(2\varphi_k)}}=\frac{P_{up}(k)}{\sqrt{P^2_{up}(k)+1/4}}\, .
\eea
Maximal (anti-)correlation is obtained for $\mathcal{P}_{up}\to\pm\infty$. A necessary condition for this is $\vert r_k\vert\gg 1$. This condition is however not sufficient if $\vert \sin(2\varphi_k)\vert \ll 1$, \ie if $\varphi_k\simeq  n(\pi/2)$ with $n\in\mathbb{Z}$. Without loss of generality, let us consider the case where $\varphi_k\simeq 0$. Expanding \Eq{eq:up:correlator} in the limit $\vert r_k\vert\gg 1$ and $\vert \varphi_k \vert\ll 1$, one finds that maximal (anti-)correlation is obtained when $\vert \varphi_k\vert \gg \ee^{-2r_k}$.\footnote{It is interesting to notice that in \Refa{Martin:2016tbd}, the condition found for Bell inequality violation reads $\vert \varphi_k \vert < 0.34 \ee^{-r_k}$, which is looser than the condition for maximal (anti-)correlation. Hence Bell inequality violation can be obtained even for low correlations between phase-space variables.}\\

Let us finally note that, although $2\times2$ hermitic matrices have four degrees of freedom in general, $\boldsymbol{\Sigma}_k$ is determined by two parameters only, namely $r_k$ and $\varphi_k$. A first degree of freedom is removed by the commutation relation (\ie the Wronskian condition), which explains why $\boldsymbol{\Sigma}_k$ depends on three power spectra only. A second degree of freedom can be removed thanks to the invariance by rotation of the vacuum state. This entails that the three power spectra are related through one constraint, namely\footnote{This relation~(\ref{eq:PuuPpp}) implies that Heisenberg's uncertainty principle is saturated, $P_{uu}(k)P_{pp}(k)=1/4$, when $u$ and $p$ are uncorrelated, that is when $P_{up}(p)=0$. From \Eq{eq:Pup}, this implies that $\varphi_k=0$ (unless $r_k=0)$, hence the claim made above that minimisation of Heisenberg's uncertainty principle fixes the squeezing angle.\label{footnote:Heisenberg:Uncertainty:Principle}}
\bea
\label{eq:PuuPpp}
P_{uu}(k)P_{pp}(k)=P^2_{up}(k)+\frac{1}{4}\, ,
\eea
which can be easily found from \Eqs{eq:Puu}-\eqref{eq:Pup}, and which was already used in the derivation of \Eq{eq:up:correlator}. Note also that \Eqs{eq:Puu}-\eqref{eq:Pup} can be inverted to give the squeezing parameters in terms of the power spectra,
\bea
\cosh(2r_k)&=&kP_{uu}(k)+k^{-1}P_{pp}(k)\\
\tan(2\varphi_k)&=&\frac{2P_{up}(k)}{kP_{uu}(k)-k^{-1}P_{pp}(k)}
\eea
while their signs are derived from the sign of $P_{up}(k)$ and using that $\sinh(2r_k)\cos(2\varphi_k)=kP_{uu}(k)-k^{-1}P_{pp}(k)$.  Hence, the tow-point correlation matrix of any two-mode squeezed states reads
\bea
	\boldsymbol{\Sigma}_k(\eta)=P_{uu}(k)\boldsymbol{I}_++P_{pp}(k)\boldsymbol{I}_-\pm\sqrt{P_{uu}(k)P_{pp}(k)-\frac{1}{4}}\boldsymbol{J}_x-\frac{1}{2}\boldsymbol{J}_y\, ,
\eea
where $(u,p)$ should be understood as any set of canonical variables. 
\subsubsection{Two-point correlations in the ${\alpha}$-vacua}
\label{sec:2pt:alphavac}
Let us now perform the same calculation but starting from the vacuum state selected out by the operators $\widehat{\boldsymbol{d}}_{\vec{k}}$. The correlation matrix is still given by \Eq{eq:Sigma:uvacuum:def}, 
\bea
\label{eq:Sigma:vvacuum:def}
	\widetilde{\boldsymbol{\Sigma}}_k(\eta)=\boldsymbol{D}^{-1}_k\boldsymbol{U}^\dag\boldsymbol{\mathcal{G}_u}(\eta,\eta_0)\left<\slashed{0}_v\right|\widehat{\boldsymbol{a}}_{\vec{k}}(\eta_0)\widehat{\boldsymbol{a}}^\dag_{\vec{k}}(\eta_0)\left|\slashed{0}_v\right>\boldsymbol{\mathcal{G}_u}^\dag(\eta,\eta_0)\boldsymbol{U}\left(\boldsymbol{D}^{-1}_k\right)^\dag\, ,
\eea
where the vacuum state $\left|\slashed{0}_v\right>$ is now such that $\widehat{d}_{\vec{k}}(\eta_0)\left|\slashed{0}_v\right>=0$ for all $\vec{k}$. This can be calculated as follows. One first notes that $\left|\slashed{0}_v\right>=\widehat{\mathcal{M}}_{(0)}\left|\slashed{0}_u\right>$, where $\widehat{\mathcal{M}}_{(0)}$ is the quantum operator generating the canonical transformation defined by the $\boldsymbol{\mathcal{M}}^{(0)}_k$'s, see \Eq{eq:canonicalquantum}. We introduce subscripts and superscripts $(0)$ to highlight that the quantities associated to the canonical transformation are evaluated at the initial time $\eta_0$. Making use of the fact that the transformation is unitary, one has 
\bea
\left<\slashed{0}_v\right|\widehat{\boldsymbol{a}}_{\vec{k}}(\eta_0)\widehat{\boldsymbol{a}}^\dag_{\vec{k}}(\eta_0)\left|\slashed{0}_v\right>=\left<\slashed{0}_u\right|\left[\widehat{\mathcal{M}}^\dag_{(0)}\widehat{\boldsymbol{a}}_{\vec{k}}(\eta_0)\widehat{\mathcal{M}}_{(0)}\right]\left[\widehat{\mathcal{M}}^\dag_{(0)}\widehat{\boldsymbol{a}}^\dag_{\vec{k}}(\eta_0)\widehat{\mathcal{M}}_{(0)}\right]\left|\slashed{0}_u\right>.
\eea
Since the above is simply the application of the canonical transformation to the creation and annihilation operators, it can be written as
\bea
	\left<\slashed{0}_v\right|\widehat{\boldsymbol{a}}_{\vec{k}}(\eta_0)\widehat{\boldsymbol{a}}^\dag_{\vec{k}}(\eta_0)\left|\slashed{0}_v\right>&=&\boldsymbol{\mathcal{M}}^{(0)}_{\vec{k}}\left<\slashed{0}_u\right|\widehat{\boldsymbol{a}}_{\vec{k}}{(0)}\widehat{\boldsymbol{a}}^\dag_{\vec{k}}{(0)}\left|\slashed{0}_u\right>\boldsymbol{\mathcal{M}}^{{(0)}\dag}_{\vec{k}}  \\
	&=&\left(\begin{array}{cc}
	\cosh^2\left(d^{(0)}_k\right) & \frac{1}{2}e^{2i\varpi^{(0)}_k}\sinh\left(2d^{(0)}_k\right) \\
	\frac{1}{2}e^{-2i\varpi^{(0)}_k}\sinh\left(2d^{(0)}_k\right) & \sinh^2\left(d^{(0)}_k\right)
\end{array}\right), 
\eea
where \Eq{eq:aa:u} has been used, together with the left-polar decomposition~\eqref{eq:bloch:left:right} $\boldsymbol{\mathcal{M}}_k=\boldsymbol{\mathcal{S}}_k(d_k,\varpi_k)\boldsymbol{\mathcal{R}}_k(\vartheta_k)$ and \Eq{eq:polar:RS:explicit}. Note that since the vacuum state is rotationally invariant, the rotation angle $\vartheta_k$ does not appear in the above result. Let us now introduce the vector made of the Bogolyubov coefficients of the matrix $\boldsymbol{\mathcal{S}}(d_k,\varpi_k)$ given in \Eq{eq:polar:RS:explicit},
\bea
	\boldsymbol{\alpha}_M(\eta_0)=\left(\begin{array}{c}
		\alpha^{(0)}_M \\
		\beta^{(0)^*}_M
	\end{array}\right)=\left(\begin{array}{c}
		\cosh d^{(0)}_k\\
		e^{-2i\varpi^{(0)}_k}\sinh d^{(0)}_k
	\end{array}\right),
\eea
using the notations of \Eq{eq:G:Bogoliubov}. The expectation value $\left<\slashed{0}_v\right|\widehat{\boldsymbol{a}}_{\vec{k}}{(0)}\widehat{\boldsymbol{a}}^\dag_{\vec{k}}{(0)}\left|\slashed{0}_v\right>$ can then be written as the tensor product $\left<\slashed{0}_v\right|\widehat{\boldsymbol{a}}_{\vec{k}}{(0)}\widehat{\boldsymbol{a}}^\dag_{\vec{k}}{(0)}\left|\slashed{0}_v\right>=\boldsymbol{\alpha}_M(\eta_0)\boldsymbol{\alpha}^\dag_M(\eta_0)$ (note that the same formula could be obtained for the $\boldsymbol{u}$-vacuum by simply setting $\alpha^{(0)}_M=1$ and $\beta^{(0)*}_M=0$). The two-point correlation matrix evaluated on the $\boldsymbol{v}$-state, \Eq{eq:Sigma:vvacuum:def}, then reads as the tensor product of the evolved vectors, $\widetilde{\boldsymbol{\Sigma}}_k(\eta)=\boldsymbol{u}^{(\alpha)}_{\vec{k}}(\eta)\boldsymbol{u}^{(\alpha)\dag}_{\vec{k}}(\eta)$ with
\bea
\label{eq:u:Gu:alpha}
	\boldsymbol{u}^{(\alpha)}_{\vec{k}}(\eta)&=&
	\boldsymbol{D}^{-1}_k\boldsymbol{U}^\dag\boldsymbol{\mathcal{G}_u}(\eta,\eta_0)\boldsymbol{\alpha}_M(\eta_0)
	\\ &=&
	\frac{1}{\sqrt{2}}\left(\begin{array}{c}
		\ds\frac{\alpha^{(u)}_k+\beta^{(u)*}_k}{\sqrt{k}}\alpha^{(0)}_M+\frac{\alpha^{(u)*}_k+\beta^{(u)}_k}{\sqrt{k}}\beta^{(0)^*}_M \\
		\ds-i\sqrt{k}\left(\alpha^{(u)}_k-\beta^{(u)*}_k\right)\alpha^{(0)}_M+i\sqrt{k}\left(\alpha^{(u)*}_k-\beta^{(u)}_k\right)\beta^{(0)^*}_M
	\end{array}\right),
\eea
where we recall that $\alpha_k^{(u)}(\eta)$ and $\beta^{(u)}_k(\eta)$ are the two Bogolyubov coefficients of $\boldsymbol{\mathcal{G}_u}(\eta,\eta_0)$. These Bogolyubov coefficients can be expressed in terms of the power spectra expected in the original $\boldsymbol{u}$-vacuum, and the two phases $\varphi^{(\pm)}_k$ of the complex numbers $\alpha^{(u)}_k\pm\beta^{(u)*}_k$. One obtains
\bea
	\boldsymbol{u}^{(\alpha)}_{\vec{k}}(\eta)=\left(\begin{array}{c}
		\ds\sqrt{P_{uu}(k)}\left(e^{i\varphi_k^{(+)}}\alpha^{(0)}_M+e^{-i\varphi_k^{(+)}}\beta^{(0)^*}_M\right) \\
		\ds-i\sqrt{P_{pp}(k)}\left(e^{i\varphi_k^{(-)}}\alpha^{(0)}_M-e^{-i\varphi_k^{(-)}}\beta^{(0)^*}_M\right)
	\end{array}\right),
\eea
where the phases can be expressed in terms of the squeezing parameters using \Eqs{eq:alpha:squeezing} and~(\ref{eq:beta:squeezing}) and read 
\bea
	\varphi_k^{(\pm)}=\theta_k\mp\arctan\left[\frac{\sinh(r_k)\sin(2\varphi_k)}{\cosh(r_k)\pm\sinh(r_k)\cos(2\varphi_k)}\right].
\eea 
We note that $\varphi^{(+)}_k+\varphi^{(-)}_k=2\theta_k$ and that $\varphi^{(+)}_k-\varphi^{(-)}_k\equiv 2\Delta\varphi_k$ (the notation $\Delta\varphi_k$ is introduced for later use) is a function of $r_k$ and $\varphi_k$ only. 

The power spectra can then be obtained by projecting the correlation matrix on the basis $\left\{\boldsymbol{J}_a\right\}$ by computing $\frac{1}{2}\mathrm{Tr}\left[\widetilde{\boldsymbol{\Sigma}}_k\boldsymbol{J}_a\right]=\frac{1}{2}\boldsymbol{u}^{(\alpha)\dag}_{\vec{k}}(\eta)\boldsymbol{J}_a\boldsymbol{u}^{(\alpha)}_{\vec{k}}(\eta)$, as explained in footnote~\ref{footnote:decomposition:IJxJyJz}. One obtains
\bea
	\widetilde{P}_{uu}(k)&=&P_{uu}(k)\left[\cosh\left(2d^{(0)}_k\right)+\cos\left(2\varpi^{(0)}_k+2\varphi^{(+)}_k\right)\sinh\left(2d^{(0)}_k\right)\right], \label{eq:alphapuu} \\
	\widetilde{P}_{pp}(k)&=&P_{pp}(k)\left[\cosh\left(2d^{(0)}_k\right)-\cos\left(2\varpi^{(0)}_k+2\varphi^{(-)}_k\right)\sinh\left(2d^{(0)}_k\right)\right], \label{eq:alphappp} \\
	\widetilde{P}_{up}(k)&=&\sqrt{P_{uu}(k)P_{pp}(k)}
	\nonumber \\ & & \times
	\left[\sinh\left(2d^{(0)}_k\right)\sin\left(2\varpi^{(0)}_k+2\theta_k\right)
	-\cosh\left(2d^{(0)}_k\right)\sin\left(2\Delta\varphi_k\right)\right]. \label{eq:alphapup}
\eea
One can finally check that $\mathrm{Tr}\left[\widetilde{\boldsymbol{\Sigma}}_k\boldsymbol{J}_y\right]/2=-1/2=\mathrm{Tr}\left[{\boldsymbol{\Sigma}}_k\boldsymbol{J}_y\right]/2$ since canonical transformations (which generate the $\boldsymbol{v}$-vacuum) preserve the canonical commutation relations.\footnote{This can be shown either by a direct calculation or as follows. One first plugs \Eq{eq:u:Gu:alpha} and the fact that $\boldsymbol{J}_y=-i\boldsymbol{\Omega}$ into $\mathrm{Tr}\left[\widetilde{\boldsymbol{\Sigma}}_k\boldsymbol{J}_y\right]=\boldsymbol{u}^{(\alpha)\dag}_{\vec{k}}(\eta)\boldsymbol{J}_y\boldsymbol{u}^{(\alpha)}_{\vec{k}}(\eta)$, which leads to
\bea
\mathrm{Tr}\left[\widetilde{\boldsymbol{\Sigma}}_k\boldsymbol{J}_y\right]=(-i)\boldsymbol{\alpha}_M^\dag(\eta_0)\boldsymbol{\mathcal{G}_u}^\dag(\eta,\eta_0)\boldsymbol{U}\boldsymbol{D}^{-1\dag}_k\boldsymbol{\Omega}\boldsymbol{D}^{-1}_k\boldsymbol{U}^\dag\boldsymbol{\mathcal{G}_u}(\eta,\eta_0)\boldsymbol{\alpha}_M(\eta_0).
\eea
One then successively makes use of $\boldsymbol{D}^{-1\dag}_k\boldsymbol{\Omega}\boldsymbol{D}^{-1}_k=\boldsymbol{\Omega}$ since $\boldsymbol{D}_k\in\mathrm{Sp}(2,\mathbb{R})$, $\boldsymbol{U}\boldsymbol{\Omega}\boldsymbol{U}^\dag=\boldsymbol{\mathcal{J}}$ because of \Eq{eq:J:def}, and $\boldsymbol{\mathcal{G}_u}^\dag(\eta,\eta_0)\boldsymbol{\mathcal{J}}\boldsymbol{\mathcal{G}_u}(\eta,\eta_0)=\boldsymbol{\mathcal{J}}$ since $\boldsymbol{\mathcal{G}_u}\in\mathrm{SU}(1,1)$. From the expression of $\boldsymbol{\mathcal{J}}$ this gives $\mathrm{Tr}\left[\widetilde{\boldsymbol{\Sigma}}_k\boldsymbol{J}_y\right]=-\left|\alpha^{(0)}_M\right|^2+\left|\beta^{(0)}_M\right|^2$ which equals $-1$ since $\alpha^{(0)}_M$ and $\beta^{(0)}_M$ are Bogolyubov coefficients.} \\

 From \Eqs{eq:alphapuu}-\eqref{eq:alphapup}, one can see that the three tilted power spectra can be expressed in terms of the untilted ones, the squeezing parameters $d^{(0)}_k$ and $\varpi_k^{(0)}$ of the canonical transform $\boldsymbol{\mathcal{M}}^{(0)}_{\vec{k}}$ (and not its rotation angle, the vacuum being invariant under rotation), and the dynamical rotation angle $\theta_k$. The fact that, contrary to \Eqs{eq:Puu}-\eqref{eq:Pup}, the dynamical rotation angle is involved, is because the $\alpha$-vacua are not vacua for the $\widehat{\boldsymbol{a}}_{\vec{k}}(\eta_0)$ operators, hence they are not invariant under rotation in that representation. One can also check that a sufficient condition for the entire set of power spectra to remain unchanged is to set $d^{(0)}_k=0$. In this case the canonical transformation reduces to a rotation in the phase space which leaves the vacuum state unchanged. Evaluating \Eq{eq:alphapup} in this situation, one finds that $\sin(2\Delta\varphi_k)=-P_{up}/\sqrt{P_{uu}P_{pp}}$. The angle  $\Delta\varphi_k$, which depends only on the dynamical squeezing parameters, thus quantifies the correlation between the configuration variable and the momentum. Since it is nothing but the relative phase between the configuration mode function $u_k(\eta)$ and its associated momentum $p_k(\eta)$, the correlation between these two variables is thus entirely encoded in their relative phase. Let us finally mention that the presence of circular functions in \Eqs{eq:alphapuu}-\eqref{eq:alphapup} usually leads to superimposed oscillations in the power spectra, which is a typical feature of non Bunch-Davies initial states.\\

Before closing this subsection, let us notice that the above results can also be obtained by making use of the composition laws of $\mathrm{SU}(1,1)$ derived in \Sec{sec:SU11:Composition}. Since the initial state is given by $\left|\slashed{0}_v\right>=\widehat{\mathcal{M}}_{(0)}\left|\slashed{0}_u\right>$, the final state reads $\widehat{\mathcal{G}}_u(\eta,\eta_0)\left|\slashed{0}_v\right>=\widehat{\mathcal{G}}_u(\eta,\eta_0)\widehat{\mathcal{M}}_{(0)}\left|\slashed{0}_u\right>$. Using the composition laws of $\mathrm{SU}(1,1)$, one can find the squeezing parameters and the rotation angle of the product $\widehat{\mathcal{G}}_u(\eta,\eta_0)\widehat{\mathcal{M}}_{(0)}$, \ie the numbers $\widetilde{r}_k$,  $\widetilde{\varphi}_k$ and $\widetilde{\theta}_k$ such that $\boldsymbol{\mathcal{G}_u}(\eta,\eta_0)\boldsymbol{\mathcal{M}}_k(\eta_0)=\boldsymbol{\mathcal{S}}(\widetilde{r}_k,\widetilde{\varphi}_k)\boldsymbol{\mathcal{R}}(\widetilde{\theta}_k)$. The power spectra are then given by \Eqs{eq:Puu}-\eqref{eq:Pup}, where $(r_k,\varphi_k)$ is replaced by $(\widetilde{r}_k,\widetilde{\varphi}_k)$.

\subsubsection{Two-point correlations in the \textit{\textbf{u}}- and \textit{\textbf{v}}-vacua}
\label{sec:2pt:vvac}
Let us now specify the above considerations to the $\boldsymbol{u}$- and $\boldsymbol{v}$-vacua in an inflationary universe. For simplicity, we consider the case of de Sitter where the scale factor reads $a(\eta)=-1/H\eta$. The equation of motion for the classical mode functions can be obtained from \Eqs{eq:eom:Fourrier} and~\eqref{eq:Hu}, and is given by
\bea
\label{eq:mode:equation:u}
	u''_{\vec{k}}+\left(k^2-\frac{\nu^2-\frac{1}{4}}{\eta^2}\right)u_{\vec{k}}=0,
\eea
where $\nu=\frac{3}{2}\sqrt{1-\left(\frac{2m}{3H}\right)^2}$. We restrict the analysis to $m<3H/2$, in order to ensure that $\nu$ is real-valued. Two independent solutions, normalised with the Klein-Gordon product [see \Eq{eq:Wronskian} and footnote~\ref{footnote:KleinGordonProduct}], are given by
\bea
\label{eq:uk:BD}
u^{(1)}_{\vec{k}}(\eta)=\frac{1}{2}\sqrt{-\pi\eta}H_\nu^{(1)}(-k\eta)
\eea
and $u^{(2)}_{\vec{k}}=u^{(1)*}_{\vec{k}}$, with $H_\nu^{(1)}$ the Hankel function of the first kind and of order $\nu$. The momenta associated to these mode functions are
\bea
\label{eq:pk:BD}
	p^{(1)}_{\vec{k}}(\eta)=\frac{1}{2}\sqrt{-\pi\eta}\left[\frac{1}{\eta}\left(\frac{3}{2}-\nu\right)H_\nu^{(1)}(-k\eta)-kH_{\nu-1}^{(1)}(-k\eta)\right]
\eea
and $p^{(2)}_{\vec{k}}=p^{(1)*}_{\vec{k}}$
They reduce to the Bunch-Davies vacuum in the sub-Hubble (or equivalently infinite-past) limit, \ie $u^{(1)}_{\vec{k}}(-k\eta\to\infty)\sim \ee^{-ik\eta-\frac{i}{2}\nu\pi-i\frac{\pi}{4}}/\sqrt{2k}$ and $p^{(1)}_{\vec{k}}(-k\eta\to\infty)\sim-i\sqrt{k/2} ~\ee^{-ik\eta-\frac{i}{2}\nu\pi-i\frac{\pi}{4}}$.\footnote{For large arguments, the asymptotic behaviour of the Hankel functions is given by~\cite{abramowitz+stegun}
\bea
\label{eq:Hankel:large:arg}
H_\nu^{(1)}(z)\underset{\vert z \vert \gg 1}{\simeq} \sqrt{\frac{2}{\pi z}}\ee^{iz-\frac{i}{2}\nu\pi-i\frac{\pi}{4}}\left[1+\frac{4\nu^2-1}{8}\frac{i}{z}-\frac{\left(4\nu^2-1\right)\left(4\nu^2-9\right)}{128 z^2}+\order{\frac{1}{z^3}}\right]\, .
\eea}

Other choices for the mode functions are of course possible. Since \Eq{eq:mode:equation:u} is a linear differential equation, they are related to the ones introduced above by
\bea
\label{eq:u:u1:u2}
	u_{\vec{k}}(\eta)&=&A_ku^{(1)}_{\vec{k}}(\eta)+B_ku^{(2)}_{\vec{k}}(\eta), \\
	p_{\vec{k}}(\eta)&=&A_kp^{(1)}_{\vec{k}}(\eta)+B_kp^{(2)}_{\vec{k}}(\eta),
\label{eq:p:p1:p2}
\eea
where $\left|A_k\right|^2-\left|B_k\right|^2=1$ for the Wronskian condition~(\ref{eq:Wronskian}) to be satisfied. Different choices for the mode functions correspond to different vacua, related one to each other by a canonical transformation. Indeed, because of the Wronskian condition, the matrix
\bea
	\boldsymbol{\mathcal{M}}_k=\left(\begin{array}{cc}
		A_k & B^*_k \\
		B_k & A^*_k
	\end{array}\right)
\eea
defines an element of $\mathrm{SU}(1,1)$, see \Eq{eq:su11mat}, from which one can define a new set of creation and annihilation operators, $\boldsymbol{\widehat{d}}_{\vec{k}}$, via $\boldsymbol{\widehat{a}}_{\vec{k}}=\boldsymbol{\mathcal{M}}_k\boldsymbol{\widehat{d}}_{\vec{k}}$.

 Introducing the operators $\boldsymbol{\widehat{d}}_{\vec{k}}$ into the field decomposition~(\ref{eq:field:decomposition:quantum:z:a}), one gets
\bea
	\boldsymbol{\widehat{z}}(\vec{x},\eta)&=&\ds\int_{\mathbb{R}^{3+}}\frac{\dd^3k}{(2\pi)^{3/2}}\left[\widehat{d}_{\vec{k}}(\eta_0)A_k\boldsymbol{z}_k(\eta)e^{-i\vec{k}\cdot\vec{x}}+\widehat{d}_{-\vec{k}}(\eta_0)B_k\boldsymbol{z}^*_k(\eta)e^{i\vec{k}\cdot\vec{x}}\right. \nonumber \\
	&&\left.+\widehat{d}^\dag_{\vec{k}}(\eta_0)A^*_k\boldsymbol{z}^*_k(\eta)e^{i\vec{k}\cdot\vec{x}}+\widehat{d}^\dag_{-\vec{k}}(\eta_0)B^*_k\boldsymbol{z}_k(\eta)e^{-i\vec{k}\cdot\vec{x}}\right].
\eea
After performing the change of integration variable $\vec{k}\to-\vec{k}$ in the second and the fourth terms, the field decomposition reads
\bea
\kern-4em
	\boldsymbol{\widehat{z}}(\vec{x},t) \kern-0.2em = \kern-0.2em
	\ds\int_{\mathbb{R}^{3+}} \kern-0.2em \frac{\dd^3k}{(2\pi)^{3/2}}\left\{\widehat{d}_{\vec{k}}(\eta_0)\left[A_k\boldsymbol{z}_k(\eta)+B_k\boldsymbol{z}^*_k(\eta)\right]e^{-i\vec{k}\cdot\vec{x}}+\widehat{d}^\dag_{\vec{k}}(\eta_0)\left[A^*_k\boldsymbol{z}^*_k(\eta)+B^*_k\boldsymbol{z}_k(\eta)\right]e^{i\vec{k}\cdot\vec{x}}\right\} \kern-0.2em , \nonumber \kern-5em \\
\eea
which is nothing but the field decomposition that would be obtained from the mode functions  $\boldsymbol{\widetilde{z}}_k(\eta)=A_k\boldsymbol{z}_k(\eta)+B_k\boldsymbol{z}^*_k(\eta)$. Notice that the same Green's matrix generates the dynamics of the mode functions $\boldsymbol{z}_k(\eta)$ and $\boldsymbol{\widetilde{z}}_k(\eta)$.

We now perform the calculation of the power spectra in the super-Hubble regime, \ie at times such that $\left|k\eta\right|\ll1$, if initial conditions are set in different vacua.

\paragraph{The Bunch-Davies vacuum--} We start by considering the case of the well-known Bunch-Davies vacuum, which will serve as a reference, and which is obtained by setting $A_k=1$ and $B_k=0$. The power spectra at the end of inflation are obtained by taking the super-Hubble limit, $\left| k\eta\right|\ll 1$, of \Eqs{eq:uk:BD} and~\eqref{eq:pk:BD}, which gives rise to\footnote{For small arguments and real $\nu$, the behaviour of the Hankel functions is given by~\cite{abramowitz+stegun}
\bea
H_\nu^{(1)}(z)\underset{\vert z \vert \ll 1}{\simeq} 
\begin{cases}
-\frac{i}{\pi}\Gamma\left(\nu\right) \left(\frac{z}{2}\right)^{-\nu}\quad \text{for}\quad \nu>0\\
\frac{2 i}{\pi}\log(z)\quad \text{for}\quad \nu=0\\
-\frac{i}{\pi}\Gamma\left(-\nu\right)\ee^{-i \nu\pi} \left(\frac{z}{2}\right)^{\nu}\quad \text{for}\quad\nu<0
\end{cases}
\, .
\eea}
\bea
\label{eq:u1:superH}
	u^{(1)}_{\vec{k}}(-k\eta\to0)\sim\left\{\begin{array}{ll}
		\ds\frac{-i}{\sqrt{2k\pi}}\Gamma\left(\nu\right) \left(-\frac{k\eta}{2}\right)^{\frac{1}{2}-\nu} & \mathrm{for}~\nu>0 \\
		\ds\frac{i}{\sqrt{\pi k}}\sqrt{-k\eta}\log\left(-k\eta\right) & \mathrm{for}~\nu=0
	\end{array}\right.,
\eea
and
\bea
\label{eq:p1:superH}
	p^{(1)}_{\vec{k}}(-k\eta\to0)\sim\left\{\begin{array}{ll}
		\ds \frac{i}{2} \sqrt{\frac{k}{2\pi}}\Gamma\left(\nu\right) \left(\frac{3}{2}-\nu\right)\left(-\frac{k\eta}{2}\right)^{-\frac{1}{2}-\nu} & \mathrm{for}~\nu>0\ \mathrm{and}\ \nu\neq 3/2 \\
		\ds i \sqrt{\frac{k}{2}}  & \mathrm{for}~\nu=3/2 \\
		\ds-\frac{3}{2}i\sqrt{\frac{k}{\pi}}\frac{\log\left(-k\eta\right)}{\sqrt{-k\eta}} & \mathrm{for}~\nu=0
	\end{array}\right. .
\eea
The value $\nu=3/2$ corresponds to a massless field and is singular, since in this case the leading-order result vanishes and one needs to perform the expansion in $-k\eta$ one order beyond. Notice that for $\nu=1/2$, which corresponds to $m=\sqrt{2}H$, the modulus of $u^{(1)}_{\vec{k}}$ remains frozen. This is because \Eq{eq:mode:equation:u} reduces to $u''_{\vec{k}}+k^2u_{\vec{k}}=0$ in that case, \ie simply describes a free field in the Minkowski space time (notice however that since the conjugate momentum receives a contribution involving $a'/a$, see Table~\ref{tab:variables}, it differs from the Minkowski space-time case). 

For $\nu$ different from $0$ and $3/2$, the power spectra are then given by
\bea
\label{eq:Puu:BD}
	P^\mathrm{BD}_{uu}(k, -k\eta \to0)&=&\frac{\Gamma^2\left(\nu\right)}{2k\pi}\left(\frac{-k\eta}{2}\right)^{1-2\nu}, \\
\label{eq:Ppp:BD}
	P^\mathrm{BD}_{pp}(k, -k\eta \to0)&=&k\frac{\Gamma^2\left(\nu\right)}{8\pi}\left(\frac{3}{2}-\nu\right)^2\left(\frac{-k\eta}{2}\right)^{-1-2\nu}, \\
\label{eq:Pup:BD}
	P^\mathrm{BD}_{up}(k,-k\eta \to0)&=&\mathrm{sgn}(\nu-3/2)\sqrt{P^\mathrm{BD}_{uu}(k, -k\eta \to0)P^\mathrm{BD}_{pp}(k,-k\eta \to0)},
\eea
where the last expression indicates that in the super-Hubble limit, the configuration and momentum variables are maximally correlated or anti-correlated if $\nu>3/2$ or $\nu<3/2$ respectively. This is because, in the super-Hubble limit, $p^{(1)}_{\vec{k}}=(3/2-\nu)u^{(1)}_{\vec{k}}/\eta$. The relative phase between the configuration and momentum variables thus freezes to $0$ or $\pi$ (depending on the sign of $\nu-3/2$), which indicates maximal (anti-)correlation. Let us notice that, from the definition of $\nu$ given below \Eq{eq:mode:equation:u}, one always has $\nu<3/2$ in the situation considered here, hence the configuration and momentum variables are found to be always maximally anti-correlated in the super-Hubble limit. This result plays an important role in the so-called ``quantum-to-classical transition'', see \Refs{Polarski:1995jg, Lesgourgues:1996jc, Martin:2015qta, Grain:2017dqa} and \Sec{sec:QuantumToClassical} below.

If $\nu=3/2$, \Eq{eq:Puu:BD} is still valid, but the power spectra involving the conjugate momentum read $P^\mathrm{BD}_{pp} = k/2$ and $P^\mathrm{BD}_{up}=-\sqrt{P^\mathrm{BD}_{uu}P^\mathrm{BD}_{pp}}$, hence the configuration and momentum variables are still maximally anti-correlated. If $\nu=0$, one finds $P^\mathrm{BD}_{uu} =\eta\log^2(-k\eta)/\pi$, $P^\mathrm{BD}_{pp} = 9 \log^2(-k\eta)/(-4\pi\eta)$ and $P^\mathrm{BD}_{up}=-\sqrt{P^\mathrm{BD}_{uu}P^\mathrm{BD}_{pp}}$, hence the configuration and momentum variables are maximally anti-correlated in that case as well.
\paragraph{The \textit{\textbf{u}}-vacuum--} Let us then consider the case of the $\boldsymbol{u}$-vacuum. The two coefficients $A_k$ and $B_k$ are determined by the condition $\boldsymbol{e}_\alpha=\boldsymbol{UD}_k\boldsymbol{u}_k(\eta_0)$ where $\boldsymbol{D}_k=\mathrm{diag}\left(\sqrt{k},1/\sqrt{k}\right)$, see \Eq{eq:initialD}, which leads to
\bea
	A_k&=&\sqrt{\frac{k}{2}}u^{(2)}_{\vec{k}}(\eta_0)-\frac{i}{\sqrt{2k}} {p^{(2)}_{\vec{k}}(\eta_0)}\, , \label{eq:Auvac}\\
	B_k&=&-\sqrt{\frac{k}{2}}u^{(1)}_{\vec{k}}(\eta_0)+\frac{i}{\sqrt{2k}}{p^{(1)}_{\vec{k}}(\eta_0)}\, . \label{eq:Buvac}
\eea
Here we have used that the Wronskian $W$ between $u^{(1)}_{\vec{k}}$ and $u^{(2)}_{\vec{k}}$ equals $i$, otherwise, the above expressions would have to be multiplied by $i/W$. Note that if $\eta_0$ is sent to the infinite past, $\eta_0\to -\infty$, then  $A_k=1$ and $B_k=0$ and the Bunch-Davies vacuum is recovered. However, as we shall now see, the Bunch-Davies vacuum and the $\boldsymbol{u}$-vacuum have vanishing overlap, even for arbitrarily large value of $\vert\eta_0\vert$. This can be shown by noticing that, since the two vacua are related by a canonical transformation, the $\boldsymbol{u}$-vacuum is a two-modes squeezed state of Bunch-Davies quanta.  Combining \Eqs{eq:vacuum:NumberOfParticle:def} and~\eqref{eq:TMSS}, the overlap between the two vacua is thus given by
\bea
	\left<\slashed{0}_u\right|\left.\slashed{0}_\mathrm{BD}\right>=
	 \prod_{\vec{k}\in\mathbb{R}^{3+}} \frac{\ee^{i \vartheta_k}}{\cosh d_k}
	 = \exp\left\lbrace \int_{\mathbb{R}^{3+}}\dd^3k\left[i\vartheta_k-\ln\left(\cosh d_k\right)\right]\right\rbrace ,
\eea
where $d_k$ and $\vartheta_k$ are the squeezing amplitude and rotation angle of the canonical transformation relating the two vacua. The rotation angle only describes a pure phase between the two vacua, while the squeezing amplitude can be related to the constants $A_k$ and $B_k$ from \Eq{eq:alpha:squeezing}, which leads to $\cosh(d_k)=\vert A_k\vert$. This gives rise to
\bea
\label{eq:vacua:overlap}
	\left\vert \left<\slashed{0}_u\right|\left.\slashed{0}_\mathrm{BD}\right>\right\vert =
	 \exp\left[- \int_{\mathbb{R}^{3+}}\dd^3k \ln\left(\vert A_k\vert\right)\right] ,
\eea
where one recovers that, when $A_k=1$ for all $\vec{k}$'s, the two vacua are identical. Let us now consider the two cases mentioned in the discussion below \Eq{eq:vartheta:u:v}. 

First, if $\eta_0$ is set independently of $\vec{k}$, one can, for simplicity, bound \Eq{eq:vacua:overlap} by the contribution of the modes that are sub-Hubble at the time $\eta_0$. For these modes, after a lengthy but straightforward calculation, \Eqs{eq:Auvac}-\eqref{eq:Buvac}, \eqref{eq:uk:BD}-\eqref{eq:pk:BD} and~\eqref{eq:Hankel:large:arg} give 
\bea
\label{eq:Ak:large:keta0}
A_k &\simeq& \ee^{ik\eta_0+i \nu\frac{\pi}{2}+i\frac{\pi}{4}} \left\lbrace1+\frac{5-4\nu^2}{8}\frac{i}{-k\eta_0} - \frac{\left(4\nu^2-1\right)\left(4\nu^2-9\right)}{128 \left(-k\eta_0\right)^2}+\mathcal{O}\left[\left(-k\eta_0\right)^{-3}\right]\right\rbrace ,\\
B_k &\simeq & \ee^{-ik\eta_0-i \nu\frac{\pi}{2}-i\frac{\pi}{4}} \left\lbrace \frac{i}{2 k\eta_0}+\mathcal{O}\left[\left(-k\eta_0\right)^{-3}\right]\right\rbrace ,
\label{eq:Bk:large:keta0}
\eea
where the large $k\eta_0$ expansion for $B_k$ is also given for future use. This leads to $\vert A_k\vert \simeq 1+(k\eta_0)^{-2}/8+\mathcal{O}[(-k\eta_0)^{-3}]$ (which, quite remarkably, does not depend on $\nu$), and as a consistency check, one can verify that the Wronskian condition $\left|A_k\right|^2-\left|B_k\right|^2=1$ is satisfied at the order at which the above expressions are written. One then has
\bea
\label{eq:vacua:overlap:1}
	\left\vert \left<\slashed{0}_u\right|\left.\slashed{0}_\mathrm{BD}\right>\right\vert \simeq
	\exp\left[- \int_{-1/(\sigma \eta_0)}^\infty \dd k 4 \pi k^2 \frac{8}{k^2\eta_0^2} \right]
	 \exp\left[- \int_{0}^{-1/(\sigma \eta_0)}\dd k 4\pi k^2 \ln\left(\vert A_k\vert\right)\right] .
\eea
In this expression, we have split the integral between deep sub-Hubble modes, for which $k>-1/(\sigma \eta_0)$ (where $\sigma\ll 1$) and the formula given above for $\vert A_k\vert$ applies, and super-Hubble modes for which $k<-1/(\sigma \eta_0)$. It is indeed enough to notice that the integral has a UV divergence in the sub-Hubble limit to conclude that $\left\vert \left<\slashed{0}_u\right|\left.\slashed{0}_\mathrm{BD}\right>\right\vert=0$.\footnote{One can also check that there is no additional IR divergence in the super-Hubble limit. Making use of \Eqs{eq:u1:superH}, \eqref{eq:p1:superH} and \eqref{eq:Auvac}, one indeed finds that $\vert A_k\vert \simeq \vert 3/2-\nu\vert \Gamma(\nu)(-k\eta_0/2)^{-1/2-\nu}/(4\sqrt{\pi})$ if $\nu \neq 0$ and $\nu\neq 3/2$, $\vert A_k\vert \simeq -1/(2k\eta_0)$ if $\nu=3/2$, and $\vert A_k\vert \simeq 3\log(-k\eta_0)/(-2\sqrt{2\pi}k\eta_0)$ if $\nu=0$. This means that, in \Eq{eq:vacua:overlap:1}, one has to perform integrals of the type $\int_0 k^2 \log(k) \dd k$, which are IR-convergent.}

Second, let us consider the situation where initial conditions are set at a fixed sub-Hubble physical scale, such that $-k \eta_0(k) = 1/\sigma$ is constant, where $\sigma \ll 1$. In this case, the formula given above for $\vert A_k\vert$ applies for all modes, but now gives $\vert A_k\vert \simeq 1+[k\eta_0(k)]^{-2}/8+\mathcal{O}\lbrace [-k\eta_0(k)]^{-3}\rbrace = 1+\sigma^2/8+\order{\sigma^3}$. The integral appearing in \Eq{eq:vacua:overlap} is therefore still divergent (and even more so than in the previous case), leading to the same conclusion that the two vacua have zero overlap.

This result remains true for the overlap between any other states in the Bunch-Davies Hilbert space because the term $\exp\left[-\int\dd^3k\ln\left(\cosh d_k\right)\right]$ is systematically present. This is a generic property of unitarily inequivalent representations in quantum field theory (see \Refa{haag1955} for an original formulation in the context of interacting quantum fields, and \eg \Refa{PhysRevD.7.2850} for a realisation in the context of quantum-field theory in curved spaces).\\

Let us now compute the power spectra. In the super-Hubble limit, they are simply related to the Bunch-Davies ones. Indeed, \Eq{eq:u:u1:u2} gives rise to $P_{uu}=\left(\vert A_k\vert^2 + \vert B_k\vert ^2\right) P^\mathrm{BD}_{uu} + 2 \Rea\left(A_k B_k^* {u^{(1)2}_{\vec{k}}} \right)$. In the super-Hubble limit however, $u^{(1)}_{\vec{k}}$ is pure imaginary, see \Eq{eq:u1:superH}, so $u^{(1)2}_{\vec{k}}=-\vert u^{(1)}_{\vec{k}} \vert^2 = - P^\mathrm{BD}_{uu}$. This gives rise to
\bea
P\left(k,-k\eta\to 0\right) & \simeq & \left[\vert A_k\vert^2 + \vert B_k\vert ^2 - 2 \Rea\left(A_k B_k^*\right) \right]P^\mathrm{BD} \left(k,-k\eta\to 0\right) \nonumber \\
& = & \left\vert A_k - B_k \right\vert^2 P^\mathrm{BD} \left(k,-k\eta\to 0\right).
\label{eq:PowerPspectrum:u:BD:superH}
\eea
For $P_{pp}$, since $p^{(1)}_{\vec{k}}$ is also pure imaginary in the super-Hubble limit, see \Eq{eq:p1:superH}, one obtains from \Eq{eq:p:p1:p2} a similar expression. For $P_{up}$, \Eqs{eq:u:u1:u2} and~\eqref{eq:p1:superH} lead to $P_{up}=\left(\vert A_k\vert^2 + \vert B_k\vert ^2\right) P^\mathrm{BD}_{up} + 2 \Rea\left(A_k B_k^* {u^{(1)}_{\vec{k}}} {p^{(1)}_{\vec{k}}} \right)$. Since both ${u^{(1)}_{\vec{k}}}$ and ${p^{(1)}_{\vec{k}}}$ are pure imaginary, one has $P^\mathrm{BD}_{up} = \Rea \left({u^{(1)}_{\vec{k}}} {p^{(1)}_{\vec{k}}} ^*\right) = -{u^{(1)}_{\vec{k}}} {p^{(1)}_{\vec{k}}} $, hence the same expression~\eqref{eq:PowerPspectrum:u:BD:superH} is again recovered. This explains why $P$ carries no subscript in \Eq{eq:PowerPspectrum:u:BD:superH}, since it applies to all three power spectra. Combined with \Eqs{eq:uk:BD}, \eqref{eq:pk:BD}, \eqref{eq:Auvac} and \eqref{eq:Buvac}, this provides an explicit expression for the power spectra in the $\boldsymbol{u}$-vacuum. Let us also note that \Eq{eq:PowerPspectrum:u:BD:superH} guarantees that the position and momentum variables are still maximally anti-correlated, as in the Bunch-Davies vacuum, \ie \Eq{eq:Pup:BD} is also valid in the $\boldsymbol{u}$-vacuum.\footnote{Since $A_k$ and $B_k$ are left unspecified at this stage, maximal anti-correlation of phase-space variables thus generalises to any $\alpha$-vacuum. The only exception is the vacuum choice for with $ A_k= B_k$. In that case, one needs to expand the mode functions one order beyond in the super-Hubble limit, and the argument does not apply anymore. This case is further discussed in \Sec{sec:QuantumToClassical}.}

\begin{figure}[t]
\begin{center}
\includegraphics[width=0.49\textwidth]{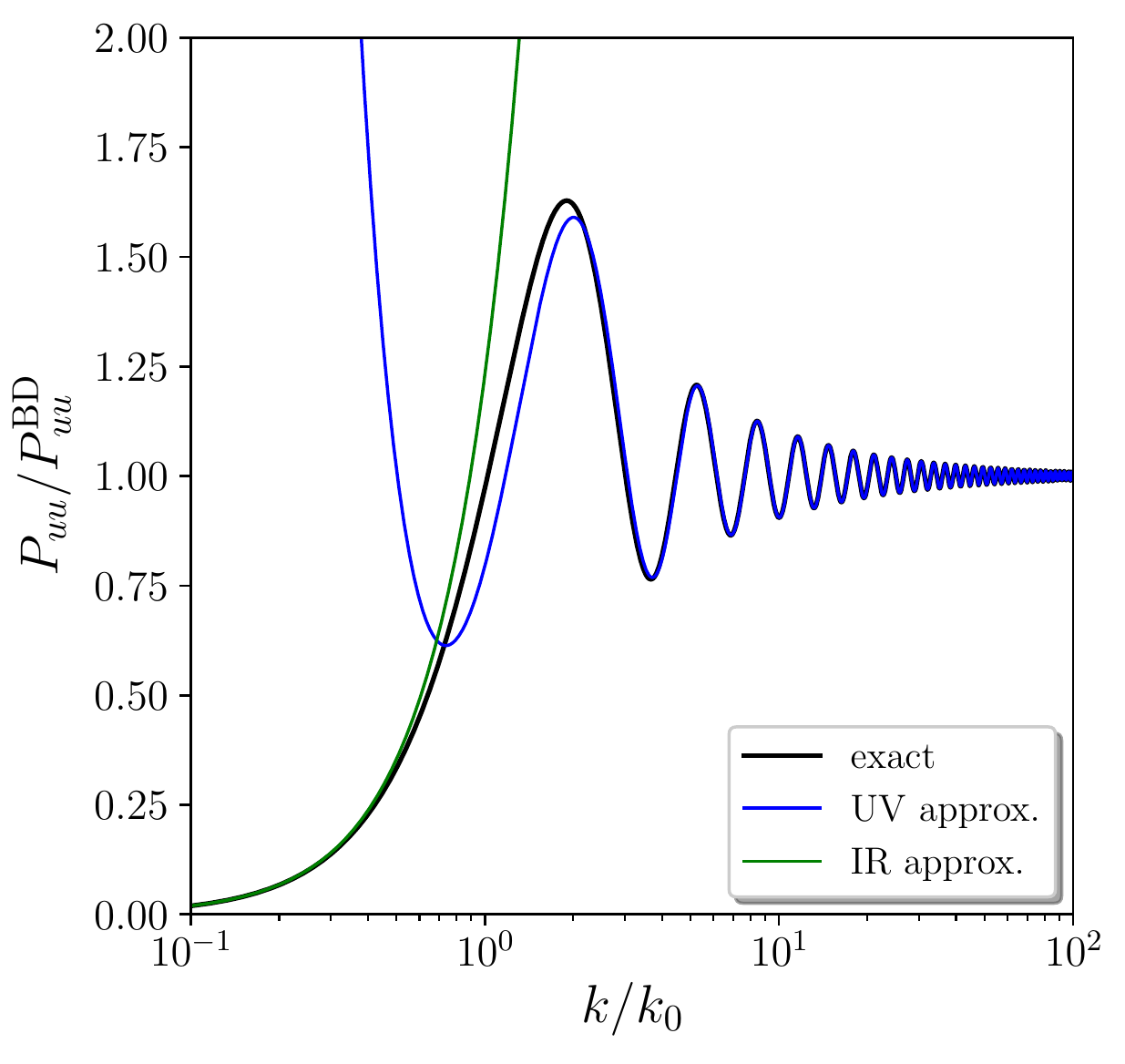}
\includegraphics[width=0.49\textwidth]{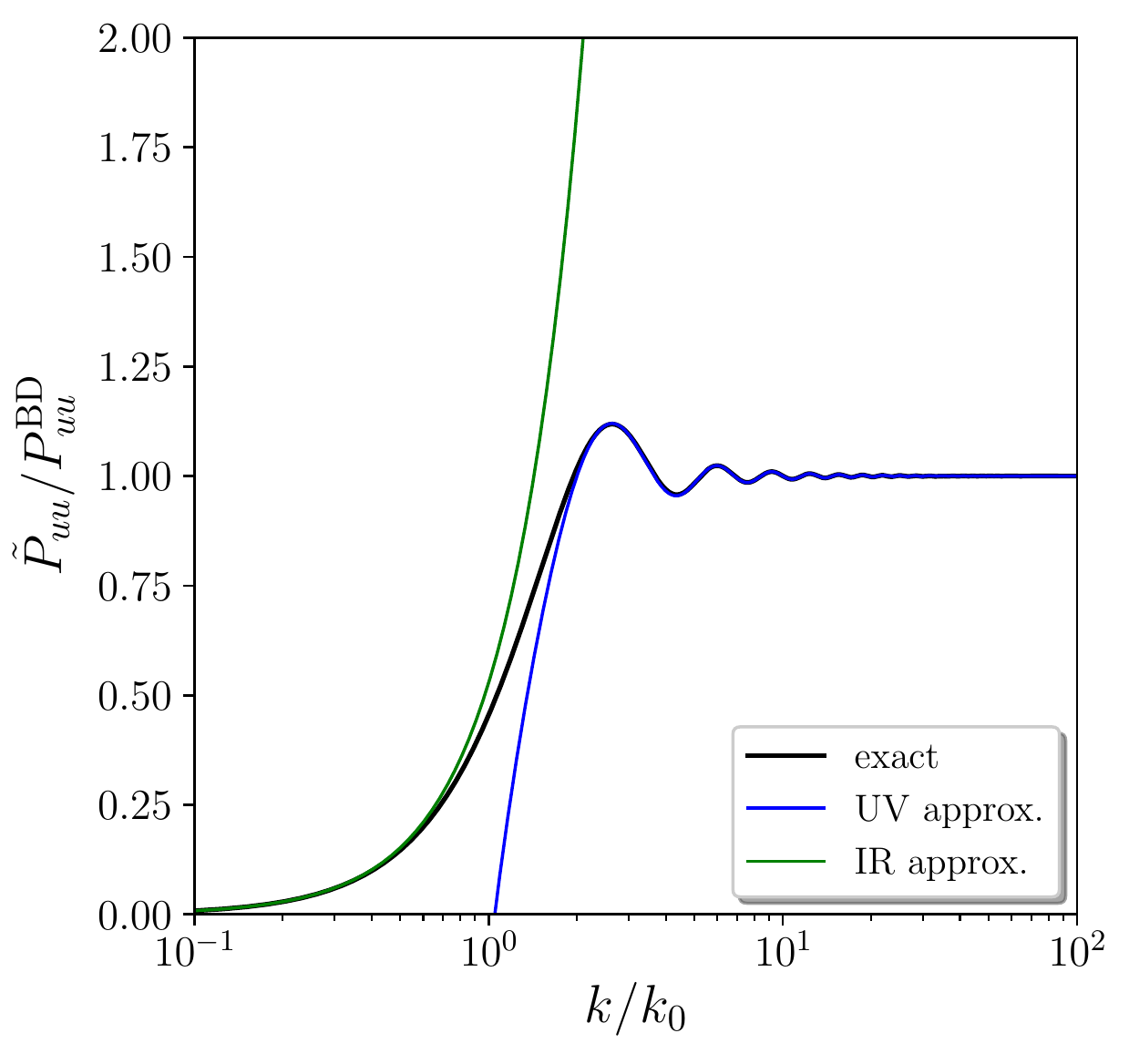}
\caption{Power spectrum in the $\boldsymbol{u}$-vacuum (left panel) and in the $\boldsymbol{v}$-vacuum (right panel), normalised to the one in the Bunch-Davies vacuum, as a function of $k/k_0$ where $k_0=-1/\eta_0$, for $\nu=3/2-0.1$. The black line corresponds to the exact result, while the blue line stands for the large-$k/k_0$ approximation, \Eq{eq:PowerPspectrum:u:BD:superH:subHeta0} in the left panel and \Eq{eq:PowerPspectrum:v:BD:superH:subHeta0} in the right panel, and the green line stands for the small-$k/k_0$ approximation, \Eq{eq:PowerPspectrum:u:BD:superH:superHeta0} in the left panel and \Eq{eq:PowerPspectrum:v:BD:superH:superHeta0} in the right panel.}  
\label{fig:modified:Spectrum}
\end{center}
\end{figure}

In order to gain some further insight, one can approximate this result in the limit where initial conditions are set in the sub-Hubble regime, $-k\eta_0 \gg 1$. In that limit, one can make use of \Eqs{eq:Ak:large:keta0} and~\eqref{eq:Bk:large:keta0}, and \Eq{eq:PowerPspectrum:u:BD:superH} gives rise to 
\bea
\label{eq:PowerPspectrum:u:BD:superH:subHeta0}
P \kern-0.2em \left(k,-k\eta\to 0\right)&\simeq & \left\lbrace  1 + \frac{\cos\left(\nu\pi+2k\eta_0\right)}{-k\eta_0}+\frac{4+\left(4\nu^2-5\right)\sin\left(\nu\pi+2k\eta_0\right)}{8\left(-k\eta_0\right)^2} 
\right\rbrace P^\mathrm{BD} \kern-0.2em\left(k,-k\eta\to 0\right) .
\nonumber \\
\eea
Let us comment again on the two cases mentioned in the discussion below \Eq{eq:vartheta:u:v}. If the initial time $\eta_0$ is set independently of $\vec{k}$, the above expression indicates that oscillations are added to the power spectrum, with an amplitude that decays with $k$. This can be checked in \Fig{fig:modified:Spectrum} where the power spectrum of $u_{\vec{k}}$ has been displayed as a function of $k/k_0$ where $k_0=-1/\eta_0$. In the situation where initial conditions are set at a fixed physical scale however, \ie such that $-k \eta_0(k)$ is constant, the power spectra are simply multiplied by an overall, $k$-independent constant. In this case, no additional features are present in the power spectra, which are simply rescaled. One should however note that this is a consequence of the de Sitter invariance: in a space time inflating in the slow-roll regime, scale-dependent corrections would be obtained. Finally, in the opposite limit, \ie when initial conditions are set in the super-Hubble regime, $-k\eta_0 \ll 1$, one finds\footnote{One needs to use a higher-order expansion of the Hankel function, namely~\cite{abramowitz+stegun}
\bea
H_{\nu}^{(1)}(z)\underset{\vert z \vert \ll 1}\simeq \frac{- i}{\pi}\left[\Gamma(\nu)\left(\frac{z}{2}\right)^{-\nu}+\Gamma\left(\nu-1\right)\left(\frac{z}{2}\right)^{2-\nu}+\Gamma(-\nu)e^{-i\nu\pi}\left(\frac{z}{2}\right)^{\nu}
\right]
\eea
which is valid for $0<\nu<2$ (for $\nu<1$, the ordering between the two last terms needs to be flipped). One then makes use of \Eqs{eq:uk:BD} and~({\ref{eq:pk:BD}}), and~(\ref{eq:Auvac}) and~(\ref{eq:Buvac}), in order to derive \Eq{eq:PowerPspectrum:u:BD:superH:subHeta0}.\label{footnote:IR:lim}
}
\bea
\label{eq:PowerPspectrum:u:BD:superH:superHeta0}
P \kern-0.2em \left(k,-k\eta\to 0\right)&\simeq & 
\left(\nu+\frac{3}{2}\right)^2\frac{\Gamma^2\left(-\nu\right)}{4\pi} \sin^2\left(\nu\pi\right)\left(\frac{k}{2k_0}\right)^{2\nu-1}
P^\mathrm{BD} \kern-0.2em\left(k,-k\eta\to 0\right) ,
\nonumber \\
\eea
where this expression applies if $1<\nu<2$. As a consequence, on large scales, if the initial time $\eta_0$ is set independently of $\vec{k}$, the power spectrum is suppressed compared to the Bunch-Davies one.

\paragraph{The  \textit{\textbf{v}}-vacuum--} 
Let us finally consider the case of the $\boldsymbol{v}$-vacuum. We proceed in a similar way as what was done for the $\boldsymbol{u}$-vacuum. The two coefficients $\tilde{A}_k$ and $\tilde{B}_k$ are still determined by the condition $\boldsymbol{e}_\alpha=\boldsymbol{U\tilde{D}}_k\boldsymbol{u}_k(\eta_0)$, see \Eq{eq:initialD}, except that instead of working with $\boldsymbol{D}_k=\mathrm{diag}\left(\sqrt{k},1/\sqrt{k}\right)$, one uses $\boldsymbol{\tilde{D}}_k=\boldsymbol{M}\boldsymbol{D}_k$, where $\boldsymbol{M}=\boldsymbol{U}^\dag\boldsymbol{\mathcal{M}}\boldsymbol{U}$, see \Eq{eq:calM:def}, and $\boldsymbol{\mathcal{M}}$ is given by \Eq{eq:canomatapp}. This expression for $\boldsymbol{\tilde{D}}_k$ has been derived making use of \Eq{eq:twovacua:def}, and is such that $\widehat{\boldsymbol{d}}_{\vec{k}}=\boldsymbol{U} \boldsymbol{\tilde{D}}_k \widehat{\boldsymbol{u}}_{\vec{k}}$. It gives rise to
\bea
\boldsymbol{\tilde{D}}_k = \left(\begin{array}{cc}
\sqrt{k} & 0 \\
\dfrac{aH}{\sqrt{k}} & \dfrac{1}{\sqrt{k}}
\end{array}\right),
\eea
which leads to
\bea
	\tilde{A}_k&=&\sqrt{\frac{k}{2}}u^{(2)}_{\vec{k}}(\eta_0)\left(1+\frac{i}{k\eta_0}\right)-\frac{i}{\sqrt{2k}} {p^{(2)}_{\vec{k}}(\eta_0)}\, , \label{eq:Avvac}\\
	\tilde{B}_k&=&-\sqrt{\frac{k}{2}}u^{(1)}_{\vec{k}}(\eta_0)\left(1+\frac{i}{k\eta_0}\right)+\frac{i}{\sqrt{2k}}{p^{(1)}_{\vec{k}}(\eta_0)}\, . \label{eq:Bvvac}
\eea
Combined with \Eqs{eq:uk:BD}, \eqref{eq:pk:BD} and \eqref{eq:PowerPspectrum:u:BD:superH} (where $A_k$ and $B_k$ have to be replaced with $\tilde{A}_k$ and $\tilde{B}_k$ respectively), this provides an explicit expression for the power spectra in the $\boldsymbol{v}$-vacuum. In the situation where initial conditions are set in the sub-Hubble regime, $k\eta_0\gg 1$, an approximated form of the result can be derived. In this limit indeed, making use of \Eq{eq:Hankel:large:arg}, one has
\bea
\label{eq:tildeAk:large:keta0}
\tilde{A}_k &\simeq& \ee^{ik\eta_0+i \nu\frac{\pi}{2}+i\frac{\pi}{4}} \left\lbrace1+\frac{1-4\nu^2}{8}\frac{i}{-k\eta_0} - \frac{\left(1-4\nu^2\right)^2}{128 \left(-k\eta_0\right)^2}+\mathcal{O}\left[\left(-k\eta_0\right)^{-3}\right]\right\rbrace ,\\
B_k &\simeq & \ee^{-ik\eta_0-i \nu\frac{\pi}{2}-i\frac{\pi}{4}} \left\lbrace \frac{1-4\nu^2}{16\left(- k\eta_0\right)^2}+\mathcal{O}\left[\left(-k\eta_0\right)^{-3}\right]\right\rbrace .
\label{eq:tildeBk:large:keta0}
\eea
Plugging these expressions into \Eq{eq:PowerPspectrum:u:BD:superH}, one obtains
\bea
\label{eq:PowerPspectrum:v:BD:superH:subHeta0}
\tilde{P} \left(k,-k\eta\to 0\right)&\simeq & \left\lbrace  1 +
\frac{1-4\nu^2}{8\left(-k\eta_0\right)^2}\sin\left(\nu\pi+2k\eta_0\right)
+\frac{\left(1-4\nu^2\right)^2}{64\left(-k\eta_0\right)^3}\cos\left(\nu\pi+2k\eta_0\right)
\right\rbrace
\nonumber \\ & & \quad \times
 P^\mathrm{BD} \left(k,-k\eta\to 0\right) .
\nonumber \\
\eea
This result has a similar structure as what was obtained in the $\boldsymbol{u}$-vacuum, see \Eq{eq:PowerPspectrum:u:BD:superH:subHeta0}, with the noticeable difference that the oscillations are now modulated by $(-k\eta_0)^{-2}$ instead of $(-k\eta_0)^{-1}$. They are therefore more suppressed, as can be checked in \Fig{fig:modified:Spectrum}. In the opposite situation where initial conditions are set in the super-Hubble regime, $-k\eta_0\ll 1$, following a similar approach as the one underlined in footnote~\ref{footnote:IR:lim}, one obtains
\bea
\label{eq:PowerPspectrum:v:BD:superH:superHeta0}
P \kern-0.2em \left(k,-k\eta\to 0\right)&\simeq & 
\left(\nu+\frac{1}{2}\right)^2\frac{\Gamma^2\left(-\nu\right)}{4\pi} \sin^2\left(\nu\pi\right)\left(\frac{k}{2k_0}\right)^{2\nu-1}
P^\mathrm{BD} \kern-0.2em\left(k,-k\eta\to 0\right) ,
\nonumber \\
\eea
for $1<\nu<2$, which is very similar to \Eq{eq:PowerPspectrum:u:BD:superH:superHeta0}. The two vacua therefore share the same suppression at large scales.

These results illustrate that, if initial conditions are set at a finite time in the past, not only different predictions that those derived in the Bunch-Davies vacuum are obtained, but these predictions depend on the choice of canonical variables (here $\boldsymbol{u}$ or $\boldsymbol{v}$) used to describe the system. The amplitude of the superimposed oscillations is found to be suppressed by some power (which depends on the choice of canonical variables) of the ratio between the physical wavelength of the mode under consideration and the Hubble radius at the initial time, as is often encountered in trans-Planckian effects~\cite{Martin:2003kp}.
\subsection{Quantum-to-classical transition}
\label{sec:QuantumToClassical}
In the cosmological context, the squeezing that operates for perturbations on super-Hubble scales if often described as playing a role in the so-called ``quantum-to-classical'' transition of these fluctuations~\cite{Polarski:1995jg, Lesgourgues:1996jc, Kiefer:2008ku}. Indeed, in the two-point correlation matrix of the squeezed state, see \Eq{eq:2ptcorr:2msq}, the term proportional to $\boldsymbol{J}_y$ comes from the non-vanishing commutator between the two phase-space variables. It is therefore natural to define ``classicality'' as being the regime where this term provides a negligible contribution to the correlation matrix. More precisely, when the expectation value of the anti-commutator, $P_{up}(k)$ (which provides the contribution proportional to $\boldsymbol{J}_x$), is much larger than the expectation value of the commutator, $1/2$ (which provides the contribution proportional to $\boldsymbol{J}_y$), the quantum imprint onto the correlation matrix can be neglected and the system is therefore ``classical''. Such a criterion translates into $\vert P_{up}\vert \gg 1$, such that a ``classical'' system is one for which maximal (anti-)correlation between the phase space variables is reached, see the discussion below \Eq{eq:up:correlator}. 

From \Eq{eq:Pup}, a necessary condition for this is $\vert r_k\vert \gg 1$, which is why large squeezing is usually associated with classicality. For instance, in \Refs{Martin:2015qta, Grain:2017dqa}, it is explicitly shown that expectation values of hermitian quartic combinations of phase-space variables do not depend on the ordering of these phase-space variables (as one would expect if, instead of quantum operators, they could be described by classical stochastic quantities) in the large-squeezing limit. This ``classicalisation'' should however be taken with a grain of salt for the following reason.

For a given physical state, the squeezing parameters are only defined up to a choice of canonical variables. Therefore, a criterion based on the value of the squeezing parameters is not invariant under reparameterisation of phase space. For instance, below \Eq{eq:up:correlator}, it is explained that maximal (anti-)correlation is in fact obtained when $\vert \varphi_k\vert \gg \ee^{-2r_k}$. One can always operate a change of canonical variables that absorbs the squeezing angle and sets it to $0$ (this simply means working with the major and minor axes of the Wigner ellipse as the canonical variables). In that case, the two phase-space variables are entirely uncorrelated, see \Eq{eq:Pup}, and according to the above criterion the system would have to be said to be highly quantum. In that case indeed, the expectation values of hermitian quartic combinations of phase-space variables does strongly depend on the ordering of these phase-space variables. Another striking example is the invariant representation introduced in \Sec{ssec:invariant}, where the squeezing amplitude $r_k$ vanishes at any time, hence the two phase-space variables are again uncorrelated. A genuine property of a quantum system cannot depend on the choice of phase-space variables used to describe it, which is why the above criterion only provides partial insight.

Since a two-mode squeezed state features a non-negative Wigner function, it is also often said that \Eq{eq:expectationweyl} can be interpreted as the  Wigner function playing the role of a quasi-distribution function in phase space, and endowing it with classical properties (note that this is independent of the amount of squeezing). This is however the case only if the operator $\widehat{\mathcal{O}}$ in \Eq{eq:expectationweyl} is such that its Weyl transform, $\mathcal{O}_{(W)}$, takes values with the set of eigenvalues of $\widehat{\mathcal{O}}$. Such operators are called ``proper'', but there are ``improper'' operators~\cite{2005PhRvA..71b2103R} for which this property is not satisfied, and of which the expectation value can never be reproduced by a stochastic theory, even in the large-squeezing limit. 

Improper operators can, for instance, give rise to explicit Bell inequality violations~\cite{Martin:2016tbd, Martin:2017zxs}. Let us note that, in the setup discussed in \Refa{Martin:2016tbd}, Bell inequalities are found to be violated if $\vert r_k \gg 1\vert$ and $\varphi_k<0.34 e^{-\left|r_k\right|}$. So if one works with canonical variables such that $r_k\gg 1$ and $e^{-2\left|r_k\right|}<\varphi_k<0.34 e^{-\left|r_k\right|}$, one both has maximal (anti-)correlation and violation of Bell inequalities, implying that even maximal (anti-)correlation (which, we recall, is anyway not a canonically invariant concept) does not guarantee the erasure of all quantum imprints.

The above discussion should however be tempered with the following remark. Although it is true that large squeezing does not remove quantum correlations in principle, it can make these quantum correlations more difficult to observe in practice. To illustrate this statement, let us consider cosmological perturbations during inflation, described with the canonical variables $u_{\vec{k}}$ and $p_{\vec{k}}$, or with $v_{\vec{k}}$ and $\Pi_{\vec{k}}$ (see table~\ref{tab:variables}). As perturbations cross out the Hubble radius, the squeezing amplitude increases, and quantum correlations are indeed not suppressed, they even strongly increase~\cite{Martin:2016tbd}. However, in the standard theory of cosmological perturbations, the stretched and squeezed directions of phase space are associated to a ``growing'' and ``decaying'' mode respectively, and although the non-vanishing commutator between these two modes can give rise to, \eg, Bell inequality violations, their experimental confirmation would require to measure the decaying mode, which is made smaller as squeezing proceeds. In this case, large squeezing is also responsible for hiding the decaying mode, hence preventing the observational detection of quantum correlations. Of course, this description holds within certain choices of canonical variables only (\eg $u_{\vec{k}}$ and $p_{\vec{k}}$ or $v_{\vec{k}}$ and $\Pi_{\vec{k}}$), and one can find some for which the squeezing amplitude remains constant, or even decreases on large scales. But when the system is parametrised in terms of quantities that can be concretely measured, large squeezing does signal the practical difficulty in measuring some of them, and in this sense, it can be associated with effective classicalisation.
\section{Conclusion}
\label{sec:conclusion}
In this work we have investigated various aspects of canonical transformations. After noting that they are endowed with a symplectic structure, we have studied mathematical properties of the symplectic group. We have then discussed how canonical transformations generate the dynamics of classical and quantum scalar fields, making the connection with the squeezing formalism. We have also highlighted the fact that different canonical variables select out different vacuum states. In the context of inflationary cosmology, where initial conditions are often set in the vacuum, under the assumption that all pre-existing classical fluctuations are red-shifted by the accelerated expansion, this results into an ambiguity: different initial conditions are obtained by working with different canonical variables, and we have shown explicitly how this translates into obtaining different observational predictions if initial conditions are set at a finite time in the past. The formalism developed in this work also allowed us to shed new light on issues related to the quantum-to-classical transition of cosmological perturbations.

Let us note that the concept of quantum vacuum also depends on the choice of space-time coordinates. In quantum field theory on the flat Minkowski space-time, there is a distinguished Lorentz frame, so a particular space-time splitting. This is not the case in general relativity, where different slicings of de Sitter space time would further yield different vacuum states~\cite{Brandenberger:1984wt}. 

Note that other approaches, such as setting initial conditions in the ground state of the Hamiltonian (\ie the eigenstate of the Hamiltonian with minimum eigenvalue) rather than in the vacuum state (\ie the state annihilated by annihilation operators), have been proposed~\cite{Handley:2016ods}. However, under a time-dependent canonical transformation, the Hamiltonian changes according to \Eq{eq:Hamiltonian:canonical:transform} and the ground state is mapped onto a state which is generically not an eigenstate of the new Hamiltonian, hence is not its ground state. As a consequence, ``ground state'' is not a canonically-invariant concept, and cannot be used to uniquely define initial conditions either. 

Ultimately, the issue of initial conditions in cosmology is a delicate one. If inflation is preceded by another cosmological epoch, such as a contracting phase followed by a bounce for instance, the state of the universe at the onset of inflation is inherited from this earlier epoch. If inflation lasts arbitrarily long, the Bunch-Davies vacuum does feature some specific properties, such as \eg being de Sitter invariant, but it is also the case for any state obtained from the Bunch-Davies vacuum through a $k$-independent Bogolyubov transformation~\cite{PhysRevD.32.3136}. It is also sometimes argued that the Bunch-Davies vacuum is a local attractor~\cite{Brandenberger:1985fc, Kaloper:2018zgi}, which is mainly due to the explosive particle production mechanism taking place on super-Hubble scales: for any state that differs from the Bunch-Davies vacuum by a finite number of particles, this difference is quickly overtaken by the vast number of particles created on super-Hubble scales, and becomes negligible. However, this remains true for other reference states, and we have seen explicit examples where substantial differences with the Bunch-Davies predictions remain even at late time. If inflation lasts arbitrarily long, the trans-Planckian problem~\cite{Brandenberger:2012aj} also arises, and one may have to trace the details of initial conditions all the way back to the quantum gravity regime.

Let us finally mention that the formalism presented in this work could be further extended to more generic situations, such as those containing multiple fields. If two scalar fields contribute to cosmological perturbations for instance, they can be decomposed into an adiabatic and an entropic degree of freedom, and canonical transformations on this extended phase space belong to the group $\mathrm{Sp}(4,\mathbb{R})$. It would be interesting to study the properties of this group, using similar techniques as the ones employed here to investigate $\mathrm{Sp}(2,\mathbb{R})$, and to use them to analyse quantum correlations in multiple-field inflationary models. If gravitational non linearities were to be included, in order to study non-free fields, or cosmological perturbations at higher order, quantum states would not remain separable in Fourier space, and one would also have to generalise the present setup to non-linear canonical transformations, which would be technically more challenging. 
\begin{acknowledgments}
V.~V. acknowledges funding from the European Union's
Horizon 2020 research and innovation programme under the Marie Sk\l odowska-Curie grant agreement N${}^0$ 750491.
\end{acknowledgments}
\appendix
\section{Composition law for $\mathrm{SU}(1,1)$}
\label{app:composition}
In this appendix, we derive the composition law of matrices in $\mathrm{SU}(1,1)$ that are presented in \Sec{sec:SU11:Composition}. Considering two matrices with parameters $(\gamma_1,d_1,\gamma'_1)$, and $(\gamma_2,d_2,\gamma'_2)$, respectively, we want to obtain the parameters $(\gamma,d,\gamma')$ of the product matrix between these two, see \Eq{eq:com:su11:formal}. Defining $\gamma_+$, $\gamma_-$ and $\gamma_{1,2}$ according to \Eqs{eq:gamma+-12:def}, in \Sec{sec:SU11:Composition} it is shown that the system that needs to be inverted is given by \Eqs{eq:c1}-(\ref{eq:c4}) that we reproduce here for convenience: 
\bea
	\cosh d&=&\cos(\gamma_+)\cos(\gamma_{1,2})\cosh(d_1+d_2)-\sin(\gamma_+)\sin(\gamma_{1,2})\cosh(d_1-d_2), \label{eq:c1:app}\\
	\sinh d&=&\cos(\gamma_-)\cos(\gamma_{1,2})\sinh(d_1+d_2)+\sin(\gamma_-)\sin(\gamma_{1,2})\sinh(d_1-d_2), \label{eq:c2:app}\\
	0&=&\sin(\gamma_-)\cos(\gamma_{1,2})\sinh(d_1+d_2)-\cos(\gamma_-)\sin(\gamma_{1,2})\sinh(d_1-d_2), \label{eq:c3:app}\\
	0&=&\sin(\gamma_+)\cos(\gamma_{1,2})\cosh(d_1+d_2)+\cos(\gamma_+)\sin(\gamma_{1,2})\cosh(d_1-d_2). \label{eq:c4:app}
\eea
The unknowns are the squeezing parameter $d$ and the two angles $\gamma_\pm$, to be expressed as functions of the two squeezing parameters $d_1$ and $d_2$ and the angle $\gamma_{1,2}$. The system is singular if either $\sin\gamma_{1,2}=0$ or $\cos\gamma_{1,2}=0$, for which we first derive the solution, before turning to the case where neither $\sin\gamma_{1,2}$  nor $\cos\gamma_{1,2}$ vanishes.

\subsection{The case $\sin\gamma_{1,2}=0$}
\label{sec:singamma12_eq_0}
This case corresponds to $\gamma'_1+\gamma_2=n\pi$ with $n\in\mathbb{Z}$, meaning that $\cos\gamma_{1,2}=(-1)^n$. The set of constraints reduces to
\bea
	\cosh d&=&(-1)^n\cos(\gamma_+)\cosh(d_1+d_2), \\
	\sinh d&=&(-1)^n\cos(\gamma_-)\sinh(d_1+d_2), \\
	0&=&\sin(\gamma_-)\sinh(d_1+d_2), \\
	0&=&\sin(\gamma_+)\cosh(d_1+d_2).
\eea
One should then consider two sub-cases depending on whether $d_1+d_2$ is vanishing or not.

First, if $d_1+d_2\neq0$, then the two last equations give $\gamma_\pm=m_\pm\pi$. Since the hyperbolic cosine is strictly positive, the first constraint leads to $m_+=n$ modulo 2, and $d=\pm\left(d_1+d_2\right)$. From the second constraint, one can set either (i) $d=d_1+d_2$ and $m_-=n$ modulo 2, or (ii) $d=-d_1-d_2$ and $m_-=n+1$ modulo 2. Let us note that these two possible solutions in fact lead to the very same matrix thanks to the symmetry~(\ref{eq:sym}).

Second, if $d_1+d_2=0$, then the second constraint leads to $d=0$. The first constraint subsequently gives $\gamma_+=m_+\pi$ with $m_+=n$ modulo 2, while $\gamma_-$ remains unconstrained. This is because, in this case, $\boldsymbol{\mathcal{M}} = \boldsymbol{\mathcal{R}}(\gamma+\gamma')= \boldsymbol{\mathcal{R}}(\gamma_1+\gamma_2'-\gamma_+)$ (see footnote~\ref{footnote:gamma:gammapm}), in which $\gamma_-$ does not appear.

\subsection{The case $\cos\gamma_{1,2}=0$}
This case corresponds to $\gamma'_1+\gamma_2=(n+1/2)\pi$ with $n\in\mathbb{Z}$, meaning that $\sin\gamma_{1,2}=(-1)^n$. The set of constraints simplifies to
\bea
	\cosh d&=&(-1)^{n+1}\sin(\gamma_+)\cosh(d_1-d_2), \\
	\sinh d&=&(-1)^{n}\sin(\gamma_-)\sinh(d_1-d_2), \\
	0&=&\cos(\gamma_-)\sinh(d_1-d_2), \\
	0&=&\cos(\gamma_+)\cosh(d_1-d_2).
\eea
Two sub-cases need again to be distinguished, depending on wether $d_1-d_2$ is vanishing or not. If $d_1-d_2=0$, then one obtains $d=0$, $\gamma_+=m_+\pi$ with $m_+=n+1$ modulo 2, and $\gamma_-$ remains unconstrained. If $d_1-d_2\neq0$, one obtains $\gamma_\pm=(m_\pm+1/2)\pi$ from the last two constraints. The first constraint then gives $d=\pm\left(d_1-d_2\right)$ and $m_+=n+1$ modulo 2.  The second constraint finally yields either (i) $d=d_1-d_2$ and $m_-=n$ modulo 2, or (ii) $d=-d_1+d_2$ and $m_-=n+1$ modulo 2. As is the case for $\sin\gamma_{1,2}=0$, see \Sec{sec:singamma12_eq_0}, these two solutions give the same final matrix because of the symmetry~(\ref{eq:sym}). 

\subsection{The general case}
In the generic situation where both $\cos\gamma_{1,2}$ and $\sin\gamma_{1,2}$ are different from zero, one has to solve for the full set of constraints. We proceed as follows: first we determine $\gamma_+$ and $\cosh d$, and second we determine $\gamma_-$ and $\sinh d$. By determining  $\gamma_\pm$, we mean determining both their sine and cosine.
In order to lighten the expressions, we introduce the following notations: $x_\pm=\cos\gamma_\pm$, $y_\pm=\sin\gamma_{\pm}$, and $d_\pm=d_1\pm d_2$.

Let us start with $\gamma_+$ and $\cosh d$. We determine first the squares of $x_+$  and of $\cosh d$. To this end, we add the squares of \Eqs{eq:c1:app} and~(\ref{eq:c4:app}), which allows one to get rid of the terms $\propto\sin\gamma_+\cos\gamma_+$,
\bea
	\cosh^2d&=& \cos^2\gamma_{1,2}\cosh^2 d_+ + \sin^2\gamma_{1,2}\cosh^2 d_-\, .
\eea
One can then derive an alternative expression for $\cosh^2d$ by making use of \Eq{eq:c4:app} to express $\sin\gamma_+$ as a function of $\cos\gamma_+$, and by plugging the resulting expression into \Eq{eq:c1:app} to obtain
\bea
	\cosh^2d&=&x_+^2\left(\frac{\cos^2\gamma_{1,2}\cosh^2 d_+ +\sin^2\gamma_{1,2}\cosh^2 d_-}{\cos\gamma_{1,2}\cosh d_+}\right)^2.
\eea
This leads to expressions for $x_+^2$ and $\cosh^2d$ as functions of $f_+=\cos^2\gamma_{1,2}\cosh^2 d_+$ and $f_-=\sin^2\gamma_{1,2}\cosh^2 d_-$ (which are both positive-valued),
\bea
	\cosh^2d&=&f_++f_- \label{eq:coshdapp}\, , \\
	x^2_+&=&\frac{f_+}{f_++f_-} \label{eq:x+^2}\, .
\eea
It is straightforward to check that the right-hand side of the above are positive valued, and for \Eq{eq:x+^2}, that it is smaller than one. 

Let us now go from $\cosh^2d$ and $x^2_+$, to $\cosh d$ and $x_+$. This is done by noticing that hyperbolic cosines are positive valued so we can use the positive square root of $\cosh^2d$ as given by \Eq{eq:coshdapp}, 
\bea
	\cosh d=\sqrt{f_++f_-}\, . 
\eea
From \Eq{eq:c4:app}, one can write $x_+/y_+=-\mathrm{sgn}(\cos\gamma_{1,2})\mathrm{sgn}(\sin\gamma_{1,2})\sqrt{f_+/f_-}$, which, once plugged back into \Eq{eq:c1:app}, gives rise to
\bea
	x_+&=&\mathrm{sgn}\left(\cos\gamma_{1,2}\right) \sqrt{\frac{f_+}{f_++f_-}}\, .
\eea
In the above, $\mathrm{sgn}(f)$ is the sign function, meaning $+1$ for $f>0$, $-1$ for $f<0$, and $0$ if $f=0$. 

To fully determine $\gamma_+$, one finally needs to derive its sine (remind that $x_+=\cos\gamma_+$). This is easily obtained using \Eq{eq:c4:app}, which leads to
\bea
	y_+&=&-\mathrm{sgn}\left(\sin\gamma_{1,2}\right)\sqrt{\frac{f_-}{f_++f_-}}\, .
\eea
It is straightforward to check that our solution satisfies $x^2_++y^2_+=1$. This closes the determination of $\cosh d$ and $\gamma_+$. \\

The second step consists in proceeding similarly for deriving $x_-$ and $\sinh d$. The first expression for $\sinh^2d$ is obtained by adding the squared of \Eqs{eq:c2:app} and~(\ref{eq:c3:app}), leading to
\bea
\label{eq:sinhdapp}
	\sinh^2d&=&g_+ + g_-\, ,
\eea
where we have introduced $g_+=\cos^2\gamma_{1,2}\sinh^2 d_+$ and $g_-=\sin^2\gamma_{1,2}\sinh^2 d_-$. Unlike the case of $\cosh d$ and $\gamma_+$, one should now consider both the positive and the negative roots of \Eq{eq:sinhdapp}. It is however sufficient to consider \eg the positive root, since the two roots are mapped one to each other by sending $d\to -d$, which leads to the same final matrix $\boldsymbol{\mathcal{M}}$ provided the angles are mapped according to \Eq{eq:sym}. We thus take
\bea
	\sinh d&=&\sqrt{g_+ + g_-}\, .
\eea
Solving \Eqs{eq:c2:app} and~(\ref{eq:c3:app}) as a linear system in $x_-$ and $y_+$ then leads to 
\bea
	x_-&=&\mathrm{sgn}\left(\cos\gamma_{1,2}\right)\sqrt{\frac{g_+}{g_++g_-}}\, ,\\
	y_-&=&\mathrm{sgn}\left(\sin\gamma_{1,2}\right)\sqrt{\frac{g_-}{g_++g_-}}\, .
\eea
We note that apart from the minus sign in the expression for $y_-$, the results are the same as the ones for $(\cosh d,x_+,x_-)$ if one replaces $f_\pm$ by $g_\pm$ (\ie if one replaces $\cosh d_\pm$ by $\sinh d_\pm$).

\section{Quantum representations of $\frak{su}(1,1)$ and $\frak{sp}(2,\mathbb{R})$}
\label{app:representation}
In this appendix, we derive quantum representations of the Lie algebra of $\mathrm{Sp}(2,\mathbb{R})$ and $\mathrm{SU}(1,1)$, first in terms of the field operators (working with $\mathrm{Sp}(2,\mathbb{R})$), and then in terms of the creation and annihilation operators (working with $\mathrm{SU}(1,1)$).
\subsection{Field operators}
\label{sec:QuantumRepresentation:FieldOperators}
Let us start with the field operators $\widehat{\boldsymbol{z}}_{\vec{k}}$ satisfying the commutation relation~(\ref{eq:commutator:hatz}). We study the action of the symplectic group $\mathrm{Sp}(2,\mathbb{R})$, which is a Lie Group with a two-dimensional representation. Let $\boldsymbol{M}^{(k)}$ be a matrix of $\mathrm{Sp}(2,\mathbb{R})$, and consider the matrix generators $\boldsymbol{K}_a$ with $a=1,2,3$ introduced in \Eq{eq:K123} (note that the generators do not depend on the wavenumber), which satisfy the algebra $\left[\boldsymbol{K}_a,\boldsymbol{K}_b\right]=f_{ab}^c\boldsymbol{K}_c$ given below \Eq{eq:K123}.\footnote{From the relations given below \Eq{eq:K123}, the non-vanishing structure constants are $f_{12}^3=-f_{21}^3 = f_{13}^2=-f_{31}^2=f_{23}^1=f_{32}^1=-2$.} 

Close to the identity, one can expand 
\bea
\label{eq:Mk:linearisation}
\boldsymbol{M}^{(k)}\simeq\boldsymbol{I}+\ds\sum_{a=1}^3\epsilon^a_{(k)}\boldsymbol{K}_a
\eea 
with $\epsilon^a_{(k)}\in\mathbb{R}^3$, and $|\sum_a\epsilon^a_{(k)}\epsilon^a_{(k)}|\ll1$. As recalled below \Eq{eq:K123}, since $\boldsymbol{M}^{(k)}$ satisfies \Eq{eq:symp}, one has $\left(\boldsymbol{\Omega K}_a\right)^\mathrm{T}=\boldsymbol{\Omega K}_a$. We also note that since we restrict our study to isotropic canonical transformations, the infinitesimal vector $\epsilon^a_{(k)}$ is labelled by the wavenumber only (extension to non-isotropic transformations is straightforward).

Let us then consider the following linear transformation of the field operator,
\bea
	\widehat{\boldsymbol{z}}_{\vec{k},\mu}\to\widehat{\boldsymbol{z}}'_{\vec{k},\mu}=\ds\sum_{\nu=1}^2\boldsymbol{M}^{(k)}_{\mu\nu}\widehat{\boldsymbol{z}}_{\vec{k},\nu} . \label{eq:lintransfo}
\eea
Since the transformation is for all $\vec{k}$-modes, there should exist a unitary transformation, $\widehat{{M}}$, representing the action of all the $\boldsymbol{M}^{(k)}$'s, such that operators transform according to 
\bea
\label{eq:Lie:Transf:operator}
\widehat{{O}}\to\widehat{{O}}'=\widehat{{M}}^\dag\widehat{{O}}\widehat{{M}}.
\eea
One can design a systematic way to derive such a unitary operator following approaches developed in quantum mechanics and quantum optics \cite{PhysRevA.36.3868,SIMON1987223,Arvind:1995ab}. The first step consists in finding the representation of the generator using the phase-space operator $\left\{\hat{\boldsymbol{z}}_{\vec{k}}\right\}_{\vec{k}\in\mathbb{R}^3}$. Then one can either use the polar decomposition or the Bloch-Messiah decomposition~(\ref{eq:bloch:left:right}) for any element of the symplectic group, expressed in terms of exponentials of the generators, see \Eq{eq:polar:calK}. The unitary representation $\widehat{{M}}$ is finally obtained by replacing the generator $\boldsymbol{K}_a$ by their operational representation in the exponential map.

Close to the identity, one can expand 
\bea
\label{eq:hatMk:linearisation}
\widehat{{M}}\simeq\widehat{{I}}+i\ds\int_{\mathbb{R}^3}\dd^3k\epsilon^a_{(k)}\widehat{{K}}^{(\vec{k})}_a ,
\eea
where the $\widehat{{K}}^{(\vec{k})}_a$'s are hermitic operators. Compared to the matrix linearisation~(\ref{eq:Mk:linearisation}), there is an additional $i$ factor since the $\widehat{{K}}^{(\vec{k})}_a$'s are hermitic operators and $\widehat{{M}}$ is unitary.\footnote{Alternatively, one could impose the $\widehat{{K}}^{(\vec{k})}_a$ to be {\it anti-hermitic}, \ie $\left(\widehat{{K}}^{(\vec{k})}_a\right)^\dag=-\widehat{{K}}^{(\vec{k})}_a$, and remove the factor $i$ for the transformation to be unitary. This is just a matter of convention.} The infinitesimal version of the operator transformation rule~(\ref{eq:Lie:Transf:operator}) then reads
\bea
\label{eq:Lie:Transf:operator:infinitesimal}
	\widehat{{O}}'\simeq\widehat{{O}}+i\left[\widehat{{O}},\ds\int_{\mathbb{R}^3}\dd^3k\epsilon^a_{(k)}\widehat{{K}}^{(\vec{k})}_a\right].
\eea
The above applies to the specific case of the field operators $\widehat{\boldsymbol{z}}_{\vec{k},\mu}$, and should be identical to the infinitesimal transformation deduced from \Eq{eq:lintransfo}, \ie $\widehat{\boldsymbol{z}}'_{\mu,\vec{k}}=\widehat{\boldsymbol{z}}_{\mu,\vec{k}}+\epsilon^a_{(k)}\ds\sum_{\nu=1}^2\boldsymbol{K}_{a,\mu\nu}\widehat{\boldsymbol{z}}_{\vec{k},\nu}$. For this to be true, one notes that $\widehat{{K}}_a^{(\vec{k})}$ has to be a quadratic form of $\widehat{\boldsymbol{z}}_{\vec{k}}$, \ie 
\bea
\label{eq:hatK:Q}
\widehat{{K}}^{(\vec{k})}_a=\widehat{\boldsymbol{z}}^\dag_{\vec{k}}\boldsymbol{Q}^a\widehat{\boldsymbol{z}}_{\vec{k}}
\eea 
with $\boldsymbol{Q}^a$ some matrices that are independent of the wavevector. This ensures the commutator $\left[\widehat{\boldsymbol{z}}_{\vec{k},\mu},\widehat{{K}}^{(\vec{q})}_a\right]$ appearing in the right-hand side of \Eq{eq:Lie:Transf:operator:infinitesimal} to be linear in the phase-space operators. Moreover, the hermiticity of the $\widehat{{K}}^{(\vec{k})}_a$'s imposes that the matrices $\boldsymbol{Q}_a$ are also hermitic. One can then explicitly compute the commutator describing the infinitesimal transformation, \ie
\bea
	\left[\widehat{\boldsymbol{z}}_{\mu,\vec{k}},\ds\int_{\mathbb{R}^3}\dd^3q\epsilon^a_{(q)}\widehat{{K}}^{(\vec{q})}_a\right]&=&\ds\int_{\mathbb{R}^3}\dd^3q\epsilon^a_{(q)}\ds\sum_{\nu,\kappa=1}^2\left[\widehat{\boldsymbol{z}}_{\mu,\vec{k}},\widehat{\boldsymbol{z}}^\dag_{\vec{q},\nu}\boldsymbol{Q}^a_{\nu\kappa}\widehat{\boldsymbol{z}}_{\vec{q},\kappa}\right], 
\eea
for all $\mu$ and $\vec{k}$. By using the commutation relation~(\ref{eq:commutator:hatz}) and the fact that $\widehat{\boldsymbol{z}}^\dag_{\vec{q},\nu}=\widehat{\boldsymbol{z}}_{-\vec{q},\nu}$ [which follows from the scalar field being real-valued, as explained above \Eq{eq:Hamiltonian:Fourrier}], one obtains
\bea
	\left[\widehat{\boldsymbol{z}}_{\mu,\vec{k}},\ds\int_{\mathbb{R}^3}\dd^3q\epsilon^a_{(q)}\widehat{{K}}^{(\vec{q})}_a\right]&=&\epsilon^a_{(k)}\ds\sum_{\nu,\kappa=1}^2i\boldsymbol{\Omega}_{\mu\nu}\left(\boldsymbol{Q}^a_{\nu\kappa}+\boldsymbol{Q}^a_{\kappa\nu}\right)\widehat{\boldsymbol{z}}_{\vec{k},\kappa}\, .
\eea
Only the symmetric part of the matrix $\boldsymbol{Q}^a$ is needed in the above, meaning that we can restrict these hermitic matrices to be real and symmetric matrices. By using $\boldsymbol{\Omega}^\mathrm{T}=\boldsymbol{\Omega}^{-1}=-\boldsymbol{\Omega}$, one finally obtains
\bea
\label{eq:Q:sol}
	\boldsymbol{Q}^a=-\frac{1}{2}\boldsymbol{\Omega}\boldsymbol{K}_a\, ,
\eea
which is obviously real and symmetric, as recalled below \Eq{eq:Mk:linearisation}. We finally note that $\widehat{{K}}^{(-\vec{k})}_a=\widehat{{K}}^{(\vec{k})}_a$ since $\boldsymbol{Q}^a$ is real and symmetric, and given that $\widehat{\boldsymbol{z}}^\dag_{\vec{k},\mu}=\widehat{\boldsymbol{z}}_{-\vec{k},\mu}$.\footnote{The integral over $\mathbb{R}^3$ in the linearisation~(\ref{eq:hatMk:linearisation}) of $\widehat{{M}}$ can therefore be restricted to half of $\mathbb{R}^{3}$, \ie
\bea
	\widehat{{M}}\simeq\widehat{{I}}-i\ds\int_{\mathbb{R}^{3+}}\dd^3k\epsilon^a_{(k)}\left(\widehat{\boldsymbol{z}}_{\vec{k}}^\dag\boldsymbol{\Omega K}_a\widehat{\boldsymbol{z}}_{\vec{k}}\right).
\eea}

One can then check that the infinitesimal operators appearing in the right-hand side of \Eq{eq:hatMk:linearisation} satisfy the same commutation relations as the ones appearing in the right-hand side of \Eq{eq:Mk:linearisation}, \ie
\bea
	\left[i\ds\int_{\mathbb{R}^3}\dd^3k\epsilon^a_{(k)}\widehat{{K}}^{(\vec{k})}_a,i\ds\int_{\mathbb{R}^3}\dd^3q\epsilon'^b_{(q)}\widehat{{K}}^{(\vec{q})}_b\right]=i\ds\int_{\mathbb{R}^3}\dd^3kf_{ab}^c\epsilon^a_{(k)}\epsilon'^b_{(k)}\widehat{{K}}^{(\vec{k})}_c .
\eea
This can be verified in a straightforward (though rather cumbersome) way from the expression $\widehat{{K}}^{(\vec{k})}_a=-\frac12\widehat{\boldsymbol{z}}^\dag_{\vec{k}}\boldsymbol{\Omega}\boldsymbol{K}_a\widehat{\boldsymbol{z}}_{\vec{k}}$ that arises from combining \Eqs{eq:hatK:Q} and~(\ref{eq:Q:sol}). This is precisely what is needed for the $\widehat{{K}}^{(\vec{k})}_a$'s to be a representation of the generators of $\mathrm{Sp}(2,\mathbb{R})$. 

The explicit expressions for the generators is finally
\bea
\kern-2em
	\int_{\mathbb{R}^3}\dd^3k\widehat{{K}}^{(\vec{k})}_1&=&-\frac{1}{2}\ds\int_{\mathbb{R}^3}\dd^3k\left(\widehat{\phi}_{\vec{k}}^\dag\widehat{\phi}_{\vec{k}}-\widehat{\pi}_{\vec{k}}^\dag\widehat{\pi}_{\vec{k}}\right)=-\frac{1}{2}\ds\int_{\mathbb{R}^{3+}}\dd^3k\left(\left\{\widehat{\phi}_{\vec{k}},\widehat{\phi}^\dag_{\vec{k}}\right\}_{\kern-0.3em+}-\left\{\widehat{\pi}_{\vec{k}},\widehat{\pi}^\dag_{\vec{k}}\right\}_{\kern-0.3em+}\right), \\
\kern-2em
	\int_{\mathbb{R}^3}\dd^3k\widehat{{K}}^{(\vec{k})}_2&=&\frac{1}{2}\ds\int_{\mathbb{R}^3}\dd^3k\left(\widehat{\phi}_{\vec{k}}^\dag\widehat{\phi}_{\vec{k}}+\widehat{\pi}_{\vec{k}}^\dag\widehat{\pi}_{\vec{k}}\right)=\frac{1}{2}\ds\int_{\mathbb{R}^{3+}}\dd^3k\left(\left\{\widehat{\phi}_{\vec{k}},\widehat{\phi}^\dag_{\vec{k}}\right\}_{\kern-0.3em+}+\left\{\widehat{\pi}_{\vec{k}},\widehat{\pi}^\dag_{\vec{k}}\right\}_{\kern-0.3em+}\right),  \\
\kern-2em
	\int_{\mathbb{R}^3}\dd^3k\widehat{{K}}^{(\vec{k})}_3&=&\frac{1}{2}\ds\int_{\mathbb{R}^3}\dd^3k\left(\widehat{\phi}_{\vec{k}}^\dag\widehat{\pi}_{\vec{k}}+\widehat{\pi}_{\vec{k}}^\dag\widehat{\phi}_{\vec{k}}\right)=\frac{1}{2}\ds\int_{\mathbb{R}^{3+}}\dd^3k\left(\left\{\widehat{\phi}_{\vec{k}},\widehat{\pi}^\dag_{\vec{k}}\right\}_{\kern-0.3em+}+\left\{\widehat{\phi}^\dag_{\vec{k}},\widehat{\pi}_{\vec{k}}\right\}_{\kern-0.3em+}\right), 
\eea
where $\left\{\widehat{O}_1,\widehat{O}_2\right\}_{\kern-0.3em+}=\widehat{O}_1\widehat{O}_2+\widehat{O}_2\widehat{O}_1$ stands for the anti-commutator. The second equalities are obtained by first splitting the integral over $\mathbb{R}^{3+}$ and $\mathbb{R}^{3-}$, and then by using the fact that $\widehat{\boldsymbol{z}}_{\vec{k},\mu}^\dag=\widehat{\boldsymbol{z}}_{-\vec{k},\mu}$ since one deals with a real-valued scalar field. This allows one to express the integral over $\mathbb{R}^{3-}$ as an integral over $\mathbb{R}^{3+}$, and introduce the anti-commutator.
\subsection{Creation and annihilation operators}
\label{sec:QuantumRepresentation:CreationAnnihilationOperators}
A similar study can be performed by using instead the creation and annihilation operators. In this case, one is working with the group $\mathrm{SU}(1,1)$. From the considerations presented in \Sec{sec:LieAlgebra}, the linearisation of the elements of this group is of the form
\bea
	\boldsymbol{\mathcal{M}}^{(k)}\simeq\boldsymbol{I}-2i\epsilon^x\boldsymbol{\mathcal{K}}_x-2i\epsilon^y\boldsymbol{\mathcal{K}}_y+2i\epsilon^z\boldsymbol{\mathcal{K}}_z,
\eea
with $\epsilon^a\in\mathbb{R}^3$ for $a=x$, $y$ and $z$, and $\boldsymbol{\mathcal{K}}_x$, $\boldsymbol{\mathcal{K}}_y$ and $\boldsymbol{\mathcal{K}}_z$ have been introduced below \Eq{eq:Lie:commutation}. The conditions~(\ref{eq:eqdef:SU11}) imply that $\boldsymbol{\mathcal{K}}^\dag_a\boldsymbol{\mathcal{J}}-\boldsymbol{\mathcal{J}}\boldsymbol{\mathcal{K}}_a=0$, and since $\boldsymbol{\mathcal{J}}$ is anti-hermitian, see \Eq{eq:J:def}, one obtains that $\boldsymbol{\mathcal{J K}}_a$ is anti-hermitian too, \ie $\left(\boldsymbol{\mathcal{J K}}_a\right)^\dag=-\boldsymbol{\mathcal{J K}}_a$.

For the creation and annihilation operator, the ``phase-space" operators satisfy the canonical commutation relation~(\ref{eq:commutator:hat:a}), and evolve under the action of $\boldsymbol{\mathcal{M}}^{(k)}$ according to
\bea
\label{eq:a:evolv:calK:xyz:infinitesimal}
	\widehat{\boldsymbol{a}}'_{\vec{k},\mu}&\simeq&\widehat{\boldsymbol{a}}_{\vec{k},\mu}+\ds\sum_{\nu}\left(-2i\epsilon^x\boldsymbol{\mathcal{K}}_x-2i\epsilon^y\boldsymbol{\mathcal{K}}_y+2i\epsilon^z\boldsymbol{\mathcal{K}}_z\right)_{\mu\nu}\widehat{\boldsymbol{a}}_{\vec{k},\nu}\, ,
\eea
where we remind that $\mu,~\nu$ indices should read as $\widehat{\boldsymbol{a}}_{\vec{k},\mu}=\widehat{a}_{\vec{k}},~\widehat{a}^\dag_{-\vec{k}}$ for $\mu=1,~2$ respectively.

On the other hand, similarly to what was done for the field operators, one can introduce the unitary transformation $\widehat{\mathcal{M}}$ that represents the action of all the $\boldsymbol{\mathcal{M}}^{(k)}$, and with which the creation and annihilation operators transform as $\widehat{\boldsymbol{a}}'_{\vec{k},\mu}=\widehat{\mathcal{M}}^\dag\widehat{\boldsymbol{a}}_{\vec{k},\mu}\widehat{\mathcal{M}}$. Linearising this relation, one obtains 
\bea
\label{eq:a:evolv:calK:xyz:infinitesimal:representation}
	\widehat{\boldsymbol{a}}'_{\vec{k},\mu}&\simeq&\widehat{\boldsymbol{a}}_{\vec{k},\mu}+\left[\widehat{\boldsymbol{a}}_{\vec{k},\mu},\ds\int_{\mathbb{R}^{3}}\dd^3q\left(-2i\epsilon^x\widehat{\mathcal{K}}^{(\vec{q})}_x-2i\epsilon^y\widehat{\mathcal{K}}^{(\vec{q})}_y+2i\epsilon^z\widehat{\mathcal{K}}^{(\vec{q})}_z\right)\right],
\eea
where the $\widehat{\mathcal{K}}^{(\vec{q})}_a$'s have to be hermitic operators for $\widehat{\mathcal{M}}$ to be unitarity. Identifying the right-hand sides of \Eqs{eq:a:evolv:calK:xyz:infinitesimal} and~(\ref{eq:a:evolv:calK:xyz:infinitesimal:representation}), one finds that these operators are quadratic in the creation and annihilation operators, \ie $\widehat{\mathcal{K}}^{(\vec{q})}_a=\widehat{\boldsymbol{a}}_{\vec{q}}^\dag\boldsymbol{\mathcal{Q}}^a\widehat{\boldsymbol{a}}_{\vec{q}}$ with $\boldsymbol{\mathcal{Q}}^a$ a hermitic matrix, and an explicit calculation of the commutator gives
\bea
	\boldsymbol{\mathcal{Q}}^a=\frac{i}{2}\boldsymbol{\mathcal{J K}}_a\, .
\eea
From the expressions given for $\boldsymbol{\mathcal{J K}}_a$ below \Eq{eq:Lie:commutation}, one can readily check that the above solutions are indeed hermitic matrices.
This gives rise to
\bea
	\int_{\mathbb{R}^3}\dd^3k\,\widehat{\mathcal{K}}^{(\vec{k})}_x&=&\frac{i}{2}\ds\int_{\mathbb{R}^{3+}}\dd^3k\left(\widehat{a}^\dag_{\vec{k}}\widehat{a}^\dag_{-\vec{k}}-\widehat{a}_{\vec{k}}\widehat{a}_{-\vec{k}}\right), \\
	\int_{\mathbb{R}^3}\dd^3k\,\widehat{\mathcal{K}}^{(\vec{k})}_y&=&\frac{1}{2}\ds\int_{\mathbb{R}^{3+}}\dd^3k\left(\widehat{a}^\dag_{\vec{k}}\widehat{a}^\dag_{-\vec{k}}+\widehat{a}_{\vec{k}}\widehat{a}_{-\vec{k}}\right), \\
	\int_{\mathbb{R}^3}\dd^3k\,\widehat{\mathcal{K}}^{(\vec{k})}_z&=&\frac{1}{2}\ds\int_{\mathbb{R}^{3+}}\dd^3k\left(\widehat{a}^\dag_{\vec{k}}\widehat{a}_{\vec{k}}+\widehat{a}^\dag_{-\vec{k}}\widehat{a}_{-\vec{k}}+1\right),
\eea
and we also define $\widehat{\mathcal{K}}^{(\vec{k})}_\pm=\widehat{\mathcal{K}}^{(\vec{k})}_x\pm i\widehat{\mathcal{K}}^{(\vec{k})}_y$. This leads to \Eqs{eq:K+:a}-(\ref{eq:Kz:a}) quoted in the main text.

\section{Canonical transformations for the one-mode variables}
\label{app:realvariables}
The use of the Wigner-Weyl function in \Sec{sec:RelatingObservationalPredictions} requires to introduce the hermitian and one-mode coordinates of the phase space $(\mathfrak{q}_{\vec{k}},\mathfrak{q}_{-\vec{k}},\mathfrak{p}_{\vec{k}},\mathfrak{p}_{-\vec{k}})$~\cite{Martin:2015qta} defined in \Eqs{eq:Wigner:Canonical:Transform:phi}-\eqref{eq:Wigner:Canonical:Transform:pi}. In this Appendix, we study how applications of canonical transformations in the two-mode representation $(\phi_{\vec{k}},\phi_{-\vec{k}},\pi_{\vec{k}},\pi_{-\vec{k}})$ used throughout this article, translate into canonical transformations in the one-mode representation needed to use the Wigner-Weyl function.  

This is done in the classical phase space, but since only linear transformations are needed, there are no issues of operator ordering when extending to the quantum phase space. Therefore, the following results can be directly translated to the quantum phase space by replacing field variables by field operators, and helicity basis variables by creation and annihilation operators. The helicity basis~\eqref{eq:a(z)} will also be systematically defined with $\boldsymbol{D}_k=\mathrm{diag}\left(\sqrt{k},1/\sqrt{k}\right)$, since any other choice for $\boldsymbol{D}_k$ can be recast that way by introducing a canonical transformation, which is precisely the purpose of this Appendix. The following considerations are therefore generic. 
\subsection{Definitions, notations, and posing the problem}
The initial phase space is given by $\Gamma=\bigotimes_{\vec{k}\in\mathbb{R}^{3+}}\left(\phi_{\vec{k}},\pi_{\vec{k}}\right)$. Each subspace $\left(\phi_{\vec{k}},\pi_{\vec{k}}\right)$ contains two modes since it is unequivocally mapped to the space $\left(a_{\vec{k}},a^*_{-\vec{k}}\right)$ by using the unitary transform~\eqref{eq:a(z)} generated by $\boldsymbol{U}$, \ie $\boldsymbol{a}_{\vec{k}}=\boldsymbol{UD}_k\boldsymbol{z}_{\vec{k}}$. The one-mode and hermitian representation of the phase space is obtained using $\Gamma=\bigotimes_{\vec{k}\in\mathbb{R}^{3}}\left(\mathfrak{q}_{\vec{k}},\mathfrak{p}_{\vec{k}}\right)$ with
\bea
\label{eq:Wigner:Canonical:Transform:phi:App}
 \mathfrak{q}_{\vec{k}} &=& \frac{1}{2}\left[\phi_{\vec{k}}+\phi_{-\vec{k}}+\frac{i}{k}\left(\pi_{\vec{k}}-\pi_{-\vec{k}}\right)\right]\\
 \mathfrak{p}_{\vec{k}} &=& \frac{1}{2i}\left[k\left(\phi_{\vec{k}}-\phi_{-\vec{k}}\right)+i\left(\pi_{\vec{k}}+\pi_{-\vec{k}}\right)\right],
 \label{eq:Wigner:Canonical:Transform:pi:App}
 \eea
see \Eqs{eq:Wigner:Canonical:Transform:phi} and~\eqref{eq:Wigner:Canonical:Transform:pi}.
Since $\phi_{\vec{k}}^*=\phi_{-\vec{k}}$ and $\pi_{\vec{k}}^*=\pi_{-\vec{k}}$, one can check that $\mathfrak{q}_{\vec{k}}^*=\mathfrak{q}_{\vec{k}}$ and $\mathfrak{p}_{\vec{k}}^*=\mathfrak{p}_{\vec{k}}$, hence the one-mode representation is indeed hermitian. The corresponding helicity basis is made of one single mode. Indeed, introducing
\bea
	\left(\begin{array}{c}
		\mathfrak{a}_{\vec{k}} \\
		\mathfrak{a}^*_{\vec{k}}
	\end{array}\right)&:=&\boldsymbol{UD}_k\left(\begin{array}{c}
		\mathfrak{q}_{\vec{k}} \\
		 \mathfrak{p}_{\vec{k}}
	\end{array}\right), 
\eea
one can readily check from \Eq{eq:a(z)} that
\bea
	\left(\begin{array}{c}
		\mathfrak{a}_{\vec{k}} \\
		\mathfrak{a}^*_{\vec{k}}
	\end{array}\right)=\left(\begin{array}{c}
		{a}_{\vec{k}} \\
		{a}^*_{\vec{k}}
	\end{array}\right).
\eea
Unlike $\boldsymbol{a}_{\vec{k}}=\left(a_{\vec{k}},a^*_{-\vec{k}}\right)^\mathrm{T}$, which involves the modes $\vec{k}$ and $-\vec{k}$, the above helicity basis $\boldsymbol{\mathfrak{a}}_{\vec{k}}=\left(a_{\vec{k}},a^*_{\vec{k}}\right)^\mathrm{T}$ is made of one single mode $\vec{k}$. \\

The canonical transformations used in the core of this paper act on the two-mode subspaces $\left(\phi_{\vec{k}},\pi_{\vec{k}}\right)$ without mixing them [in particular, they do not mix $(\phi_{\vec{k}},\pi_{\vec{k}})$ and $(\phi_{-\vec{k}},\pi_{-\vec{k}})$]. The one-mode subspaces $\left(\mathfrak{q}_{\vec{k}},\mathfrak{p}_{\vec{k}}\right)$ and $\left(\mathfrak{q}_{-\vec{k}},\mathfrak{p}_{-\vec{k}}\right)$ are however mixed within $\left(\phi_{\vec{k}},\pi_{\vec{k}}\right)$, so studying those canonical transformations in the one-mode representation requires to work in a four-dimensional subspace that contains the degrees of freedom of opposite wavevectors. 

To this end, we introduce the four-vectors containing the dimensionless configuration and momentum variables, 
\bea
	\vec{\boldsymbol{\phi}}_{\vec{k}}=\left(\begin{array}{c}
		\sqrt{k}\phi_{\vec{k}} \\
		\sqrt{k}\phi_{-\vec{k}} \\
		\pi_{\vec{k}}/\sqrt{k} \\
		\pi_{-\vec{k}}/\sqrt{k}
	\end{array}\right) & ~~~\mathrm{and}~~~ & \vec{\boldsymbol{q}}_{\vec{k}}=\left(\begin{array}{c}
		\sqrt{k}\mathfrak{q}_{\vec{k}} \\
		\sqrt{k}\mathfrak{q}_{-\vec{k}} \\
		\mathfrak{p}_{\vec{k}}/\sqrt{k} \\
		\mathfrak{p}_{-\vec{k}}/\sqrt{k}
	\end{array}\right),
\eea
where the upper arrows refer to four-dimensional structures in phase space. The corresponding helicity basis is built by first introducing the block-wise matrix
\bea
	\vec{\boldsymbol{U}}=\frac{1}{\sqrt{2}}\left(\begin{array}{cc}
		\boldsymbol{I} & i\boldsymbol{I} \\
		\boldsymbol{I} & -i \boldsymbol{I}
	\end{array}\right),
\eea
which is nothing but the Kronecker product between the matrix $\boldsymbol{U}$, defined in \Eq{eq:U:def}, and the identity $\boldsymbol{I}$, \ie $\vec{\boldsymbol{U}} = \boldsymbol{U}\otimes\boldsymbol{I} $.
This is a unitary matrix, \ie $\vec{\boldsymbol{U}}\vec{\boldsymbol{U}}^\dag=\vec{\boldsymbol{I}}$, with $\vec{\boldsymbol{I}}$ the $4\times4$ identity matrix. The helicity variables are then defined by
\bea
	\vec{\boldsymbol{a}}_{\vec{k}}=\vec{\boldsymbol{U}}\vec{\boldsymbol{\phi}}_{\vec{k}} & ~~~\mathrm{and}~~~ & \vec{\boldsymbol{\mathfrak{a}}}_{\vec{k}}=\vec{\boldsymbol{U}}\vec{\boldsymbol{q}}_{\vec{k}}\, .
\eea	
(Note that the application of the matrix $\boldsymbol{D}_k$ is already accounted for by considering the dimensionless configuration and momentum variables.) We stress that $\vec{\boldsymbol{a}}_{\vec{k}}\neq\vec{\boldsymbol{\mathfrak{a}}}_{\vec{k}}$: they have the same entries but ordered differently, since
\bea
	\vec{\boldsymbol{a}}_{\vec{k}}=\left(\begin{array}{c}
		a_{\vec{k}} \\
		a_{-\vec{k}} \\
		a^*_{-\vec{k}} \\
		a^*_{\vec{k}} 
	\end{array}\right) & ~~~\mathrm{and}~~~ & \vec{\boldsymbol{\mathfrak{a}}}_{\vec{k}}=\left(\begin{array}{c}
		a_{\vec{k}} \\
		a_{-\vec{k}} \\
		a^*_{\vec{k}} \\
		a^*_{-\vec{k}} 
	\end{array}\right).
\eea
\\

In the following, we show how canonical transformations in the two-mode representation translate into canonical transformations in the one-mode representation. For completeness, this is done using both the field variables and the helicity variables, though as it will be clear, it is much easier in the helicity basis. We stress that such a mapping for canonical transformations directly applies to the evolution of those variables under the Green's matrix, since both the Green's matrices and the canonical transformations are generated by the same group, either $\mathrm{Sp}(2,\mathbb{R})$ for the field variables or $\mathrm{SU}(1,1)$ for the helicity variables.
\subsection{Relating the two-mode and one-mode representations}
\label{subapp:Vmatrix}
\subsubsection{Field variables} 
The relation~(\ref{eq:Wigner:Canonical:Transform:phi:App})-(\ref{eq:Wigner:Canonical:Transform:pi:App}) between the one- and the two-mode variables can be recast in matricial form 
\bea
	\vec{\boldsymbol{q}}_{\vec{k}}=\vec{\boldsymbol{V}}\vec{\boldsymbol{\phi}}_{\vec{k}},
\eea
where the $4\times4$ matrix $\vec{\boldsymbol{V}}$ is given by
\bea
	\vec{\boldsymbol{V}}=\left(\begin{array}{cc}
		 \boldsymbol{V}_+ & i\boldsymbol{V}_- \\
		-i\boldsymbol{V}_- & \boldsymbol{V}_+
	\end{array}\right),
\eea
and where we have introduced the $2\times2$ matrices 
\bea
	\boldsymbol{V}_{\pm}&=&\frac{1}{2}\left(\begin{array}{cc}
		1 & \pm1 \\
		\pm1 & 1
	\end{array}\right)=\frac{1}{2}\left(\boldsymbol{I}\pm \boldsymbol{J}_x\right).
\eea

Let us discuss the properties of these matrices. First, the matrix $\vec{\boldsymbol{V}}$ is both involutory (hence invertible) and hermitian, \ie $\vec{\boldsymbol{V}}^{-1}=\vec{\boldsymbol{V}}$ and $\vec{\boldsymbol{V}}^\dag=\vec{\boldsymbol{V}}$.

Second, the matrices $\boldsymbol{V}_\pm$ are projectors since $\left(\boldsymbol{V}_\pm\right)^2=\boldsymbol{V}_\pm$. They are thus singular though hermitian. Moreover, they project onto two {\it orthogonal} subspaces since $\boldsymbol{V}_\pm\boldsymbol{V}_\mp=\boldsymbol{0}$, which can be readily shown by noticing that $\left(\boldsymbol{J}_x\right)^2=\boldsymbol{I}$. This relation also allows one to show that $\boldsymbol{J}_x \boldsymbol{V}_{\pm}=\boldsymbol{V}_{\pm}\boldsymbol{J}_x=\pm\boldsymbol{V}_\pm$. Since $\boldsymbol{V}_+$ and $\boldsymbol{V}_-$ are projectors, their eigenvalues are $0$ and $1$. One can thus diagonalise \eg$\boldsymbol{V}_+$, 
\bea
	 \boldsymbol{V}_+=\boldsymbol{P}{\left(\begin{array}{cc}
	 	1 & 0 \\
		0 & 0
	\end{array}\right)}\boldsymbol{P}^{-1},
\eea
where the matrix $\boldsymbol{P}$ built from the eigenvectors of $\boldsymbol{V}_+$. The matrix $\boldsymbol{V}_-$ can then be written as 
\bea
	 \boldsymbol{V}_-=\boldsymbol{P}{\left(\begin{array}{cc}
	 	0 & 0 \\
		0 & 1
	\end{array}\right)}\boldsymbol{P}^{-1}.
\eea
The matrix $\boldsymbol{V}_-$ is indeed diagonalised in the same basis as $\boldsymbol{V}_+$ (but with reversed eigenvalues), since these two matrices are orthogonal projectors on a two-dimensional vector space. An explicit computation of the eigenvectors shows that $\boldsymbol{P}=\frac{1}{\sqrt{2}}\left(\boldsymbol{J}_x+\boldsymbol{J}_z\right)$, which is also an involutory matrix, \ie $\boldsymbol{P}^{-1}=\boldsymbol{P}$.

A last useful property is the following one. Consider two matrices which are linear combinations of $\boldsymbol{V}_\pm$, \ie $\boldsymbol{V}(\lambda,\mu)=\lambda\boldsymbol{V}_++\mu\boldsymbol{V}_-$ and $\boldsymbol{V}(\lambda',\mu')=\lambda'\boldsymbol{V}_++\mu'\boldsymbol{V}_-$ with $(\lambda,\mu,\lambda',\mu')\in\mathbb{C}^4$. These two matrices commute and their product is also a linear combination of $\boldsymbol{V}_\pm$ given by
\bea
	\boldsymbol{V}(\lambda,\mu)\boldsymbol{V}(\lambda',\mu')=\boldsymbol{V}(\lambda',\mu')\boldsymbol{V}(\lambda,\mu)=\boldsymbol{V}(\lambda\lambda',\mu\mu'). \label{eq:Vproduct}
\eea
This results from the fact that the matrices $\boldsymbol{V}_\pm$ form a set of two orthogonal projectors.\footnote{This property guarantees that the linear combinations of $\boldsymbol{V}_\pm$ form a two-dimensional abelian group over (and isomorphic to) ${\mathbb{C}^*}^2$. The identity is obtained from the parameters $\lambda=\mu=1$, and the inverse of $\boldsymbol{V}(\lambda,\mu)$ is $\boldsymbol{V}(\lambda^{-1},\mu^{-1})$ (note that the product law~(\ref{eq:Vproduct}) is valid on $\mathbb{C}^2$, working on ${\mathbb{C}^*}^2$ is mandatory only in order to be able to define the inverse.}\\

Let us now show that the transformation generated by $\vec{\boldsymbol{V}}$ preserves the symplectic structure. To this end, one first notes that the Poisson bracket in the two-mode representation reads
\bea
	\left\{\vec{\boldsymbol{\phi}}_{\vec{k}},\vec{\boldsymbol{\phi}}_{\vec{k}}^\dag\right\}=\vec{\boldsymbol{\Omega}}\, ,
\eea
with
\bea
	\vec{\boldsymbol{\Omega}}=\left(\begin{array}{cc}
		\boldsymbol{0} & \boldsymbol{I} \\
		-\boldsymbol{I} & \boldsymbol{0}
	\end{array}\right)
\eea
and where $\boldsymbol{0}$ is the $2\times2$ zero matrix. The matrix $\vec{\boldsymbol{\Omega}}$ is obtained by taking the Kronecker product of $\boldsymbol{\Omega}$ with the identity, \ie $\vec{\boldsymbol{\Omega}}=\boldsymbol{\Omega}\otimes\boldsymbol{I}$. In the one-mode representation, the Poisson bracket is thus
\bea
	\left\{\vec{\boldsymbol{q}}_{\vec{k}},\vec{\boldsymbol{q}}_{\vec{k}}^\dag\right\}=\vec{\boldsymbol{V}}\vec{\boldsymbol{\Omega}}\vec{\boldsymbol{V}}.
\eea
The right-hand side can be obtained using block-wise matrix multiplication, and reads
\bea
\label{eq:VOmegaV}
	\vec{\boldsymbol{V}}\vec{\boldsymbol{\Omega}}\vec{\boldsymbol{V}}=\left(\begin{array}{c|c}
		-i\left(\boldsymbol{V}_+\boldsymbol{V}_-+\boldsymbol{V}_-\boldsymbol{V}_+\right) & \left(\boldsymbol{V}_+\right)^2+ \left(\boldsymbol{V}_-\right)^2 \\ \hline
	-\left(\boldsymbol{V}_+\right)^2-\left(\boldsymbol{V}_-\right)^2 & \-i\left(\boldsymbol{V}_+\boldsymbol{V}_-+\boldsymbol{V}_-\boldsymbol{V}_+\right)
	\end{array}\right).
\eea
The diagonal blocks are vanishing since $\boldsymbol{V}_\pm\boldsymbol{V}_\mp=\boldsymbol{0}$. For the off-diagonal blocks, since $\left(\boldsymbol{V}_\pm\right)^2=\boldsymbol{V}_\pm$, they are given by $\pm\left(\boldsymbol{V}_++\boldsymbol{V}_-\right)=\pm\boldsymbol{I}$. One finally obtains 
\bea
	\left\{\vec{\boldsymbol{q}}_{\vec{k}},\vec{\boldsymbol{q}}_{\vec{k}}^\dag\right\}=\vec{\boldsymbol{\Omega}}\, ,
\eea
showing that the symplectic structure is indeed preserved when going from the two-mode representation to the one-mode representation.\footnote{At the quantum level, one replaces the Poisson brackets by the commutators, and in the right-hand side, $\vec{\boldsymbol{\Omega}}$ by $i\vec{\boldsymbol{\Omega}}$. It is then straightforward to see that the canonical commutation relations are preserved in exactly the same way, since all the transformations are linear.} \\
\subsubsection{Helicity basis}
Let us now show that the symplectic structures is preserved, but using the helicity basis instead. In this basis, the Poisson bracket is 
\bea
	\left\{\vec{\boldsymbol{a}}_{\vec{k}},\vec{\boldsymbol{a}}^\dag_{\vec{k}}\right\}=\vec{\boldsymbol{\mathcal{J}}}
\eea
where we introduce $\vec{\boldsymbol{\mathcal{J}}}=\boldsymbol{\mathcal{J}}\otimes\boldsymbol{I}$ with $\boldsymbol{\mathcal{J}}$ defined in \Eq{eq:J:def}. The mapping from $\vec{\boldsymbol{a}}_{\vec{k}}$ to $\vec{\boldsymbol{\mathfrak{a}}}_{\vec{k}}$ reads
\bea
\label{eq:a:fraka}
	\vec{\boldsymbol{\mathfrak{a}}}_{\vec{k}}=\underbrace{\left(\begin{array}{cc}
		\boldsymbol{I} & \boldsymbol{0} \\
		\boldsymbol{0} & \boldsymbol{J}_x
	\end{array}\right)}_{\vec{\boldsymbol{\mathcal{V}}}}\vec{\boldsymbol{a}}_{\vec{k}}\, ,
\eea
where $\vec{\boldsymbol{\mathcal{V}}}=\vec{\boldsymbol{U}}\vec{\boldsymbol{V}}\vec{\boldsymbol{U}}^\dag$. This matrix inherits the properties of involution and unitarity. The Poisson bracket then gives $\left\{\vec{\boldsymbol{\mathfrak{a}}}_{\vec{k}},\vec{\boldsymbol{\mathfrak{a}}}^\dag_{\vec{k}}\right\}=\left(\begin{array}{cc}
		\boldsymbol{I} & \boldsymbol{0} \\
		\boldsymbol{0} & \boldsymbol{J}_x
	\end{array}\right)\vec{\boldsymbol{\mathcal{J}}}\left(\begin{array}{cc}
		\boldsymbol{I} & \boldsymbol{0} \\
		\boldsymbol{0} & \boldsymbol{J}_x
	\end{array}\right)^\dag$. A direct calculation of the right-hand side using $\left(\boldsymbol{J}_x\right)^2=\boldsymbol{I}$ shows that it is equal to $\vec{\boldsymbol{\mathcal{J}}}$. 
\subsection{Canonical transformations}
\subsubsection{Field variables}
Let us now consider an isotropic canonical transformation in the two-mode field-variables representation, $\boldsymbol{M}_k\in\mathrm{Sp}(2,\mathbb{R})$, such that $\left(\Phi_{\vec{k}},\Pi_{\vec{k}}\right)^\mathrm{T}=\boldsymbol{M}_k\left(\phi_{\vec{k}},\pi_{\vec{k}}\right)^\mathrm{T}$. The entries of this matrix are denoted
\bea
	\boldsymbol{M}_k=\left(\begin{array}{cc}
		M^{(\Phi,\phi)}_k & M^{(\Phi,\pi)}_k \\
		M^{(\Pi,\phi)}_k & M^{(\Pi,\pi)}_k
	\end{array}\right),
\eea
where the determinant constraint (see below \Eq{eq:Omega:2D:def}) imposes $M^{(\Phi,\phi)}_kM^{(\Pi,\pi)}_k-M^{(\Phi,\pi)}_kM^{(\Pi,\phi)}_k=1$. Since we consider isotropic transformations, the same matrix applies to the $-\vec{k}$ subspace, \ie$\left(\Phi_{-\vec{k}},\Pi_{-\vec{k}}\right)^\mathrm{T}=\boldsymbol{M}_k\left(\phi_{-\vec{k}},\pi_{-\vec{k}}\right)^\mathrm{T}$. Hence in the four-dimensional subspace, the canonical transformation reads $\vec{\boldsymbol{\Phi}}_{\vec{k}}=\vec{\boldsymbol{M}}_k\vec{\boldsymbol{\phi}}_{\vec{k}}$,
with
\bea
	\vec{\boldsymbol{M}}_k=\left(\begin{array}{c|c}
		M^{(\Phi,\phi)}_k\boldsymbol{I} & M^{(\Phi,\pi)}_k\boldsymbol{I} \\ \hline
		M^{(\Pi,\phi)}_k\boldsymbol{I} & M^{(\Pi,\pi)}_k\boldsymbol{I}
	\end{array}\right),
	\label{eq:vecM:entries}
\eea
which is nothing but the Kronecker product $\boldsymbol{M}_k\otimes\boldsymbol{I}$. It is straightforward to check that if $\boldsymbol{M}_k\in\mathrm{Sp}(2,\mathbb{R})$, then $\vec{\boldsymbol{M}}_k\vec{\boldsymbol{\Omega}}\vec{\boldsymbol{M}}_k=\vec{\boldsymbol{\Omega}}$, hence the canonical transformation preserves the symplectic structure in the four-dimensional subspace. The Kronecker product preserves other important properties. For instance, since $\det(\boldsymbol{M}_k)=\det(\boldsymbol{I})=1$, one has $\det(\vec{\boldsymbol{M}}_k)=1$. This means that $\vec{\boldsymbol{M}}_k$ is invertible and one can check that its inverse is simply $\vec{\boldsymbol{M}}_k^{-1}=\boldsymbol{M}_k^{-1}\otimes\boldsymbol{I}$. Also, if $\boldsymbol{M}_k=\boldsymbol{I}$, then $\vec{\boldsymbol{M}}_k=\vec{\boldsymbol{I}}$. Finally, it can be shown that for two symplectic matrices $\boldsymbol{M}_k$ and $\boldsymbol{N}_k$, one has $\left(\boldsymbol{M}_k\otimes\boldsymbol{I}\right)\left(\boldsymbol{N}_k\otimes\boldsymbol{I}\right)=\left(\boldsymbol{M}_k\boldsymbol{N}_k\right)\otimes\boldsymbol{I}$. In other words, the matrices $\vec{\boldsymbol{M}}_{\vec{k}}$ form a group and feature the symplectic structure. They are thus elements of the four-dimensional symplectic group $\mathrm{Sp}(4,\mathbb{R})$.\footnote{These properties result from the following formulas: consider $\boldsymbol{A}$, $\boldsymbol{C}$ to be $n\times n$ matrices, and $\boldsymbol{B}$, $\boldsymbol{D}$ to be $m\times m$ matrices, then 
\bea
	&&\det(\boldsymbol{A}\otimes\boldsymbol{B})=\det(\boldsymbol{A})^n\det(\boldsymbol{B})^m,  \label{eq:Kronecker:P1}\\
	&&\left(\boldsymbol{A}\otimes\boldsymbol{B}\right)^{-1}=\boldsymbol{A}^{-1}\otimes\boldsymbol{B}^{-1}, \label{eq:Kronecker:P2} \\
	&&\left(\boldsymbol{A}\otimes\boldsymbol{B}\right)\left(\boldsymbol{C}\otimes\boldsymbol{D}\right)=\left(\boldsymbol{AC}\otimes\boldsymbol{BD}\right) \label{eq:Kronecker:P3},\\
	& &  \boldsymbol{A}\otimes \left(\boldsymbol{B}+\boldsymbol{D}\right)= \boldsymbol{A}\otimes\boldsymbol{B}+\boldsymbol{A}\otimes\boldsymbol{D}\label{eq:Kronecker:P4},\\
	& & \left(\boldsymbol{A}+\boldsymbol{C}\right)\otimes \boldsymbol{B} = \boldsymbol{A}\otimes\boldsymbol{B}+\boldsymbol{C}\otimes\boldsymbol{B}\label{eq:Kronecker:P5},
\eea
where \Eq{eq:Kronecker:P2} assumes $\boldsymbol{A}$ and $\boldsymbol{B}$ to be invertible. The preservation of the unit determinant comes from \Eq{eq:Kronecker:P1} by taking $\boldsymbol{A}=\boldsymbol{M}_k$ and $\boldsymbol{B}=\boldsymbol{I}$, and the preservation of the group composition law derives from \Eq{eq:Kronecker:P3} by setting $\boldsymbol{A}$ and $\boldsymbol{C}$ to be elements of $\mathrm{Sp}(2,\mathbb{R})$ and $\boldsymbol{B}=\boldsymbol{D}=\boldsymbol{I}$.} In the following, any $4\times4$ matrix that is built from the Kronecker product of $\boldsymbol{M}\in\mathrm{Sp}(2,\mathbb{R})$ with $\boldsymbol{I}$ will be denoted with an upper arrow, $\vec{\boldsymbol{M}}$.

The one-mode representation associated to $\vec{\boldsymbol{\Phi}}_{\vec{k}}$ can be studied by introducing $\vec{\boldsymbol{Q}}_{\vec{k}}=\vec{\boldsymbol{V}}\vec{\boldsymbol{\Phi}}_{\vec{k}}$. In the one-mode basis, the canonical transformation reads
\bea
	\vec{\boldsymbol{Q}}_{\vec{k}}=\left(\vec{\boldsymbol{V}}\vec{\boldsymbol{M}}_k\vec{\boldsymbol{V}}\right)\vec{\boldsymbol{q}}_{\vec{k}}\, ,
\eea
where we use that $\vec{\boldsymbol{V}}^\dag=\vec{\boldsymbol{V}}$. Let us briefly study the matrix $\vec{\boldsymbol{\mathfrak{M}}}_{{k}}:=\vec{\boldsymbol{V}}\vec{\boldsymbol{M}}_k\vec{\boldsymbol{V}}$, where hereafter we systematically use the font $\vec{\boldsymbol{\mathfrak{M}}}$ to denote the result of sandwiching a matrix $\vec{\boldsymbol{M}}$ with $\vec{\boldsymbol{V}}$. Since $\vec{\boldsymbol{V}}$ is involutory, $\det(\vec{\boldsymbol{V}}^2)=1$ (one can check explicitly that $\det (\vec{\boldsymbol{V}}) =-1$) and as noted below \Eq{eq:vecM:entries}, $\det(\vec{\boldsymbol{M}}_k)=1$, which implies that $\det(\vec{\boldsymbol{\mathfrak{M}}}_{{k}})=1$.  Using block-wise multiplications, one can then show that
\bea
	\vec{\boldsymbol{\mathfrak{M}}}_{{k}}=\left(\begin{array}{c|c}
		M^{(\Phi,\phi)}_k\boldsymbol{V}_++M^{(\Pi,\pi)}_k\boldsymbol{V}_- & M^{(\Phi,\pi)}_k\boldsymbol{V}_+-M^{(\Pi,\phi)}_k\boldsymbol{V}_- \\ \hline
		-M^{(\Phi,\pi)}_k\boldsymbol{V}_-+M^{(\Pi,\phi)}_k\boldsymbol{V}_+ & M^{(\Phi,\phi)}_k\boldsymbol{V}_-+M^{(\Pi,\pi)}_k\boldsymbol{V}_+
	\end{array}\right),
\eea 
where we made use of $\boldsymbol{V}_\pm\boldsymbol{V}_\mp=\boldsymbol{0}$ and $\left(\boldsymbol{V}_\pm\right)^2=\boldsymbol{V}_\pm$. This matrix is real-valued and all its $2\times2$ blocks are symmetric, though the full matrix is not (note also that each $2\times2$ block is a linear combination of $\boldsymbol{V}_\pm$.)

Showing that such a transformation is indeed canonical and preserves the symplectic structure requires to prove that 
\bea
\label{eq:VMV:symplectic:structure}
	\vec{\boldsymbol{\mathfrak{M}}}_{{k}}\vec{\boldsymbol{\Omega}}\vec{\boldsymbol{\mathfrak{M}}}_{k}^\dag=\vec{\boldsymbol{\Omega}}\, .
\eea
One can first note that, using the peculiar anti-symmetric structure of $\vec{\boldsymbol{\Omega}}$ as well as \Eq{eq:Vproduct}, the left-hand side of \Eq{eq:VMV:symplectic:structure} is of the form
\bea
	\vec{\boldsymbol{\mathfrak{M}}}_{{k}}\vec{\boldsymbol{\Omega}}\vec{\boldsymbol{\mathfrak{M}}}_{{k}}^\dag=\left(\begin{array}{cc}
		\boldsymbol{0} & \boldsymbol{\mathfrak{M}}_{k} \\
		-\boldsymbol{\mathfrak{M}}_{k}  & \boldsymbol{0}
	\end{array}\right) ,
\eea
where 
\bea
	\boldsymbol{\mathfrak{M}}_k&=&\det\left(\boldsymbol{M}_k\right)\left(\boldsymbol{V}_++\boldsymbol{V}_-\right).
\eea
Since $\boldsymbol{M}_k\in\mathrm{Sp}\left(2,\mathbb{R}\right)$, its determinant is one, and given that $\boldsymbol{V}_++\boldsymbol{V}_-=\boldsymbol{I}$, this finishes to prove \Eq{eq:VMV:symplectic:structure}.

The matrix $\vec{\boldsymbol{\mathfrak{M}}}_{{k}}$ thus generates a canonical transformation, and from any isotropic canonical transformation in the two-mode representation, one obtains a canonical transformation in the one-mode representation. We stress however that this requires $\boldsymbol{M}_{-\vec{k}}=\boldsymbol{M}_{\vec{k}}$ (for which isotropy is a sufficient condition).
\subsubsection{Helicity basis}
Let us now reproduce these results in the helicity basis, where an isotropic canonical transformation is generated by the matrix $\boldsymbol{\mathcal{M}}_k\in\mathrm{SU}(1,1)$. It acts in the two-dimensional subspace as $\boldsymbol{a}_{\pm\vec{k}}\to\boldsymbol{A}_{\pm\vec{k}}=\boldsymbol{\mathcal{M}}_k\boldsymbol{a}_{\pm\vec{k}}$. In the four-dimensional subspace of the helicity basis, the corresponding canonical transformation is obtained from the Kronecker product of $\boldsymbol{\mathcal{M}}_k$ with the identity, and is thus given by 
\bea
	\vec{\boldsymbol{A}}_{\vec{k}}=\left(\boldsymbol{\mathcal{M}}_k\otimes\boldsymbol{I}\right)\vec{\boldsymbol{a}}_{\vec{k}}\, .
\eea
Using the Bogolyubov coefficients notation of \Eq{eq:su11mat}, the four-dimensional canonical transformation in the two-mode representation is thus
\bea	
	\vec{\boldsymbol{\mathcal{M}}}_k:=\left(\boldsymbol{\mathcal{M}}_k\otimes\boldsymbol{I}\right)=\left(\begin{array}{cc}
		\alpha_k\boldsymbol{I} & \beta_k\boldsymbol{I} \\
		\beta^*_k\boldsymbol{I} & \alpha^*_k\boldsymbol{I}
	\end{array}\right).
\eea
Switching to the one-mode representation, similarly to \Eq{eq:a:fraka}, one introduces $\vec{\boldsymbol{\mathfrak{A}}}_{\vec{k}}:=\vec{\boldsymbol{\mathcal{V}}} \vec{\boldsymbol{A}}_{\vec{k}}=\left(\vec{\boldsymbol{\mathcal{V}}}\vec{\boldsymbol{\mathcal{M}}}_k\vec{\boldsymbol{\mathcal{V}}}\right)\vec{\boldsymbol{\mathfrak{a}}}_{\vec{k}}$. The matrix generating the transformation thus reads
\bea
	\vec{\boldsymbol{\mathcal{V}}}\vec{\boldsymbol{\mathcal{M}}}_k\vec{\boldsymbol{\mathcal{V}}}=\left(\begin{array}{cc}
		\alpha_k\boldsymbol{I} & \beta_k\boldsymbol{J}_x \\
		\beta^*_k\boldsymbol{J}_x & \alpha^*_{k}\boldsymbol{I}
	\end{array}\right).
\eea
Checking that such a transformation preserves the symplectic structure proceeds by a direct calculation of $\left(\vec{\boldsymbol{\mathcal{V}}}\vec{\boldsymbol{\mathcal{M}}}_k\vec{\boldsymbol{\mathcal{V}}}\right)\vec{\boldsymbol{\mathcal{J}}}\left(\vec{\boldsymbol{\mathcal{V}}}\vec{\boldsymbol{\mathcal{M}}}_k\vec{\boldsymbol{\mathcal{V}}}\right)^\dag$ to prove that it equals $\vec{\boldsymbol{\mathcal{J}}}$. This can be done straightforwardly by block-wise matrix multiplication, making use of $\left(\boldsymbol{J}_x\right)^2=\boldsymbol{I}$ and $\left|\alpha_k\right|^2-\left|\beta_k\right|^2=1$, see \Eq{eq:su11mat}.
\subsection{Mapping the two-mode representation to the one-mode representation}
\label{subapp:mappingcomplexreal}
Let us summarise the previous considerations. Any matrix $\boldsymbol{M}\in\mathrm{Sp}(2,\mathbb{R})$ (either a Green's matrix or a canonical transformation) acting on the two-dimensional subspace $\left(\phi_{\vec{k}},\pi_{\vec{k}}\right)$ (and identically on the subspace with opposite wavevector), can be mapped onto a $4\times4$ matrix acting on the subspace $\left(\mathfrak{q}_{\vec{k}},\mathfrak{q}_{-\vec{k}},\mathfrak{p}_{\vec{k}},\mathfrak{p}_{-\vec{k}}\right)$ by first taking its Kronecker product with $\boldsymbol{I}$, and then by sandwiching it with the involutory (hence unitary) matrix $\vec{\boldsymbol{V}}$, \ie $\vec{\boldsymbol{\mathfrak{M}}}=\vec{\boldsymbol{V}}\left(\boldsymbol{M}\otimes\boldsymbol{I}\right)\vec{\boldsymbol{V}}$. All the group properties (composition law and inverse) are preserved through this mapping, \ie for three matrices of $\mathrm{Sp}(2,\mathbb{R})$ such that $\boldsymbol{M}_{1,2}=\boldsymbol{M}_{1}\boldsymbol{M}_{2}$, one has $\vec{\boldsymbol{\mathfrak{M}}}_{1,2}=\vec{\boldsymbol{\mathfrak{M}}}_{1}\vec{\boldsymbol{\mathfrak{M}}}_{2}$. As a consequence, $\vec{\boldsymbol{\mathfrak{M}}}_{{k}}$ has a unit determinant, and the symplectic structure is also preserved, \ie $\vec{\boldsymbol{\mathfrak{M}}}\vec{\boldsymbol{\Omega}}\vec{\boldsymbol{\mathfrak{M}}}^\dag=\vec{\boldsymbol{\Omega}}$.  \\

In fact, many other properties satisfied by matrices in $\mathrm{Sp}(2\mathbb{R})$ are preserved through such a mapping. First, any polynomial relation satisfied by a set of matrices in $\mathrm{Sp}(2,\mathbb{R})$ is directly mapped into the same polynomial relation for their imaged matrices, in the space of matrices acting on $\left(\mathfrak{q}_{\vec{k}},\mathfrak{q}_{-\vec{k}},\mathfrak{p}_{\vec{k}},\mathfrak{p}_{-\vec{k}}\right)$. Suppose indeed that we have 
\bea
	\ds\sum_{n=0}^N\boldsymbol{M}_{1,n}\boldsymbol{M}_{2,n}\cdots\boldsymbol{M}_{\ell_n,n}=\boldsymbol{0}\, , \label{eq:app:polynom}
\eea
where $\boldsymbol{M}_{1,n}\boldsymbol{M}_{2,n}\cdots\boldsymbol{M}_{\ell_n,n}$ is the product of $n$ arbitrary matrices of $\mathrm{Sp}(2,\mathbb{R})$. One has
\bea
	\vec{\boldsymbol{V}}\left\{\left[\ds\sum_{n=0}^N\boldsymbol{M}_{1,n}\boldsymbol{M}_{2,n}\cdots\boldsymbol{M}_{\ell_n,n}\right]\otimes\boldsymbol{I}\right\}\vec{\boldsymbol{V}}=\vec{\boldsymbol{V}}\left(\boldsymbol{0}\otimes\boldsymbol{I}\right)\vec{\boldsymbol{V}}=\vec{\boldsymbol{0}}\, ,
\eea
where $\vec{\boldsymbol{0}}$ is the $4\times4$ zero matrix. Making use of \Eq{eq:Kronecker:P5}, one has $\left\{\left[\ds\sum_{n=0}^N\boldsymbol{M}_{1,n}\boldsymbol{M}_{2,n}\cdots\boldsymbol{M}_{\ell_n,n}\right]\otimes\boldsymbol{I}\right\}=\ds\sum_{n=0}^N\left(\boldsymbol{M}_{1,n}\boldsymbol{M}_{2,n}\cdots\boldsymbol{M}_{\ell_n,n}\right)\otimes\boldsymbol{I}$. Recalling that the Kronecker product preserves the group composition law, see the discussion below \Eq{eq:vecM:entries}, one has $\left(\boldsymbol{M}_{1,n}\boldsymbol{M}_{2,n}\cdots\boldsymbol{M}_{\ell_n,n}\right)\otimes\boldsymbol{I} = \left(\boldsymbol{M}_{1,n}\otimes\boldsymbol{I}\right)\left( \boldsymbol{M}_{2,n}\otimes\boldsymbol{I} \right)\cdots\left(\boldsymbol{M}_{\ell_n,n}\otimes\boldsymbol{I} \right) = \vec{\boldsymbol{M}}_{1,n}\vec{\boldsymbol{M}}_{2,n}\cdots\vec{\boldsymbol{M}}_{\ell_n,n}$. Since $\vec{\boldsymbol{V}}$ is involutory, it is then straightforward to show that $\vec{\boldsymbol{V}} \vec{\boldsymbol{M}}_{1,n}\vec{\boldsymbol{M}}_{2,n}\cdots\vec{\boldsymbol{M}}_{\ell_n,n} \vec{\boldsymbol{V}}  =\left(\vec{\boldsymbol{V}} \vec{\boldsymbol{M}}_{1,n} \vec{\boldsymbol{V}}\right)\left(\vec{\boldsymbol{V}} \vec{\boldsymbol{M}}_{2,n} \vec{\boldsymbol{V}}\right)\cdots \left(\vec{\boldsymbol{V}} \vec{\boldsymbol{M}}_{\ell_n,n} \vec{\boldsymbol{V}}\right) =\vec{\boldsymbol{\mathfrak{M}}}_{1,n}\vec{\boldsymbol{\mathfrak{M}}}_{2,n}\cdots \vec{\boldsymbol{\mathfrak{M}}}_{\ell_n,n}$. This finally proves that, if \Eq{eq:app:polynom} holds, then 
\bea
\ds\sum_{n=0}^N\vec{\boldsymbol{\mathfrak{M}}}_{1,n}\vec{\boldsymbol{\mathfrak{M}}}_{2,n}\cdots \vec{\boldsymbol{\mathfrak{M}}}_{\ell_n,n}=\vec{\boldsymbol{0}}
\eea 
holds too.

The second important preserved property is exponentiation. Consider a matrix of $\mathrm{Sp}(2,\mathbb{R})$ obtained from the exponential map of $\mathfrak{sp}(2,\mathbb{R})$, \ie
\bea
\label{eq:2mrep-1mrep:map:exp}
	\boldsymbol{M}=\exp\left(i\boldsymbol{m}\right).
\eea
The Kronecker product of exponentials of matrices is given by
\bea
	\exp(A)\otimes \exp(B)=\exp\left(A\otimes I_m + I_n\otimes B\right),
\eea
with $A$ a $n\times n$ matrix, $B$ a $m\times m$ matrix, and $I_n$ the identity of dimension $n$. Since $\boldsymbol{I}$ is the exponential of the zero matrix, one obtains
\bea
	\boldsymbol{M}\otimes\boldsymbol{I}=\exp\left(i\boldsymbol{m}\otimes\boldsymbol{I}\right).
\eea
Since the matrix $\vec{\boldsymbol{V}}$ is involutory, this gives rise to
\bea
	\vec{\boldsymbol{V}}\left(\boldsymbol{M}\otimes\boldsymbol{I}\right)\vec{\boldsymbol{V}}=\exp\left[i\vec{\boldsymbol{V}}\left(\boldsymbol{m}\otimes\boldsymbol{I}\right)\vec{\boldsymbol{V}}\right],
\eea
which implies that, if \Eq{eq:2mrep-1mrep:map:exp} holds, then
\bea
\vec{\boldsymbol{\mathfrak{M}}}=\exp\left(i\vec{\boldsymbol{\mathfrak{m}}}\right)
\eea
also holds.\footnote{Notice that the two above properties can be combined to show that, if $\ds\sum_{n=0}^N \exp(i\boldsymbol{m}_{1,n}) \exp(i\boldsymbol{m}_{2,n}) \cdots \exp(i\boldsymbol{m}_{\ell_n,n}) = \boldsymbol{0}$ holds, 
then the relation $\allowbreak \ds\sum_{n=0}^N \exp(i\boldsymbol{\mathfrak{m}}_{1,n})\allowbreak \exp(i\boldsymbol{\mathfrak{m}}_{2,n}) \allowbreak \cdots \allowbreak \exp(i\boldsymbol{\mathfrak{m}}_{\ell_n,n}) \allowbreak=\allowbreak \boldsymbol{0}$ holds too.}

Finally, the property $\left(\boldsymbol{M}\otimes\boldsymbol{I}\right)\left(\boldsymbol{N}\otimes\boldsymbol{I}\right)=\left(\boldsymbol{M}\boldsymbol{N}\right)\otimes\boldsymbol{I}$ derived below \Eq{eq:vecM:entries} also applies to the generators of $\mathrm{Sp}(2,\mathbb{R})$. Denoting $\vec{\boldsymbol{\mathfrak{K}}}_i\equiv \vec{\boldsymbol{V}}\left(\boldsymbol{K}_i\otimes\boldsymbol{I}\right)\vec{\boldsymbol{V}}$, with $\boldsymbol{K}_i$ the generators of $\mathrm{Sp}(2,\mathbb{R})$, then the set of $4\times4$ matrices $\left\{\vec{\boldsymbol{\mathfrak{K}}}_i\right\}$ satisfies the same algebraic relations than the $\boldsymbol{K}_i$'s, given below \Eq{eq:K123}. \\

All the above is directly translated to the helicity basis variables by using the following replacements. For the phase-space variables, they read
\bea
	\vec{\boldsymbol{\phi}}_{\vec{k}}\to\vec{\boldsymbol{a}}_{\vec{k}}~~~\mathrm{and}~~~\vec{\boldsymbol{q}}_{\vec{k}}\to\vec{\boldsymbol{\mathfrak{a}}}_{\vec{k}}\, .
\eea
The dynamics and the canonical transformations in the 4-dimensional two-mode representation are then obtained by the replacements 
\bea
	\left[\boldsymbol{G}_k(t,t_0)\otimes\boldsymbol{I}\right]\to\left[\boldsymbol{\mathcal{G}}_k(t,t_0)\otimes\boldsymbol{I}\right] &~~~\mathrm{and}~~~& \left[\boldsymbol{M}_k(t)\otimes\boldsymbol{I}\right]\to\left[\boldsymbol{\mathcal{M}}_k(t)\otimes\boldsymbol{I}\right],
\eea
where we remind that $\boldsymbol{\mathcal{G}}_k(t,t_0)=\boldsymbol{UG}_k(t,t_0)\boldsymbol{U}^\dag$, see below \Eq{eq:calM:def}, and $\boldsymbol{\mathcal{M}}=\boldsymbol{U}\boldsymbol{M}\boldsymbol{U}^\dag$, see \Eq{eq:calM:def}.\footnote{Note that for any matrix $\boldsymbol{M}$ in $\mathrm{Sp}(2,\mathbb{R})$, one also has the relation $\boldsymbol{\mathcal{M}}\otimes\boldsymbol{I}=\left(\boldsymbol{UMU}^\dag\right)\otimes\boldsymbol{I}=\vec{\boldsymbol{U}}\left(\boldsymbol{M}\otimes\boldsymbol{I}\right)\vec{\boldsymbol{U}}^\dag$, so $\vec{\boldsymbol{\mathcal{M}}} = \vec{\boldsymbol{U}} \vec{\boldsymbol{M}}\vec{\boldsymbol{U}}^\dag$. \label{footnote:rel:calMvec}} The resulting dynamics and canonical transformations for the one-mode representation are then obtained in the helicity basis with the replacement
\bea
	&&\vec{\boldsymbol{V}}\left[\boldsymbol{G}_k(t,t_0)\otimes\boldsymbol{I}\right]\vec{\boldsymbol{V}}\to\vec{\boldsymbol{\mathcal{V}}}\left[\boldsymbol{\mathcal{G}}_k(t,t_0)\otimes\boldsymbol{I}\right]\vec{\boldsymbol{\mathcal{V}}}\, , \\
	&&\vec{\boldsymbol{V}}\left[\boldsymbol{M}_k(t)\otimes\boldsymbol{I}\right]\vec{\boldsymbol{V}}\to\vec{\boldsymbol{\mathcal{V}}}\left[\boldsymbol{\mathcal{M}}_k(t)\otimes\boldsymbol{I}\right]\vec{\boldsymbol{\mathcal{V}}}\, .
\eea
As is the case for the field variables, all the properties satisfied by matrices in $\mathrm{SU}(1,1)$ (being either interpreted as Green's matrices or as canonical transformations) are preserved through the set operations $\boldsymbol{\mathcal{M}}\to\boldsymbol{\mathcal{M}}\otimes\boldsymbol{I}$ and $\boldsymbol{\mathcal{M}}\to\vec{\boldsymbol{\mathcal{V}}}\left(\boldsymbol{\mathcal{M}}\otimes\boldsymbol{I}\right)\vec{\boldsymbol{\mathcal{V}}}$. This can be proven by direct calculations or by using the unitary transform sending the 4-dimensional one-mode field variables to the helicity basis, \ie $\vec{\boldsymbol{\mathcal{V}}}\left(\boldsymbol{\mathcal{M}}\otimes\boldsymbol{I}\right)\vec{\boldsymbol{\mathcal{V}}}=\vec{\boldsymbol{U}}\vec{\boldsymbol{\mathfrak{M}}}\vec{\boldsymbol{U}}^\dag$ (where footnote~\ref{footnote:rel:calMvec} has been used).
\section{Solving for the dynamics in the invariant representation}
\label{app:rho}
In this section, we show how the differential equation implementing the dynamics in the invariant representation, \Eq{eqrho} can be solved in terms of the so-called ``mode functions'' that are solutions of \Eq{eqqcl}. We follow the approach presented in \Refs{doi:10.1063/1.1664532,Lewis:1968tm,doi:10.1063/1.524625} and extend it to complex-valued mode functions. We then apply this procedure to the cosmological context by considering a phase of inflation followed by a radiation era. We finally discuss the choice of initial conditions when implementing the invariant representation as a canonical transformation.
\subsection{Solution in terms of mode functions}
Given a set of two linearly independent solutions of \Eq{eqqcl}, $f_1$ and $f_2$, a general solution of \Eq{eqrho} reads
\bea
	\rho(\eta)=\frac{\gamma_1}{W}\left[Af^2_1(\eta)+Bf^2_2(\eta)+2\gamma_2\left(AB-W^2\right)^{1/2}f_1(\eta)f_2(\eta)\right]^{1/2},
\eea
where $A$ and $B$ are arbitrary constants, the $\gamma_i$'s have to be such that $(\gamma_1)^4=(\gamma_2)^2=1$, and $W=f_1f'_2-f_2f'_1$ is the Wronskian of the solutions of \Eq{eqqcl} (note that $W$ is conserved, \ie $W'=0$, and it is necessarily non zero since it is constructed from two linearly independent solutions). 

Since the coefficients appearing in the differential equation~\eqref{eqqcl} are real, if $f_1$ is a solution of \Eq{eqqcl}, so is $f_1^*$, and these two solutions are independent as soon as $f_1$ is neither real nor pure imaginary. One can therefore take $f_2=f_1^*$. The Wronskian then becomes pure imaginary, and without loss of generality, by multiplying $f_1$ with the required constant, one can set it to $W=i$ so as to represent the commutation relation (though any other choice works equally well).  
The function $\rho$ being real-valued, one can then take $\gamma_1=i$ and constrain $B=A^*$, leading to 
\bea
\label{eq:rho:sol:gen}
	\rho(\eta)=\sqrt{2\mathrm{Re}\left[Af^2_1(\eta)\right]+2\gamma_2\left(\left|A\right|^2+1\right)^{1/2}\left|f_1(\eta)\right|^2},
\eea
where we remind that $\gamma_2=\pm 1$ is real.
\subsection{Connecting two asymptotic harmonic oscillators}
\label{sec:connect:harmonic:oscillators}
For concreteness, let us assume that the time-dependent frequency tends to a real-valued constant both in the infinite past, $\omega(\eta\to-\infty)=\omega_{(-)}$, and in the infinite future, $\omega(\eta\to+\infty)=\omega_{(+)}$, where the two asymptotic values need not be equal. Let us see how the integration constants appearing in \Eq{eq:rho:sol:gen} can be set. For a constant frequency, $\omega_{(\pm)}$, a general solution of \Eq{eqrho} is \cite{Lewis:1968tm}
\bea
	\rho=\frac{\delta^{(\pm)}_1}{\sqrt{\omega_{(\pm)}}}\left[\cosh{\epsilon^{(\pm)}}+\delta^{(\pm)}_2\sinh{\epsilon^{(\pm)}}~\sin{\left(2\omega_{(\pm)}\eta+\varphi^{(\pm)}\right)}\right]^{1/2}, \label{eqapp:rhocst}
\eea
where the $\delta^{(\pm)}_i$'s are such that $(\delta^{(\pm)}_1)^4=(\delta^{(\pm)}_2)^2=1$, $\varphi^{(\pm)}$ is a real constant phase, and ${\epsilon^{(\pm)}}$ is another integration constant. These parameters can be chosen in such a way that, in the infinite past, the set of operators $(\widehat{d},\widehat{d}^\dag)$ and $(\widehat{a},\widehat{a}^\dag)$ coincide. In the infinite past indeed, the Hamiltonian reduces to the one of a harmonic oscillator, reading $\widehat{H}_{(q,p)}=[\widehat{p}^2+\omega_{(-)}^2\widehat{q}^2]/2$. The natural definition for the creation and annihilation operator is thus 
\bea
\label{eq:invariant:a}
	\widehat{a}&=&\frac{1}{\sqrt{2}}\left(\sqrt{\omega_{(-)}}\widehat{q}+i\frac{\widehat{p}}{\sqrt{\omega_{(-)}}}\right),  \\
	\widehat{a}^\dag&=&\frac{1}{\sqrt{2}}\left(\sqrt{\omega_{(-)}}\widehat{q}-i\frac{\widehat{p}}{\sqrt{\omega_{(-)}}}\right),
\label{eq:invariant:adag}
\eea
leading to $\widehat{H}_{(q,p)}=\omega_{(-)}(\widehat{a}^\dag\widehat{a}+1/2)$. Using instead the $(Q,P)$ variables introduced above \Eq{eq:I:P:Q}, the creation and annihilation operators $\widehat{d}$ and $\widehat{d}^\dagger$ defined in \Eqs{eq:d:QP:def}-(\ref{eq:d:dagger:QP:def}) match $\widehat{a}$ and $\widehat{a}^\dagger$ provided 
\bea
\label{eq:rho:asympt:omega}
\rho(\eta\to-\infty)=1/\sqrt{\omega_{(-)}}\, ,
\eea
which is obtained by setting $\delta^{(-)}_1=1$ and $\epsilon^{(-)}=0$. In that case, in the infinite past, the two sets of creation and annihilation operators define the same basis states for the Hilbert space. On can further check that, in that limit, since $\widehat{H}_{(Q,P)}=(1/\rho^2)(\widehat{d}^\dag\widehat{d}+1/2)$ as derived below \Eq{eq:I:P:Q}, the two Hamiltonians also coincide, $\widehat{H}_{(Q,P)}=\widehat{H}_{(q,p)}$.

This prescription allows one to fix the values of $A$ and $\gamma_2$ in \Eq{eq:rho:sol:gen}, by requiring that the general solution~(\ref{eq:rho:sol:gen}) matches the asymptotic solution~(\ref{eq:rho:asympt:omega}) in the infinite past. One first notes that for a constant frequency, a solution $f_1$ to \Eq{eqqcl} such that the Wronskian between $f_1$ and $f_2=f_1^*$ equals $i$ is given by
\bea
	f^{(\pm)}_1(\eta)=\frac{e^{-i\omega_{(\pm)}\eta}}{\sqrt{2\omega_{(\pm)}}}\, . 
\eea
The matching is obtained by setting $f_{1(2)}(\eta\to-\infty)=f^{(-)}_{1(2)}$ in the asymptotic past, leading to
\bea
	\sqrt{2}=\sqrt{A~e^{-2i\omega_{(-)}\eta}+A^*~ e^{2i\omega_{(-)}\eta}+2\gamma_2(\left|A\right|^2+1)^{1/2}}.
\eea
This imposes that $A=0$ and $\gamma_2=1$. Hence the solution for $\rho$ that ensures the invariant to be  initially equal to the Hamiltonian is
\bea
\label{eq:rho:f1:f2}
	\rho(\eta)=\sqrt{2}\left\vert f_1(\eta)\right\vert\, ,
\eea
where $f_1$ is chosen such that $f_{1}(\eta\to-\infty)\to f^{(-)}_1(\eta)$. 

This solution is valid throughout the entire evolution and can be evaluated in the infinite future. In full generality, the evolution is not adiabatic, hence there is no reason for the mode functions, $f_{i}$, which initially equate $f^{(-)}_{i}$, to evolve towards $f^{(+)}_i$. However, both the $f_i$'s and the $f^{(+)}_i$'s form a set of linearly independent solutions in the asymptotic future, and share the same Wronskian. They are thus related by a Bogolyubov transform such as \Eqs{eq:Bogoljubov:a:1} and~(\ref{eq:Bogoljubov:a:2}), namely
\bea
\label{eq:Bogoljubov:f1:f2}
	\lim_{\eta\to+\infty}{\left(\begin{array}{c}
		f_1(\eta) \\
		f_2(\eta)
	\end{array}\right)}=\left(\begin{array}{cc}
		\alpha^{(+)} & \beta^{(+)} \\
		\beta^{(+)*} & \alpha^{(+)*}
	\end{array}\right)\left(\begin{array}{c}
		f^{(+)}_1(\eta) \\
		f^{(+)}_2 (\eta)
	\end{array}\right)
\eea
with $\left\vert \alpha^{(+)}\right\vert^2-\left\vert \beta^{(+)}\right\vert^2=1 $.
Plugging \Eq{eq:Bogoljubov:f1:f2} into \Eq{eq:rho:f1:f2} leads to the following asymptotic behaviour for $\rho$,
\bea
	\rho(\eta)&\underset{{\eta\to+\infty}}{=}&\frac{1}{\sqrt{\omega_{(+)}}}\left\lbrace\left|\alpha^{(+)}\right|^2+\left|\beta^{(+)}\right|^2+2\mathrm{Re}\left[\alpha^{(+)}\beta^{(+)*}\right]\cos{(2\omega_{(+)}\eta)}\right. \nonumber \\
	&&\left.+2\mathrm{Im}\left[\alpha^{(+)}\beta^{(+)*}\right]\sin{(2\omega_{(+)}\eta)}\right\rbrace^{1/2}.
	\label{eq:rho:Bogoljubov:asympt}
\eea
This is equivalent to the general solution for a constant frequency~(\ref{eqapp:rhocst}), with the identification 
\bea
	&&\cosh\epsilon^{(+)}=\left|\alpha^{(+)}\right|^2+\left|\beta^{(+)}\right|^2, \\
	&&\sinh\epsilon^{(+)}\sin\varphi^{(+)}=2\mathrm{Re}\left[\alpha^{(+)}\beta^{(+)*}\right], \\
	&&\sinh\epsilon^{(+)}\cos\varphi^{(+)}=2\mathrm{Im}\left[\alpha^{(+)}\beta^{(+)*}\right],
\eea
and setting $\delta^{(+)}_1=\delta^{(+)}_2=1$. When $\beta^{(+)}\neq 0$, \ie when $\lim_{\eta\to +\infty }f_1\neq f_1^{(+)}$, $\epsilon^{(+)}\neq 0$, $\lim_{\eta\to +\infty }\rho\neq 1/\sqrt{\omega^{(+)}}$ and the invariant deviates from the Hamiltonian.
\subsection{Application to inflation followed by a radiation phase}
\label{sec:InvariantRep:inf:plus:rad}
Let us now illustrate the above procedure in a specific example belonging to the cosmological standard model, and dealing with the cosmological fluctuations introduced in \Sec{sec:inflation}. We consider a scenario starting with an inflationary era, well approximated by a de Sitter phase, and followed by a radiation-dominated period. Working with conformal time, the scale factor for this model reads
\bea
\label{eq:a(eta):inf:rad}
a\left(\eta\right) =
\begin{cases}
-\dfrac{1}{H\eta} & \text{for}\quad -\infty<\eta<\eta_{\mathrm{r}}\, ,\\
& \\
\dfrac{\eta-2\eta_{\mathrm{r}}}{H\eta^2_{\mathrm{r}}}& \text{for}\quad \eta_{\mathrm{r}}<\eta\, ,
\end{cases}
\eea
with $\eta_{\mathrm{r}}<0$ being the time when the universe exits the de Sitter phase and enters the radiation-dominated era. The above expression ensures both the scale factor and the Hubble parameter to be continuous at the transition time. From \Eq{eq:a(eta):inf:rad}, one can infer the time dependence of the frequency entering the Hamiltonian density~(\ref{eq:Hv:inf}), for the scalar field described with the  canonical variables $\boldsymbol{v}_{\vec{k}}$,

\bea
\label{eq:omegasquareapp}
\omega^2_k\left(\eta\right) =
\begin{cases}
k^2-\dfrac{2}{\eta^2} & \text{for}\quad -\infty<\eta<\eta_{\mathrm{r}}\, ,\\
k^2& \text{for}\quad \eta_{\mathrm{r}}<\eta\, ,
\end{cases}
\eea
where we have further assumed that the mass of the field is vanishing. One can note that, contrary to the scale factor and the Hubble parameter, $\omega^2_k(\eta)$ is discontinuous at the transition. However, the approach sketched above remains valid for such sudden transitions, since one deals with a test field~\cite{Deruelle:1995kd}. The frequency is constant both in the infinite past and the infinite future, with the same asymptotic value $\omega_{(-)}=\omega_{(+)}=k$.

During the de Sitter (dS) inflationary era and the radiation (r) dominated era, two independent solutions of the mode functions written as $f_1(\eta)=f(\eta)$ and $f_2(\eta)=f^*(\eta)$, with a properly normalised Wronskian, are obtained from
\bea
		f_{\mathrm{dS}}(\eta)&=&\frac{1}{\sqrt{2k}}\left(1-\frac{i}{k\eta}\right)e^{-ik\eta}\, , \\
		f_{\mathrm{r}}(\eta)&=&\frac{1}{\sqrt{2k}}e^{-ik(\eta-\eta_r)}\, .
\eea
It is easy to check that $f_{\mathrm{dS}}$ asymptotes $f^{(-)}_1=e^{-ik\eta}/\sqrt{2k}$ in the remote past, while $f_{\mathrm{r}}$ obviously asymptotes $f^{(+)}_1=e^{-ik\eta}/\sqrt{2k}$ (up to an irrelevant global phase, $e^{ik\eta_{\mathrm{r}}}$) in the infinite future. With such solutions, during the de Sitter epoch the function $\rho_k$ given by \Eq{eq:rho:f1:f2} reads 
\bea
	\rho_k^{\mathrm{dS}}(\eta)=\sqrt{\frac{1}{k}\left(1+\frac{1}{k^2\eta^2}\right)}\, .
\eea
During the radiation era, introducing the Bogolyubov decomposition $f_1=\alpha_k f_{\mathrm{r}}+\beta_k f^*_{\mathrm{r}}$ as in \Eq{eq:Bogoljubov:f1:f2}, one instead finds the solution~(\ref{eq:rho:Bogoljubov:asympt}), 
\bea
	\rho^{\mathrm{r}}_k(\eta)=\frac{1}{\sqrt{k}}\left[\left|\alpha_k\right|^2+\left|\beta_k\right|^2+2\mathrm{Re}\left[\alpha_k\beta^*_k\right]\cos{(2k\eta)}+2\mathrm{Im}\left[\alpha_k\beta^*_k\right]\sin{(2k\eta)}\right]^{1/2}.
\eea
In this specific example, the time-dependent frequency is constant during the entire radiation era (and not only in the asymptotic future). Hence the above is valid during the whole radiation epoch $\eta>\eta_{\mathrm{r}}$. The last step consists in deriving the Bogolyubov coefficients, which is easily done by matching the de Sitter mode functions with the radiation mode functions at $\eta_{\mathrm{r}}$ yielding  $\alpha_k=1-i/k\eta_{\mathrm{r}}-1/2k^2\eta^2_r$ and $\beta_k=e^{2ik\eta_{\mathrm{r}}}/2k^2\eta^2_{\mathrm{r}}$. 
For modes becoming super-Hubble during the de Sitter era well before the transition, $(k\eta_{\mathrm{r}})^{-1}\gg1$, and the function $\rho_k$ during the radiation era is well approximated by 
\bea
	\rho^{\mathrm{r}}_k(\eta)\simeq\frac{1}{\sqrt{k}}\frac{\left|\sin{\left[k(\eta-\eta_{\mathrm{r}})\right]}\right|}{k^2\eta^2_{\mathrm{r}}}.
\eea
On the contrary, for modes remaining sub-Hubble during the entire evolution, $(k\eta_r)^{-1}\gg1$, and the function $\rho_k$ during the radiation epoch is given by
\bea
	\rho^{\mathrm{r}}_k(\eta)\simeq\frac{1}{\sqrt{k}}\left[1+\mathcal{O}(k\eta_r)^{-2}\right] .
\eea
In this limit, it satisfies the condition~\eqref{eq:rho:asympt:omega} (more precisely the dual of that condition in the asymptotic future) for the invariant to match the Hamiltonian, which is a consequence of the fact that sub-Hubble modes evolve adiabatically.
\subsection{Initial conditions for the invariant canonical transformation}
\label{app:iniinv}
In this section, we follow the considerations of \Sec{sec:inflation:invariant:representation} to implement the invariant representation through a canonical transformation. This requires to properly set the initial conditions for the two representations (the original one and the invariant one) to be initially identical. We remind that the canonical transformation generating the invariant representation is given by the matrix
\bea
\label{eq:invariant:canonical:transform}
\boldsymbol{M}_k(\eta)=\boldsymbol{R}(\vartheta_{\vec{k}})\underbrace{\left(\begin{array}{cc}
		1/\rho_{\vec{k}} & 0 \\
		-\rho'_{\vec{k}} & \rho_{\vec{k}}
	\end{array}\right)}_{\boldsymbol{S}_{\vec{k}}(\rho_k)}\, ,
\eea
see \Eq{eq:Q:S}.
Starting from a given representation, where the creation and annihilation operators are constructed from \Eq{eq:a(z)} with a given matrix $\boldsymbol{D}_k(\eta)$, the invariant representation, generated by \Eq{eq:invariant:canonical:transform}, yields the same initial set of creation and annihilation operators (hence the same initial Hilbert space) provided 
\bea
\label{eq:invariant:canonical:transform:initial:constraint}
	\boldsymbol{D}_{\vec{k}}\left(\eta_0\right)=\boldsymbol{R}\left(\vartheta^{(0)}_{\vec{k}}\right)\boldsymbol{S}_{\vec{k}}\left(\rho^{(0)}_k\right)\, ,
\eea
where the superscript $(0)$ means that the functions $\rho_k$ and the angle $\vartheta_k$ are to be evaluated at $\eta_0$. In \App{sec:connect:harmonic:oscillators}, this condition was studied in the case where $\boldsymbol{D}_k(\eta\to -\infty)=\mathrm{diag}\left(\sqrt{\omega}_{(-)},1/\sqrt{\omega_{(-)}}\right)$, see \Eqs{eq:invariant:a} and~\eqref{eq:invariant:adag}. Here we generalise these considerations to an arbitrary choice of $\boldsymbol{D}_k$ and an arbitrary initial time $\eta_0$. \\

The constraint~\eqref{eq:invariant:canonical:transform:initial:constraint} is more easily solved by going to the helicity basis for the matrices, leading to $\boldsymbol{\mathcal{D}}_k(\eta_0)\boldsymbol{\mathcal{S}}^{-1}_k(\rho^{(0)}_k)\boldsymbol{\mathcal{R}}(-\vartheta^{(0)}_{\vec{k}})=\boldsymbol{I}$. In the helicity basis, defined by \Eq{eq:calM:def}, each of these matrices are elements of $\mathrm{SU}(1,1)$ and they are respectively given by
\bea
	\boldsymbol{\mathcal{D}}_k(\eta_0)=\left(\begin{array}{cc}
		A^{(D)}_k & B^{(D)}_k \\
		B^{(D)*}_k & A^{(D)*}_k
	\end{array}\right)
\eea
with $A^{(D)}_k $ and the $B^{(D)}_k$ the Bogolyubov coefficients defining the canonical transformation generated by $\boldsymbol{D}_k$;
\bea
\label{eq:invariant:calS:inverse:A:B}
	\boldsymbol{\mathcal{S}}^{-1}_k(\rho^{(0)}_k)=\left(\begin{array}{cc}
		A^{(\rho)}_k & B^{(\rho)}_k \\
		B^{(\rho)*}_k & A^{(\rho)*}_k
	\end{array}\right)
\eea
with the Bogolyubov coefficients reading $A^{(\rho)}_k=(\rho_k+\rho^{-1}_k+i\rho'_k)/2$ and $B^{(\rho)}_k=(\rho_k-\rho^{-1}_k+i\rho'_k)/2$,\footnote{This expression is valid at any time, although here we only use it at time $\eta_0$.} which can be easily obtained by plugging \Eq{eq:invariant:canonical:transform} into \Eq{eq:calM:def}; and 
\bea
	\boldsymbol{\mathcal{R}}(-\vartheta^{(0)}_{\vec{k}})=\left(\begin{array}{cc}
		e^{-i\vartheta^{(0)}_k} & 0 \\
		0 & e^{i\vartheta^{(0)}_k}
	\end{array}\right),
\eea
see \Eq{eq:polar:RS:explicit}. The constraints $\boldsymbol{\mathcal{D}}_k(\eta_0)\boldsymbol{\mathcal{S}}^{-1}_k(\rho^{(0)}_k)\boldsymbol{\mathcal{R}}(-\vartheta^{(0)}_{\vec{k}})=\boldsymbol{I}$ then leads to
\bea
\label{eq:invariant:init:problem:eq1}
	A^{(\rho)}_ke^{-i\vartheta^{(0)}_k}&=&A^{(D)*}_k, \\
	B^{(\rho)}_ke^{i\vartheta^{(0)}_k}&=&-B^{(D)}_k\, ,
\label{eq:invariant:init:problem:eq2}
\eea
where we have used that $\left|A^{(\rho)}_k\right|^2-\left|B^{(\rho)}_k\right|^2=1$, and where the above relations immediately show that this also implies that $\left|A^{(D)}_k\right|^2-\left|B^{(D)}_k\right|^2=1$, as it should. Note that the above system is neither over-constrained nor under-constrained. The right-hand side is a pair of Bogolyubov coefficients, and hence it has three degrees of freedom (two complex numbers constrained by the unit-determinant condition). The left-hand side has also three degrees of freedom: $\vartheta^{(0)}_k$, $\rho_k(\eta_0)$ and $\rho'_k(\eta_0)$. \\

The easiest way to solve the above system is to express the Bogolyubov coefficients in terms of squeezing parameters, $A^{(D)}_k=e^{i\vartheta^{(D)}_k}\cosh(d_k^{(D)})$ and $B^{(D)}_k=e^{-i\vartheta_k^{(D)}+2i\varpi_k^{(D)}}\sinh(d_k^{(D)})$, see \Eqs{eq:alpha:squeezing}-\eqref{eq:beta:squeezing}\footnote{Note that since we parametrise canonical transformations rather than dynamical evolutions, the squeezing parameters are denoted $d,\varpi,\vartheta$ rather than $r,\varphi,\theta$ as explained at the end of \Sec{sec:canonical:transf:as:squeezing}}, and similarly $A^{(\rho)}_k=e^{i\vartheta^{(\rho)}_k}\cosh(d_k^{(\rho)})$ and $B^{(\rho)}_k=e^{-i\vartheta_k^{(\rho)}+2i\varpi_k^{(\rho)}}\sinh(d_k^{(\rho)})$. The solution to \Eqs{eq:invariant:init:problem:eq1}-\eqref{eq:invariant:init:problem:eq2} is given by
\bea
\label{eq:invariant:rep:canonical:transfo:relate:squeezing:rho:D}
	d_k^{(\rho)}=-d_k^{(D)}, &~~~\varpi_k^{(\rho)}=\varpi_k^{(D)}-\vartheta^{(D)}_k,&~~~\mathrm{and}~~~\vartheta^{(0)}_k-\vartheta_k^{(\rho)}=\vartheta_k^{(D)}\, .
\eea
As it is, the system may appear as being not closed since one cannot determine $\vartheta^{(0)}_k$ and $\vartheta_k^{(\rho)}$ independently. However the three parameters $(d_k^{{(\rho)}},\varpi_k^{(\rho)},\vartheta_k^{(\rho)})$ describing $\boldsymbol{\mathcal{S}}^{-1}_k$ derive from only two numbers, $\rho_k(\eta_0)$ and $\rho'_k(\eta_0)$. Hence there is a constraint allowing for instance to express the rotation angle $\vartheta_k^{(\rho)}$ as a function of the squeezing parameters $d_k^{{(\rho)}}$ and $\varpi_k^{(\rho)}$. 

To derive such a constraint, we turn back to the field basis of the phase space, \ie $\boldsymbol{S}^{-1}_k=\boldsymbol{U}^\dag\boldsymbol{\mathcal{S}}^{-1}_k\boldsymbol{U}$. Starting from
the expression of $\boldsymbol{\mathcal{S}}_k^{-1}$ using the squeezing parameters and the rotation angle, one obtains an expression of $\boldsymbol{S}_k^{-1}$ in terms of $d_k^{(\rho)}$, $\varpi_k^{(\rho)}$ and $\vartheta_k^{(\rho)}$, which can be identified with its expression as a function of $\rho_k$ and $\rho'_k$, \ie \Eq{eq:invariant:rep:S:inverse}. The fact that the element $\left[\boldsymbol{S}_k^{-1}\right]_{(1,2)}$ vanishes provides the constraint relating $\vartheta_k^{(\rho)}$ to $d_k^{(\rho)}$ and $\varpi_k^{(\rho)}$,
\bea
\label{eq:invariant:rep:canonical:transfo:vartheta:d:and:varpi}
	\tan\left(\vartheta_k^{(\rho)}\right)=\frac{\sin\left(2\varpi_k^{(\rho)}\right)\tanh\left(d_k^{(\rho)}\right)}{1+\cos\left(2\varpi_k^{(\rho)}\right)\tanh\left(d_k^{(\rho)}\right)}\, .
\eea
Combining the two first relations of \Eq{eq:invariant:rep:canonical:transfo:relate:squeezing:rho:D} with \Eq{eq:invariant:rep:canonical:transfo:vartheta:d:and:varpi}, $\vartheta_k^{(\rho)}$ can then be expressed in terms of the squeezing parameters and the rotation angle of $\boldsymbol{\mathcal{D}}_k(\eta_0)$, hence, using the third relation of \Eq{eq:invariant:rep:canonical:transfo:relate:squeezing:rho:D}, an expression of $\vartheta_k^{(0)}$ in terms of the squeezing parameters and the rotation angle of $\boldsymbol{\mathcal{D}}_k(\eta_0)$ can be obtained.\footnote{We stress that \Eq{eq:invariant:rep:canonical:transfo:vartheta:d:and:varpi} applies to any canonical transformation generated  by a symplectic matrix of the lower-triangular Borel subgroup, \ie by a matrix of the form
\bea
	\boldsymbol{M}=\left(\begin{array}{cc}
		f & 0 \\
		g & f^{-1}
	\end{array}\right),
\eea
like $\boldsymbol{S}_{\vec{k}}$ and $\boldsymbol{S}_{\vec{k}}^{-1}$.
Conversely, if the symplectic matrix belongs to the upper-triangular Borel subgroup, \ie is of the form 
\bea
	\boldsymbol{M}=\left(\begin{array}{cc}
		f & g \\
		0 & f^{-1}
	\end{array}\right),
\eea
then the constraint is given by
\bea
	\tan\left(\vartheta^{(\rho)}_k\right)=\frac{-\sin\left(2\varpi^{(\rho)}_k\right)\tanh\left(d_k^{(\rho)}\right)}{1-\cos\left(2\varpi_k^{(\rho)}\right)\tanh\left(d_k^{(\rho)}\right)}.
\eea
}

The last step consists in finding $\rho_k$ and $\rho'_k$ from the squeezing parameters $d_k^{(\rho)}$ and $\varpi_k^{(\rho)}$ (keeping in mind that $\vartheta_k^{(\rho)}$ is a function of $d_k^{(\rho)}$ and $\varpi_k^{(\rho)}$ thanks to \Eq{eq:invariant:rep:canonical:transfo:vartheta:d:and:varpi}). As above, this can be done by identifying the expression of $\boldsymbol{S}_k^{-1}$ in terms of $d_k^{(\rho)}$, $\varpi_k^{(\rho)}$ and $\vartheta_k^{(\rho)}$ with the one in terms of $\rho_k$ and $\rho'_k$, \ie \Eq{eq:invariant:rep:S:inverse}. The entry $\left[\boldsymbol{S}_k^{-1}\right]_{(1,2)}$ gave rise to \Eq{eq:invariant:rep:canonical:transfo:vartheta:d:and:varpi}, here, the entry $\left[\boldsymbol{S}_k^{-1}\right]_{(1,1)}$ gives $\rho_k$ and the entry $\left[\boldsymbol{S}_k^{-1}\right]_{(2,1)}$ gives $\rho'_k$. One obtains
\bea
\label{eq:invariant:rep:rho:squeezing:of:rho}
	\rho_k&=&\cos\left(\vartheta_k^{(\rho)}\right)\cosh\left(d_k^{(\rho)}\right)+\cos\left(2\varpi_k^{(\rho)}-\vartheta_k^{(\rho)}\right)\sinh\left(d_k^{(\rho)}\right), \\
	\rho'_k&=&\sin\left(\vartheta_k^{(\rho)}\right)\cosh(d_k)+\sin\left(2\varpi_k^{(\rho)}-\vartheta_k^{(\rho)}\right)\sinh\left(d_k^{(\rho)}\right) ,
\label{eq:invariant:rep:rhoprime:squeezing:of:rho}
\eea
where these expressions have to be evaluated at $\eta_0$. In this way, the initial values of $\rho_k$, $\rho'_k$, and $\vartheta_k$ are unequivocally determined from the initial matrix $\boldsymbol{D}_k$, ensuring that the two representations select the same initial vacuum state. 

Let us note that the above formulas determine the initial canonical transformation sending to the invariant representation in terms of the initial squeezing parameters of $\boldsymbol{D}_k$. In most practical cases however, one starts by setting the Bogolyubov coefficients of $\boldsymbol{D}_k$ rather than its squeezing parameters. However, the later can be expressed in terms of the former by inverting \Eqs{eq:alpha:squeezing}-\eqref{eq:beta:squeezing}, leading to
\bea
\label{eq:d:Bogoljiubov}
	\cosh^2\left(d_k^{(D)}\right)&=&\left|A^{(D)}_k\right|^2, \\ 
\label{eq:varpi:Bogoljiubov}
	\tan\left(2\varpi_k^{(D)}\right)&=& i\left(\frac{A^{(D)*}_kB^{(D)*}_k-A^{(D)}_kB^{(D)}_k}{A^{(D)*}_kB^{(D)*}_k+A^{(D)}_kB^{(D)}_k}\right), \\
\label{eq:vartheta:Bogoljiubov} 
	\tan\left(\vartheta_k^{(D)}\right)&=&-i\left(\frac{A^{(D)}_k-A^{(D)*}_k}{A^{(D)}_k+A^{(D)*}_k}\right). 
\eea

To conclude, let us illustrate this procedure with the example discussed in \App{sec:connect:harmonic:oscillators}, where $\boldsymbol{D}_k(\eta\to -\infty)=\mathrm{diag}\left(\sqrt{k},1/\sqrt{k}\right)$. In this case, it was found that one should set $\rho_k(-\infty)=1/\sqrt{k}$ and $\rho'_k(-\infty)=0$ in order to define the same vacuum initial state, see \Eq{eq:rho:asympt:omega}. This result can be recovered as follows. Using \Eqs{eq:d:Bogoljiubov}-\eqref{eq:vartheta:Bogoljiubov}, one first obtains that $\ee^{d_k^{(D)}}=\sqrt{k}$ and $\varpi_k^{(D)}=\vartheta_k^{(D)}=0$ initially, which also follows from the Bloch-Messiah decomposition~(\ref{eq:BlochMessiah:Sp2R}). From the two first relations of \Eq{eq:invariant:rep:canonical:transfo:relate:squeezing:rho:D}, one then obtains that $\ee^{-d_k^{(\rho)}}=\sqrt{k}$ and $\varpi_k^{(\rho)}=0$. Making use of \Eq{eq:invariant:rep:canonical:transfo:vartheta:d:and:varpi}, this gives rise to $\vartheta_k^{(\rho)}=0$, such that \Eq{eq:invariant:rep:rho:squeezing:of:rho} leads to $\rho_k(\eta\to -\infty)=\ee^{d_k^{(\rho)}}=1/\sqrt{k}$ and \Eq{eq:invariant:rep:rhoprime:squeezing:of:rho} yields $\rho_k(\eta\to -\infty)=0$. Let us also notice that the third relation of \Eq{eq:invariant:rep:canonical:transfo:relate:squeezing:rho:D} gives $\vartheta_k^{(0)}=0$.
\section{Configuration representation of selected states in the invariant representation of parametric oscillators}
\label{app:wave}
In \Sec{sssec:invariant}, it was shown how a parametric oscillator, with phase-space variables $q$ and $p$ and Hamiltonian~(\ref{eq:H:parma:osc}), can be cast into the form of a harmonic oscillator
\bea
\hat{H}=\frac{1}{2\rho^2(\eta)}\left(\hat{P}^2+\hat{Q}^2\right)
\eea
for the variables 
\bea
\label{eq:Q:qp}
\hat{Q}&=&\frac{\hat{q}}{\rho}\, ,\\
\hat{P}&=&\rho\hat{p}-\rho'\hat{q}\, ,
\label{eq:P:qp}
\eea 
where $\rho(\eta)$ satisfies \Eq{eqrho}. Creation and annihilation operators can then be introduced from these variables, $\widehat{d}=(\widehat{Q}+i\widehat{P})/\sqrt{2}$ and $\widehat{d}^\dag=(\widehat{Q}-i\widehat{P})/\sqrt{2}$, see \Eqs{eq:d:QP:def} and~(\ref{eq:d:dagger:QP:def}). They define a Fock space, and in this appendix we elaborate on the results of \Refs{Pal:2011xx, Balajani:2018qnn} and derive the wavefunctions for the $n$-particles states (Fock states, see \App{sec:InvariantRep:Fock}) and for the eigenstates of the annihilation operator (quasiclassical states, see \App{sec:InvariantRep:QCS}).

\subsection{Fock states}
\label{sec:InvariantRep:Fock}
The construction of the $n$-particles states for $(\hat{d}, \hat{d}^\dagger)$, \ie the eigenstates of the invariant $\hat{I}$ introduced in \Eq{eq:I:P:Q}, is similar to the second quantisation of the harmonic oscillator. Their wavefunctions are obtained from the relation 
\bea
\label{eq:n:state:formal:from:vacuum}
\left|n,\eta\right>_I=\frac{1}{\sqrt{n!}}\left(d^\dag\right)^n\left|0,\eta\right>_I,
\eea
where the vacuum state $\left|0,\eta\right>_I$ can be obtained from writing $\hat{d}\left|0,\eta\right>_I=0$ in the configuration representation. Since $\sqrt{2} \hat{d} = (1/\rho-i\rho')\hat{q}+i\rho\hat{p}$, and given that $\hat{q}\vert\Psi\rangle = q \Psi(q)$ and $\hat{p}\vert\Psi\rangle = -i\Psi'(q)$, this leads to $(\partial/\partial q)\ln\Psi_0(q)=(-1/\rho^2+i\rho'/\rho)q$. Setting the integration constant such that the wavefunction is properly normalised, up to an overall phase, one obtains
\bea
\label{eq:InvariantRep:vacuum}
	\Psi_0(q,\eta)=\frac{1}{\left(\pi\rho^2\right)^{1/4}}\exp{\left(-\frac{1-i\rho'\rho}{2\rho^2}q^2\right)}\, .
\eea
In the configuration representation, \Eq{eq:n:state:formal:from:vacuum} can then be recast as
\bea
	\Psi_n(q,\eta)&=&\frac{1}{\sqrt{2^n(n!)}}\left[-\rho\frac{\partial}{\partial{q}}+\left(\frac{1}{\rho}+i\rho'\right)q\right]^n\Psi_0(q,t)\, .
	\label{fock-wave}
\eea
The main difference between the above expressions and the standard harmonic oscillator lies in the variance of the ground state, which is real-valued for a harmonic oscillator and complex-valued in the case of time-dependent frequencies. This is why it is convenient to introduce
\bea
	z=q\sqrt{\frac{1-i\rho\rho'}{\rho^2}}=\left|z\right|e^{i\theta_q}\, ,
\eea
with $\left|z\right|$ the modulus of $z$ and $\theta_q$ its argument,
\bea
\label{eq:InvariantRep:|z|:thetaq}
	|z|=q\left[\frac{1+(\rho\rho')^2}{\rho^4}\right]^{1/4}&\mathrm{and}&\theta_q=-\frac{1}{2}\arctan{\left(\rho\rho'\right)}. 
\eea
Because $q$ is real-valued, it only contributes to the modulus of $z$ and not to its argument. As a consequence, the creation operator can be written as
\bea
\label{eq:d:D}
	\hat{d}^\dag=\frac{\left[1+(\rho\rho')^2\right]^{1/4}}{\sqrt{2}}\underbrace{\left(-\frac{\partial}{\partial\left|z\right|}+e^{-2i\theta_q}\left|z\right|\right)}_{\hat{D}}\, ,
\eea
and the ground states takes the form
\bea
\label{eq:vacuum:wavefunction:z}
	\Psi_0(q,\eta)=\frac{1}{(\pi\rho^2)^{1/4}}\exp{\left(-\frac{\left|z\right|^2}{2}e^{2i\theta_q}\right)}.
\eea
The rest of the calculation of the wavefunction is very similar to the case of a harmonic oscillator. We first introduce a new variable, $u=|z|\cos^{1/2}(2\theta_q)$, and show recursively that
\bea
\label{eq:D:Hermite}
	\hat{D}^ne^{-\frac{\left|z\right|^2}{2}e^{2i\theta_q}}=\cos^{n/2}(2\theta_q)H_n(u)e^{-\frac{\left|z\right|^2}{2}e^{2i\theta_q}}\, ,
\eea
where $H_n(u)$ satisfies conditions that we now derive. For $n=0$, the above relation is trivially satisfied provided $H_0(u)=1$. Then, the relation at order $n$ implies the relation at order $n+1$ provided
\bea
	H_{n+1}(u)=\left(2u-\frac{\dd}{\dd u}\right)H_{n}(u)\, .
\eea
Since this is the recurrence relation verified by the Hermite polynomials, also initiated with $H_0(u)=1$, this proves \Eq{eq:D:Hermite}, where $H_n$ is the Hermite polynomial of order $n$. Combining \Eqs{eq:n:state:formal:from:vacuum}, (\ref{eq:d:D}), (\ref{eq:vacuum:wavefunction:z}) and~(\ref{eq:D:Hermite}), one obtains for the wavefunction of the $n$-th eigenstate of $\hat{I}$,
\bea
	\Psi_n(q,t) = & &\frac{1}{\sqrt{2^n(n!)}}\left\lbrace\left[1+(\rho\rho')^2\right]\cos^2(2\theta_q)\right\rbrace^{n/4}H_{n}\left[|z|\cos^{1/2}(2\theta_q)\right]
	\nonumber \\& & \times
	\dfrac{1}{\left(\pi\rho^2\right)^{1/4}}\exp{\left(-\frac{\left|z\right|^2}{2}e^{2i\theta_q}\right)}\, .
	\label{q-wave-int}
\eea
The relation~(\ref{q-wave-int}) is then simplified by noticing that $u^2=\mathrm{Re}(z^2)$ by construction, and by using \Eqs{eq:InvariantRep:|z|:thetaq} to replace $\left|z\right|$ and $\theta_q$. One obtains
\bea
\label{eq:psi:Fock:I}
	\Psi_n(q,t)=\frac{1}{\sqrt{2^n(n!)}}\frac{1}{\left(\pi\rho^2\right)^{1/4}}H_{n}\left(\frac{q}{\sqrt{\rho^2}}\right)\exp{\left(-\frac{1-i\rho\rho'}{2\rho^2}q^2\right)}.
	\label{q-wave}
\eea

Let us finally notice that the usual formula for  the $n-$particle wavefunction of a harmonic oscillator can be recovered from this expression. For a constant frequency $\omega$,  \Eq{eqrho} has a simple solution, $\rho=1/\sqrt{\omega}$. Performing this replacement in \Eq{eq:psi:Fock:I}, one obtains the standard formula
\bea
	\Psi_{\mathrm{HO},n}(q,t)=\frac{1}{\sqrt{2^n(n!)}}{\left(\frac{\omega}{\pi}\right)^{1/4}}H_{n}\left(q\sqrt{\omega}\right)\exp{\left(-\frac{\omega}{2}q^2\right)}\, .
\eea
For a parametric oscillator, the frequency $\omega$ is thus replaced by the time-dependant function $1/\rho^2$, and an imaginary part is added to the Gaussian argument of the wave-function.
 
\subsection{Quasiclassical states}
\label{sec:InvariantRep:QCS}
\subsubsection{Definition and properties}
Coherent states for parametric oscillators, also called { \it quasiclassical} or {\it time-dependant coherent} states, have been defined in \Refa{PhysRevD.25.382} in a way that is very reminiscent to the case of coherent states for harmonic oscillators. In the latter case, those states exhibit interesting properties making them the quantum counterpart of a classical point in phase space. Among those properties, the expectation values of position and momentum follow the classical motion. These states also minimise the uncertainty relation and do not spread in phase space. Moreover, a quantum field coupled to a classical source evolves to this class of states, which also constitutes the preferred basis in which quantum systems coupled to environments in thermal equilibrium decohere~\cite{PhysRevLett.70.1187}. 

\paragraph{Harmonic oscillator --}
In the case of a time-independent frequency  $\omega$, creation and annihilation operators are defined as $\hat{a}=\sqrt{\omega/2}(\hat{q}+i\hat{p}/\omega)$ and $\hat{a}^\dagger=\sqrt{\omega/2}(\hat{q}-i\hat{p}/\omega)$. Quasiclassical states $\left|\alpha\right>$ can be defined as eigenstates of the annihilation operator, $\hat{a}\left|\alpha\right>=\alpha\left|\alpha\right>$, where $\alpha$ is a complex eigenvalue. In terms of the $n$-particles states, \ie the Fock states studied in \App{sec:InvariantRep:Fock}, they thus read
\bea
	\left|\alpha\right>=e^{-\left|\alpha\right|^2/2}\displaystyle\sum^\infty_{n=0}\frac{\alpha^n}{\sqrt{n!}}\left|n\right> .
\eea
In configuration representation where $\hat{q}\vert\Psi\rangle = q \Psi(q)$ and $\hat{p}\vert\Psi\rangle = -i\Psi'(q)$, $\hat{a}=\sqrt{\omega/2}[q+(1/\omega)(\partial/{\partial q})]$, so the eigenvalue equation, $\hat{a}\Psi_\alpha(q)=\alpha\Psi_{\alpha}(q)$, gives rise to $\partial\ln\Psi_\alpha/\partial q=-\omega q+\sqrt{2\omega}\alpha$. Setting the integration constant such that the wavefunction is normalised, up to an overall constant phase factor, this gives rise to
\bea
\label{eq:coherent:state:harmonic}
	\Psi_\alpha(q)=\left(\frac{\omega}{\pi}\right)^{1/4}\exp{\left[-\frac{\omega}{2}\left(q-q_{\mathrm{cl}}\right)^2+ip_{\mathrm{cl}}q\right]},
\eea
with 
\bea
\label{eq:qcl:pcl}
\alpha=\sqrt{\frac{\omega}{2}}\left(q_{\mathrm{cl}}+\frac{i}{\omega}p_{\mathrm{cl}}\right).
\eea 
From this expression, one can easily show that $\left\langle \hat{q} \right\rangle = \int q \Psi^*_\alpha(q) \Psi_\alpha(q)\dd q = q_\ucl$ and that $\left\langle \hat{p} \right\rangle = - i \int  \Psi^*_\alpha(q) \Psi^\prime_\alpha(q)\dd q=p_\ucl$. In the same manner, one can show that 
\bea
\label{eq:Delta:q}
\left\langle \widehat{\Delta q}^2\right\rangle \equiv \left\langle \left( \hat{q}- \left\langle \hat{q} \right\rangle \right)^2\right\rangle = \frac{1}{2\omega}\\
\left\langle \widehat{\Delta p}^2\right\rangle \equiv \left\langle \left( \hat{p}- \left\langle \hat{p} \right\rangle \right)^2\right\rangle = \frac{\omega}{2}\, ,
\label{eq:Delta:p}
\eea
which proves that, as announced above, quasiclassical states minimise the uncertainty relation. From \Eq{eq:coherent:state:harmonic}, one also finds that the two phase-space variables are uncorrelated, \ie
\bea
\label{eq:Delta:q:Delta:p}
\left\langle \widehat{\Delta q} \widehat{\Delta p}+ \widehat{\Delta p} \widehat{\Delta q} \right\rangle =0\, .
\eea

\paragraph{Parametric oscillator --}
Let us now consider a time-dependent frequency $\omega(\eta)$, that asymptotes a constant in the infinite past, $\omega_{(-)}$. Quasiclassical states are defined as being the states that tend to {\it standard} coherent states in the asymptotic past~\cite{PhysRevD.25.382}. Since the states introduced in \Eq{eq:eigenstates:I:time:dependent} are solutions to the Schr\"odinger equation, and asymptote the $n$-particle states in the asymptotic past, quasiclassical states can thus be decomposed as in \Eq{ItoS}, namely
\bea
	 \left|\alpha,\eta\right>_S=e^{-\left|\alpha\right|^2/2}\displaystyle\sum^\infty_{n=0}\frac{\alpha^n}{\sqrt{n!}}e^{i\vartheta_n(\eta)}\left|n,\eta\right>_{I},
	\label{qcs-fock}
\eea
where $\left|n,\eta\right>_{I}$ are the eigenstates of the invariant $\hat{I}$ introduced in \Eq{eq:I:P:Q}, and the phases are given by \Eq{eq:vartheta:sol}, namely $\vartheta_n(\eta)=-(n+1/2)\int^\eta{\mathrm{d}\bar{\eta}}/\rho^2(\bar{\eta})$. Introducing the new parameter
\bea
\label{eq:alpha:prime}
\alpha'=\alpha\exp{\left[-i\int^\eta\frac{\mathrm{d}\bar{\eta}}{\rho^2(\bar{\eta})}\right]},
\eea
the time-dependent coherent states becomes
\bea
	 \left|\alpha,\eta\right>_S=\exp{\left[-\frac{i}{2}\int^\eta\frac{\mathrm{d}\bar{\eta}}{\rho^2(\bar{\eta})}\right]}\underbrace{e^{-\left|\alpha'\right|^2/2}\displaystyle\sum^\infty_{n=0}\frac{{\alpha'}^n}{\sqrt{n!}}\left|n,\eta\right>_{I}}_{ \left|\alpha'(\eta)\right>_{P,Q}}\, .
\eea
In the $(Q,P)$ phase space, the state $\left|\alpha'(\eta)\right>_{P,Q}$ is a coherent state and the quasiclassical states are, up to a time-dependant phase factor, coherent in this phase space.
In the original $(q,p)$ phase space, this translates into two important properties. First, according to \Eq{eq:qcl:pcl}, the expectation values of position and momentum follow the classical motion of a parametric oscillator if $\alpha'=(Q_\ucl+iP_\ucl)/\sqrt{2}$. Combining \Eq{eq:alpha:prime} with \Eqs{eq:Q:qp}-(\ref{eq:P:qp}), this gives rise to
\bea
\label{eq:InvariantRep:alpha:classical:eom}
\alpha=\frac{1}{\sqrt{2}}\exp{\left[i\int^\eta\frac{\mathrm{d}\bar{\eta}}{\rho^2(\bar{\eta})}\right]}\left[\frac{q_{\ucl}}{\rho}+i\left(\rho{p}_\ucl-\rho' q_\ucl\right)\right].
\eea
Second, quasiclassical states are eigenstates of the annihilation operator $\hat{d}$, with the associated eigenvalue $\alpha'(\eta)=\alpha{e}^{2i\vartheta_0(\eta)}$. However, according to \Eqs{eq:Delta:q}, (\ref{eq:Delta:p}) and~(\ref{eq:Delta:q:Delta:p}) one has $\langle\widehat{\Delta{Q}}^2\rangle=\langle\widehat{\Delta{P}}^2\rangle=1/2$ and $\langle \widehat{\Delta{Q}} \widehat{\Delta{P}}+\widehat{\Delta{P}} \widehat{\Delta{Q}} \rangle = 0$, so \Eqs{eq:Q:qp} and~(\ref{eq:P:qp}) lead to
\begin{eqnarray}
	\left<\Delta{q}^2\right>&=&\frac{1}{2}\rho^2, \\
	\left<\Delta{p}^2\right>&=&\frac{1}{2}\left(\dot\rho^2+\frac{1}{\rho^2}\right).
\label{eq:<Delta p^2>}
\end{eqnarray}
Quasiclassical states are therefore described by squeezed wave-packets and do not constitute minimal-uncertainty, isotropic states in the $(q,p)$ phase space.

Let us finally mention that the quasiclassical states are derived from a direct mapping with the {\it standard} coherent states associated to the invariant $\hat{I}$. Because the invariant takes the form of a harmonic oscillator in the $(Q,P)$ phase space, and since coherent states form an overcomplete basis of the Hilbert states related to $\hat{I}$, the quasiclassical states define an overcomplete basis of the Hilbert space associated to the parametric oscillator.

\subsubsection{Wavefunction}
The wavefunction of the quasiclassical states can be derived either by rewriting \Eq{qcs-fock} in the configuration representation and making use of the wavefunction~(\ref{q-wave}) of the Fock states, or by solving an eigenvalue problem. For completeness, we follow the two approaches below.

\paragraph{Using the Fock states --}
We remind that for quasiclassical states to follow the classical trajectory $\left(q_{\mathrm{cl}}(t),p_{\mathrm{cl}}(t)\right)$, $\alpha$ has to satisfy \Eq{eq:InvariantRep:alpha:classical:eom}. Plugging \Eq{q-wave} into \Eq{qcs-fock}, one then obtains that quasiclassical states are described with the gaussian wavefunction\footnote{One makes use of the identity~\cite{abramowitz+stegun}
\bea
	\displaystyle\sum_{n=0}^\infty\frac{x^n}{n!}H_n(y)=e^{-x^2+2xy}
\eea
for $x$ and $y$ real.}
\bea
	\Psi_{q_{\mathrm{cl}},p_{\mathrm{cl}}}(q,t)=\frac{e^{-\frac{i}{2}\left[q_\ucl p_\ucl+\int^\eta\frac{\dd\bar\eta}{\rho^2(\bar\eta)}\right]}}{\left(\pi\rho^2\right)^{1/4}}\exp{\left[-\frac{1-i\rho\rho'}{2\rho^2}\left(q-q_{\mathrm{cl}}\right)^2+{i}p_{\mathrm{cl}}q\right]}\, .
	\label{qcs-q}
\eea

\paragraph{Solving the eigenvalue problem in the configuration representation --}
As explained below \Eq{eq:InvariantRep:alpha:classical:eom}, quasiclassical states are eigenstates of the annihilation operator $\hat{d}$, with the associated eigenvalue $\alpha'(\eta)=\alpha{e}^{2i\vartheta_0(\eta)}$, \ie
\bea
	\hat{d} \left|\alpha,t\right>_S=\alpha{e}^{2i\vartheta_0(t)} \left|\alpha,t\right>_S\, .
\eea
In the configuration representation, this translates into a first differential equation,
\bea
	\left[\rho\frac{\partial}{\partial q}+\left(\frac{1}{\rho}-i\rho'\right)q\right]\Psi_{q_{\mathrm{cl}},p_{\mathrm{cl}}}(q)=\sqrt{2}\alpha{e}^{2i\vartheta_0(t)}\Psi_{q_{\mathrm{cl}},p_{\mathrm{cl}}}(q)\, .
\eea
Normalised solutions of the above equation, with $\alpha$ given by \Eq{eq:InvariantRep:alpha:classical:eom}, reduce to the solution~(\ref{qcs-q}) up to an overall phase, which can be determined by imposing the Schr\"odinger equation. Let us mention that the eigenstate for the zero eigenvalue corresponds to the ground state (\ie the vacuum state of the invariant), whose wavefunction is given by \Eq{qcs-q} when setting $q_{\mathrm{cl}}=p_{\mathrm{cl}}=0$, which coincides with \Eq{eq:InvariantRep:vacuum} as it should.
\bibliographystyle{JHEP}
\bibliography{SU11}
\end{document}